\documentclass[12pt]{article}%
\usepackage[latin9]{inputenc}
\usepackage[margin=1in]{geometry}
\usepackage{amsmath}
\usepackage{amssymb}
\usepackage{graphicx}
\usepackage[doublespacing]{setspace}
\usepackage{hyperref}
\usepackage{amsfonts,adjustbox}
\usepackage{natbib}
\usepackage{csquotes}
\usepackage[colorinlistoftodos]{todonotes}
\usepackage{rotating}
\usepackage{graphicx}
\usepackage{wrapfig}
\usepackage{pdflscape}
\usepackage{rotating}
\usepackage{float}
\usepackage{amsfonts}%
\usepackage{footmisc}
\usepackage{todonotes}
\usepackage{caption}
\usepackage{threeparttable}
\usepackage{tabularx}
\usepackage[normalem]{ulem}
\providecommand{\U}[1]{\protect\rule{.1in}{.1in}}
\hypersetup{colorlinks=true,citecolor=blue,linkcolor=red,filecolor=brown,urlcolor=blue}

\graphicspath{{Figures/}}

\begin{document}

\newgeometry{top=0.3in, bottom=1 in}

\title{\textbf{Empirical evidence on the Euler equation for investment in the
US}\thanks{
%\protect\doublespacing 
Ascari: University of Pavia and De Nederlandsche Bank; Haque: University of Adelaide and Centre for Applied Macroeconomic Analysis; Magnusson:  M251, 35 Stirling Highway, Department of Economics, University of Western Australia, \url{leandro.magnusson@uwa.edu.au} (corresponding author); Mavroeidis: University of Oxford. We would like to thank the Editor Marco Del Negro, three referees, Giovanni Caggiano, Firmin Doko Tchatoka, Adrian Pagan, and seminar and conference participants at Monash University, the International Association of Applied Econometrics Meeting in Rotterdam 2021, and the Econometric Society Australasian Meeting in Melbourne 2021 for very helpful comments and discussions. Leandro M. Magnusson and Qazi Haque gratefully acknowledge financial support from the Australian Research Council via grant DP170100697, and Sophocles Mavroeidis from the European Research Council via grant 647152.}} %\thanks{\protect\doublespacing Ascari: University of Oxford and University of Pavia; Haque: University of Adelaide, University of Western Australia and Center for Applied Macroeconomics Analysis; Magnusson:  M251, 35 Stirling Highway, Department of Economics,
%University of Western Australia, \url{leandro.magnusson@uwa.edu.au} (corresponding author); Mavroeidis: University of Oxford.}}
\author{Guido Ascari
\and Qazi Haque
\and Leandro M. Magnusson
\and Sophocles Mavroeidis}
%\date{}
\maketitle
% \vspace{-50pt}
\begin{abstract}
%\protect\onehalfspacing Is the typical specification of the Euler equation for investment employed in DSGE models consistent with aggregate macro data? Using state-of-the-art econometric methods that are robust to weak instruments and exploit information in possible structural changes, the answer is yes. Unfortunately, however, the structural parameters of the investment equation are generally poorly identified, that is, there is very little information about these parameters in aggregate data. This is because investment appears to be unresponsive to changes in capital utilization and the real interest rate. In DSGE models, the investment adjustment cost parameter is mainly identified by the cross-equation restrictions implied by the structure of the whole model.

%Leandro's try
\protect\onehalfspacing Is the typical specification of the Euler equation for investment employed in DSGE models consistent with aggregate macro data? Using state-of-the-art econometric methods that are robust to weak instruments and exploit information in possible structural changes, the answer is yes. Unfortunately, however, there is very little information about the values of these parameters in aggregate data because investment is unresponsive to changes in capital utilization and the real interest rate. In DSGE models, the investment adjustment cost and the persistence of the investment-specific technology shock parameters are mainly identified by, respectively, the cross-equation restrictions and the dynamics implied by the structure of the model. 
 
%The Euler equation model for investment with adjustment costs and variable capital utilization is estimated using aggregate US post-war data with econometric methods that are robust to weak instruments and exploit information in possible structural changes. Various alternative identification assumptions are considered, including external instruments, and instruments obtained from Dynamic Stochastic General Equilibrium models. Results show that the elasticity of capital utilization and investment adjustment cost parameters are very weakly identified. This is because investment appears to be unresponsive to changes in capital utilization and the real interest rate.  

\vspace{20pt}

\vspace{10pt}
\end{abstract}

% \vspace{100pt}

 \thispagestyle{empty} \setcounter{page}{0} \pagebreak

\restoregeometry

\section{Introduction}

An important component of the demand side of standard Dynamic Stochastic General Equilibrium (DSGE) models is  aggregate investment. The seminal contribution by \cite{Christiano_Eichenbaum_Evans_2005} proposes an investment-adjustment cost model coupled with the assumption of variable capital utilization to capture the inertial response of aggregate investment to monetary policy shocks. This specification for investment behavior has become standard in the DGSE literature, and the implied Euler equation for investment features in most DSGE models used for policy analysis. 
The key structural parameters of this investment block of modern DSGE models are: the investment adjustment cost parameter that denotes the inverse of the elasticity of investment with respect to the shadow price of capital, the elasticity of the capital utilization cost function, and the persistence of the investment-specific technology shock.\footnote{The results we report in the paper are for the most standard investment equation that appears in DSGE models. We also studied a specification with capital adjustment costs and the results were very similar. Appendix \ref{app: s: CAC} contains results for capital adjustment costs.}

However, estimates of these three key parameters differ greatly across papers in the literature. The next Section shows that different parameter values  entail a very different response of investment, and hence of output, to various shocks. Not surprisingly, this then translates to different implications regarding the main drivers of business cycle fluctuations, in terms of the relative importance of the various shocks. Hence, these parameters are key for our interpretation of the business cycle through the lens of the investment block of a standard DSGE model. It seems important, therefore, to investigate why the estimates of the key parameters vary greatly and how accurately they can be estimated using aggregate macro data. This is what we do in this paper. More specifically, we ask the following two questions. 

First, is the typical specification of the Euler equation for investment employed in DSGE models together with an autoregressive investment-specific shock consistent with aggregate macro data? We test the bare minimum implications of this model with state-of-the-art generalized method of moments (GMM) tests, and we find that the answer is yes. In other words, there is no evidence against this specification that can be found in aggregate macro data. 
This is a positive message that we add to the literature. We are not aware of any paper in the literature that checks the consistency of the implied Euler equation for investment with aggregate data as a single equation, rather than through the lens of Bayesian estimation of a full DSGE model. 
We use recently developed econometric methods in \cite{MM14} and in \cite{Mikusheva2021} to deal with weak identification. The former incorporates subsample information in the data arising from structural changes in the economy such as policy regime shifts. These methods also serve as parameter stability tests that are fully robust to weak instruments, and, hence, they provide reliable evidence on the stability of the parameters over the sample. The latter explores information of all potential instruments available for inference using a split-sample technique. Moreover, using the common assumption of variable capital utilization to derive our estimated equation allows us, on the one hand, to avoid issues of using proxies for unobservable variables, such as the return on capital or the Tobin's marginal Q, and, on the other hand, to use observable variables, such as the real interest rate and capacity utilization that is fully consistent with the investment block currently used in modern DSGE modeling. Finally, the single-equation approach is robust to potential misspecification in other equations of the system.\footnote{In the literature, there exists alternative approaches to dealing with misspecification in DSGE models. \cite{sargent1989two} and \cite{ireland2004technology} introduce errors in the measurement equations of the state-space model. \cite{del2004priors} use prior distributions for structural VARs that are centered at the DSGE model-implied cross-equation restrictions, generating a continuum of empirical models referred to as DSGE-VARs \citep[see also][]{del2007fit,del2009monetary}. More recently, \cite{Inoue_Kuo_Rossi_2020} propose a method for detecting and identifying misspecification in structural models and show that DSGE models can be severely misspecified. \label{footnote: dsge mis}} 
%Our work investigates the limits of the information one can expect to obtain from aggregate time series data for the behavior of aggregate investment, and, thus, provide useful guidance for future work in this area.

The second question is: how much can we learn from aggregate time series data about the values of the three key parameters mentioned above?
Unfortunately, the answer to this second question is not much. The structural parameters of the investment equation are generally very poorly identified, that is, there is very little information about these parameters in aggregate data. Using typical lagged values of the endogenous variables as instruments, the confidence sets contain almost the entire parameter space. This motivates us to consider external sets of instruments, such as oil prices, financial uncertainty measures, government expenditure shocks and monetary policy shocks; however, we find that they help very little in identifying the main structural parameters.
Then, we present the implications of this finding in terms of possible ranges of the impulse response functions and of the variance decompositions using the standard medium-scale DSGE model of  \cite{justiniano2010investment} (JPT, henceforth). Our analysis shows that the results are particularly sensitive to the value of the investment adjustment cost parameter and the persistence of the investment-specific technology shock. Hence, it seems that pinning down these two parameters is key and more important than identifying a value for the elasticity of capital utilization. 
%Next, we use as another set of instruments the structural shocks estimated from two popular medium-scale DSGE models, \cite{Smets_Wouters_2007} (SW, henceforth) and \cite{justiniano2010investment} (JPT, henceforth), in order to check whether we could obtain identification from the cross-equation restrictions of a full DSGE model that are implicit in the computation of those shocks. %These shocks result from Bayesian estimation of the full DSGE model, and, as such, derive from having imposed those cross-equation restrictions.

The above finding of weak identification contrasts with the results typically reported in the DSGE literature, leading us to investigate how DSGE models could achieve identification of these parameters (again using the JPT model). This can be achieved in two ways, either through the prior or through the joint model dynamics of the variables in the system and the related cross-equation restrictions implied by rational expectations.\footnote{\textit{``Communism of models gives rational expectations much of its empirical power and underlies the cross-equation restrictions that are used by rational expectations econometrics to identify and estimate parameters. A related perspective is that, within models that have unique rational expectations equilibria, the hypothesis of rational expectations makes agents' expectations disappear as objects to be specified by the model builder or to be estimated by the econometrician. Instead, they are equilibrium outcomes.[...] Identification is partially achieved by the rich set of cross-equation restrictions that the hypothesis of rational expectations imposes.''}\citep[][p. 194-195]{Sargent2008}\label{footnote: Sargent}}
%\footnote{In the words of Thomas Sargent, rational expectations imply \textit{``cross-equation restrictions and the disappearance of any free parameters associated with expectations. [...] In rational expectations models, people's beliefs are among the outcomes of our theorizing. They are not inputs. [...] The positive part of Lucas's critique was to urge applied macroeconomists and econometricians to develop ways to implement those cross-equation restrictions."}\citep[][p. 567]{sargent_interview_2005}\label{footnote: Sargent}} 
Both features differentiate the DSGE estimation from our method. While estimating a system of equations could help identification through cross-equation restrictions, possible misspecification in other parts of the model could lead to biased estimates.
Our analysis suggests the following. Consistent with our GMM results, the elasticity of the capital utilization cost function is not identified by the data, so identification is largely due to the prior.
%\footnote{This is also consistent with JPT, who report a lack of identification of this parameter in their model.} 
On the other hand, the investment adjustment cost parameter is mainly identified  by the cross-equation restrictions implied by the structure of the DSGE model, when using the JPT model and data set. The persistence parameter of the investment-specific shock is always high and well identified, even if we relax some of the cross-equation restrictions and use a loose prior. We conjecture this to be due to the fact that the model wants to match the very persistent dynamics of the observable macroeconomic time series. 

Finally, we estimate a semi-structural model parameterized in terms of the slope coefficients of the investment equation with respect to the capital utilization rate and the real interest rate. We show that, when the persistence of the investment-specific technology shock is large, the semi-structural parameters corresponding to the above slope coefficients are not well-identified. Weak identification arises because the change in the capital utilization rate and the real interest rate are poorly forecastable. In contrast, when the persistence of the investment-specific technology shock is low, these slope parameters can be  well-identified. However, because the confidence sets on those semi-structural parameters cover zero, and the mapping from the semi-structural to the structural parameters is ill-posed near zero, the structural parameters are weakly identified - small changes in the slope coefficients generate large changes in the capital utilization and investment adjustment cost parameters. Overall, the semi-structural analysis suggests that investment is unresponsive to the real interest rate\footnote{This is consistent with Keynes' old argument and early empirical results surveyed in \cite{Taylor1999}. See also the discussion in \cite{Mertens2010} and \cite{Brault2020}.} and to capital utilization. Therefore, at a technical level, we conclude that it is the mapping from semi-structural to structural parameters that prevents the identification of the latter. Finally, comparing our results to those in \cite{MM14}, structural change is not as informative for the identification of the Euler equation for investment as it is for the NKPC. This is in line with the results in \cite{AMM2021}, suggesting again that policy regime shifts have had more impact on nominal variables than on real variables over our sample.

%\paragraph{Related literature.} 

Our paper could be of interest to any researchers using the standard investment cost specification proposed by \cite{Christiano_Eichenbaum_Evans_2005} in their medium-scale DSGE models. The literature is immense and a small subset of influential papers are considered in Figure \ref{fig:FigureLit} below.\footnote{In a recent paper, \cite{Foronietal_2022} shows that the identification of the investment adjustment cost parameter could be biased by time aggregation, that is, by aggregating monthly data into quarterly.}
Our paper is also related to the literature that estimates individual equations employed in standard DSGE models using GMM methods and aggregate data.
Specifically, three equations have been extensively studied on their own: the New Keynesian Phillips Curve \citep{GaliGertler_1999}, the Taylor rule \citep{cgg_2000}, and the Euler equation for consumption \citep{Yogo04}. 
%Starting from \cite{GaliGertler_1999}, a large literature has estimated the New Keynesian Phillips Curve to check its ability to fit observed inflation dynamics and estimate its key structural parameters, mainly the degree of price stickiness and of forward-lookingness \citep[see][for a survey]{Mavr_plag_stoc_JEL14}. The contribution in \cite{cgg_2000} has generated another large literature on the estimation of monetary policy rules and on the debate about the conduct of monetary policy in the US \citep[see e.g.,][]{Orpha_AER01,Coib_Goro_AER11, Coib_Goro_AEJM12}. The empirical literature on the Euler equation for consumption is prominent not only in macro - studying the behavior of consumption and the degree of intertemporal substitution, but also in finance - developing capital asset pricing models. Recently, \cite{AMM2021} study the empirical fit in aggregate data of  various extensions to the baseline Euler equation  for consumption that have been employed in the DSGE literature, generalizing the work in \cite{Yogo04} by using empirical methods close to the ones we employ here. 
%The more recent studies on the NKPC \citep{KleibergenMavroeidis09jbes,MM14}, the Taylor rule \citep{Mavr_AER10} and the Euler equation for consumption \citep{AMM2021} demonstrate the importance of employing econometric methods robust to weak-instrument problems.  Such problems typically arise in these models because inflation and consumption growth are poorly forecastable, thus predetermined variables make weak instruments for them.
The present paper is the first to conduct a similar exercise for the other main component of the demand side of modern DSGE models: investment.

Finally, our contribution and results are also complementary to the literature that estimates investment functions using cross-sectional microeconomic data of firms. There is a very large literature mainly concerned with additional costs of external financing as a result of information asymmetries and agency costs. This external financing cost may increase the sensitivity of investment decisions to sources of internal finance such as cash flows for constrained firms \citep{Fazzari1988}.\footnote{However, \cite{Kaplan1997} reach the opposite conclusion by looking just at the firms classified as constrained by \cite{Fazzari1988} and they also criticise the assumption that sensitivity of investment to cash flows should increase monotonically as firms become more constrained. See also, for example, \cite{Bond1994}, \cite{Almeida2007}, \cite{agca}.} \cite{Groth_Khan_2010} and \cite{Eberly_Rebelo_Vincent_2012} are, instead, very closely related to our work because they estimate the investment adjustment-cost model of \cite{Christiano_Eichenbaum_Evans_2005} on 18 industries and on firm-level data, respectively. \cite{Groth_Khan_2010} focus on the estimation of the adjustment cost parameter and find that to be very small in U.S. manufacturing industries, implying that investment is highly sensitive to the current shadow value of capital, in contrast to the aggregate estimates from well-known DSGE models reported in Figure \ref{fig:FigureLit}, see \citet[Table 5]{Groth_Khan_2010}. \cite{Eberly_Rebelo_Vincent_2012} instead focus on investigating the importance of the lagged investment term implied by the \cite{Christiano_Eichenbaum_Evans_2005} specification. They report a strong and robust lagged-investment effect in their estimates, implying inertial dynamics of investment for the set of firms in their sample. Our approach here is very different as we want to assess the ability to identify the parameters of the investment block of DSGE models from aggregate data. As such, we also use the common assumption of variable capital utilization to derive our estimated equation. Our macroeconomic approach abstracts from heterogeneous adjustment costs or elasticities of investment, since this would require modelling the heterogeneity of firms at the micro level.\footnote{This is a common problem of linking aggregate elasticity with individual one in the presence of heterogeneity that generates a compositional effect due to the different constraints firms (or households in the case of the Euler equation for consumption) are facing. \cite{KeaneRogJel2012}, for example, make a similar point regarding the difficulty in reconciling the microeconomic and the macroeconomic estimates of the elasticity of labour supply. The relationship between these two elasticities (reduced-form aggregate one vs. microeconomic individual one) depends not only on preference parameters but also on all aspects of the economic environment households are facing: borrowing constraints, liquidity needs, family economics, bequests, taxes, and so on.}

The structure of the paper is as follows. Section \ref{s: motivation} presents the theoretical specification and it investigates the implication of using different values for the parameters of the investment equation, as taken from influential papers in the DSGE literature.  Section \ref{s:_ectmethod} describes the econometric methodology and Section \ref{s: data} presents the data. Section \ref{s: results} presents the empirical results. Section \ref{s: implications DSGE} discusses the identification of the key parameters of the investment equation in DSGE models, in light of our results. Section \ref{s: concl} concludes. Additional empirical results, data sources and econometric methods are reported in the Appendix.

\section{The Euler equation of investment and DSGE models \label{s: motivation}}
The investment equation used in our empirical investigation comes from the
most standard specification used in medium-scale DSGE models. Since the seminal
paper by \cite{Christiano_Eichenbaum_Evans_2005}, DSGE models commonly employ
the assumption of investment adjustment costs (IAC) and variable capital
utilization. Investment pertains to the demand side of the model and the
first-order conditions are commonly derived from the households' problem,
assuming that households take investment decisions, own the capital stock and
rent capital to the firms.\footnote{This is mainly for convenience in the
literature. A DSGE model yields isomorphic first-order conditions for the
investment side of the model irrespective of whether investment decisions are taken by firms or by households.} A representative household's lifetime utility,
separable in consumption, $C_{t}$, and hours worked, $L_{t}$, is expressed as
\begin{equation}
E_{0}\sum_{t=0}^{\infty}\beta^{t}U(C_{t},L_{t}), \label{eq: UF}%
\end{equation}
where $\beta$ is the discount factor. The household's period \textit{t} budget
constraint is
\begin{equation}
C_{t}+I_{t}+\dfrac{B_{t+1}}{P_{t}}\leq\dfrac{R_{t-1}B_{t}}{P_{t}}+\dfrac
{W_{t}L_{t}}{P_{t}}+\Pi_{t}+r_{t}^{k}u_{t}\hat{K}_{t}-a(u_{t})\hat{K}_{t},
\label{eq: bc}%
\end{equation}
where $I_{t}$ is investment, $B_{t}$ is the amount of risk-free bonds that pay
a nominal gross interest rate of $R_{t}$, $W_{t}$ is the nominal wage, $\Pi_{t}$
denotes firms' profits net of lump-sum taxes, $r_{t}^{k}$ is the real rental
rate of capital, $\hat{K}_{t}$ is the physical capital stock, and $a(u_{t})$
is the function that measures the cost of capital utilization per unit of
physical capital. Capital owning households choose the capital utilization
rate, $u_{t}$, that transforms physical capital $\hat{K}_{t}$ into effective
capital ${K}_{t}$ as follows
\begin{equation}
K_{t}=u_{t}\hat{K}_{t}. \label{eq: effK}%
\end{equation}
Effective capital is rented to intermediate goods producers at the rate
$r_{t}^{k}$. Standard assumptions are: i) $\bar{u}=1$ and $a(\bar{u})=0$,
where a bar over a variable denotes its steady state value; ii) the curvature
of the function $a(u)$, given by $a^{\prime\prime}(u)/a^{\prime}
(u)$, measures the elasticity of capital utilization cost and it is such that
$\zeta=a^{\prime\prime}(1)/a^{\prime}(1)>0$.

The representative household accumulates end-of-period $t$ capital according
to a standard capital accumulation equation
\begin{equation}
\hat{K}_{t+1}=\nu_{t}\left[  1-\mathcal{S}\left(  \dfrac{I_{t}}{I_{t-1}%
}\right)  \right]  I_{t}+(1-\delta)\hat{K}_{t}, \label{eq: CAE}%
\end{equation}
where $\delta$ is the depreciation rate and $\nu_{t}$ is the investment-specific technology shock, that is, a shock to the efficiency with which the final good
can be transformed into physical capital, as in JPT. 
The log of the investment shock follows the autoregressive stochastic process $\log\nu_{t}=\rho\log{\nu_{t-1}}+ \varepsilon_{t}^{v}$, where $\rho$
is the autoregressive coefficient.

The IAC is specified as
\begin{equation}
\mathcal{S}\left(  \dfrac{I_{t}}{I_{t-1}}\right)  I_{t}=\dfrac{\kappa}%
{2}\left(  \dfrac{I_{t}}{I_{t-1}}-1\right)  ^{2}I_{t}.
\label{eq: IAC function}%
\end{equation}
where the IAC function $\mathcal{S}(\cdot)$ is such that $\mathcal{S}%
(1)=\mathcal{S}^{\prime}(1)=0$ with $\kappa=\mathcal{S}^{\prime\prime}(1)>0$.
Here, $\kappa$, the adjustment cost parameter, denotes the inverse of the
elasticity of investment with respect to the shadow price of capital. There
are no adjustment costs at the steady state when $I$ is fixed.\footnote{We also derive an Investment Euler equation with Capital Adjustment Cost (CAC) in the Appendix. In this case, equation (\ref{eq: CAE}) is
\[
\hat{K}_{t+1}=\nu_{t}I_{t}+(1-\delta)\hat{K}_{t}-D(\hat{K}_{t},I_{t})\text{,}%
\]
where $D(\hat{K}_{t},I_{t})=\frac{\sigma}{2}\left(  \frac{I_{t}}{\hat{K}_{t}%
}-\delta\right)  ^{2}\hat{K}_{t}$, and $\sigma>0$ governs the magnitude of
adjustment costs to capital accumulation. We also perform the same econometric
analysis for a specification using CAC instead of
IAC. The results are similar to ones reported in
Section \ref{s: results}, and, therefore, are placed in Appendix \ref{app: s: CAC}.}

The representative household chooses ${I_{t},}$ ${u_{t},}$ ${\hat{K}_{t+1},}$ and
${B_{t+1}}$ to maximise (\ref{eq: UF}) under the period-by-period budget
constraint (\ref{eq: bc}) and capital accumulation equation (\ref{eq: CAE}).
Appendix \ref{appsec: derivation} shows how log-linearizing the
first-order conditions of this problem and rearranging them yields the
following dynamic equation for investment
\begin{equation}
\Delta\widetilde{i}_{t}=\left(  \beta+\phi_{q}\right)  E_{t}\Delta
\widetilde{i}_{t+1}-\beta\phi_{q}E_{t}\Delta\widetilde{i}_{t+2}+\dfrac
{1}{\kappa}\left[  \phi_{k}\zeta E_{t}\widetilde{u}_{t+1}-\widetilde{r}%
_{t}^{p}+\widetilde{\nu}_{t}\right]  -\dfrac{\phi_{q}}{\kappa}E_{t}%
\widetilde{\nu}_{t+1}. \label{eq: Basline_Euler_IAC_util_2}
\end{equation}
% \begin{align}
% \widetilde{i}_{t}=  &  \dfrac{1/\kappa}{1+\beta+\phi_{q}}\left[  \phi_{k}\zeta
% E_{t}\widetilde{u}_{t+1}-\widetilde{r}_{t}^{p}+\widetilde{\nu}_{t}\right]
% \nonumber\\
% &  +\dfrac{1}{1+\beta+\phi_{q}}\widetilde{i}_{t-1}+\dfrac{\beta+\phi
% _{q}(1+\beta)}{1+\beta+\phi_{q}}E_{t}\widetilde{i}_{t+1}-\dfrac{\beta\phi_{q}%
% }{1+\beta+\phi_{q}}E_{t}\widetilde{i}_{t+2}-\dfrac{\phi_{q}/\kappa}%
% {1+\beta+\phi_{q}}E_{t}\widetilde{\nu}_{t+1}.
% \label{eq: Basline_Euler_IAC_util_2}%
% \end{align}
Lowercase letters with a tilde denote the respective log deviations of
the variables from their steady state, $\widetilde{r}_{t}^{p}$ denotes the
log-deviation of the ex-ante real interest rate from steady state, and $\phi
_{q}=(1-\delta)\beta$ and $\phi_{k}=1-\phi_{q}$. Intuitively, investment depends positively on expected capital utilization ($E_{t}\widetilde{u}_{t+1}$), negatively on the
real interest rate ($\widetilde{r}_{t}^{p}$) and positively on the current
value of the investment-specific shock ($\widetilde{\nu}_{t}$). The assumed
specification of the adjustment cost makes investment depend on its own
one-period lag ($\widetilde{i}_{t-1}$) and its expected leads ($E_{t}%
\widetilde{i}_{t+1},E_{t}\widetilde{i}_{t+2}$).

\begin{figure}[htbp]
	\centering
	%{\includegraphics[scale=.6, angle=0, trim = 100 40 40 40, clip]{Literature_estimates.pdf}}\\
	{\includegraphics[width=15.5cm, height=10cm,scale=.90, angle=0, trim = 90 50 90 55, clip]{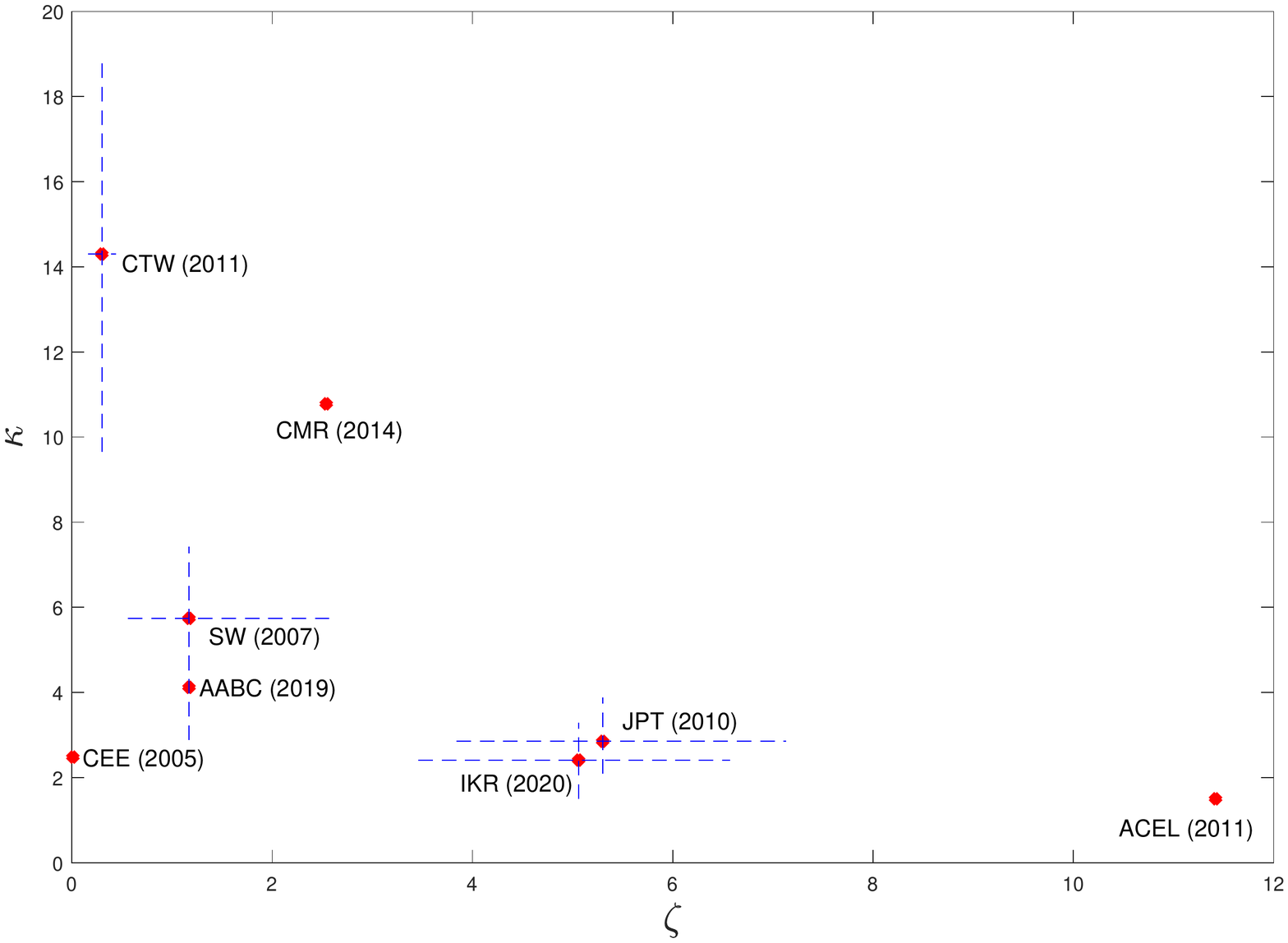}}%
	\caption{Estimates of the elasticity of capital utilization cost ($\zeta$)
		and the investment adjustment cost ($\kappa$) including the 90\% credible intervals from the literature:
		\citeauthor{Christiano_Eichenbaum_Evans_2005} -
		CEE(\citeyear{Christiano_Eichenbaum_Evans_2005}),
		\citeauthor{Smets_Wouters_2007} - SW(\citeyear{Smets_Wouters_2007}),
		\citeauthor{justiniano2010investment} -
		JPT(\citeyear{justiniano2010investment}),
		\citeauthor{Altig_Christiano_Eichenbaum_Linde_2011} -
		ACEL(\citeyear{Altig_Christiano_Eichenbaum_Linde_2011}),
		\citeauthor{Christiano_Trabandt_Walentin_2011} -
		CTW(\citeyear{Christiano_Trabandt_Walentin_2011}),
		\citeauthor{Christiano_Motto_Rostagno_2014} -
		CMR(\citeyear{Christiano_Motto_Rostagno_2014}), \citeauthor{aabc2020} -
		AABC(\citeyear{aabc2020}), \citeauthor{Inoue_Kuo_Rossi_2020} - IKR(\citeyear{Inoue_Kuo_Rossi_2020}).}%
	\label{fig:FigureLit}%
\end{figure}

Figure \ref{fig:FigureLit} shows that estimates of the two main parameters of the Euler equation for investment, i.e., the elasticity of capital utilization cost ($\zeta$)
and the investment adjustment cost ($\kappa$), vary widely across various well-known papers in the literature that estimate medium-scale DSGE models.

%%%%%%%%%%%%%%%%%%%%%%%%%%%%%%%%%%%%%%%%%%%%%%%%%%
%Figure Literature
%%%%%%%%%%%%%%%%%%%%%%%%%%%%%%%%%%%%%%%%%%%%%%%%%

% \begin{figure}[htbp]
% 	\centering
% 	{\includegraphics[width=\textwidth, height=10cm]{IRF_output_investment.png}}%
% 	\caption{Impulse responses to a one-standard deviation shock}%
% 	\label{fig:Figure_irf_out_inv}%
% \end{figure}

\begin{sidewaysfigure}[htbp]
	\centering
	{\includegraphics[width=\textwidth, height=12cm]{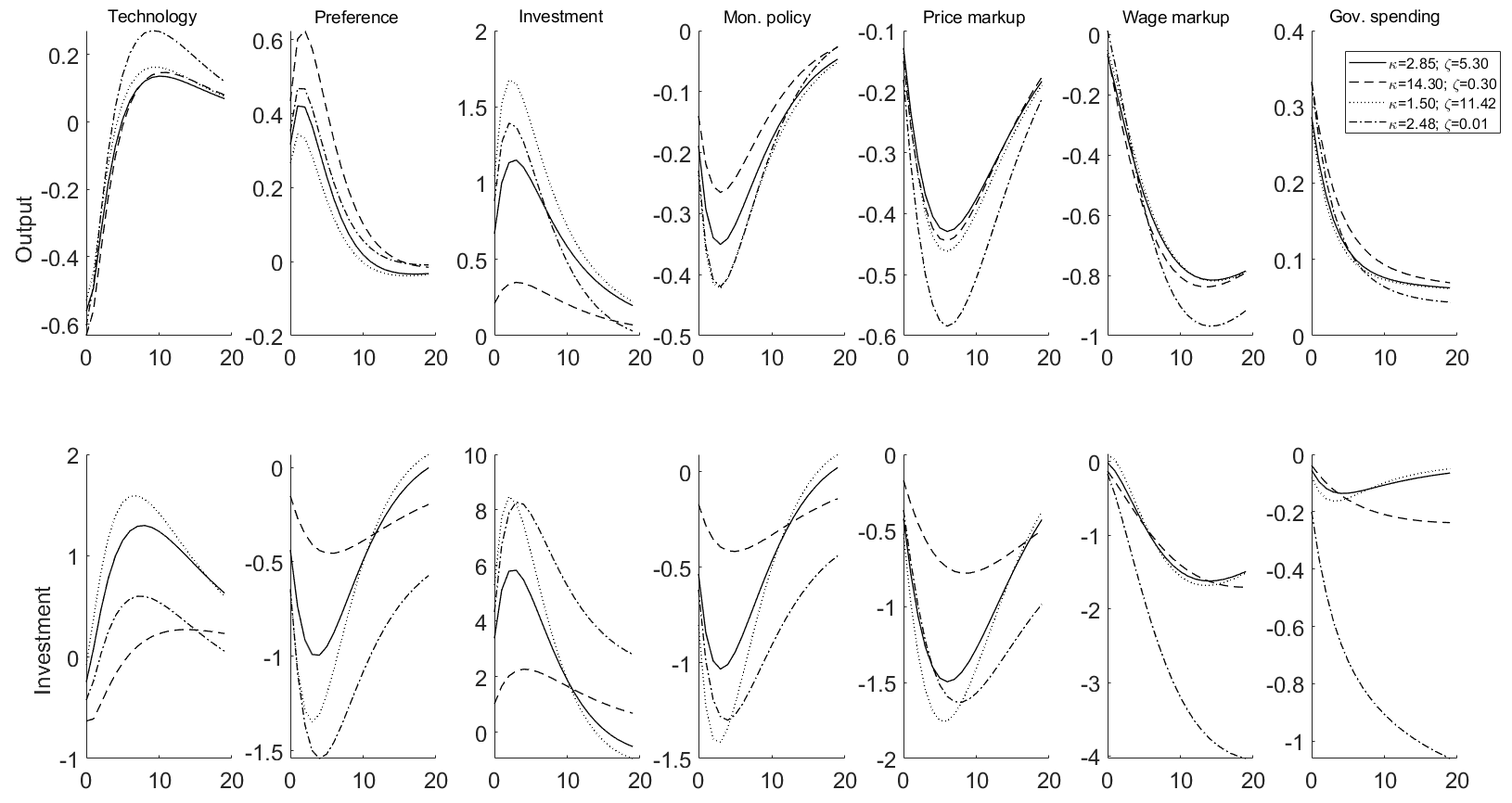}}%
	\caption{Impulse responses of output and investment to a one-standard deviation structural shock in JPT's model for alternative calibration of $\kappa$ and $\zeta$. The remaining parameters are set at the posterior median estimate of JPT.}%
	\label{fig:Figure_irf_out_inv}%
\end{sidewaysfigure}

\begin{figure}[htbhp]
\centering
%\adjustbox{min width=2.0\textwidth, height=.5\textwidth}{
\begin{tabular}{ccccc}
{\includegraphics[angle=0,width=.45\textwidth,trim=120 280 140 230 ,
totalheight=.55\textwidth]{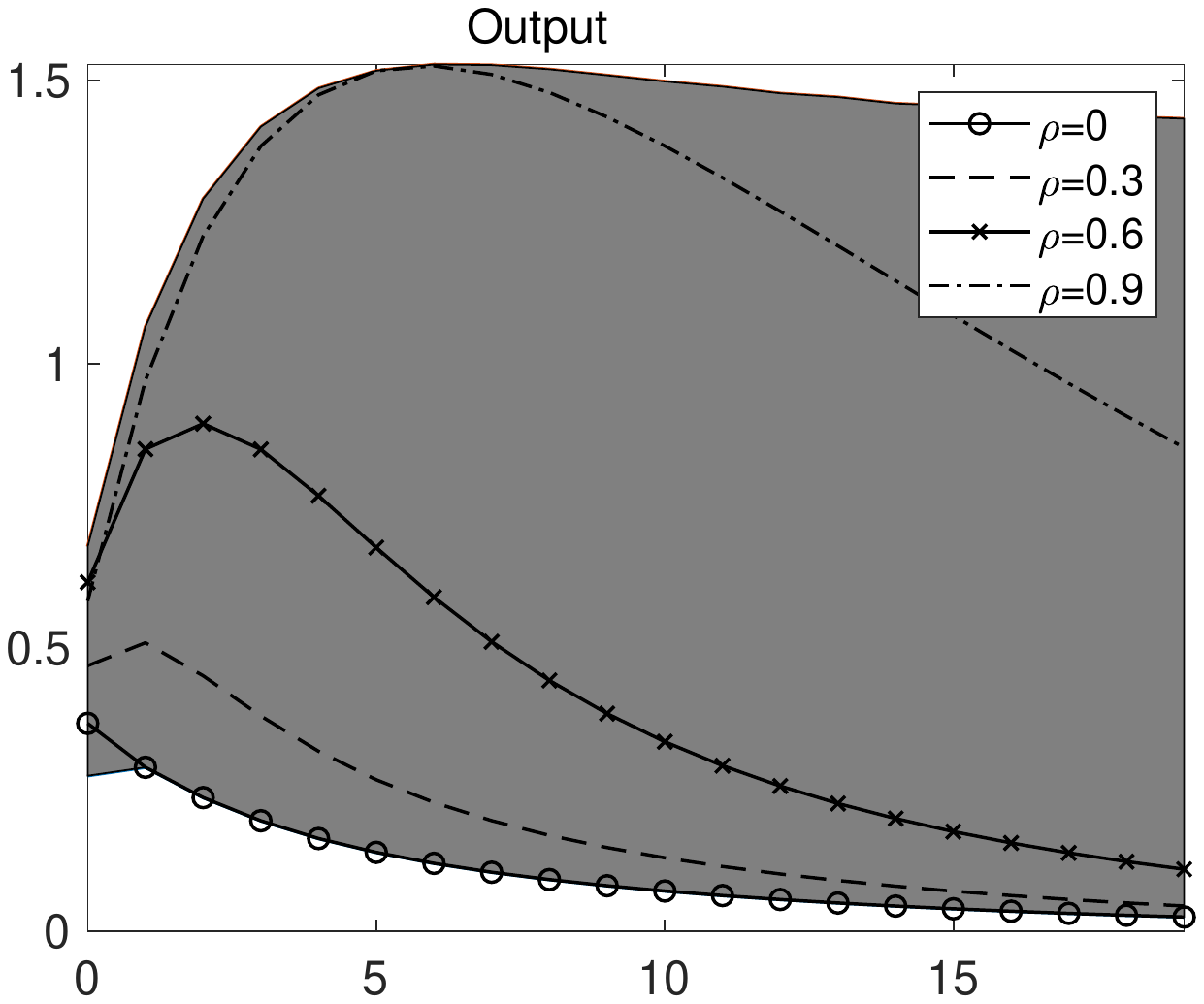}}
&& {\includegraphics[angle=0,width=.45\textwidth,trim=120 280 140 230 ,
totalheight=.55\textwidth]{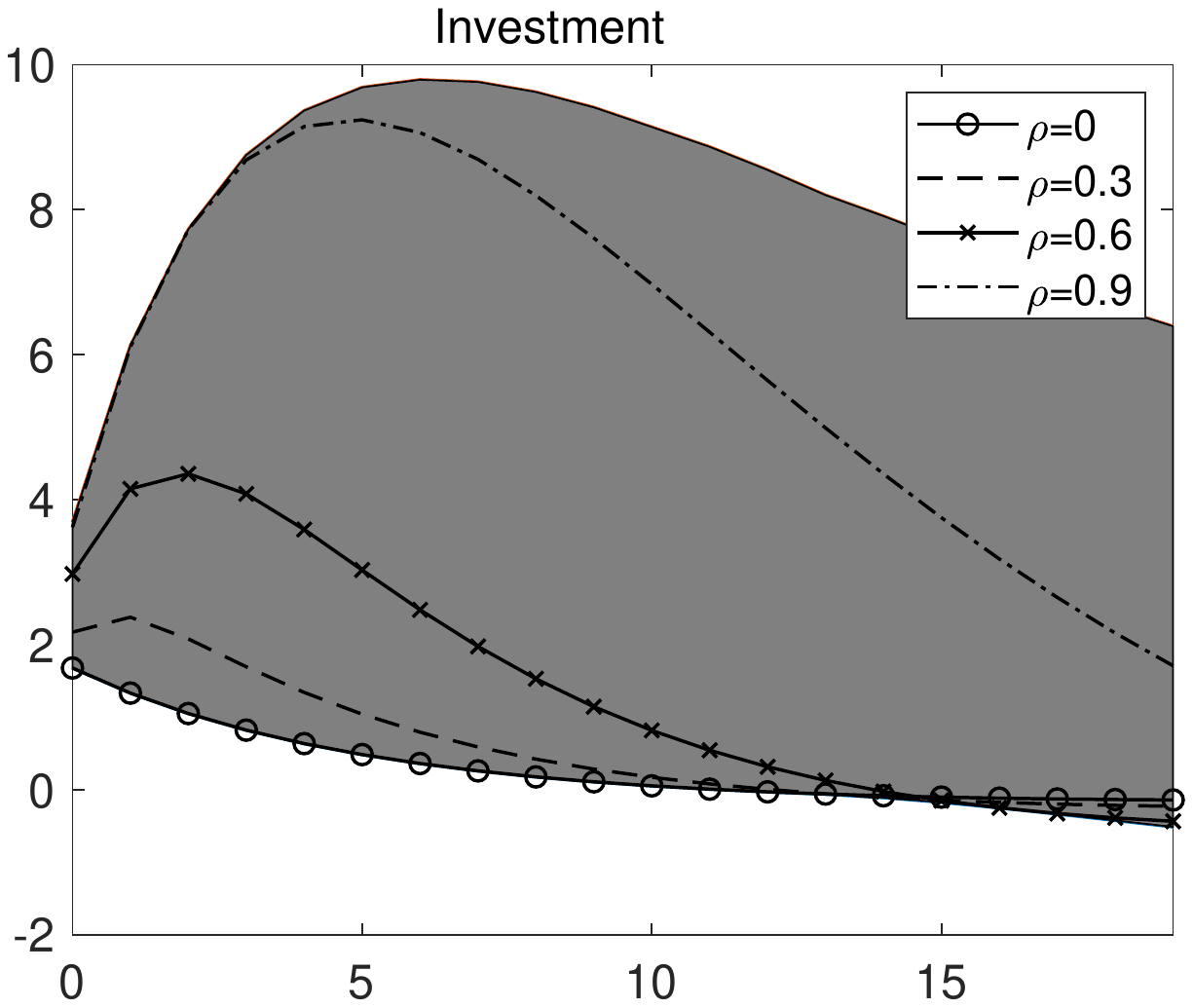}} 
\\[+25pt]
\end{tabular}
%}
\caption{Impulse responses of output and investment to a one-standard deviation investment shock in JPT's model evaluated at the posterior median of all parameters except $\rho$. The shaded area shows the range of IRFs for $\rho \in [0,1)$.}%
\label{fig:Figure_irf_rho_grid}%
\end{figure}

To appreciate the implications of the different values of the parameters characterizing the investment equation we use the JPT model. Figure \ref{fig:Figure_irf_out_inv} shows how the impulse responses of output and investment to the seven structural shocks in JPT's model change when we keep all the parameters fixed at the JPT posterior median, while we vary the value of $\kappa$ and $\zeta$ using the following four values from the literature (see Figure \ref{fig:FigureLit}) : (i) \citeauthor{justiniano2010investment} -
		JPT(\citeyear{justiniano2010investment}): $\kappa = 2.85$ and $\zeta = 5.30$; (ii) \citeauthor{Christiano_Trabandt_Walentin_2011} -
		CTW(\citeyear{Christiano_Trabandt_Walentin_2011}): $\kappa = 14.30$ and $\zeta = 0.30$; (iii) \citeauthor{Altig_Christiano_Eichenbaum_Linde_2011} -
		ACEL(\citeyear{Altig_Christiano_Eichenbaum_Linde_2011}): $\kappa = 1.50$ and $\zeta = 11.42$; (iv) \citeauthor{Christiano_Eichenbaum_Evans_2005} -
		CEE(\citeyear{Christiano_Eichenbaum_Evans_2005}): $\kappa = 2.48$ and $\zeta = 0.01.$ It is evident that the response of investment to the various shocks varies substantially from a quantitative point of view in terms of impact effect, peak response and persistence. With the exception of the technology shock, a larger value of $\kappa$ tends to dampen the response of investment. The different responses of investment then translates into different responses in output. As a consequence, these different calibrations have an impact on the variance decomposition of output growth, as shown in Table \ref{Table: VarDecom_GDPgrowth}. Not surprisingly, then a large value of $\kappa$ decreases the importance of the investment-specific shock as a driver of the business cycle. The ACEL calibration exhibits the smallest value of the investment adjustment cost parameter ($\kappa=1.5$) and the highest contribution of the investment-specific shock to the variance of output growth (69\%) among the four different calibrations. In contrast, the CTW calibration exhibits the largest value of the investment adjustment cost parameter ($\kappa=14.3$) and the lowest contribution of the investment-specific shock to the variance of output growth (8\%). The difference in the explained output growth variance is distributed mainly to the preference, technology and government spending shock. 
Hence, the main result in JPT, for example, about the investment-specific shock being the major driver of the business cycle relies on a relatively low value for $\kappa$, as also discussed by JPT. The effects of different values for $\zeta$ are more difficult to grasp from this analysis, but more will be said about it below.
Another crucial parameter for investment fluctuations is the persistence of the investment-specific shock, $\rho$. Figure \ref{fig:Figure_irf_rho_grid} shows the impulse responses of output and investment to an investment shock in JPT (2010) for grid values of $\rho$ between zero and one - using their posterior median values for $\kappa$ and $\zeta$. Not surprisingly, the larger the value of $\rho$, the larger and more persistent  the response of output and investment. Given the relevance of these parameters for our understanding and for the narrative of investment and business cycle fluctuations, it is important to understand the diversity of estimates reported in the literature and investigate whether GMM can pin down the value of these parameters more accurately.

\begin{table}[!htp]
\centering
\caption{Variance decomposition of output growth in JPT's model}
%\resizebox{\textwidth}{!}{
\begin{tabularx}{0.9\textwidth}{X|ccccccc}
 \hline \hline
\noalign{\vspace{0.5ex}}
Parameters & Tec & Pref & Inv & MP & PM & WM & Govt \\
\noalign{\vspace{0.5ex}}
\hline
\noalign{\vspace{0.5ex}}
JPT: $\kappa=2.85, \zeta =5.30$ & 0.21 & 0.10 & 0.49 & 0.04 & 0.03 & 0.05 & 0.07 \\
\\
CTW: $\kappa=14.30, \zeta =0.30$ & 0.31 & 0.29 & 0.08 & 0.03 & 0.04 & 0.10 & 0.14 \\
\\
ACEL: $\kappa=1.50, \zeta =11.42$ & 0.14 & 0.04 & 0.69 & 0.04 & 0.02 & 0.03 & 0.04 \\
\\
CEE: $\kappa=2.48, \zeta =0.01$ & 0.18 & 0.08 & 0.55 & 0.04 & 0.03 & 0.06 & 0.06 \\
\noalign{\vspace{0.5ex}}
\hline \hline
\end{tabularx}
\smallskip
		\begin{tablenotes}\setlength\labelsep{0pt} \footnotesize
			\item[] Notes: 
		\citeauthor{justiniano2010investment} -
		JPT(\citeyear{justiniano2010investment}),
		\citeauthor{Christiano_Trabandt_Walentin_2011} -
		CTW(\citeyear{Christiano_Trabandt_Walentin_2011}),
		\citeauthor{Altig_Christiano_Eichenbaum_Linde_2011} -
		ACEL(\citeyear{Altig_Christiano_Eichenbaum_Linde_2011}),
		\citeauthor{Christiano_Eichenbaum_Evans_2005} -
		CEE(\citeyear{Christiano_Eichenbaum_Evans_2005}).
		Shocks: Technology (Tec), Preference (Pref),  Investment (Inv), Monetray Policy (MP), Price markup (PM), Wage markup (WM), Governement spending (Govt).
		\end{tablenotes}
%}
\label{Table: VarDecom_GDPgrowth}
\end{table}

% \begin{figure}[htbp]
% 	\centering
% 	{\includegraphics[width=\textwidth, height=9cm]{irf_rho_grid.png}}%
% 	\caption{Impulse responses of output and investment to a one-standard deviation investment shock}%
% 	\label{fig:Figure_irf_rho_grid}%
% \end{figure}

%\section{Model \label{s: model}}

\section{Econometric Methodology\label{s:_ectmethod}}

This Section describes the methods we use to estimate the model given in equation \eqref{eq: Basline_Euler_IAC_util_2}.
Without a complete specification of a DSGE model that includes equation \eqref{eq: Basline_Euler_IAC_util_2}, the expectations on its right-hand side are unobserved. Therefore, we need to rely on a limited-information estimation method, such as single-equation GMM. The latter involves replacing the expected future terms in \eqref{eq: Basline_Euler_IAC_util_2} with their realized values and finding valid instruments for them, which are typically predetermined variables. However, in this case, we first need to quasi-difference the equation to remove the autocorrelation in $\widetilde{\nu}_t$ that would otherwise rule out using predetermined variables as instruments. Specifically, removing expectations and
quasi-differencing equation (\ref{eq: Basline_Euler_IAC_util_2}) yields
\begin{align}
\left[  1+\rho\left(  \beta+\phi_{q}\right)  \right]  \Delta\widetilde{i}_{t}=
&  \ \rho\Delta\widetilde{i}_{t-1}+\left[  \beta+\phi_{q}+\rho\beta\phi
_{q}\right]  \Delta\widetilde{i}_{t+1}-\beta\phi_{q}\Delta\widetilde{i}%
_{t+2}\nonumber\\
&  +\frac{\phi_{k}\zeta}{\kappa}\widetilde{u}_{t+1}-\frac{\rho\phi_{k}\zeta
}{\kappa}\widetilde{u}_{t}-\dfrac{1}{\kappa}\widetilde{r}_{t}^{p}+\rho
\dfrac{1}{\kappa}\widetilde{r}_{t-1}^{p}+\epsilon_{t},\label{eq: estimated}%
\end{align}
where $\epsilon_{t}$ is an error term defined in equation (\ref{app eq: error}%
) in Appendix \ref{appsec: derivation}. The time series properties of the
residual term $\epsilon_{t}$ are crucial for the selection of valid
instruments. It can be gauged from equation (\ref{app eq: error}) that, under
rational expectations, $\epsilon_{t}$ is a moving average process of order 2
that consists of current values of the investment adjustment cost shock
$\varepsilon_{t}^{\nu}$ and current and future forecast errors of inflation, investment and capital utilization. Hence, it is orthogonal to any predetermined variables. 

We estimate the structural parameters in the baseline equation (\ref{eq: estimated}) using the generalized method of moments (GMM) framework proposed by \cite{hansen1982generalized}, with orthogonality conditions obtained from the assumption that the residuals, $\epsilon_t$, in the baseline equation (\ref{eq: estimated}) are uncorrelated with any predetermined variables $Z_t$. In our estimation, we set $\beta=0.99$ and $\delta=0.025$, and compute confidence sets for the remaining structural parameters $\theta = (\rho,\kappa,\zeta)$.

Our econometric analysis relies solely on methods of inference that are robust
to the presence of potential weak instruments, while allowing for heteroskedasticity and
autocorrelation in the residuals. We estimate confidence sets based on the S
test of \cite{SW00}. The S set is constructed
as follows. We specify a grid of points within the parameter space. For each
of these points, we test whether the identifying restrictions of the model
hold using a Wald-type test \citep[][call this the S test]{SW00}. All the
points in the grid that have not been rejected by a 10\% level S test make up
the 90\% confidence S set.

In addition to the S sets, we also estimate confidence sets based on a test
proposed by \cite{MM14}, the quasi-local level S (qLL-S) test. The qLL-S set
combines the average information on the moment conditions over the sample,
which is what the S set uses, with information on the validity of the moment
conditions over subsamples. It can be thought of as using subsample
information as additional instruments. This subsample information is relevant
in two cases: (i) when the parameters of the model are unstable, or (ii) when
the parameters of the model are constant but there is time variation in other
parts of the economy, for example, monetary policy regime shifts. In case (i),
the qLL-S test can be interpreted as a structural change test that is robust
to weak identification; therefore a nonrejection is an indication of parameter
stability. In case (ii), the qLL-S can have more power than the corresponding
S test if the information that comes from structural change elsewhere in the
economy is sufficiently strong. Hence, in either case, the qLL-S sets usefully
complement the S sets. 

The orthogonality condition $E_{t-1}(\epsilon_t)=0$ in (\ref{eq: estimated}) implies that any predetermined variable could be used as an instrument. Therefore, the number of potential instruments is unbounded. However, the S and qLL-S sets may be unreliable if the number of instruments is large relative to the sample size. Therefore, we keep the number of instruments small when we compute S and qLL-S sets. This may be inefficient if information is spread over a large number of instruments, or if the most informative instruments are excluded from the set of instruments that we use. To address this possibility, we use a split-sample S set. This is a straightforward extension of a method recently proposed by \cite{Mikusheva2021} to obtain reliable inference in linear instrumental variables models with time series data and a large number of possibly weak instruments. In Section \ref{appsec: comp. details} of the Appendix we present
information about computation of the S, qLL-S and split-sample S tests.\footnote{\cite{Mikusheva2021} only explicitly discusses the case of linear moment conditions and therefore uses the Anderson-Rubin statistic. It is straightforward to extend her proposal to nonlinear GMM, see Appendix \ref{s: mikusheva} for details.}

\section{Data\label{s: data}}

We use quarterly aggregate time series data for the US over the period 1967q1
to 2019q4. We consider two proxies for Investment ($I_{t}$). One corresponds
to Fixed Private Investment as in \cite{Smets_Wouters_2007} (SW, henceforth) ,
while the other proxy is the sum of Gross Private Domestic Investment and
Personal Consumption Expenditure on Durable Goods following JPT. Both investment measures
are in real per capita terms and deflated using their respective implicit
price deflators. For the nominal interest rate $r_{t}$, we use the quarterly
average of the effective Federal Funds rate, and inflation $\pi_{t}$ is
obtained from the GDP deflator as $\pi_{t}=\log(P_{t}/P_{t-1})$. The ex-post real
interest rate $r_{t}^{p}$ is defined as $r_{t}^{p}=r_{t}-\pi_{t+1}$. For
capital utilization ($u_{t}$), we use the Federal Reserve Board's time series
on capacity utilization, which measures the intensity with which all factors
of production are used in the industrial production sector
\citep{Christiano_Eichenbaum_Evans_2005}. When estimating equation
(\ref{eq: estimated}), we use the log values of investment and capacity
utilization, that is, $i_{t}=\log(I_{t})$ and $u_{t}=\log(U_{t})$.\footnote{In equation (\ref{eq: estimated}) the variables appear with a tilde, that is, in
log-deviations from steady state. In computing the tests, we collect all
the steady state terms in the constant term included in that equation.}

Other variables included in the set of predetermined/exogenous variables
$Z_{t}$ are \citeauthor{romer2004new}'s monetary policy shock,
\citeauthor{ramey2018government}'s military news shock, oil price
inflation and (financial) uncertainty measure VXO. 
Detailed description of the data, its sources and transformations are given in Appendix \ref{appsec: data}.

\section{Results\label{s: results}}
This Section presents first the results of our baseline estimation. Then, we report results based on external instruments, followed by results obtained from combining both lagged endogenous and exogenous instruments together and employing the split-sample method proposed by \cite{Mikusheva2021} that is robust to many weak instruments. Finally, we define a semi-structural model to estimate the slope coefficients of the investment equation with respect to the capital utilization rate and the real interest rate. 

\subsection{Baseline Estimation}
\begin{figure}[htbhp]
\centering
\adjustbox{min width=2.0\textwidth, height=.5\textwidth}{
\begin{tabular}
[c]{cccccc}\hline\hline\\[-5pt]
&  & {\Large{SW Investment}} &  & {\Large{JPT Investment}} & \\\cline{2-3}\cline{5-5}\\
&  & (a)     &  & (b)     & \vspace{-.5cm}\\
& \raisebox{+20.7ex}{\rotatebox[origin=lt]{90}{S sets }}
&
\hspace{+.3cm}
{\includegraphics[angle=0,width=.5\textwidth,trim=120 280 140 230 ,
totalheight=.6\textwidth]{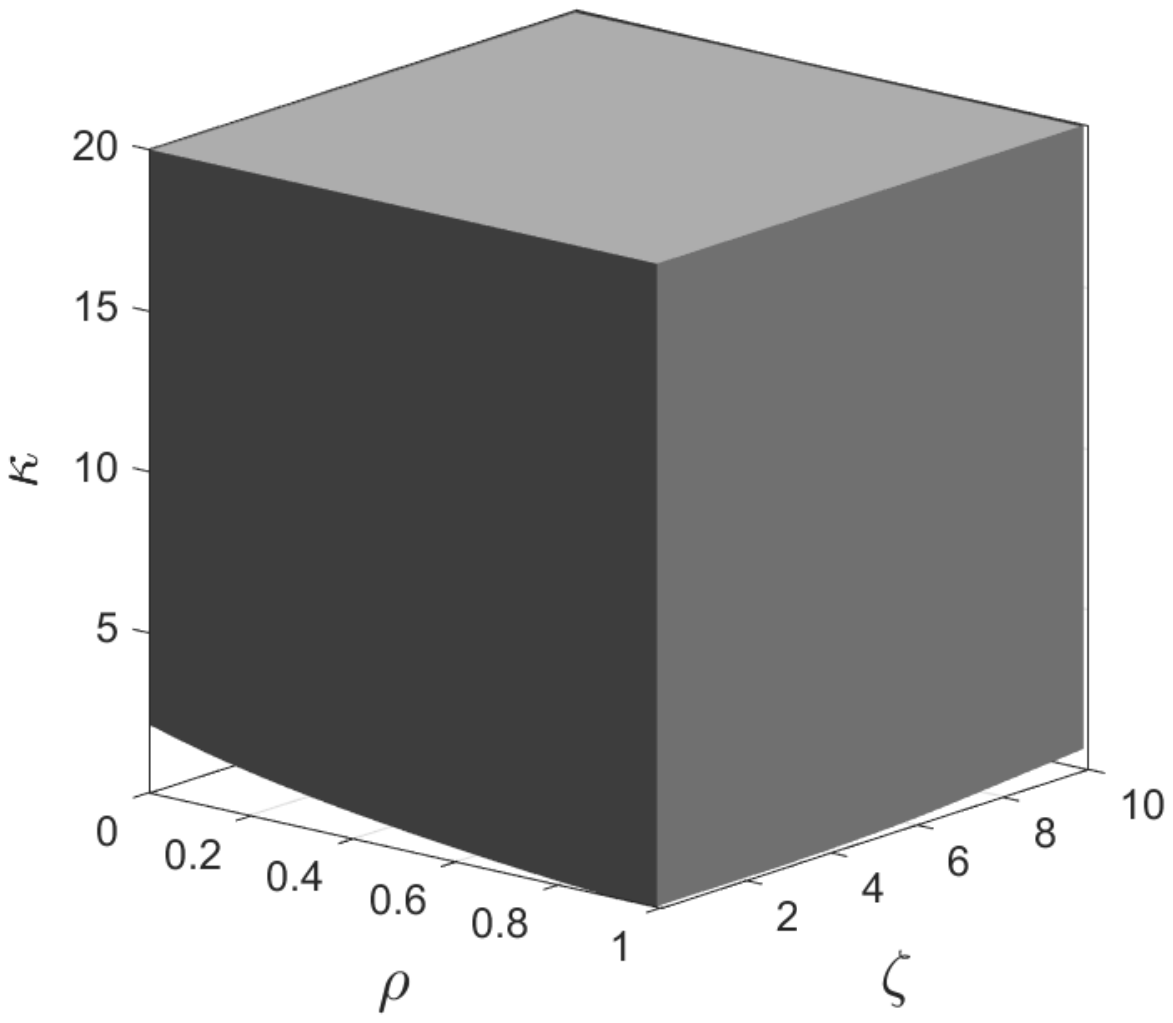}}
&
&
\hspace{+.4cm}
{\includegraphics[angle=0,width=.5\textwidth, trim=120 280 140 230 ,
totalheight=.6\textwidth]{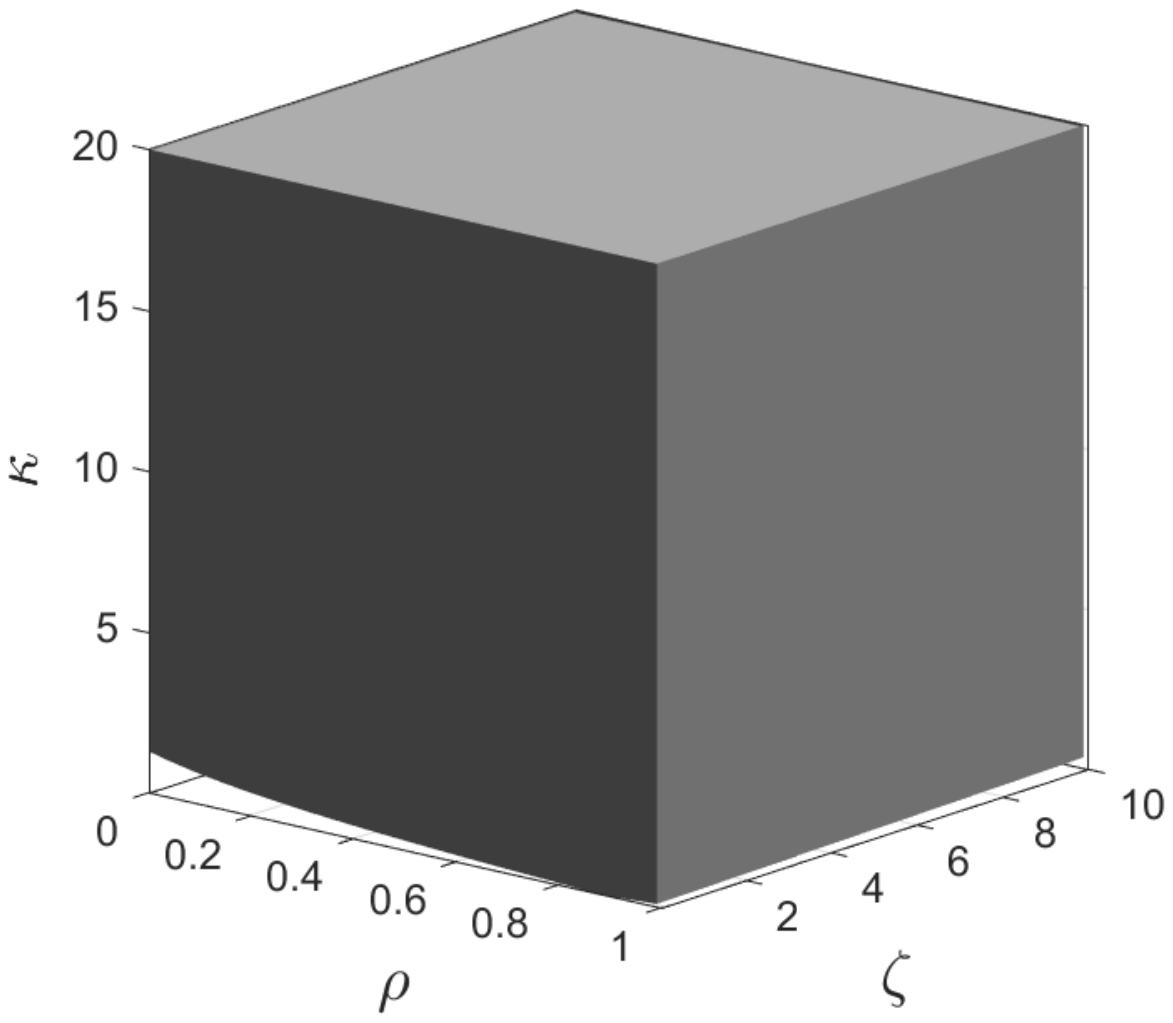}} & \\[10pt]
&  & (c)     &  &  (d)     & \vspace{-.5cm}\\
& \raisebox{+20.7ex}{\rotatebox[origin=lt]{90}{qLL-S sets }}
&
\hspace{+.3cm}
{\includegraphics[angle=0,width=.5\textwidth,trim=120 280 140 230 ,
totalheight=.6\textwidth]{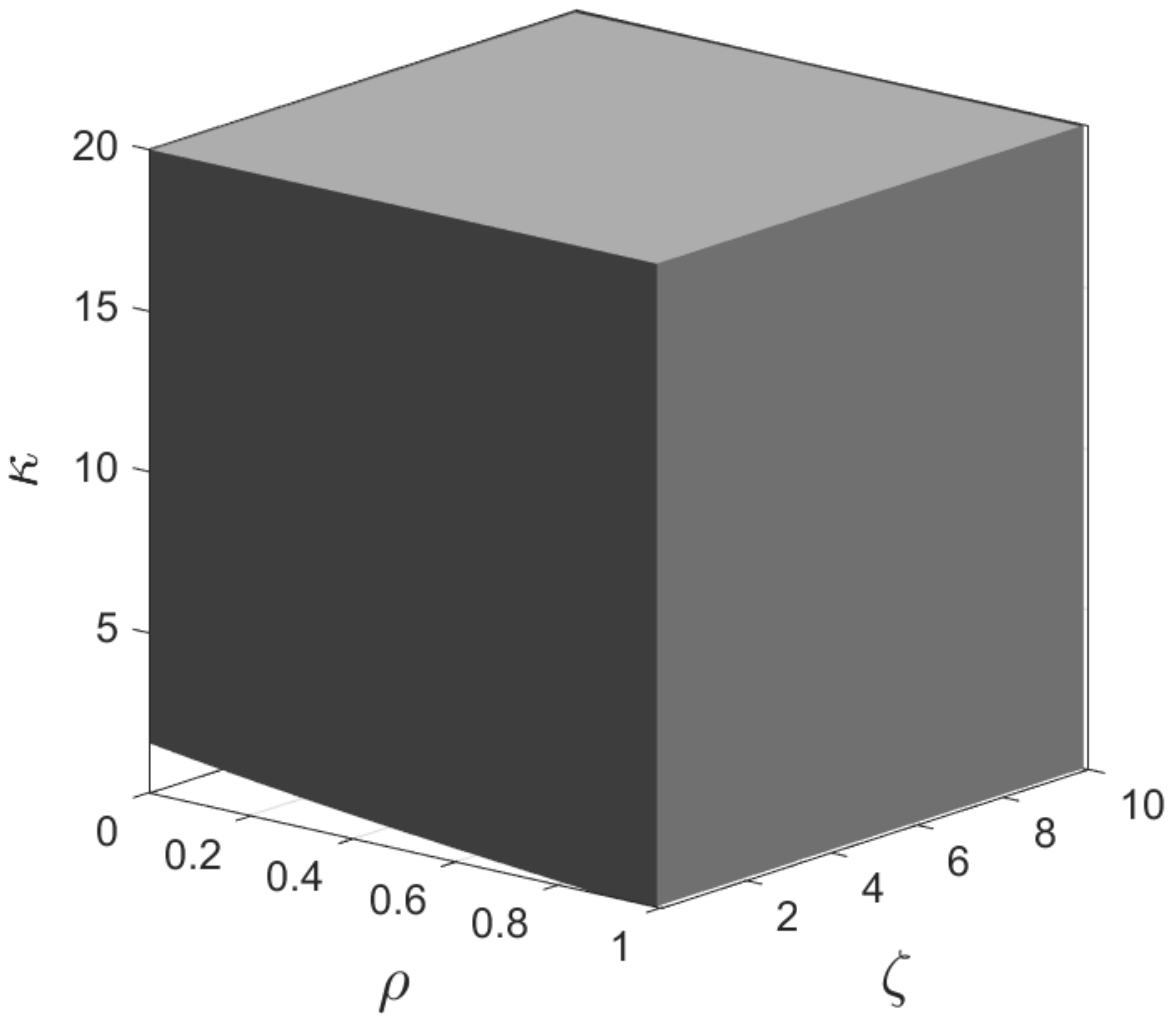}}
&
\hspace{+.5cm}
&
\hspace{+.4cm}
{\includegraphics[angle=0,width=.5\textwidth,trim=120 280 140 230, 
totalheight=.6\textwidth]
{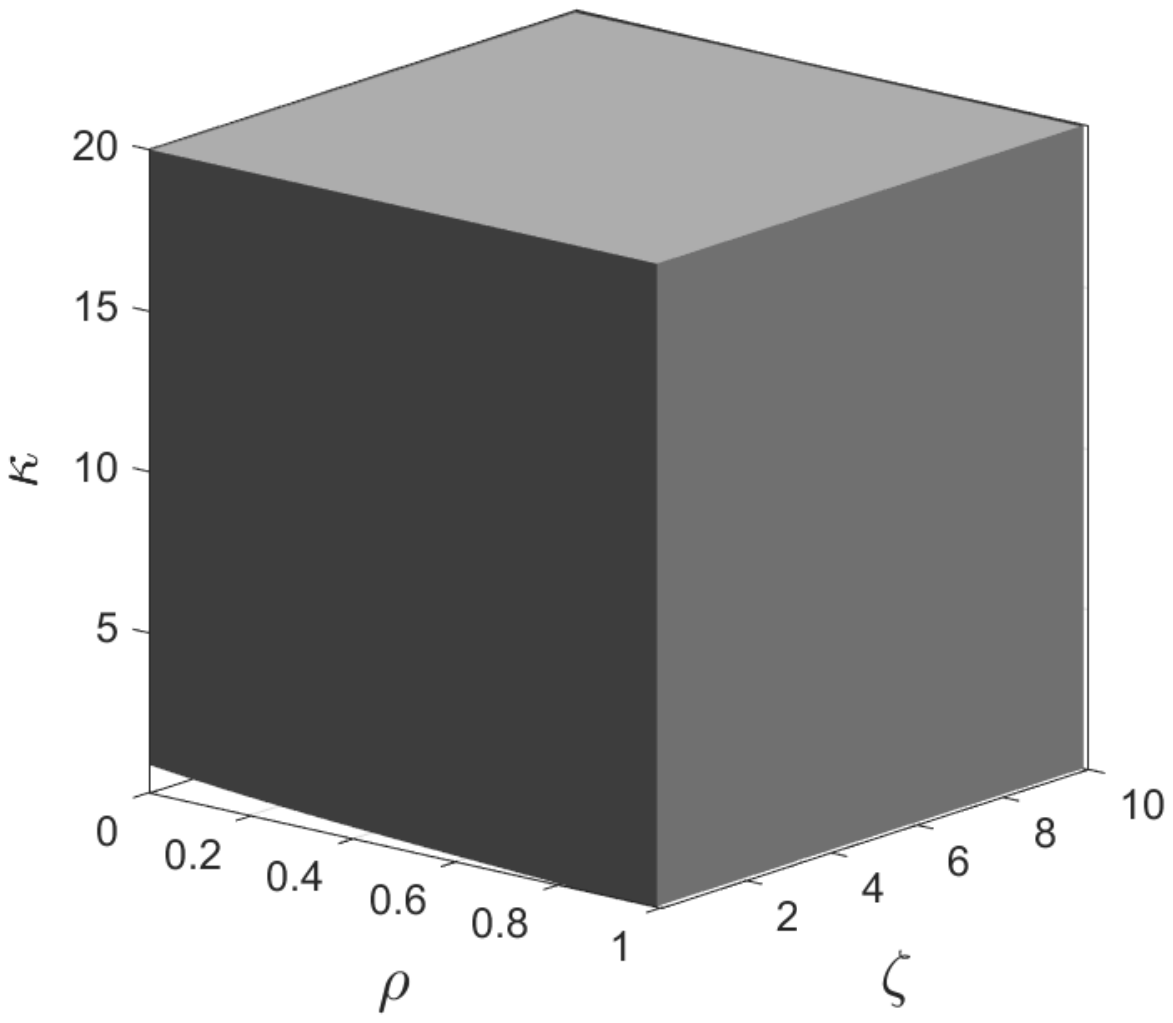}} &
\\[+15pt]
\hline\hline
\end{tabular}}
\caption{90\% S and qLL-S confidence sets for $\theta
=(\rho,\kappa,\zeta)$ derived from the investment Euler equation model
(\ref{eq: estimated}). Instruments: constant, $\Delta i_{t-1}$, $r_{t-2}^{p}$,
$u_{t-1}$. The investment proxies are Fixed Private Investment (left column) and the sum of Gross Private
Domestic Investment and Personal Consumption Expenditure on Durable Goods (right column). \cite{Newey_West_1987}
HAC. Period: 1967Q1-2019Q4.}%
\label{fig: baseline}%
\end{figure}

Following the previous studies on the estimation of the consumption Euler
equation \citep{Yogo04,AMM2021}, we begin by investigating the baseline
specification equation (\ref{eq: estimated}) using one lagged value of the variables that appear in the model as the set of instrumental variables,
namely $\Delta{i}_{t-1}$, $r_{t-2}^{p}$, and ${u}_{t-1}$. We keep the
number of instruments small to avoid problems associated with the use of many
instruments, see \cite{AndrewsStock2007}.
%\footnote{The results are very similar if we add the second lagged values of the variables in the set of instruments, see Figure \ref{appfig: Two lags} in Appendix \ref{appsec: further results}.}
We estimate the model using both the SW and the JPT definitions of investment discussed in
Section \ref{s: data}. Three-dimensional confidence sets for the parameters
$\theta=(\rho,\kappa,\zeta)$ are obtained by considering the ranges
$[0,1)\times(0,20]\times(0,10]$ in line with previous studies, see
Figure \ref{fig:FigureLit}. The results are reported in Figure
\ref{fig: baseline}.

%%%%%%%%%%%%%%%%%%%%%%%%%%%%%%%%%%%%%%%%%%%%%%%%%%
%Figure Baseline
%%%%%%%%%%%%%%%%%%%%%%%%%%%%%%%%%%%%%%%%%%%%%%%%%

Figures \ref{fig: baseline} (a) and (b) report 90\% S sets for the parameters $\rho,\kappa,\zeta$ using SW and JPT investment proxies, respectively. In both cases, the S sets comprise almost the entire parameter space. The only part rejected by the data are small values of $\kappa$ and $\zeta$ when
$\rho<1$. The good news from this result is that the investment equation is not rejected by the aggregate data. Nevertheless, the data is essentially uninformative over a very large part of the parameter space. The results remain essentially unchanged when considering two lags instead of one lag for each of the three instruments (see Figure \ref{appfig: Two lags} in the Appendix). Additionally, the qLL-S sets reported in Figures \ref{fig: baseline} (c) and (d)
are very similar to the S sets, indicating no presence of parameter
instability or violation of the moment conditions in subsamples. Hence, there is minimal information arising from the restrictions on the dynamics of the data to identify the investment equation. As a consequence of the lack of identification, all previous parameter estimates reported in Figure \ref{fig:FigureLit} are included in the confidence sets, that is, those estimates are potentially valid values of the true underlying structural parameters of the investment Euler equation (\ref{eq: estimated}).

\subsection{External Instruments}

Given the findings in Figure \ref{fig: baseline}, we explore a more extensive
set of information contained in contemporaneous external instruments. These
instruments include (i) the monetary policy shock of \cite{romer2004new}, (ii)
the military news shock of \cite{ramey2011qje,ramey2016defense} and updated by
\cite{ramey2018government}, which captures news about changes in military
spending, (iii) changes in the (log) oil price, and (iv) the (standardized)
S\&P 100 Volatility Index (VXO), which is a proxy for (financial) uncertainty
shocks.\footnote{Another related potential external instrument is overall macroeconomic
uncertainty as studied by \cite{Jurado_Ludvigson_Ng_2015}; however,
\cite{ludvigson2015uncertainty} point out that macroeconomic uncertainty responds to
business cycle fluctuations making it an endogenous variable, while
they suggest that financial uncertainty is exogenous.}
%See Section \ref{appsec: data} in the Appendix for a description of the instruments series and sources.} 
Some of the external instruments do not cover
the entire period. We, therefore, use the longest available sample when
estimating the confidence sets. Apart from oil price, the external
instruments are in levels. We use the contemporaneous
values as instruments.

\begin{figure}[ptbh]
\centering
\adjustbox{min width=\textwidth,max width=\textwidth, max height=9.5cm}{
\begin{tabular}
[c]{ccccc}\hline\hline
& \multicolumn{4}{c}{Exogenous Instruments with SW Investment Proxy }\\ \cline{1-5}
& {\small {Baseline}} & {\small {Mon. pol. shock}} & {\small {Military news}}
& {\small {Oil}}\\
& {\small {1967Q1-2019Q4}} & {\small {1969Q2-2007Q4}} &
{\small {1967Q1-2015Q4}} & {\small {1967Q1-2019Q4}}\\
& {\small (a)} & {\small (b)} & {\small (c)} & {\small (d)}\vspace{-.5cm}\\
\raisebox{+8.7ex}{\rotatebox[origin=lt]{90}{S sets }}\hspace{+.4cm} &
{\includegraphics[angle=0,width=.18\textwidth, trim=120 270 140 230 ,
totalheight=.25\textwidth]{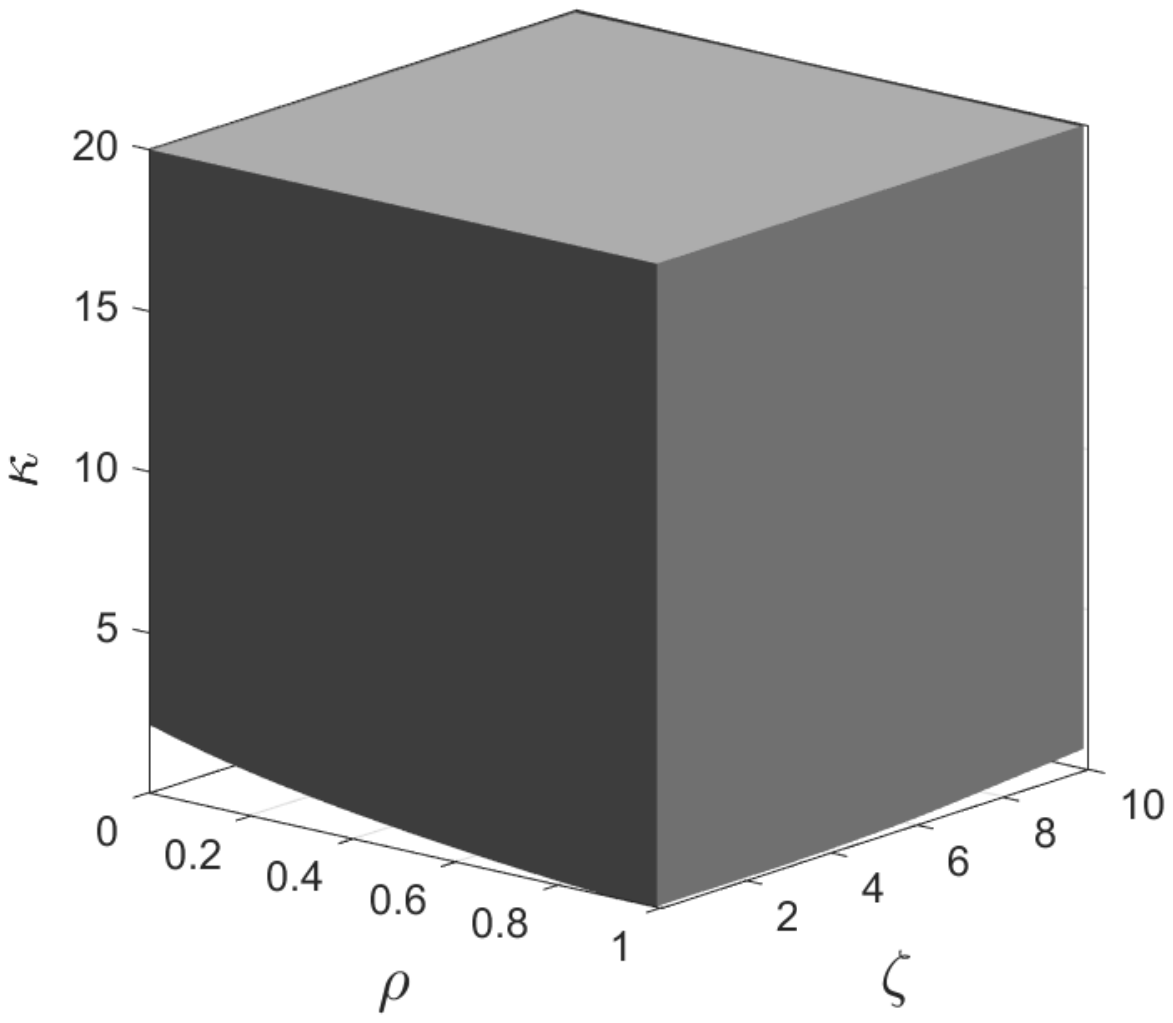}} &
\hspace{+.1cm}
{\includegraphics[angle=0,width=.20\textwidth,trim=120 280 140 230 ,
totalheight=.25\textwidth]{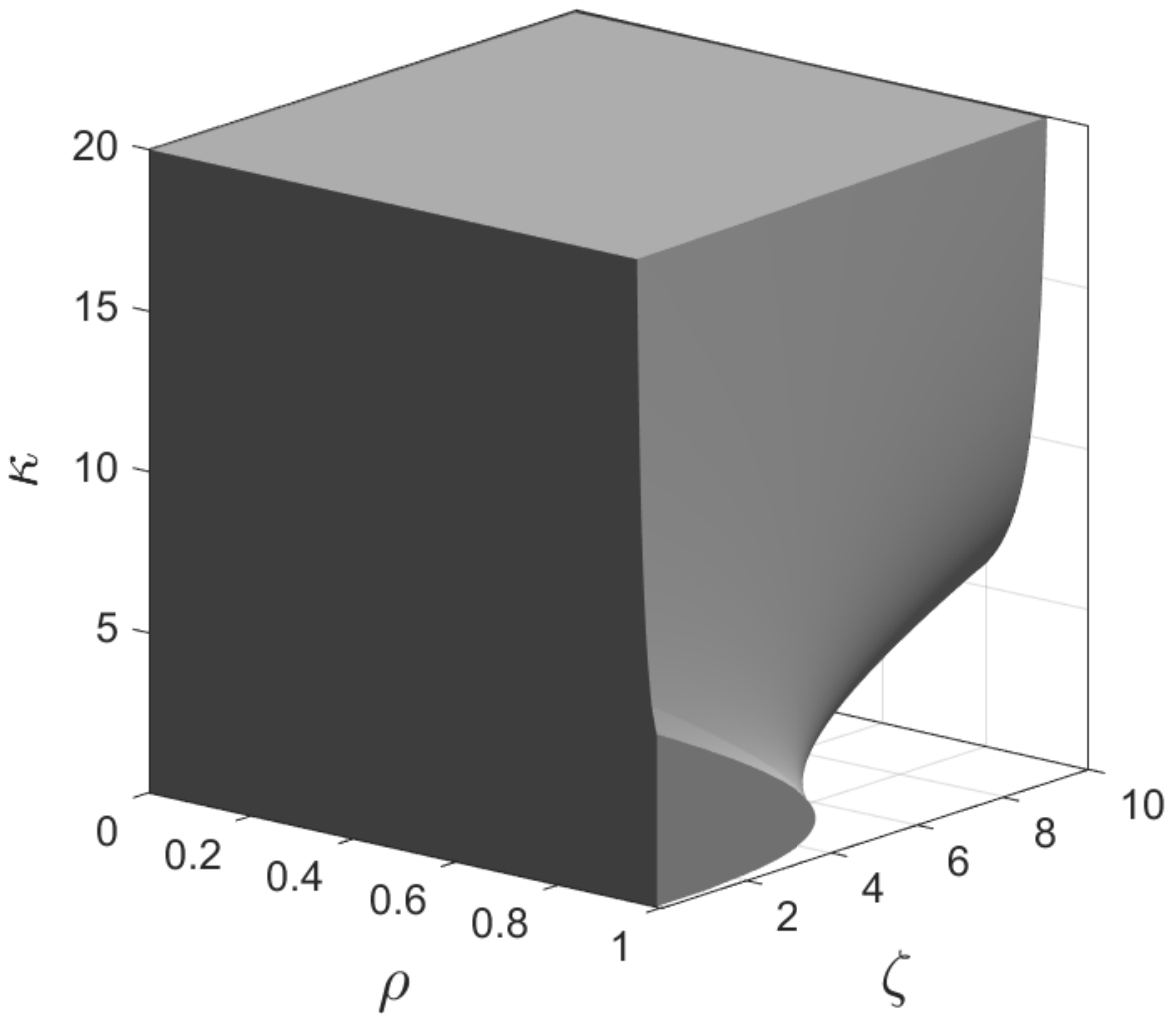}} & \hspace{+.1cm}
{\includegraphics[angle=0,width=.20\textwidth,trim=120 280 140 230 ,
totalheight=.25\textwidth]{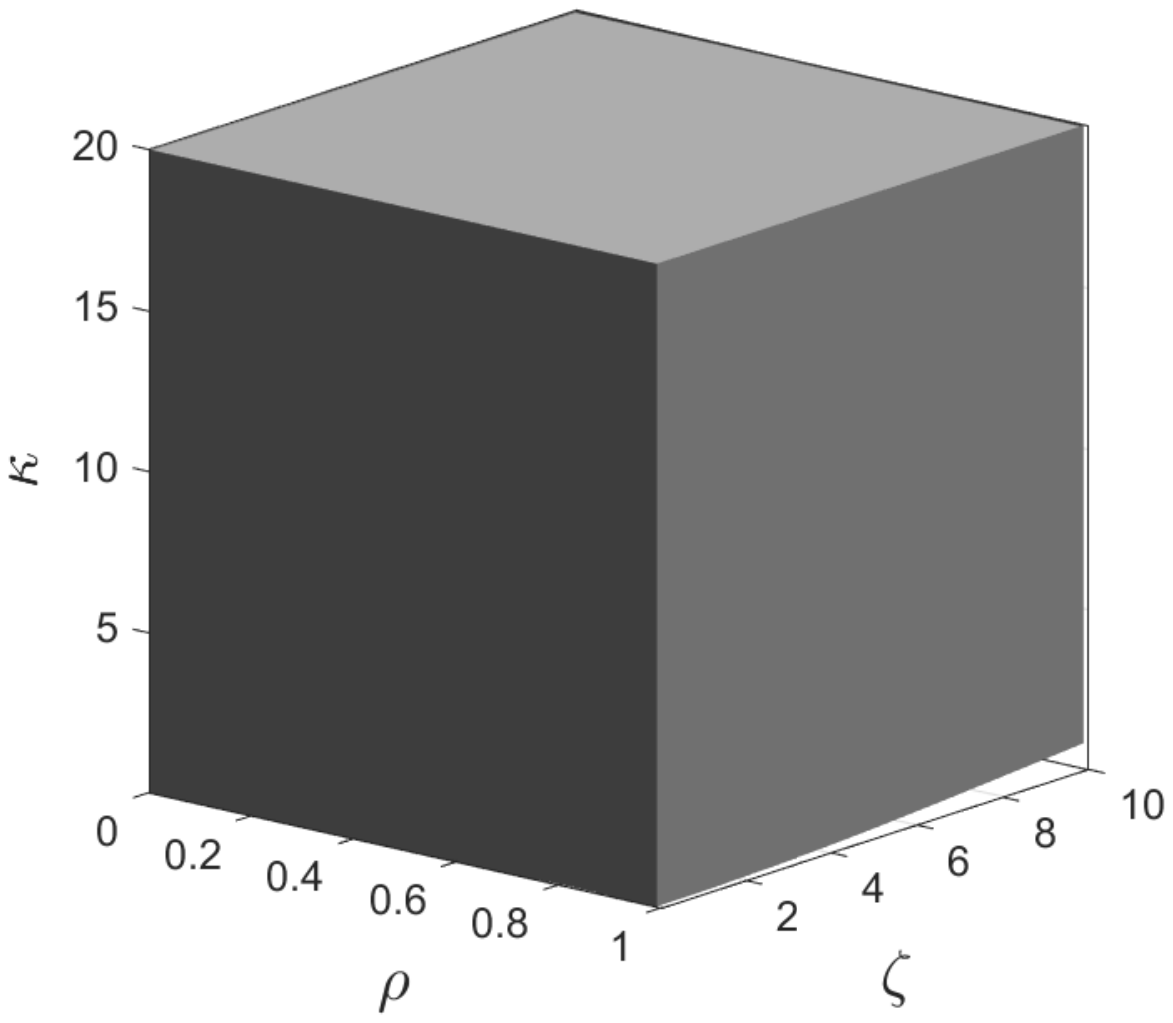}} &
\hspace{+.1cm}
{\includegraphics[angle=0,width=.20\textwidth,trim=120 280 140 230 ,
totalheight=.25\textwidth]{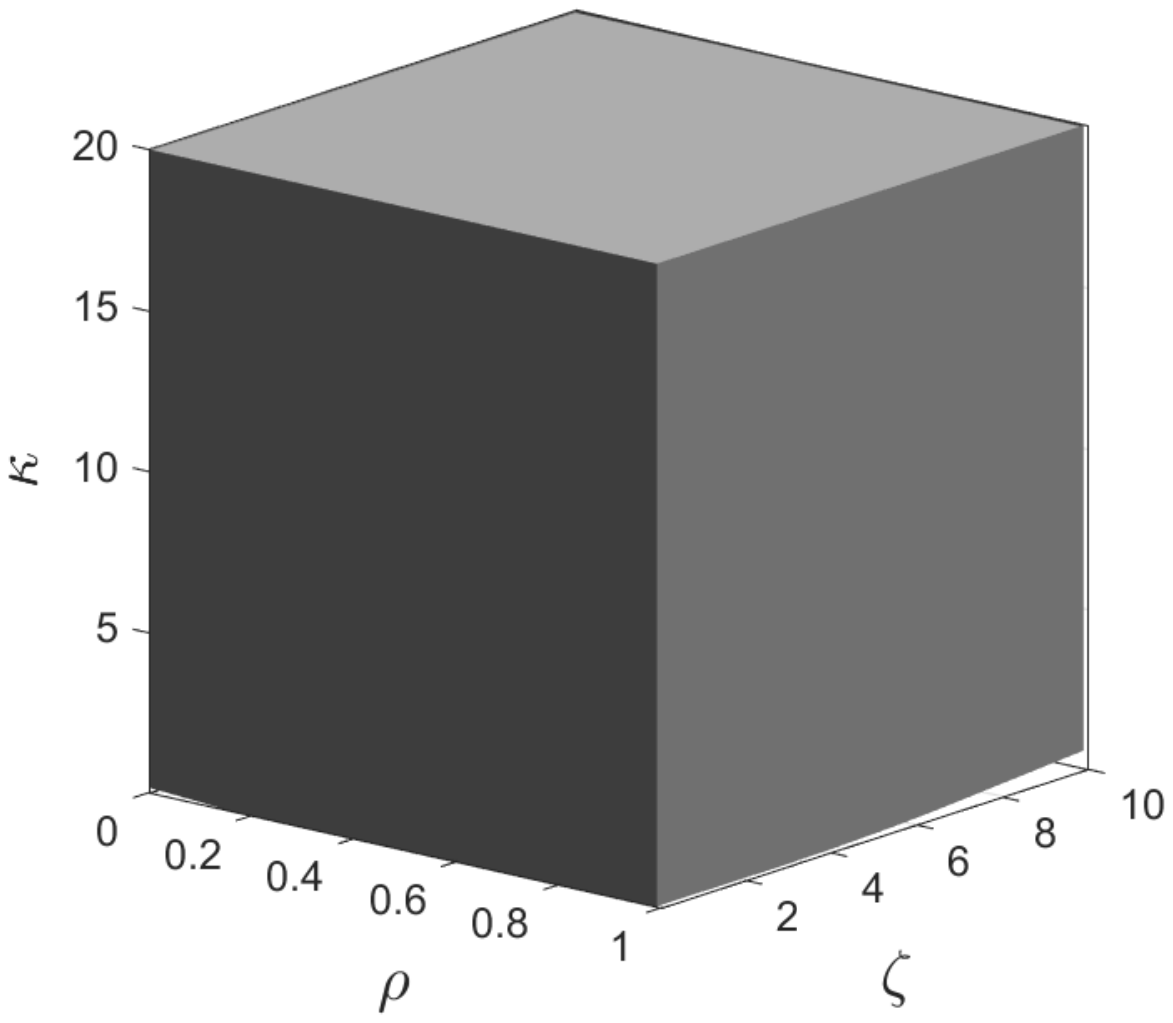}}\\[+3pt]
& {\small {VXO}} & {\small (b)+(c)} & {\small (d)+(e)} &
{\small {(b)+(c)+(d)+(e)}}\\
& {\small {1967Q1-2019Q4}} & {\small {1969Q2-2007Q4}} &
{\small {1967Q1-2019Q4}} & {\small {1969Q2-2007Q4}}\\
& {\small (e)} & {\small (f)} & {\small (g)} & {\small (h)}\vspace{-.4cm}\\
\raisebox{+8.7ex}{\rotatebox[origin=lt]{90}{S sets }}\hspace{+.4cm} &
{\includegraphics[angle=0,width=.18\textwidth, trim=120 280 140 230 ,
totalheight=.25\textwidth]{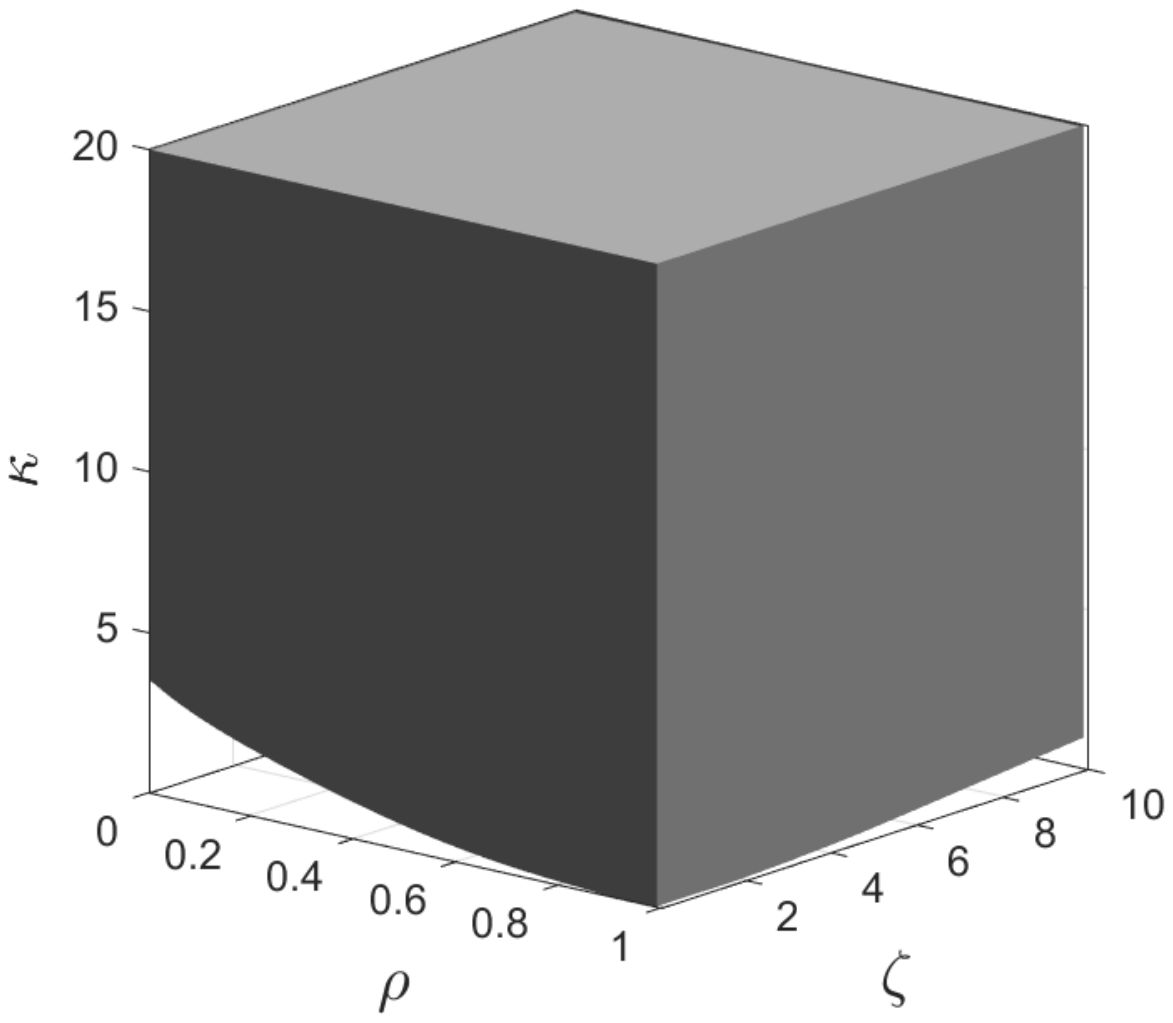}} & \hspace{+.1cm}
{\includegraphics[angle=0,width=.20\textwidth,trim=120 280 140 230 ,
totalheight=.25\textwidth]{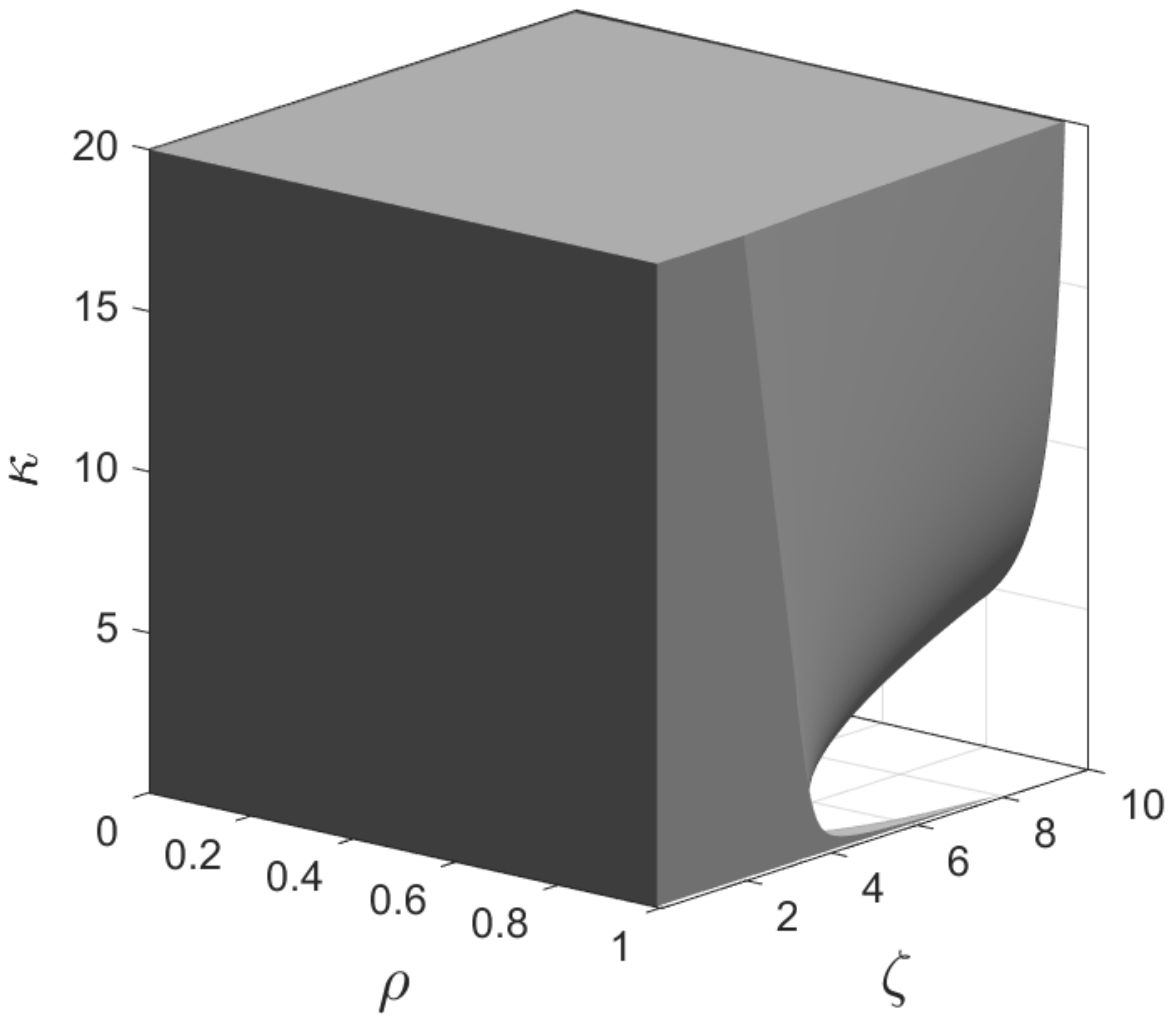}} &
\hspace{+.1cm}
{\includegraphics[angle=0,width=.20\textwidth,trim=120 280 140 230 ,
totalheight=.25\textwidth]{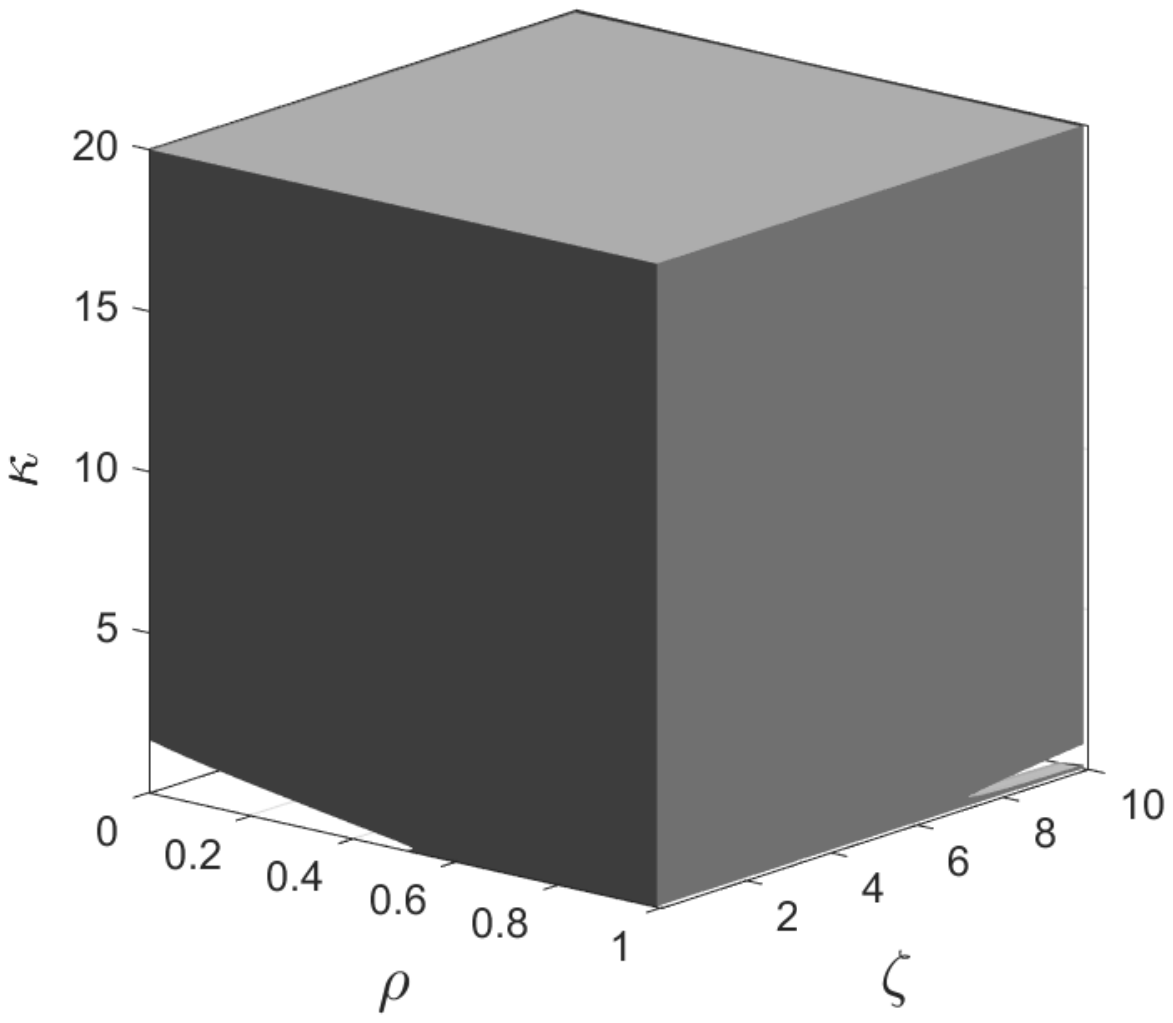}} & \hspace{+.1cm}
{\includegraphics[angle=0,width=.20\textwidth,trim=120 280 140 230 ,
totalheight=.25\textwidth]{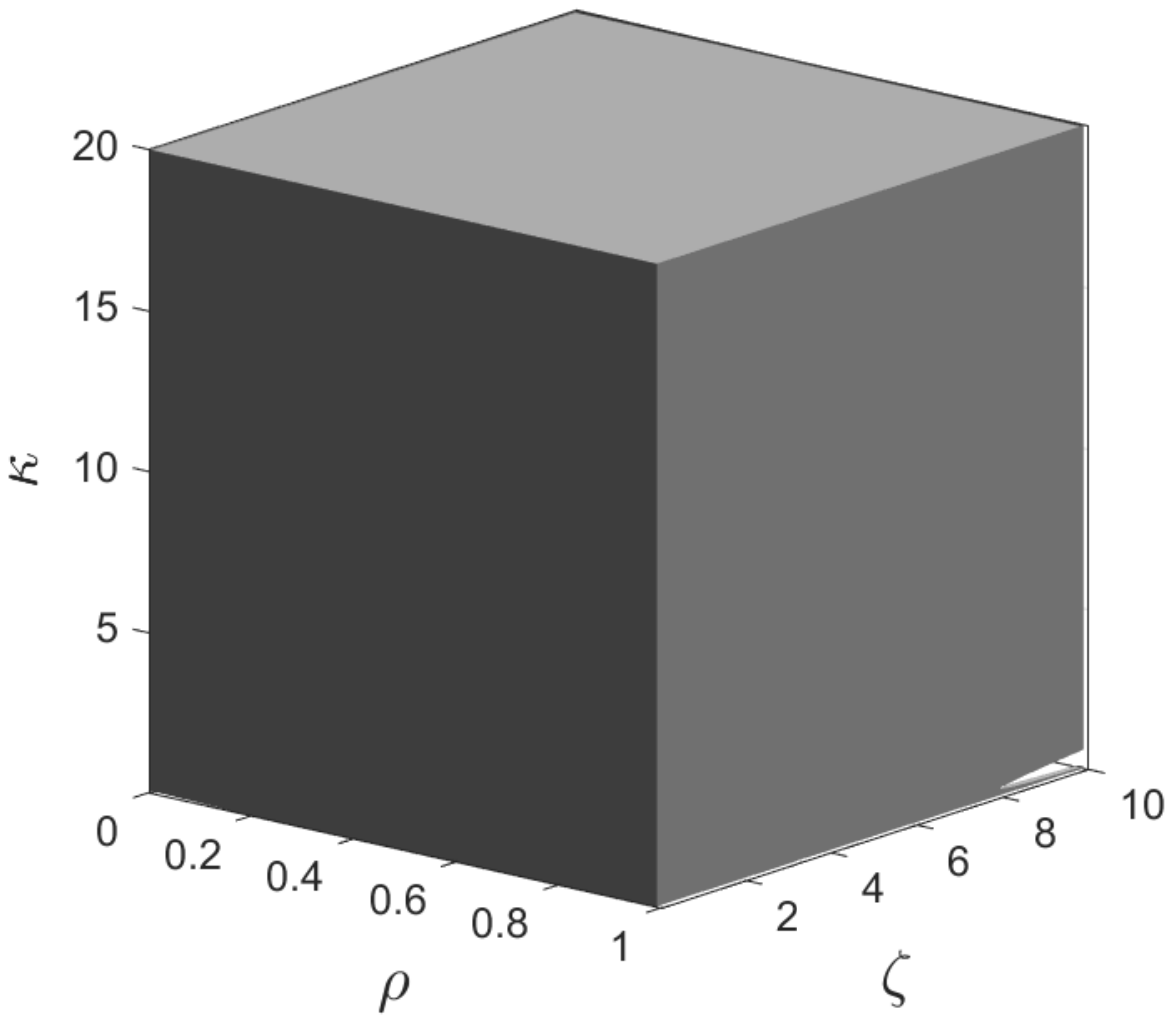}}\vspace{+10pt}\\
&  &  &  & \\
& {\small {Baseline}} & {\small {Mon. pol. shock}} & {\small {Military news}}
& {\small {Oil}}\\
& {\small {1967Q1-2019Q4}} & {\small {1969Q2-2007Q4}} &
{\small {1967Q1-2015Q4}} & {\small {1967Q1-2019Q4}}\\
& {\small (i)} & {\small (j)} & {\small (k)} & {\small (l)}\vspace{-.4cm}\\
\raisebox{+5.7ex}{\rotatebox[origin=lt]{90}{qLL-S sets }}\hspace{+.4cm} &
{\includegraphics[angle=0,width=.18\textwidth, trim=120 270 140 230 ,
totalheight=.25\textwidth]{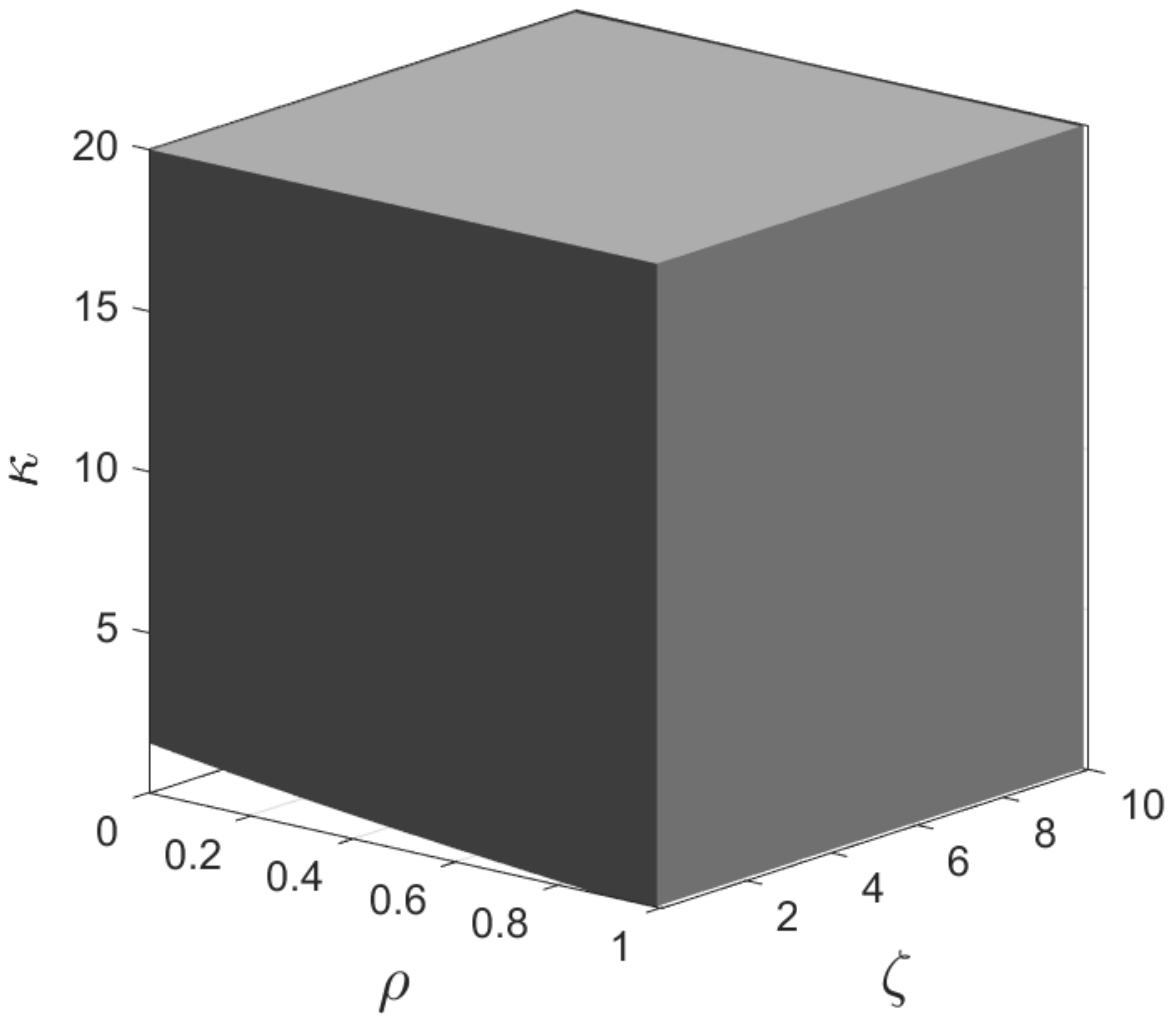}} &
\hspace{+.1cm}
{\includegraphics[angle=0,width=.20\textwidth,trim=120 280 140 230 ,
totalheight=.25\textwidth]{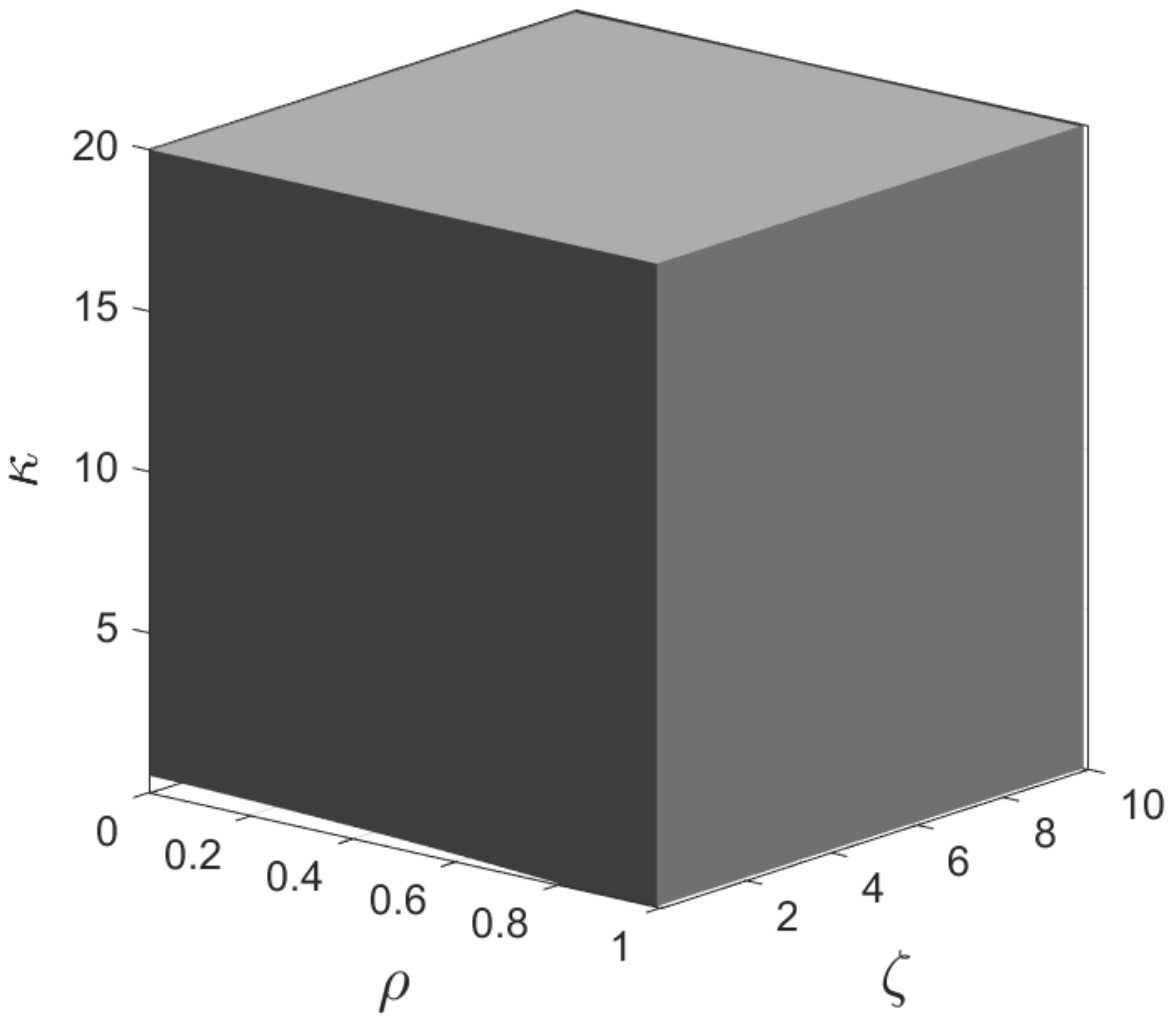}} &
\hspace{+.1cm}
{\includegraphics[angle=0,width=.20\textwidth,trim=120 280 140 230 ,
totalheight=.25\textwidth]{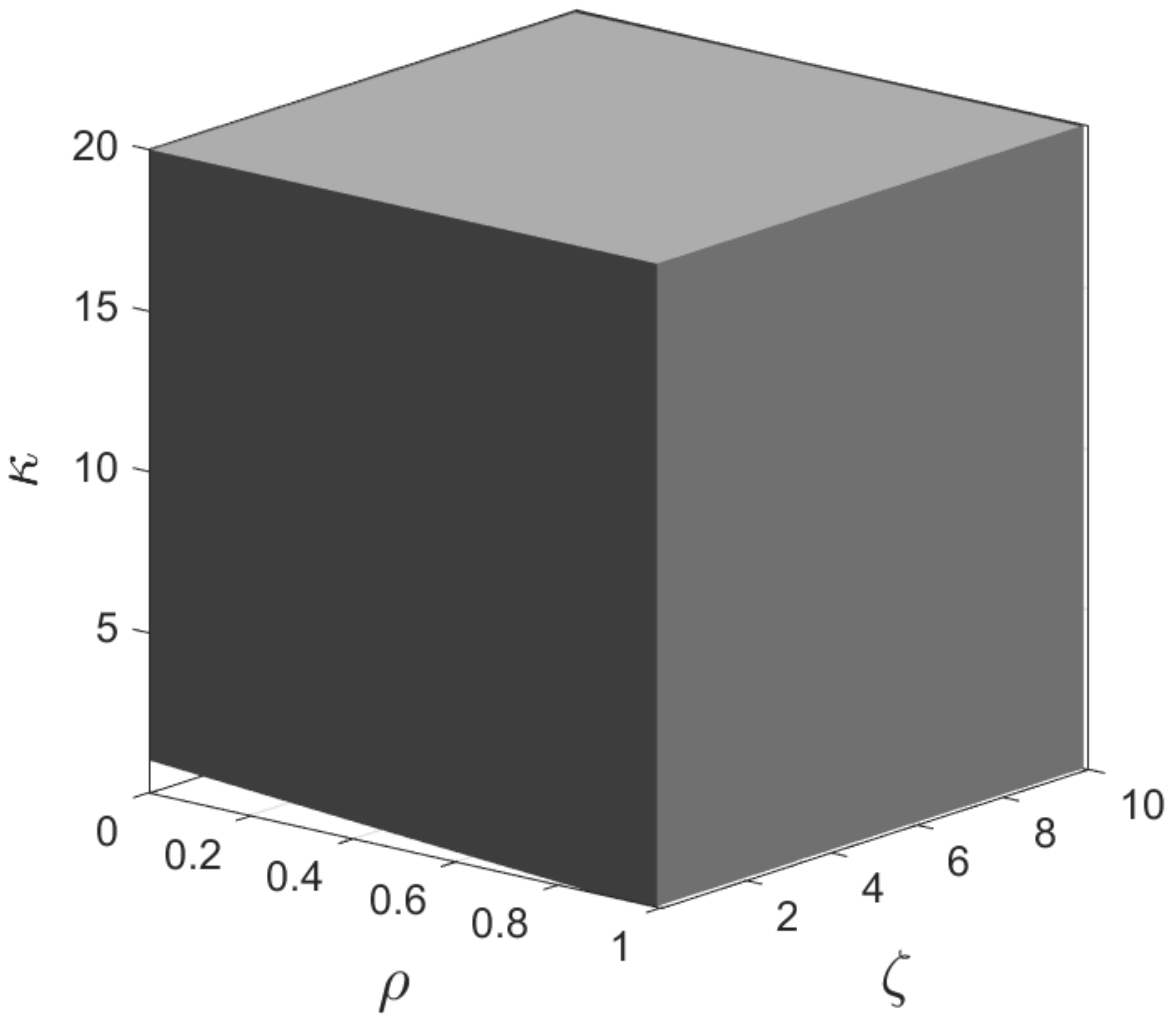}} &
\hspace{+.1cm}
{\includegraphics[angle=0,width=.20\textwidth,trim=120 280 140 230 ,
totalheight=.25\textwidth]{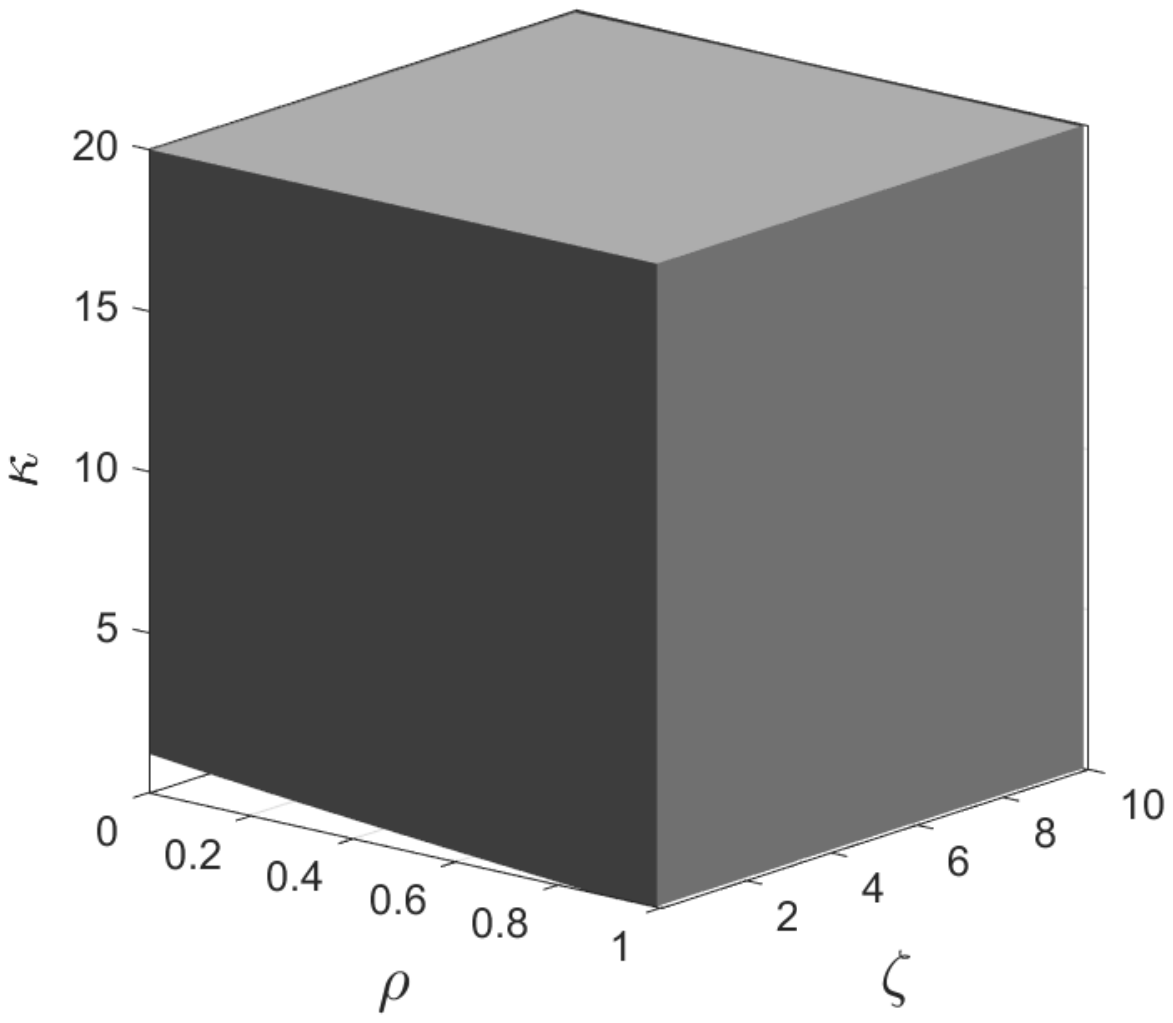}}\\[+3pt]
& {\small {VXO}} & {\small (j)+(k)} & {\small (l)+(m)} &
{\small {(j)+(k)+(l)+(m)}}\\
& {\small {1967Q1-2019Q4}} & {\small {1969Q2-2007Q4}} &
{\small {1967Q1-2019Q4}} & {\small {1969Q2-2007Q4}}\\
& {\small (m)} & {\small (n)} & {\small (o)} & {\small (p)}\vspace{-.4cm}\\
\raisebox{+5.7ex}{\rotatebox[origin=lt]{90}{qLL-S sets }}\hspace{+.4cm} &
{\includegraphics[angle=0,width=.18\textwidth, trim=120 280 140 230 ,
totalheight=.25\textwidth]{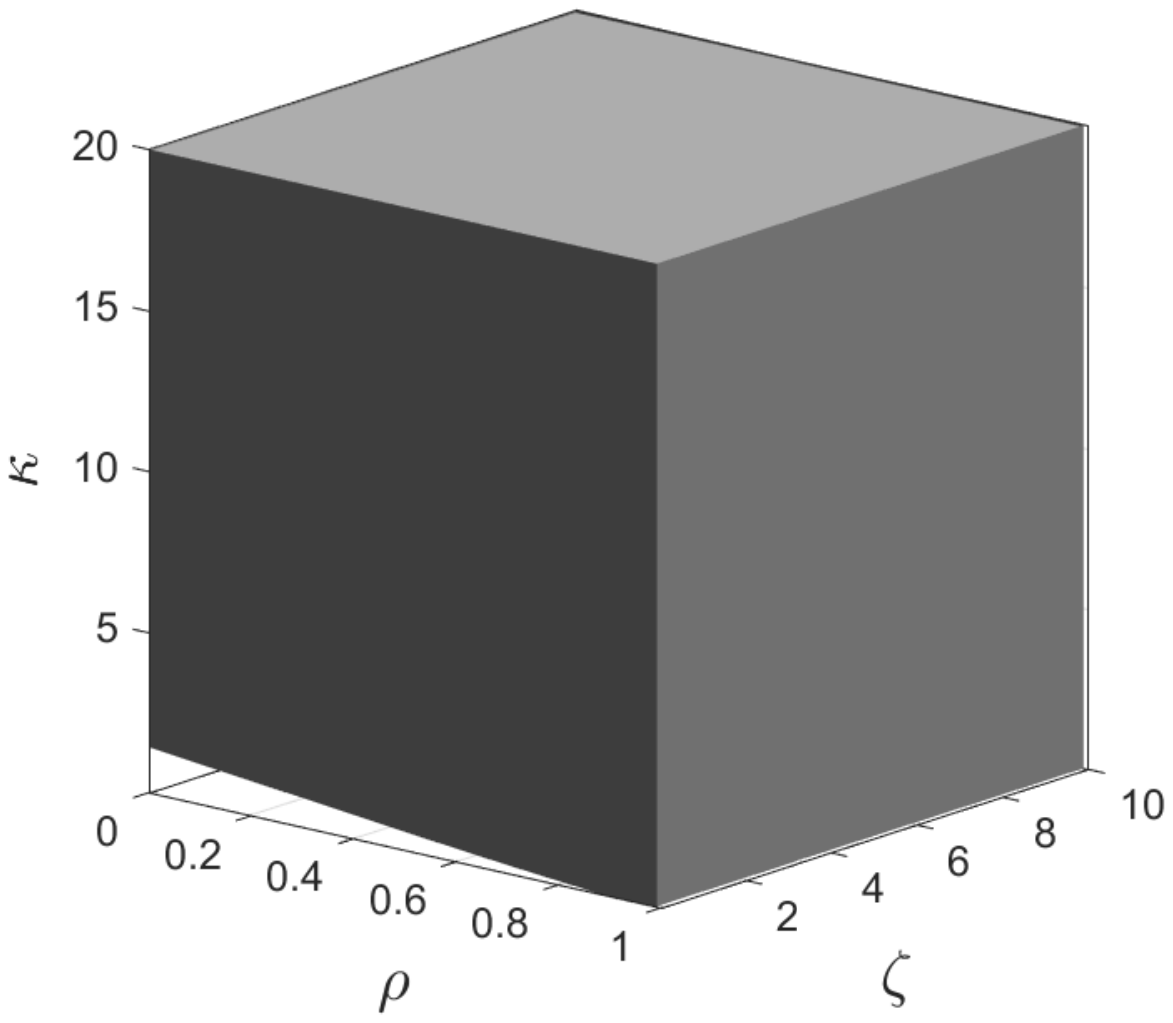}} & \hspace{+.1cm}
{\includegraphics[angle=0,width=.20\textwidth,trim=120 280 140 230 ,
totalheight=.25\textwidth]{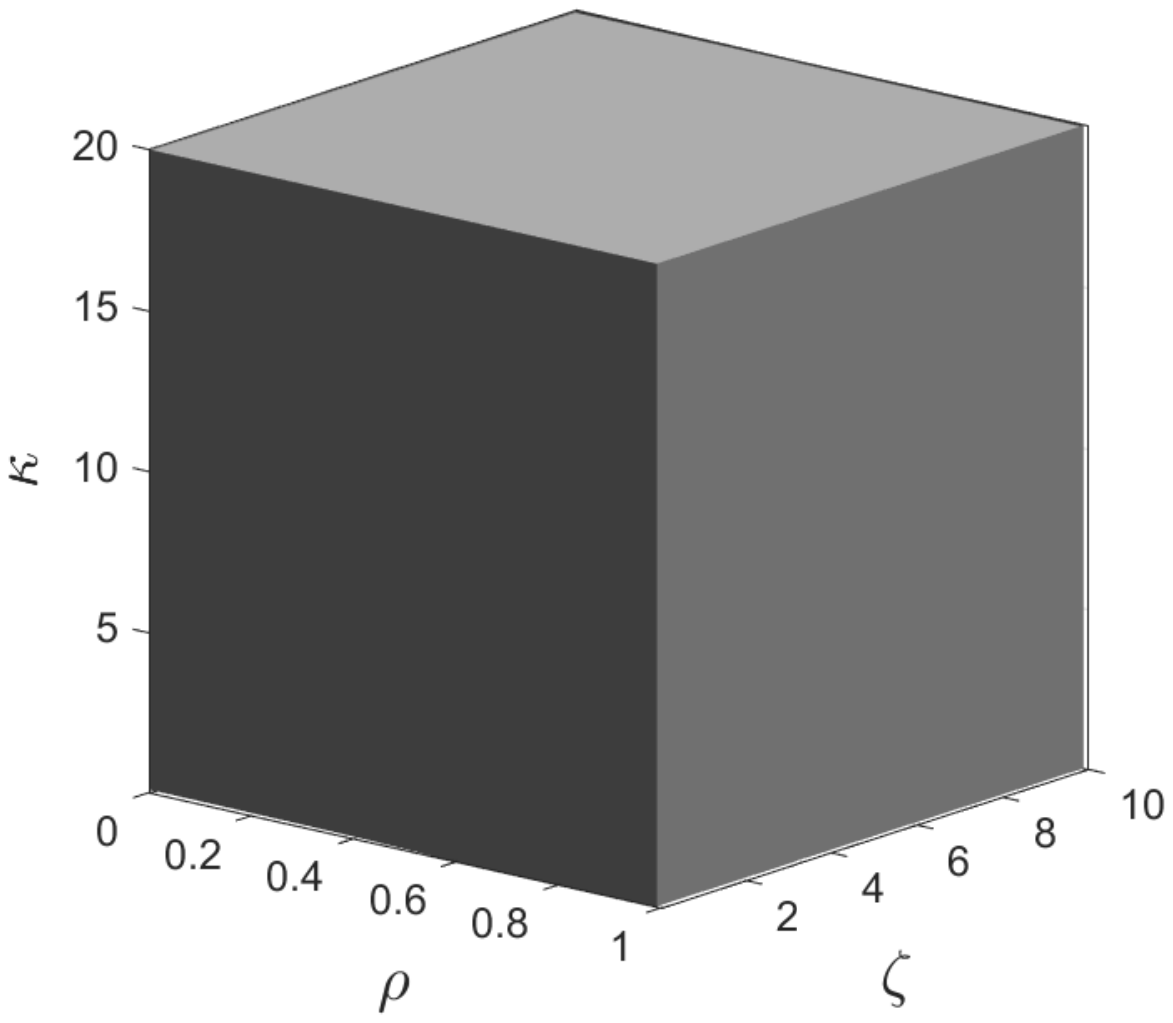}} &
\hspace{+.1cm}
{\includegraphics[angle=0,width=.20\textwidth,trim=120 280 140 230 ,
totalheight=.25\textwidth]{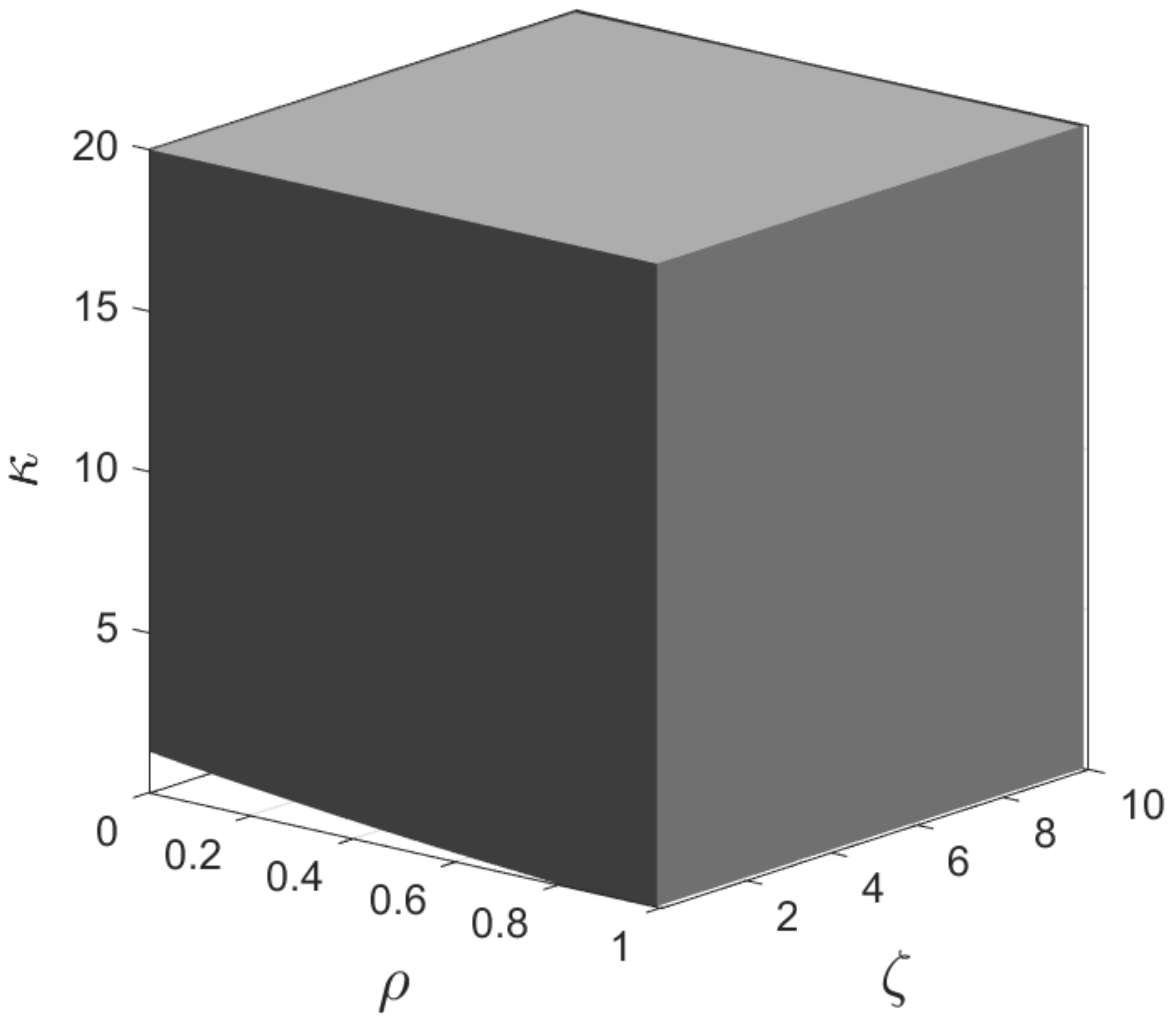}} &
\hspace{+.1cm}
{\includegraphics[angle=0,width=.20\textwidth,trim=120 280 140 230 ,
totalheight=.25\textwidth]{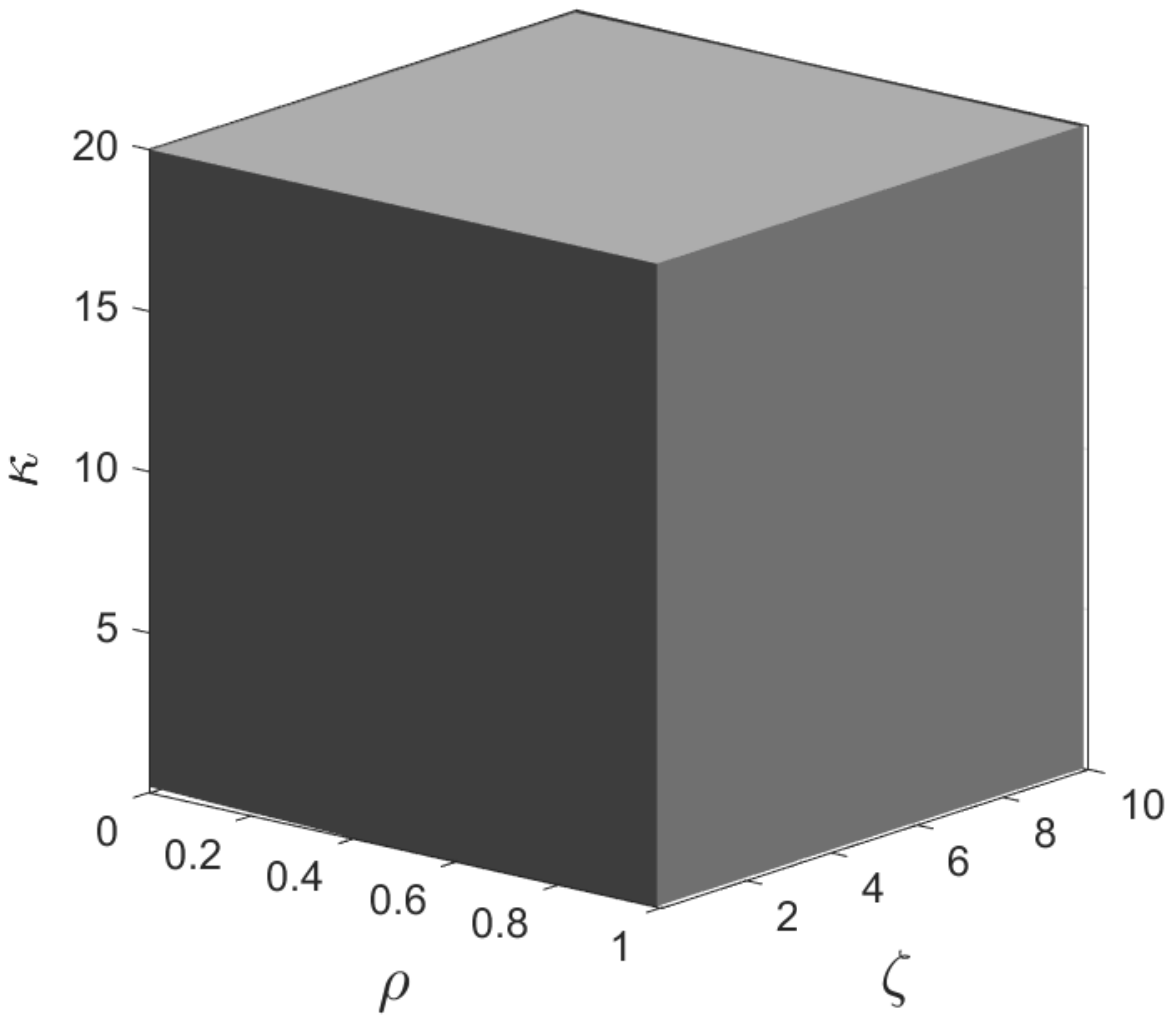}}\\\hline\hline
\end{tabular}} \caption{ 90\% S and qLL-S confidence sets for $\theta
=(\rho,\kappa,\zeta)$ derived from the investment Euler equation model
(\ref{eq: estimated}) using Fixed Private Investment as investment
proxy. A constant and $\Delta i_{t-1}$ are common instruments in all
specifications while $u_{t-1}$ enters only in specifications (a) to (e) and
(i) to (m). The additional instrument(s) by specification is (are):
\protect\underline{Baseline} $r_{t-2}^{p}$; \protect\underline{Mon. pol.
shock}: \citeauthor{romer2004new}'s \citeyearpar{romer2004new} monetary policy
shock; \protect\underline{Military news}: \citeauthor{ramey2018government}'s
\citeyearpar{ramey2018government} military news shock; \protect\underline{Oil}%
: growth rate of real oil price; \protect\underline{VXO}: financial
uncertainty. }%
\label{fig: Figure ext inst SW}%
\end{figure}

\begin{figure}[ptbh]
\centering
\adjustbox{min width=\textwidth,max width=\textwidth, max height=9.5cm}{
\begin{tabular}
[c]{ccccc}\hline\hline
& \multicolumn{4}{c}{Exogenous Instruments with JPT Investment Proxy}\\ \cline{1-5}
& {\small {Baseline}} & {\small {Mon. pol. shock}} & {\small {Military news}}
& {\small {Oil}}\\
& {\small {1967Q1-2019Q4}} & {\small {1969Q2-2007Q4}} &
{\small {1967Q1-2015Q4}} & {\small {1967Q1-2019Q4}}\\
& {\small (a)} & {\small (b)} & {\small (c)} & {\small (d)}\vspace{-.5cm}\\
\raisebox{+8.7ex}{\rotatebox[origin=lt]{90}{S sets }}\hspace{+.4cm} &
{\includegraphics[angle=0,width=.18\textwidth, trim=120 270 140 230 ,
totalheight=.25\textwidth]{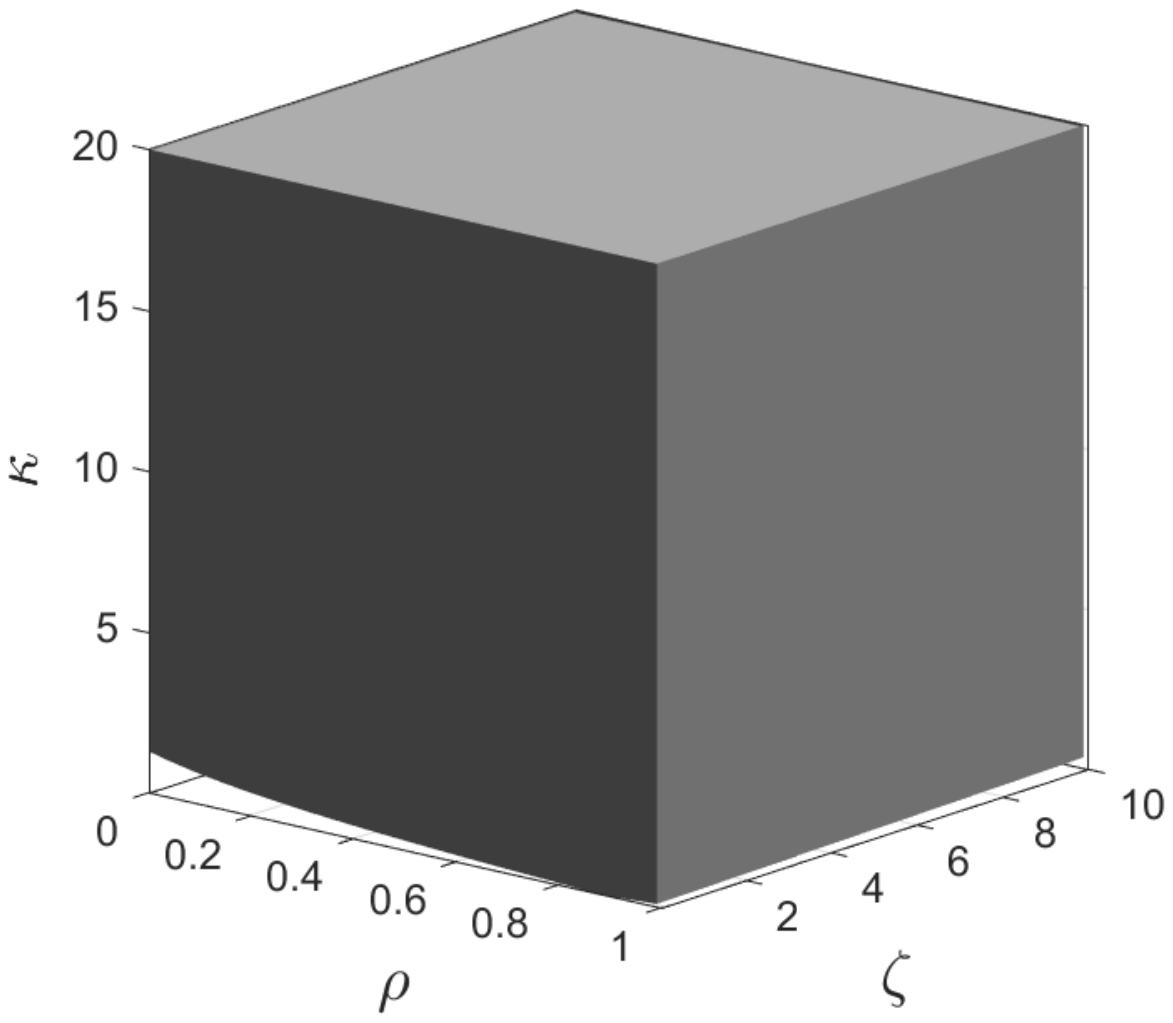}} &
\hspace{+.1cm}
{\includegraphics[angle=0,width=.20\textwidth,trim=120 280 140 230 ,
totalheight=.25\textwidth]{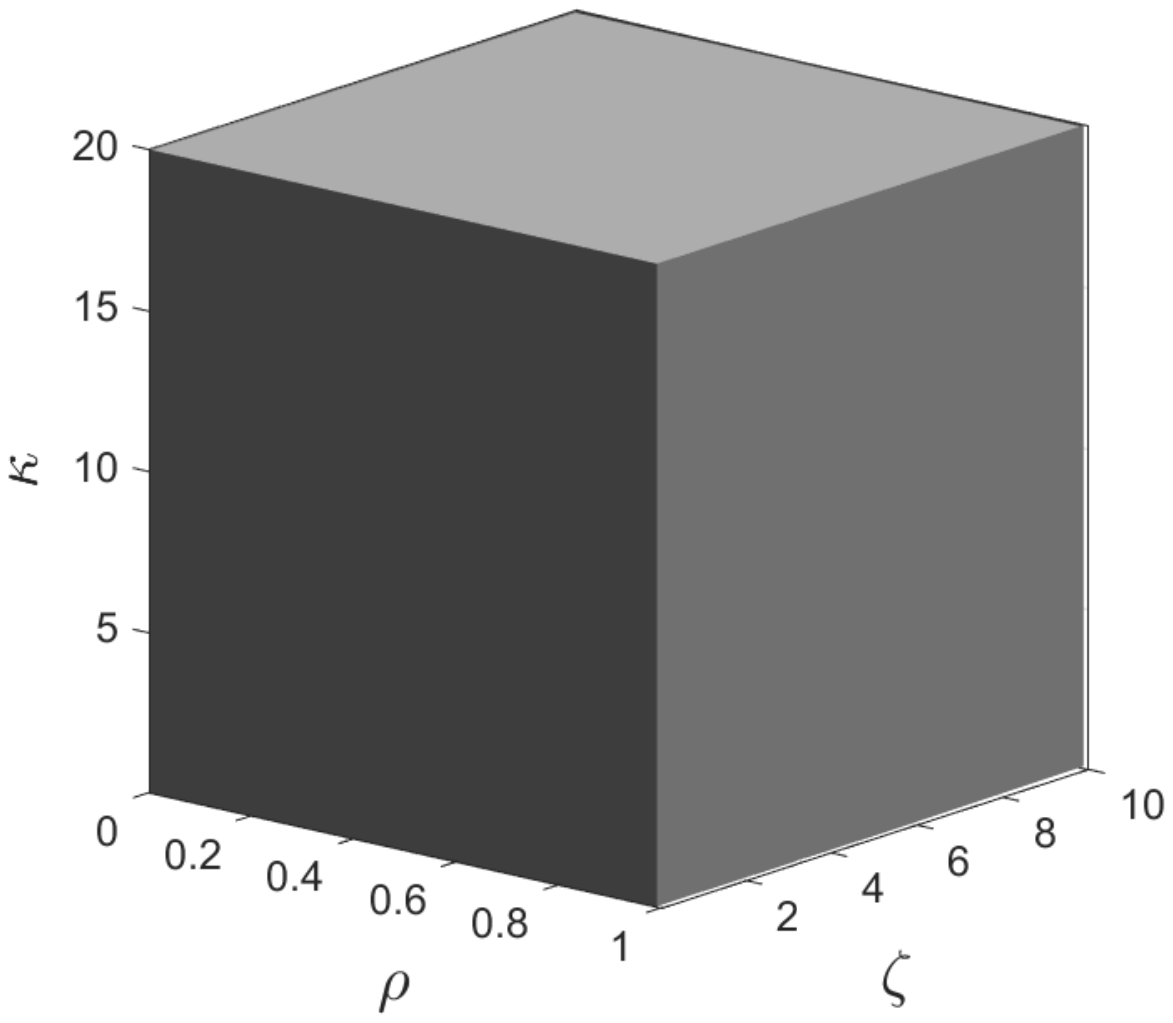}} & \hspace{+.1cm}
{\includegraphics[angle=0,width=.20\textwidth,trim=120 280 140 230 ,
totalheight=.25\textwidth]{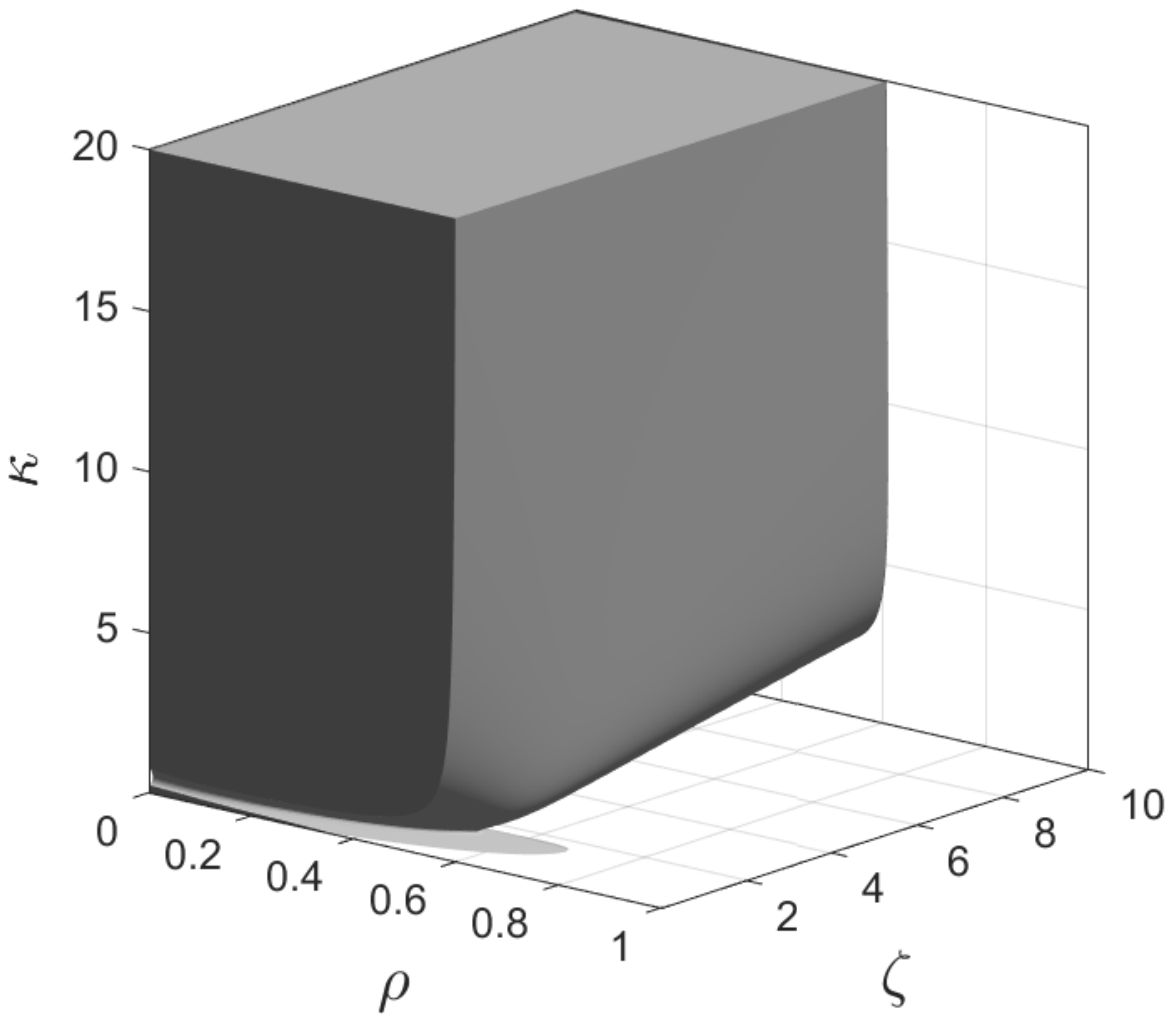}} &
\hspace{+.1cm}
{\includegraphics[angle=0,width=.20\textwidth,trim=120 280 140 230 ,
totalheight=.25\textwidth]{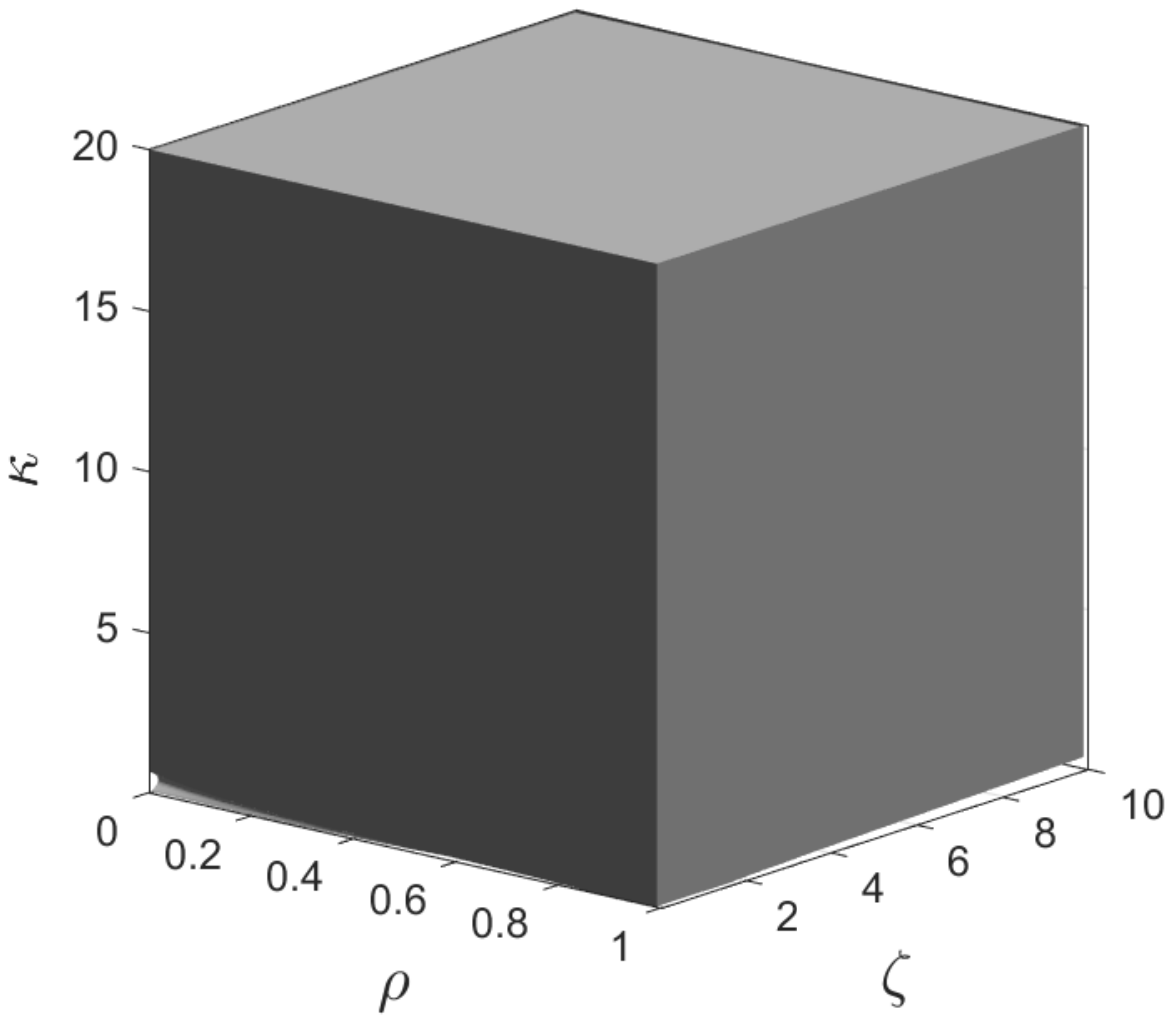}}\\[+3pt]
& {\small {VXO}} & {\small (b)+(c)} & {\small (d)+(e)} &
{\small {(b)+(c)+(d)+(e)}}\\
& {\small {1967Q1-2019Q4}} & {\small {1969Q2-2007Q4}} &
{\small {1967Q1-2019Q4}} & {\small {1969Q2-2007Q4}}\\
& {\small (e)} & {\small (f)} & {\small (g)} & {\small (h)}\vspace{-.4cm}\\
\raisebox{+8.7ex}{\rotatebox[origin=lt]{90}{S sets }}\hspace{+.4cm} &
{\includegraphics[angle=0,width=.18\textwidth, trim=120 280 140 230 ,
totalheight=.25\textwidth]{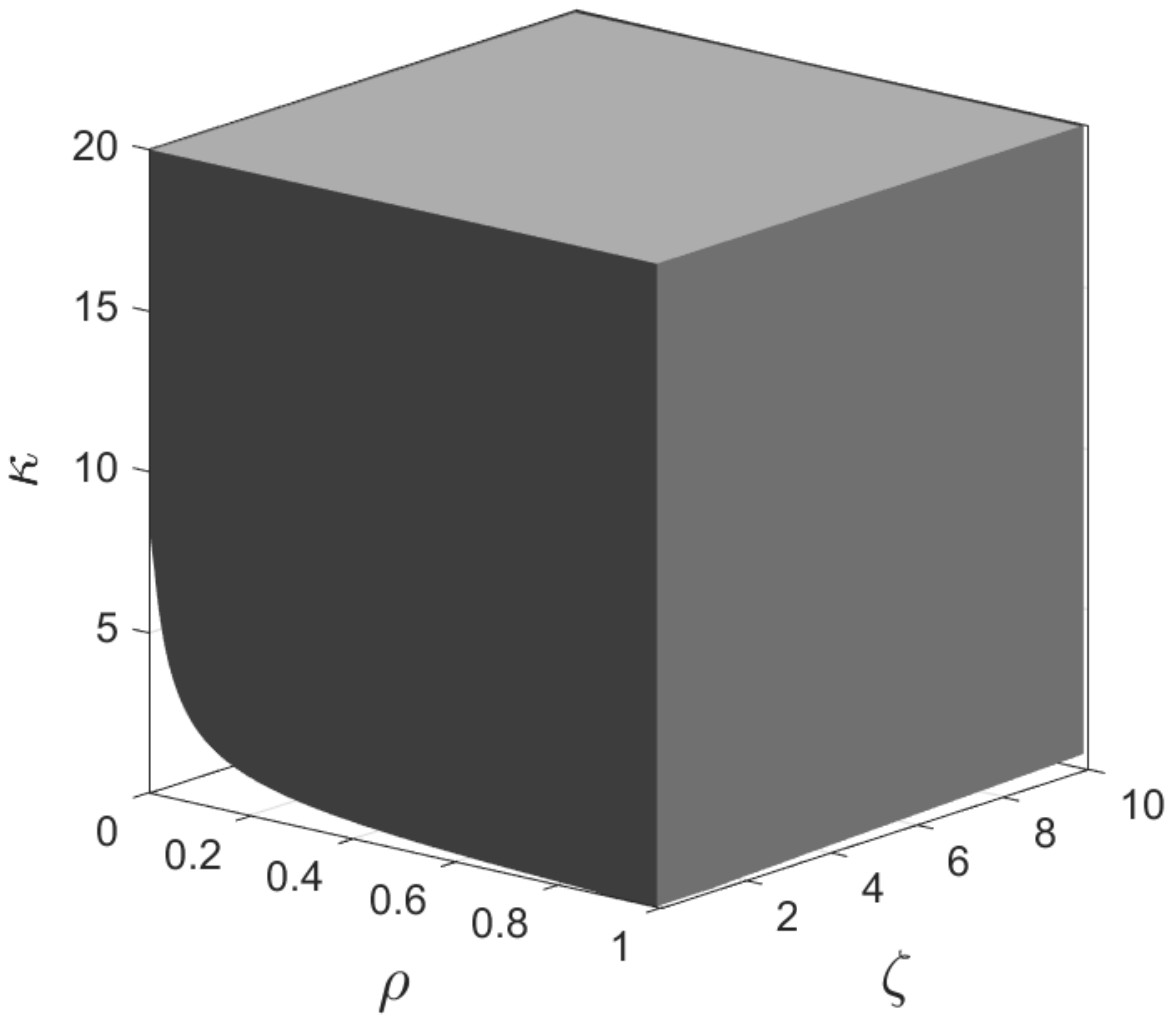}} & \hspace{+.1cm}
{\includegraphics[angle=0,width=.20\textwidth,trim=120 280 140 230 ,
totalheight=.25\textwidth]{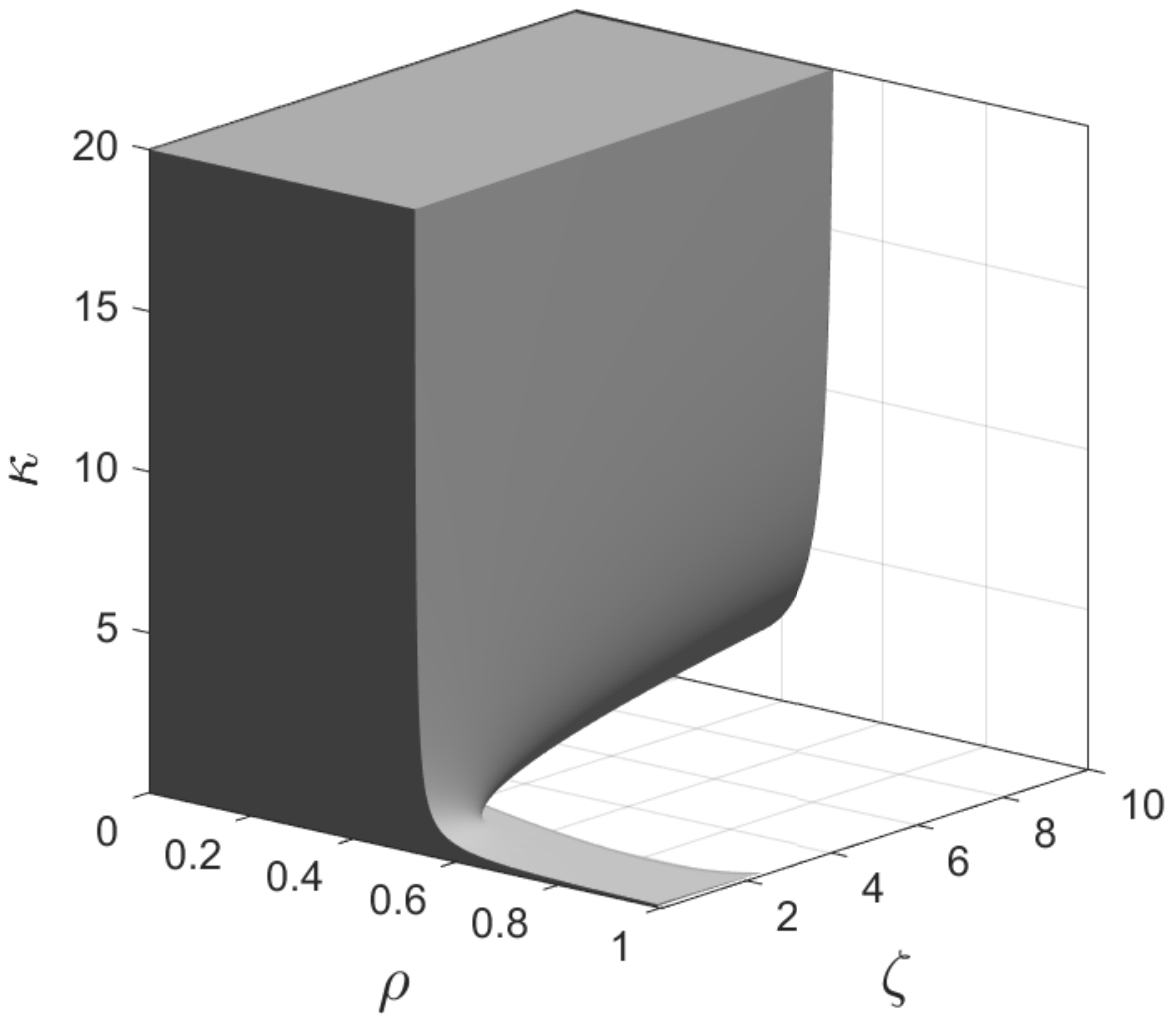}} &
\hspace{+.1cm}
{\includegraphics[angle=0,width=.20\textwidth,trim=120 280 140 230 ,
totalheight=.25\textwidth]{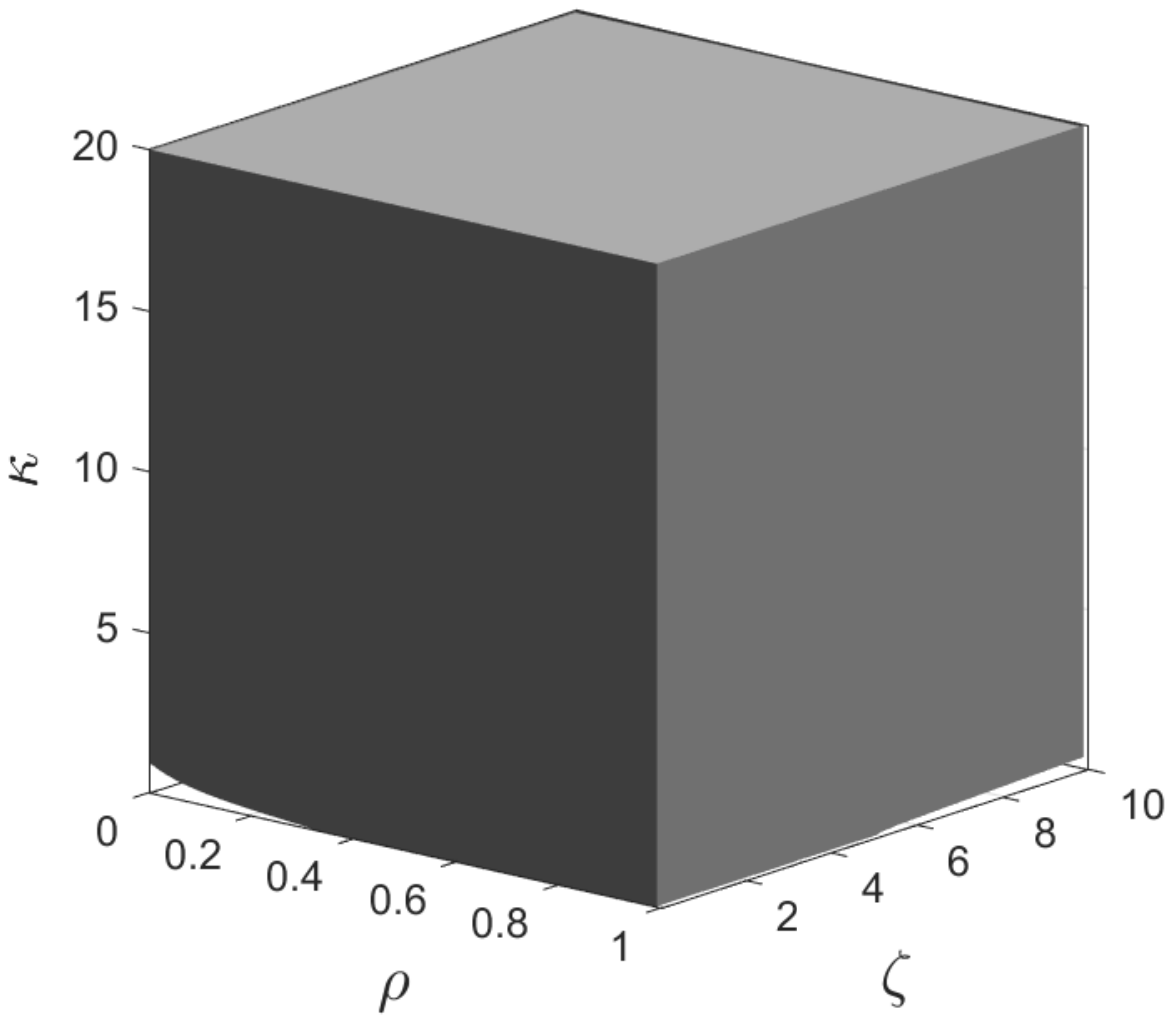}} & \hspace{+.1cm}
{\includegraphics[angle=0,width=.20\textwidth,trim=120 280 140 230 ,
totalheight=.25\textwidth]{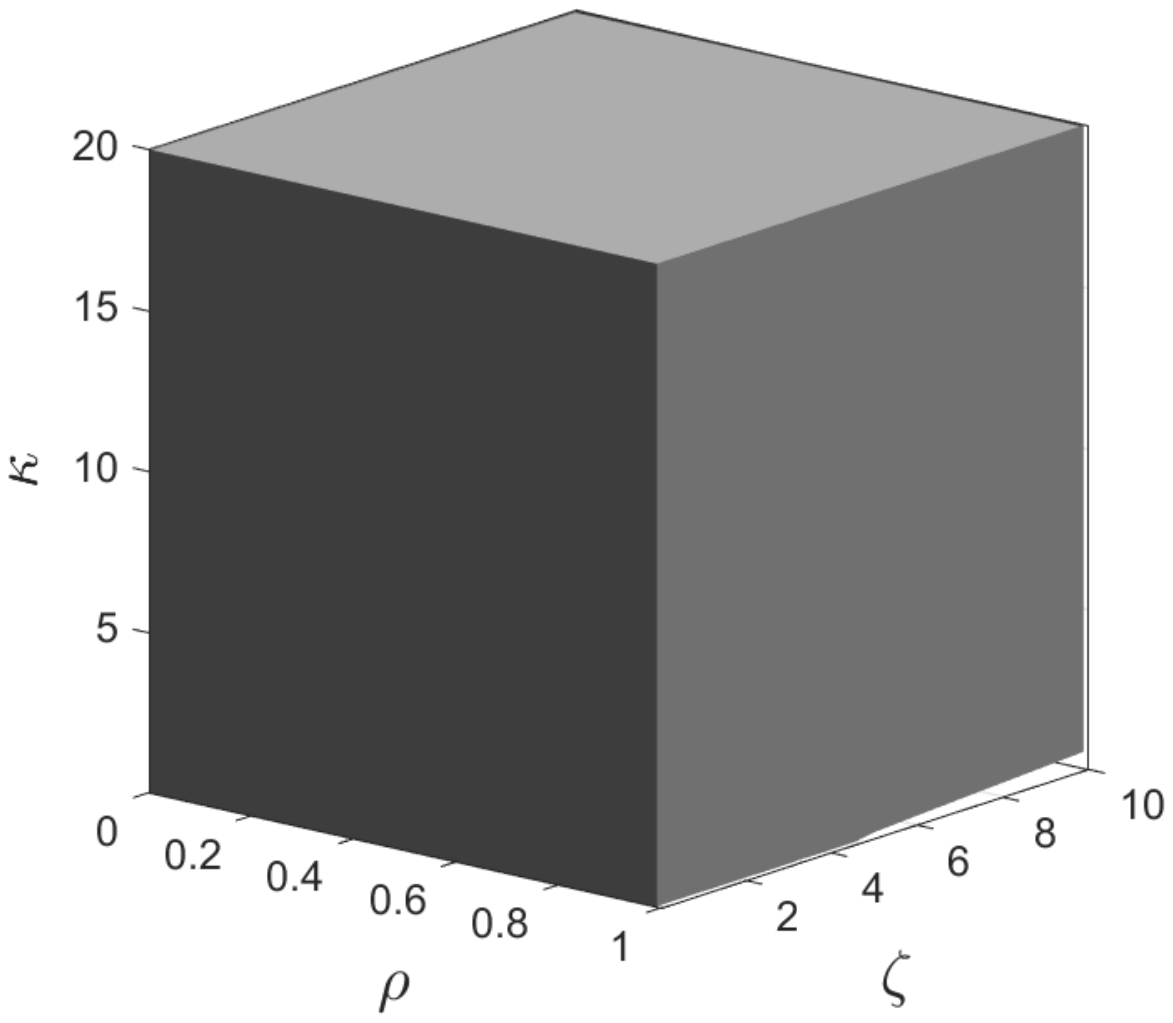}}\vspace{+10pt}\\
&  &  &  & \\
& {\small {Baseline}} & {\small {Mon. pol. shock}} & {\small {Military news}}
& {\small {Oil}}\\
& {\small {1967Q1-2019Q4}} & {\small {1969Q2-2007Q4}} &
{\small {1967Q1-2015Q4}} & {\small {1967Q1-2019Q4}}\\
& {\small (i)} & {\small (j)} & {\small (k)} & {\small (l)}\vspace{-.4cm}\\
\raisebox{+5.7ex}{\rotatebox[origin=lt]{90}{qLL-S sets }}\hspace{+.4cm} &
{\includegraphics[angle=0,width=.18\textwidth, trim=120 270 140 230 ,
totalheight=.25\textwidth]{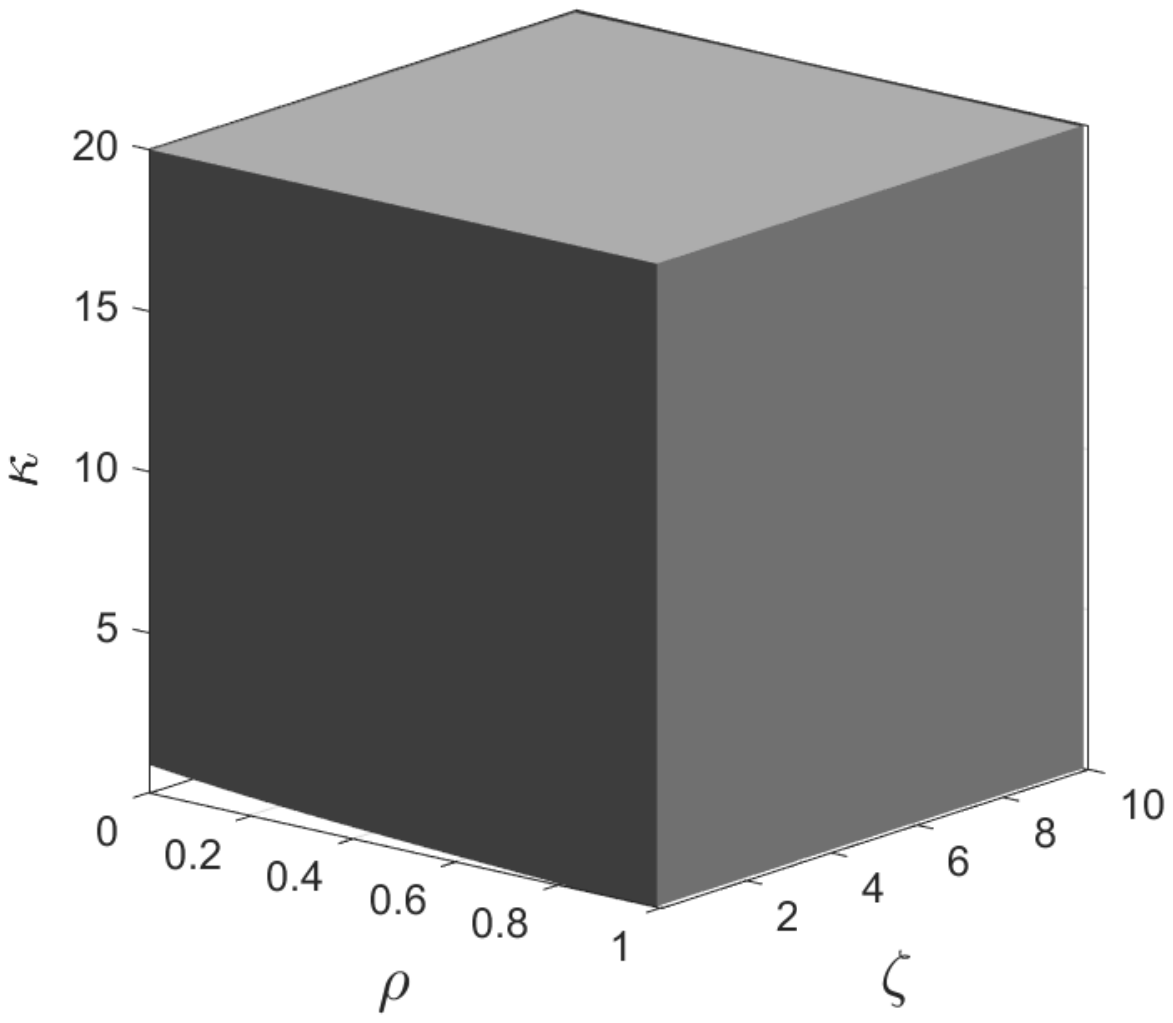}} &
\hspace{+.1cm}
{\includegraphics[angle=0,width=.20\textwidth,trim=120 280 140 230 ,
totalheight=.25\textwidth]{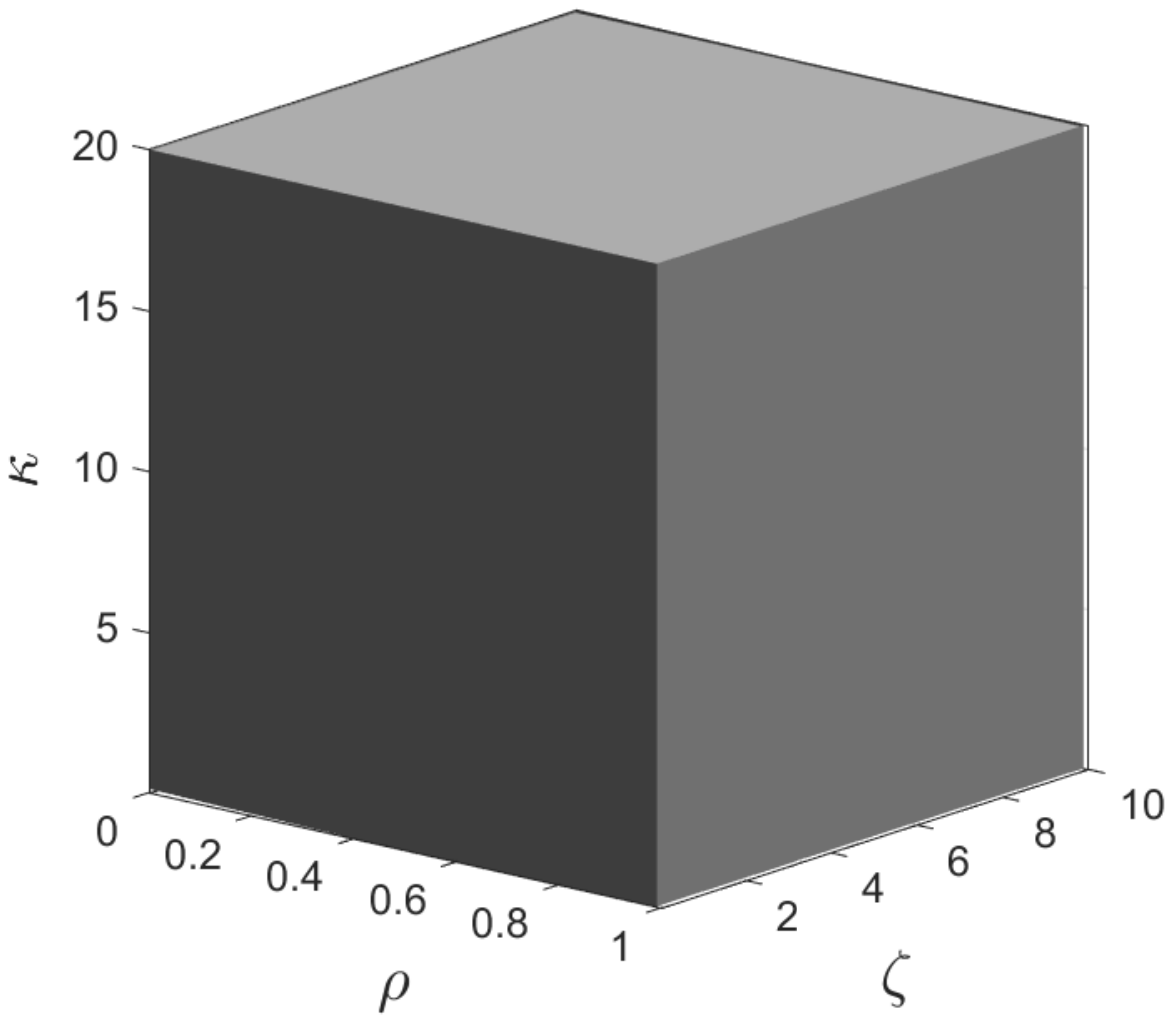}} &
\hspace{+.1cm}
{\includegraphics[angle=0,width=.20\textwidth,trim=120 280 140 230 ,
totalheight=.25\textwidth]{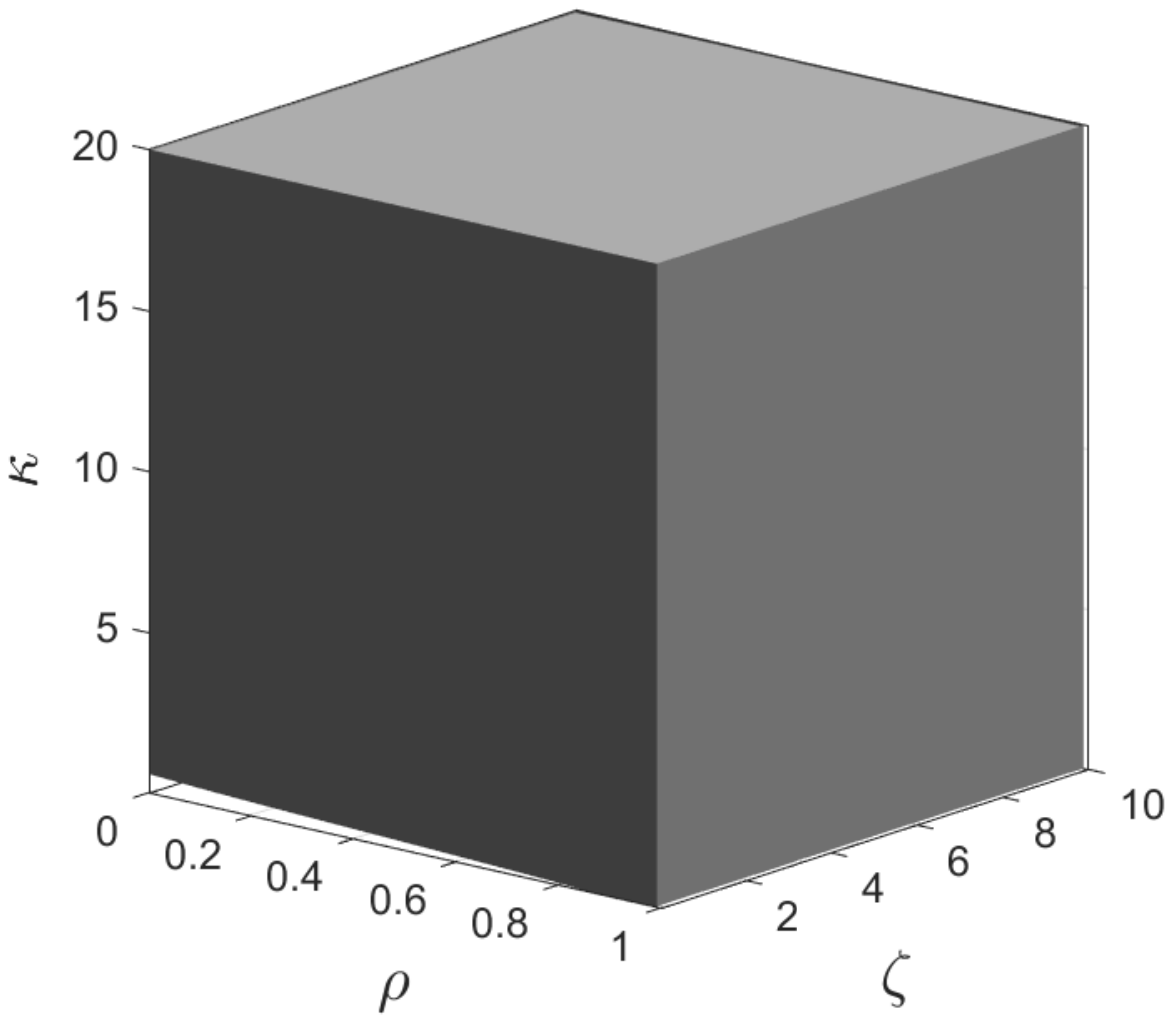}} &
\hspace{+.1cm}
{\includegraphics[angle=0,width=.20\textwidth,trim=120 280 140 230 ,
totalheight=.25\textwidth]{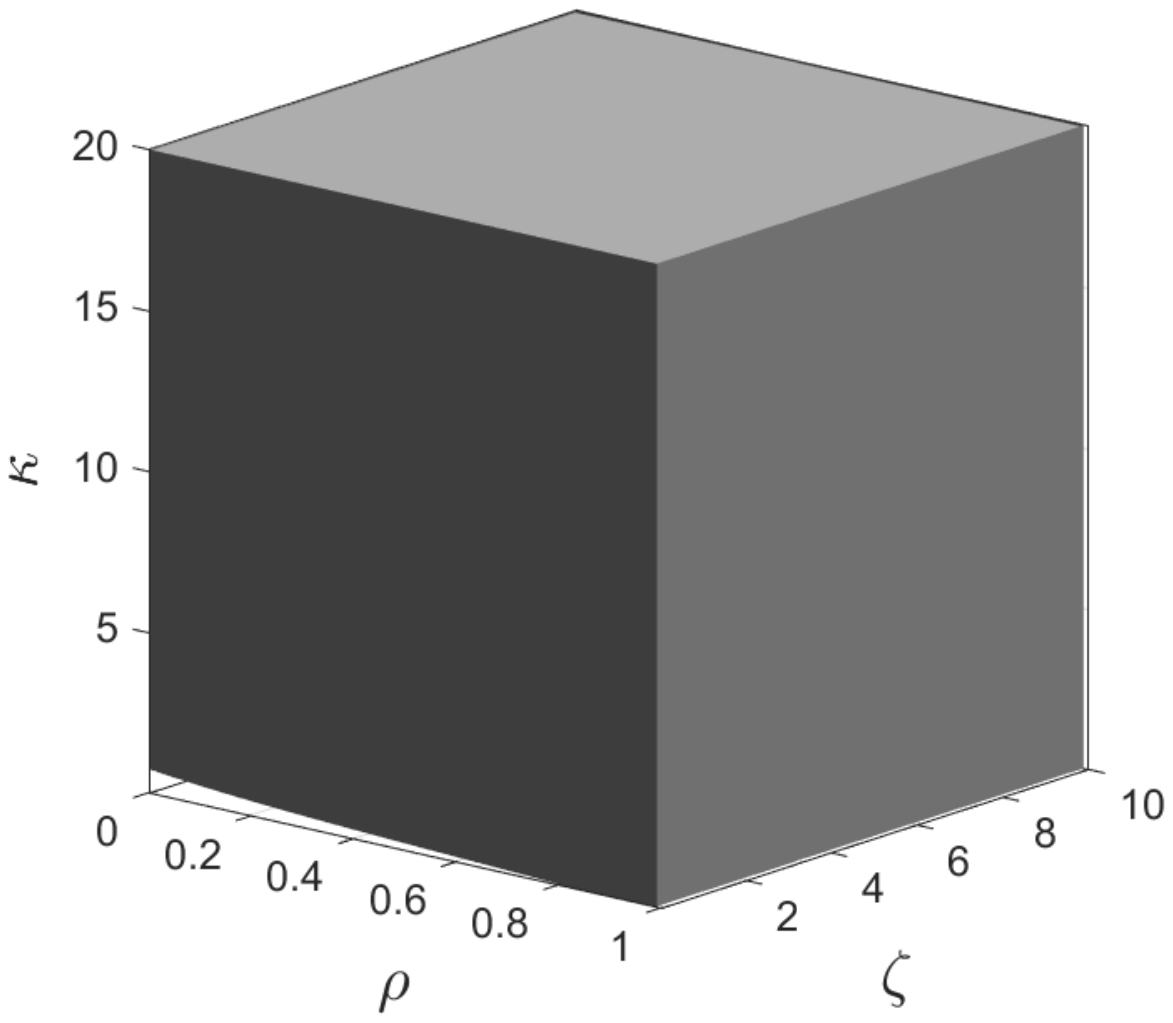}}\\[+3pt]
& {\small {VXO}} & {\small (j)+(k)} & {\small (l)+(m)} &
{\small {(j)+(k)+(l)+(m)}}\\
& {\small {1967Q1-2019Q4}} & {\small {1969Q2-2007Q4}} &
{\small {1967Q1-2019Q4}} & {\small {1969Q2-2007Q4}}\\
& {\small (m)} & {\small (n)} & {\small (o)} & {\small (p)}\vspace{-.4cm}\\
\raisebox{+5.7ex}{\rotatebox[origin=lt]{90}{qLL-S sets }}\hspace{+.4cm} &
{\includegraphics[angle=0,width=.18\textwidth, trim=120 280 140 230 ,
totalheight=.25\textwidth]{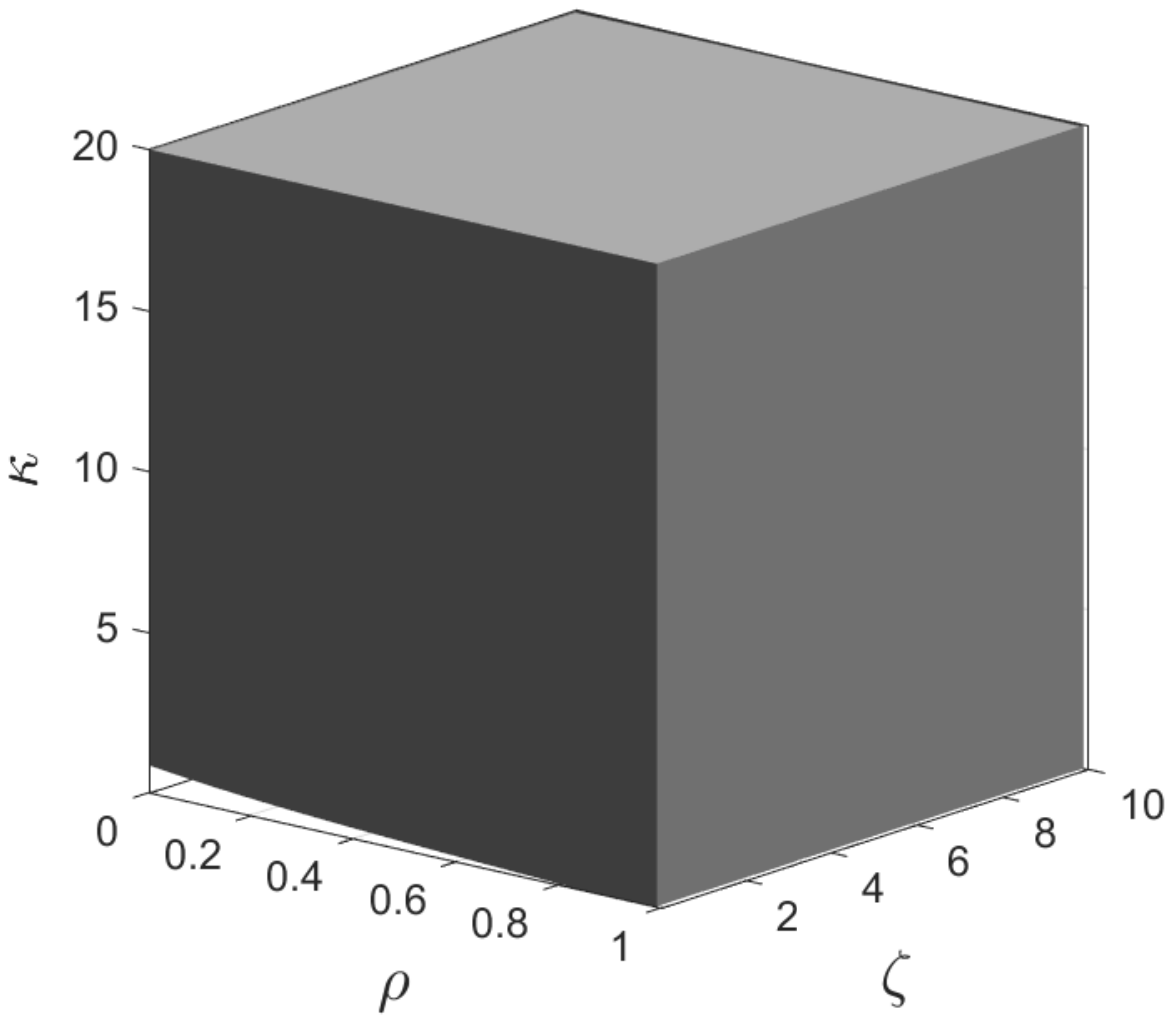}} & \hspace{+.1cm}
{\includegraphics[angle=0,width=.20\textwidth,trim=120 280 140 230 ,
totalheight=.25\textwidth]{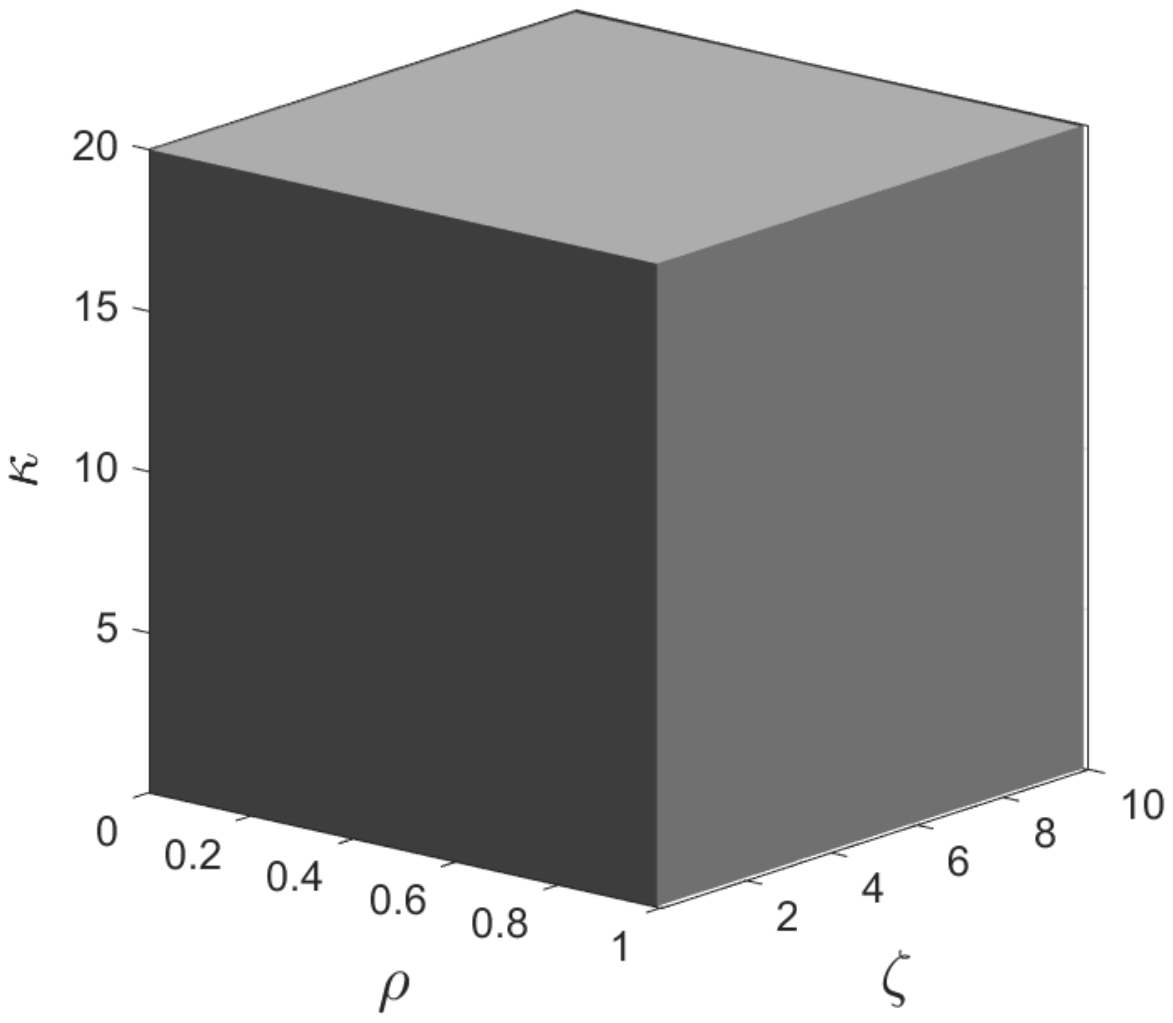}} &
\hspace{+.1cm}
{\includegraphics[angle=0,width=.20\textwidth,trim=120 280 140 230 ,
totalheight=.25\textwidth]{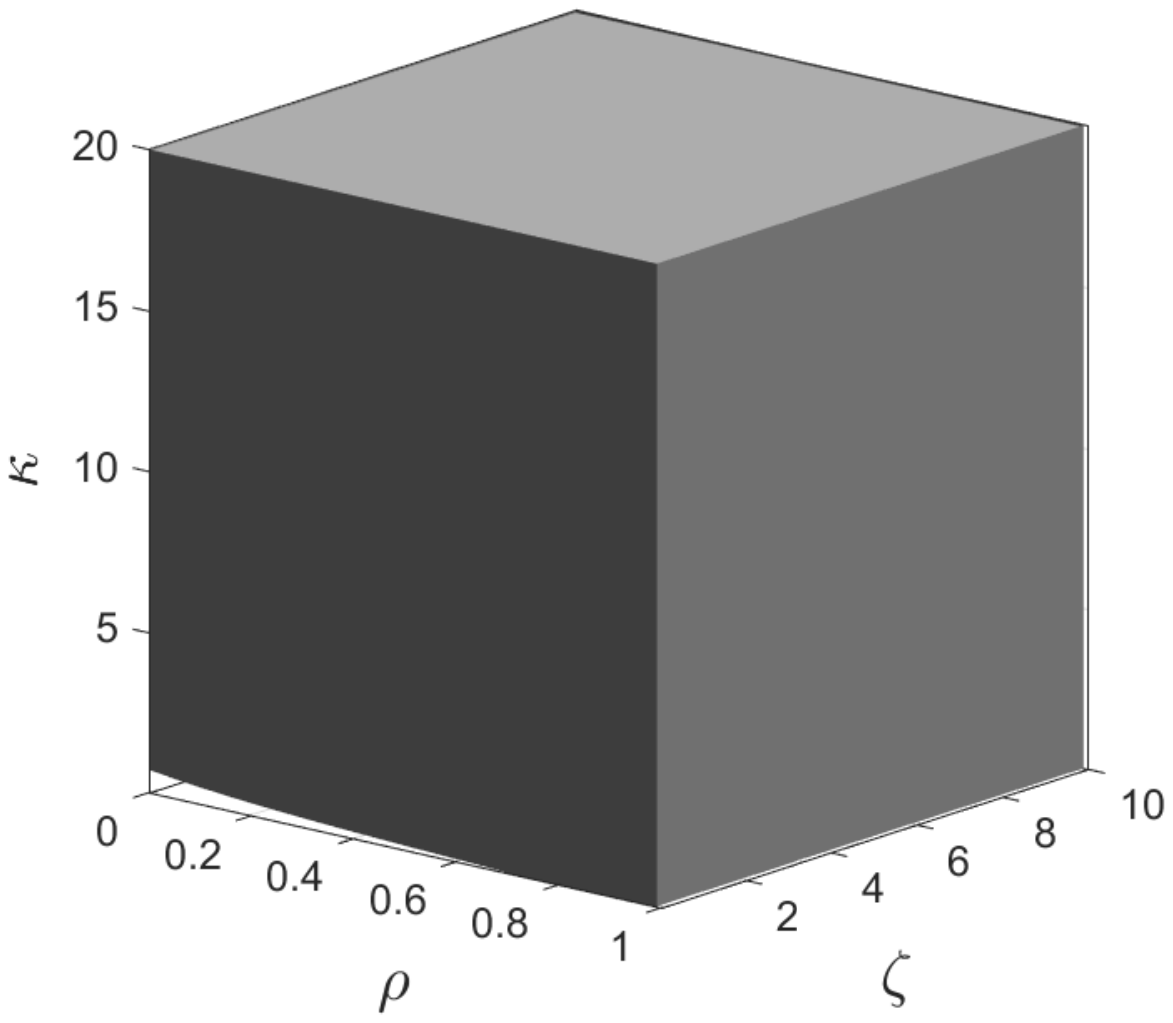}} &
\hspace{+.1cm}
{\includegraphics[angle=0,width=.20\textwidth,trim=120 280 140 230 ,
totalheight=.25\textwidth]{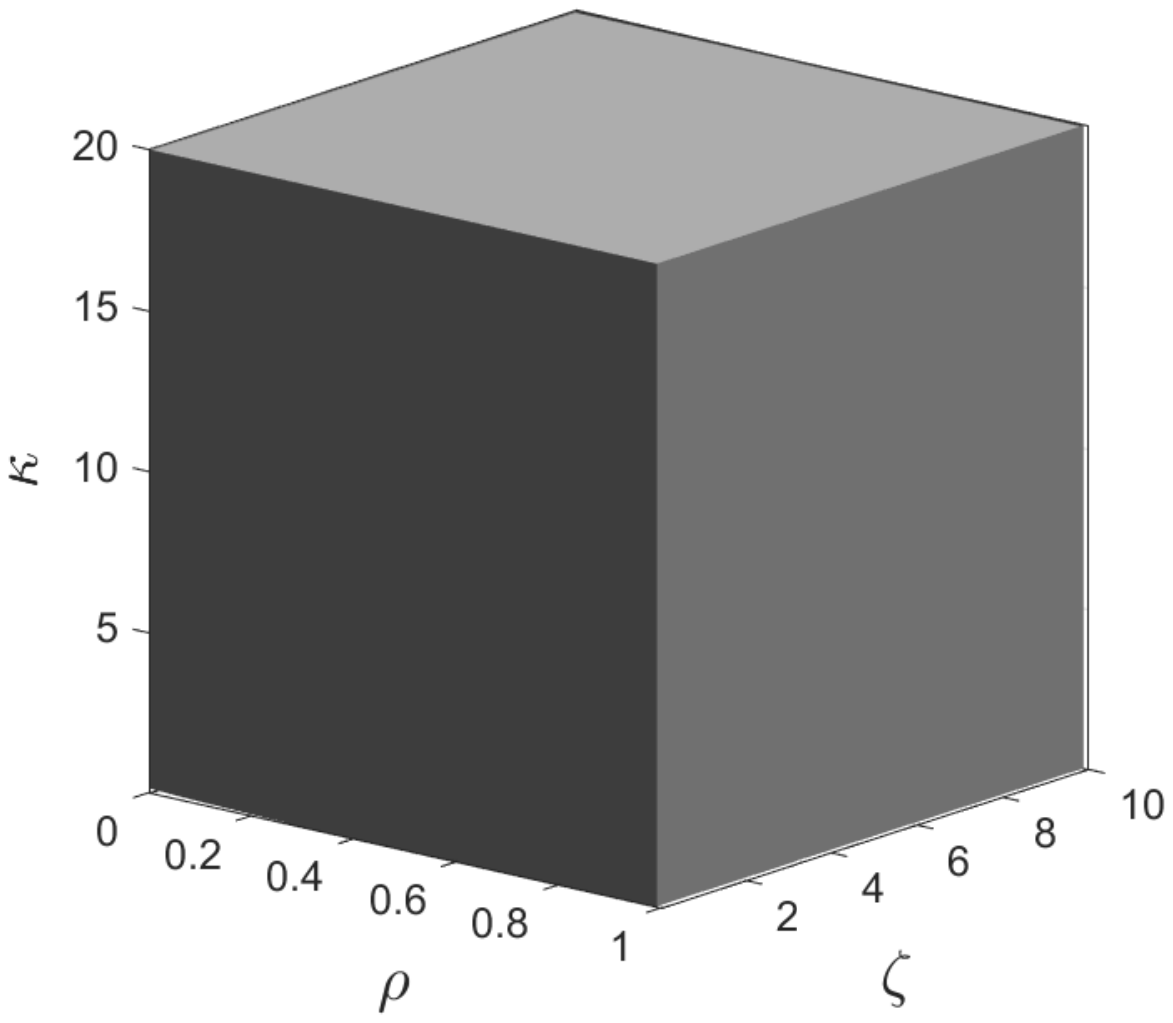}}\\\hline\hline
\end{tabular}} \caption{ 90\% S and qLL-S confidence sets for $\theta
=(\rho,\kappa,\zeta)$ derived from the investment Euler equation model
(\ref{eq: estimated}) using the sum of Gross Private
Domestic Investment and Personal Consumption Expenditure on Durable Goods as
investment proxy. A constant and $\Delta i_{t-1}$ are common instruments in
all specifications while $u_{t-1}$ enters only in specifications (a) to (e)
and (i) to (m). The additional instrument(s) by specification is (are):
\protect\underline{Baseline} $r_{t-2}^{p}$; \protect\underline{Mon. pol.
shock}: \citeauthor{romer2004new}'s \citeyearpar{romer2004new} monetary policy
shock; \protect\underline{Military news}: \citeauthor{ramey2018government}'s
\citeyearpar{ramey2018government} military news shock; \protect\underline{Oil}%
: growth rate of real oil price; \protect\underline{VXO}: financial
uncertainty. }%
\label{fig: Figure ext inst JPT}%
\end{figure}

To keep the number of instruments comparable to Figure \ref{fig: baseline}, we
report results in which $r_{t-2}^{p}$ is replaced by an external instrument,
or $r_{t-2}^{p}$ and ${u}_{t-1}$ are replaced by a pair of the external
instruments. For completeness, we also report results using all of the
external instruments together.\footnote{In Appendix \ref{appsec: further results}, we report results in which the external instruments are added together with $r_{t-2}^{p}$ and ${u}_{t-1}$ in the set of instruments. The results, which are reported in Figures \ref{appfig: SW rob inst} and \ref{appfig: JPT rob inst}, are very similar to the ones found in Figures \ref{fig: Figure ext inst SW} and \ref{fig: Figure ext inst JPT} in this section.} The resulting S and qLL-S confidence sets are
reported in Figures \ref{fig: Figure ext inst SW} and \ref{fig: Figure ext inst JPT} for the SW and JPT
investment proxies, respectively. To facilitate comparison across cases,
we report the baseline results at positions (a) and (i) in those figures. %Figures \ref{fig: Figure ext inst SW} and \ref{fig: Figure ext inst JPT}.

%%%%%%%%%%%%%%%%%%%%%%%%%%%%%%%%%%%%%%%%%%%%%%%%%%
%Figure External Instruments SW
%%%%%%%%%%%%%%%%%%%%%%%%%%%%%%%%%%%%%%%%%%%%%%%%%%

%%%%%%%%%%%%%%%%%%%%%%%%%%%%%%%%%%%%%%%%%%%%%%%%%%
%Figure External Instruments JPT
%%%%%%%%%%%%%%%%%%%%%%%%%%%%%%%%%%%%%%%%%%%%%%%%%%

In almost all cases, the results are very similar to Figure \ref{fig: baseline}
that used only lagged variables as instruments. The specifications which
include monetary policy shocks in the set of instruments for the SW investment measure, as shown in Figures
\ref{fig: Figure ext inst SW} (b) and (f), result in a slight reduction of the S confidence sets.
The reduction is somewhat more sizable for the specifications which include military news shocks
in the set of instruments when using the JPT investment proxy. In Figures \ref{fig: Figure ext inst JPT} (c)
and (f) the resulting S sets are roughly 40\% smaller
than the rest of the S sets.  Nevertheless even in this case, the
parameters $\kappa$ and $\zeta$ remain very weakly identified. The main implication of
using the military news shock is that values of $\rho>0.6$ can be rejected, which
contradicts with the findings in SW and
JPT.\footnote{The 90\% credible interval for
$\rho$ in SW and JPT is roughly between 0.60 and 0.80. The same interval is between 0.47 and 0.76 in \cite{Inoue_Kuo_Rossi_2020}.} Therefore, contemporaneous information from arguably exogenous instruments does not seem to help identify the parameters of the investment equation. Additionally, as before, there is no evidence of parameter instability or violations of the moment conditions over subsamples.

\subsection{Combining all the Instruments}\label{subsec: Mikusheva}

The methods we have used so far exploited information arising from only a handful of instruments at a time, even though equation (\ref{eq: estimated}) implies a large number of potential instruments. This was done because the S and qLL-S sets become unreliable when the number of instrument is large relative to the sample size. With the sample sizes we are dealing with here, even a dozen instruments could make results unreliable. However, use of many instruments could potentially sharpen our inference if it happens to be the case that information is spread thinly over many instruments. To study this possibility, we compute split-sample S sets, which are robust to many weak instruments. Ideally, we would like to combine all the instruments that we have used so far in a single estimation, but because of data limitations with the external instruments, we do two separate estimations instead. The first uses four lags of the three instruments $\Delta i_t, r_{t-1}^p$ and $u_t$, which are available over our full sample period. The second estimation adds to the aforementioned lagged instruments the external instruments that are available over a shorter sample. In both cases, we use approximately the first half of the sample to estimate the %optimal combination of the instruments 
first-stage regression coefficients and the second half to compute the test statistic, as explained in Appendix \ref{s: mikusheva}. 
% The results are reported in Figure \ref{fig: mikusheva}.

\begin{figure}[htbhp]
\centering
\adjustbox{min width=\textwidth,max width=\textwidth, height=.5\textwidth}{
\begin{tabular}
[c]{cccccc}\hline\hline\\[-5pt]
&  & {\Large{SW Investment}} &  & {\Large{JPT Investment}} & \\\cline{2-3}\cline{5-5}\\
&  &  1968Q1-2019Q4   &  & 1968Q1-2019Q4     & \\
&  & (a)     &  & (b)     & \vspace{-.5cm}\\
& \raisebox{+20.7ex}{\rotatebox[origin=lt]{90}{4 lags}}
&
\hspace{+.3cm}
{\includegraphics[angle=0,width=.5\textwidth,trim=120 280 140 230 ,
totalheight=.6\textwidth]{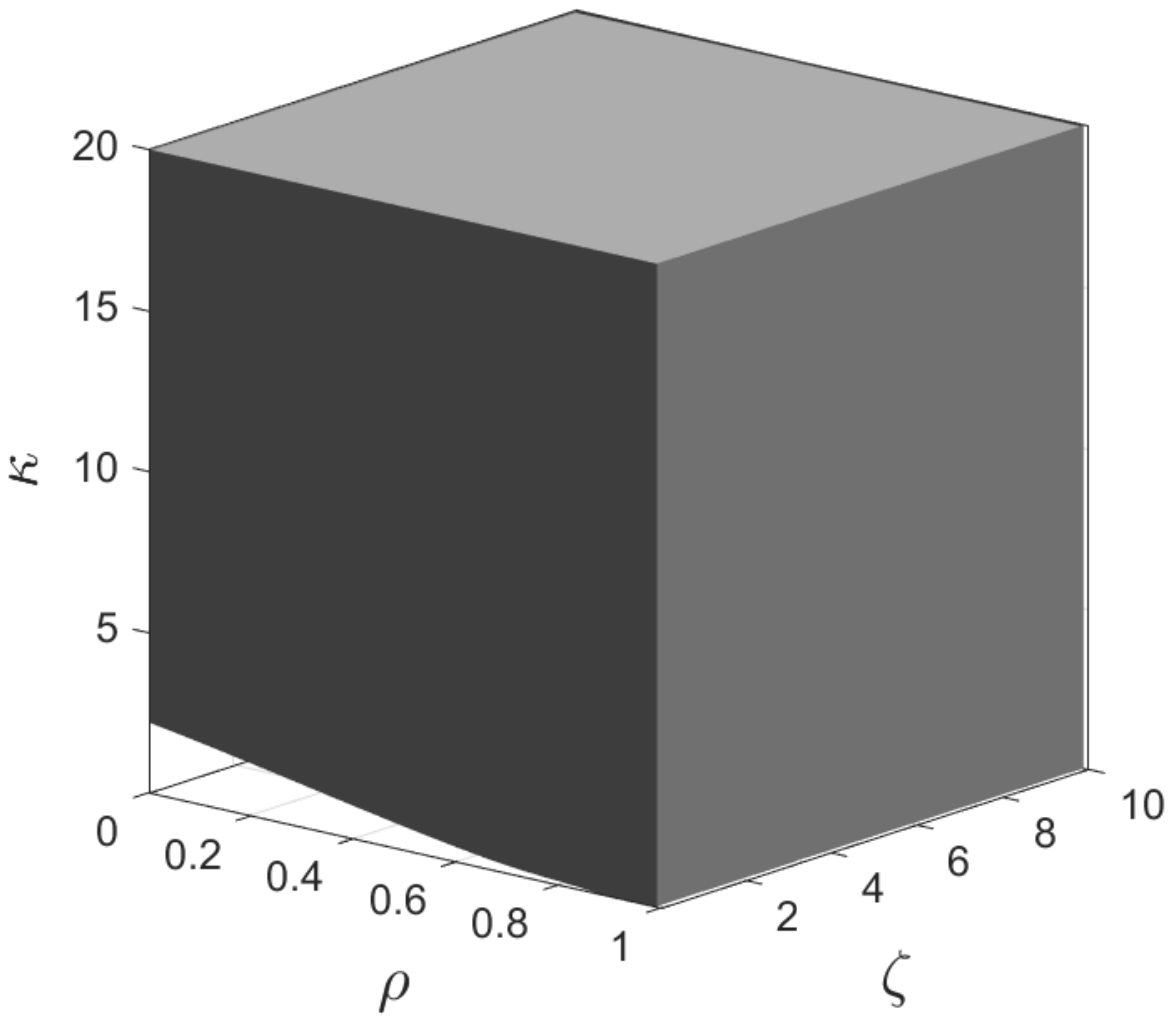}}
&
&
\hspace{+.3cm}
{\includegraphics[angle=0,width=.5\textwidth, trim=120 280 140 230 ,
totalheight=.6\textwidth]{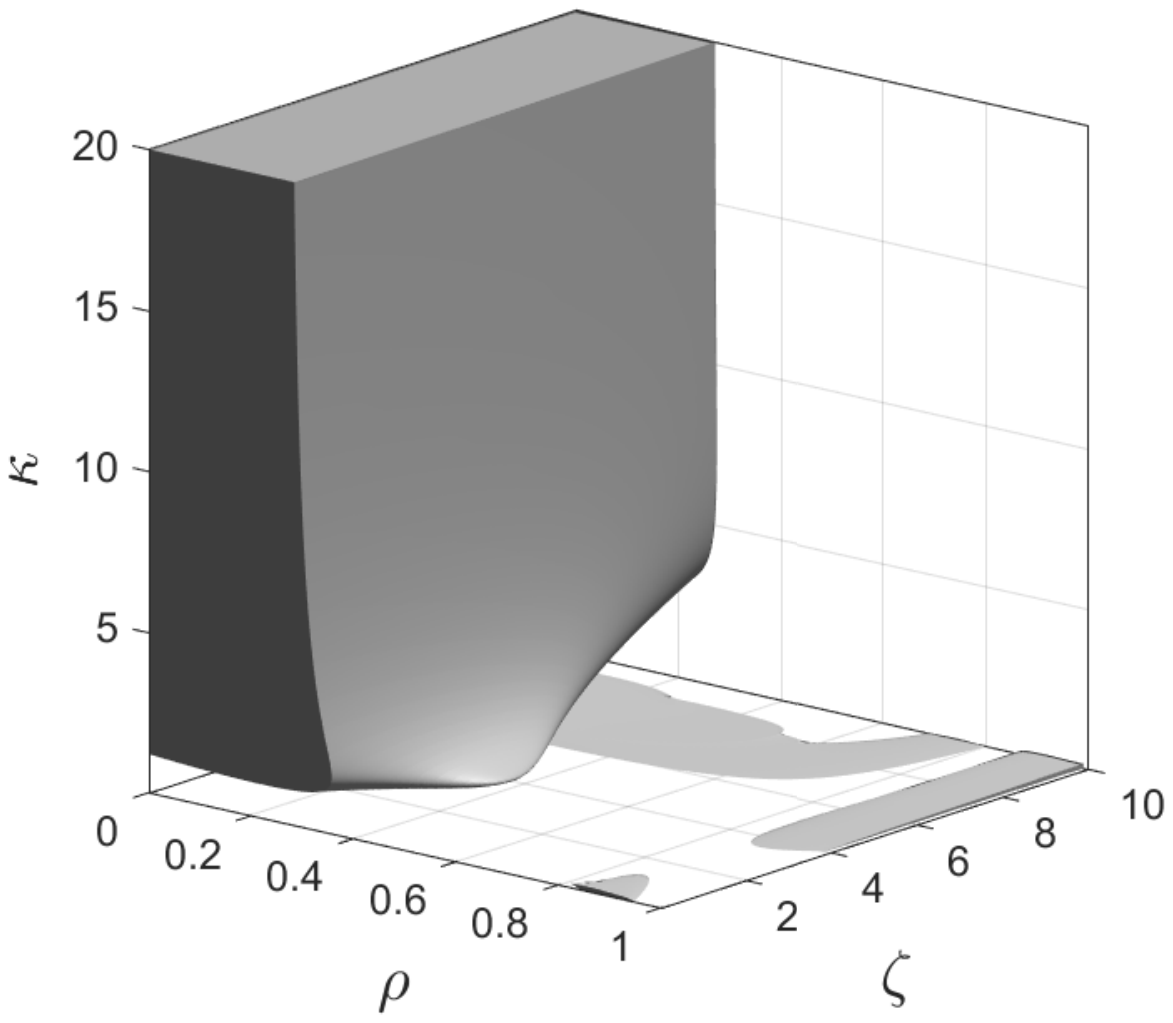}} & \\[20pt]
&  &  1970Q1-2007Q4   &  & 1970Q1-2007Q4     & \\
&  & (c)     &  & (d)     & \vspace{-.5cm}\\
& \raisebox{+14.7ex}{\rotatebox[origin=lt]{90}{4 lags + ext. instr.}}
&
\hspace{+.3cm}
{\includegraphics[angle=0,width=.5\textwidth,trim=120 280 140 230 ,
totalheight=.6\textwidth]{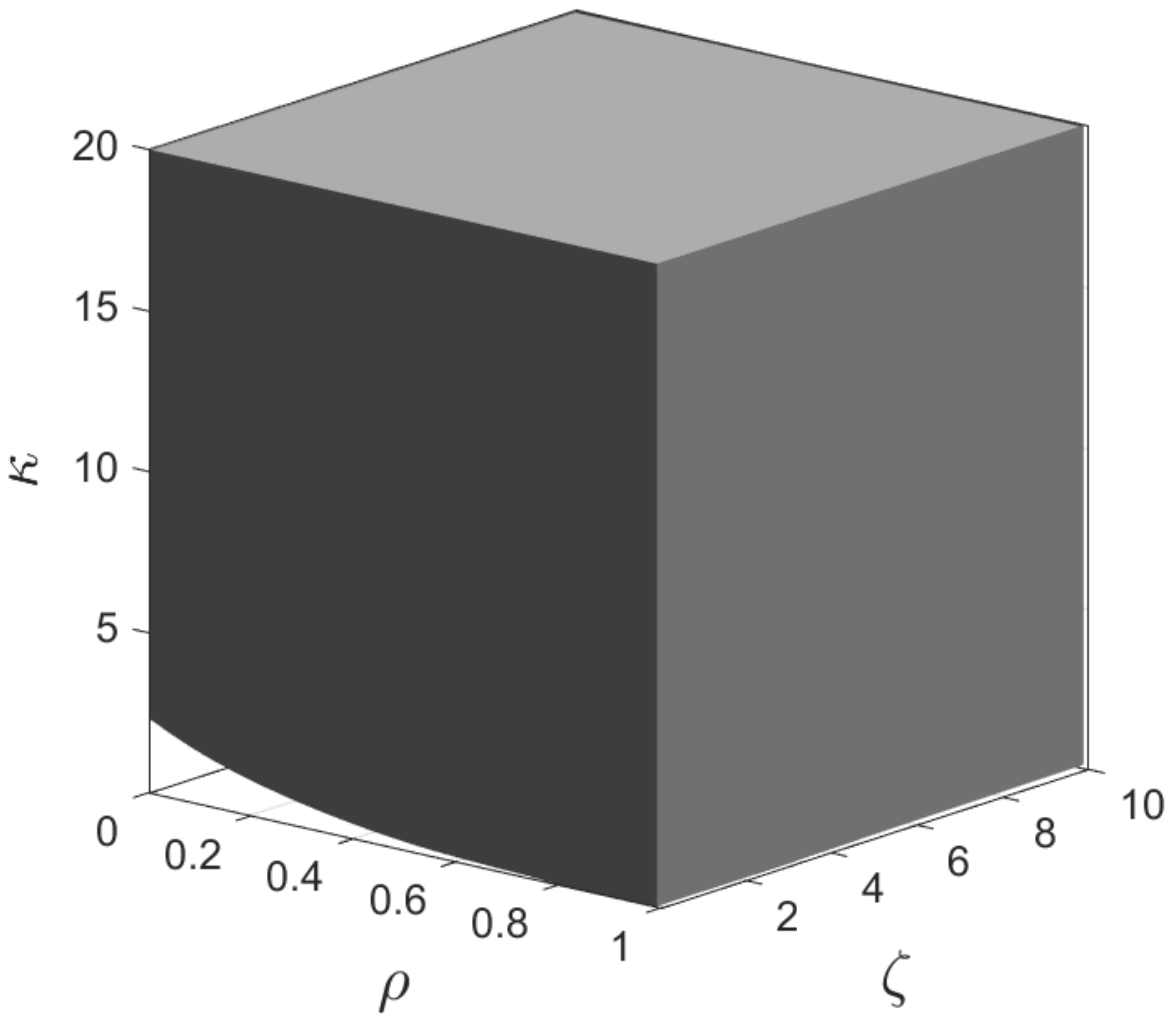}}
&
\hspace{+.5cm}
&
\hspace{+.3cm}
{\includegraphics[angle=0,width=.5\textwidth,trim=120 280 140 230, 
totalheight=.6\textwidth]
{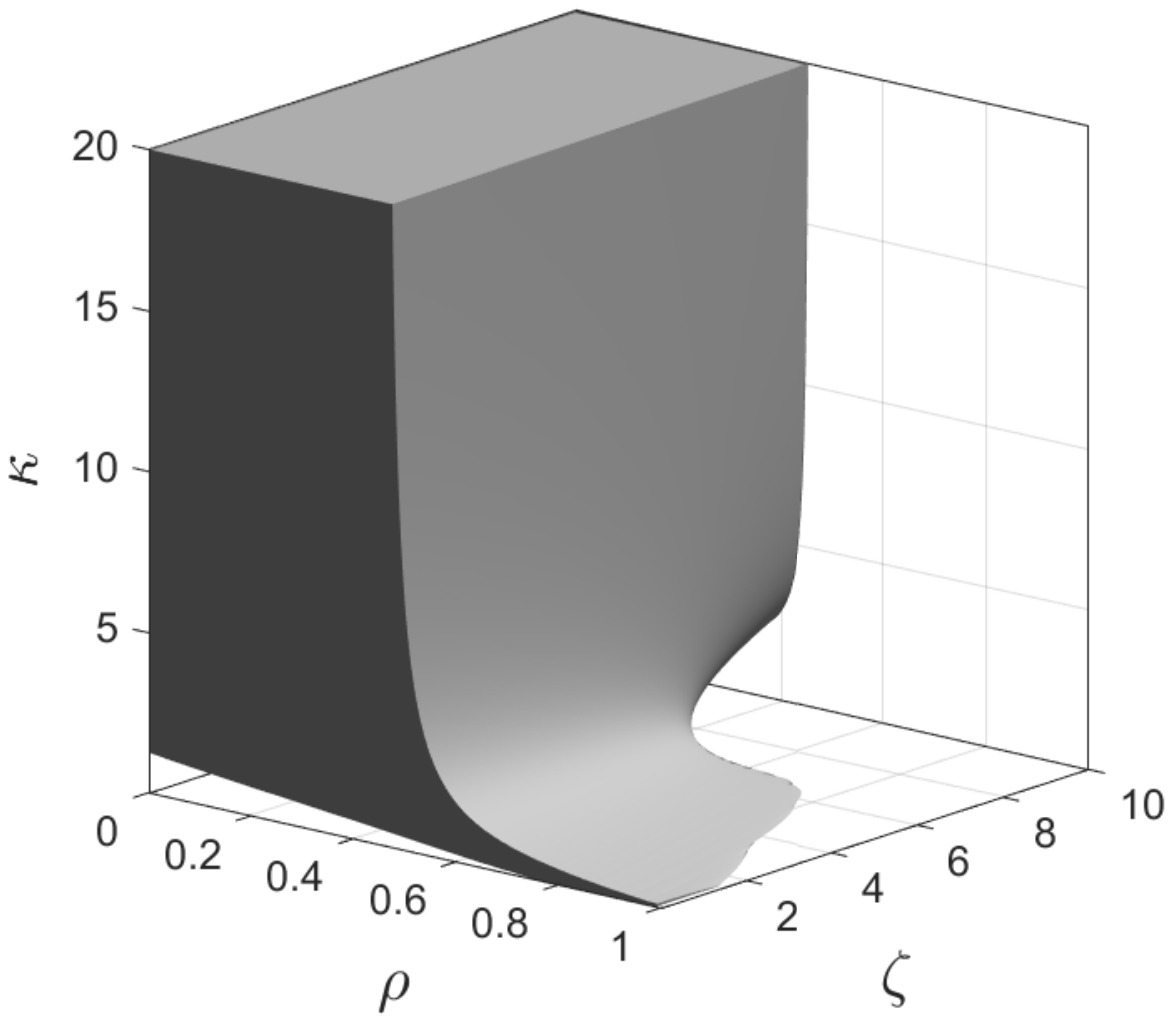}} &
\\[15pt]
\hline\hline
\end{tabular}}
\caption{90\% Mikusheva split-sample S confidence sets for $\theta
=(\rho,\kappa,\zeta)$ derived from the investment Euler equation model
(\ref{eq: estimated}). Instruments - \underline{4 lags}: $\left\{\Delta i_{t-j}, r_{t-1-j}^{p},u_{t-j} \right\}_{j=1}^{4}$.
\underline{4 lags + ext. inst.}: additional instruments are
\citeauthor{romer2004new}'s \citeyearpar{romer2004new} monetary policy
shock; \citeauthor{ramey2018government}'s
\citeyearpar{ramey2018government} military news shock, the growth rate of real oil price, and the financial uncertainty (VXO). 
A constant is included in all specifications. 
The investment proxies are Fixed Private Investment (left column) and the sum of Gross Private
Domestic Investment and Personal Consumption Expenditure on Durable Goods (right column). \cite{Newey_West_1987}
HAC.}
\label{fig: mikusheva}%
\end{figure}

Figures \ref{fig: mikusheva} (a) and (b) report the results of the split-sample S confidence sets for $(\rho,\kappa,\zeta)$ with the lagged instruments only, using the SW and JPT investement data, respectively. The results with the SW data are essentially the same as before: very weak identification of all three parameters. The results with the JPT data are somewhat more informative than before. The size of the split-sample S confidence set is less than a third of the S and qLL-S sets reported in Figures \ref{fig: baseline}(b) and (d). However, the confidence sets contain most of the values of $\kappa$ and $\zeta$, so most of the information gains affect only the parameter $\rho$: values greater than 0.5 are effectively excluded from the split-sample confidence set. This is consistent with the results reported in Figure \ref{appfig: Two lags} in Appendix \ref{appsec: further results} which compares the results using one and two lags of the instruments.
%\footnote{Appendix \ref{appsec: further results} contains other robustness checks regarding alternative combinations of instruments for the results reported in this and in the next subsection.} 
The addition of more lags shrinks the JPT confidence sets but only in terms of $\rho$, and in the same direction as in Figure \ref{fig: mikusheva}.

Figures \ref{fig: mikusheva} (c) and (d) perform the same exercise but adding the external instruments used in Figures \ref{fig: Figure ext inst SW} and \ref{fig: Figure ext inst JPT}, respectively. Note that the smaller sample size relative to Figures \ref{fig: mikusheva} (a) and (b) means that the results are not directly comparable - the combined larger instrument set can be more informative, but the smaller estimation sample makes inference less precise. Nevertheless, the pictures look very much in line with the results reported earlier. With SW investment data, the confidence set is entirely uninformative, exactly as in Figures \ref{fig: baseline} (a) and (c), and Figure \ref{fig: Figure ext inst SW} above. With JPT investment data, the split-sample S confidence set is about half the size of the sets in Figures \ref{fig: baseline} (b) and (d), with all the extra information affecting only the persistence parameter $\rho$. Interestingly, the confidence set in Figure \ref{fig: mikusheva} (d) is quite similar to the S set reported in Figure \ref{fig: Figure ext inst JPT} (c) that uses military news as a single external instrument, suggesting that military news may be the most informative of all the instruments we have considered. Still, identification of $\kappa$ and $\zeta$ remains very weak. Therefore, we conclude that the previous uninformative confidence sets were not due to the use of a limited selection of the available instruments, and that the investment equation is genuinely weakly identified.

\subsection{Semi-structural Model}\label{s: semistructural}
Finally, to better understand why the structural parameters are so poorly
identified beyond the weak instruments issues discussed above, we study a
semi-structural model. Note that we can re-write (\ref{eq: estimated}) as
\begin{align}
\left[  1+\rho\left(  \beta+\phi_{q}\right)  \right]  \Delta\widetilde{i}_{t}
-\rho\Delta\widetilde{i}_{t-1}-\left[  \beta+\phi_{q}+\rho\beta\phi
_{q}\right]  \Delta\widetilde{i}_{t+1}+\beta\phi_{q}\Delta\widetilde{i}_{t+2}\nonumber\\
=\varphi \left(\widetilde{u}_{t+1}-\rho \widetilde{u}_{t}\right)-\phi
\left(\widetilde{r}_{t}^{p}-\rho\widetilde{r}_{t-1}^{p}\right)+\epsilon_{t},
\label{eq: estimated_reduced}
\end{align}
where $\varphi=\frac{\phi_{k}\zeta}{\kappa}$ and $\phi=\frac{1}{\kappa}$ are
reduced-form parameters, for which we consider the ranges $[0,10]$ and
$[0,20]$, respectively.

\begin{figure}[ptbh]
\centering
\adjustbox{width=1.0\textwidth, max height=10cm}{
\begin{tabular}
[c]{ccccc}\hline\hline
& \multicolumn{4}{c}{Panel A: SW Investment Proxy}\\[+2pt] \cline{1-5}
& {\small {$\rho=0.0$}} & {\small {$\rho=0.6$}} & {\small {$\rho=0.8$}} & {\small {$\rho=0.9$}}\vspace{0.1cm}\\
& {\small {(a)}}&   {\small {(b)}} & {\small {(c)}} &  {\small {(d)}}\vspace{-.4cm}\\
\raisebox{+8.7ex}{\rotatebox[origin=lt]{90}{S sets }}\hspace{+.4cm}
&
{\includegraphics[angle=0,width=.23\textwidth,trim=120 280 140 230 ,
totalheight=.30\textwidth]{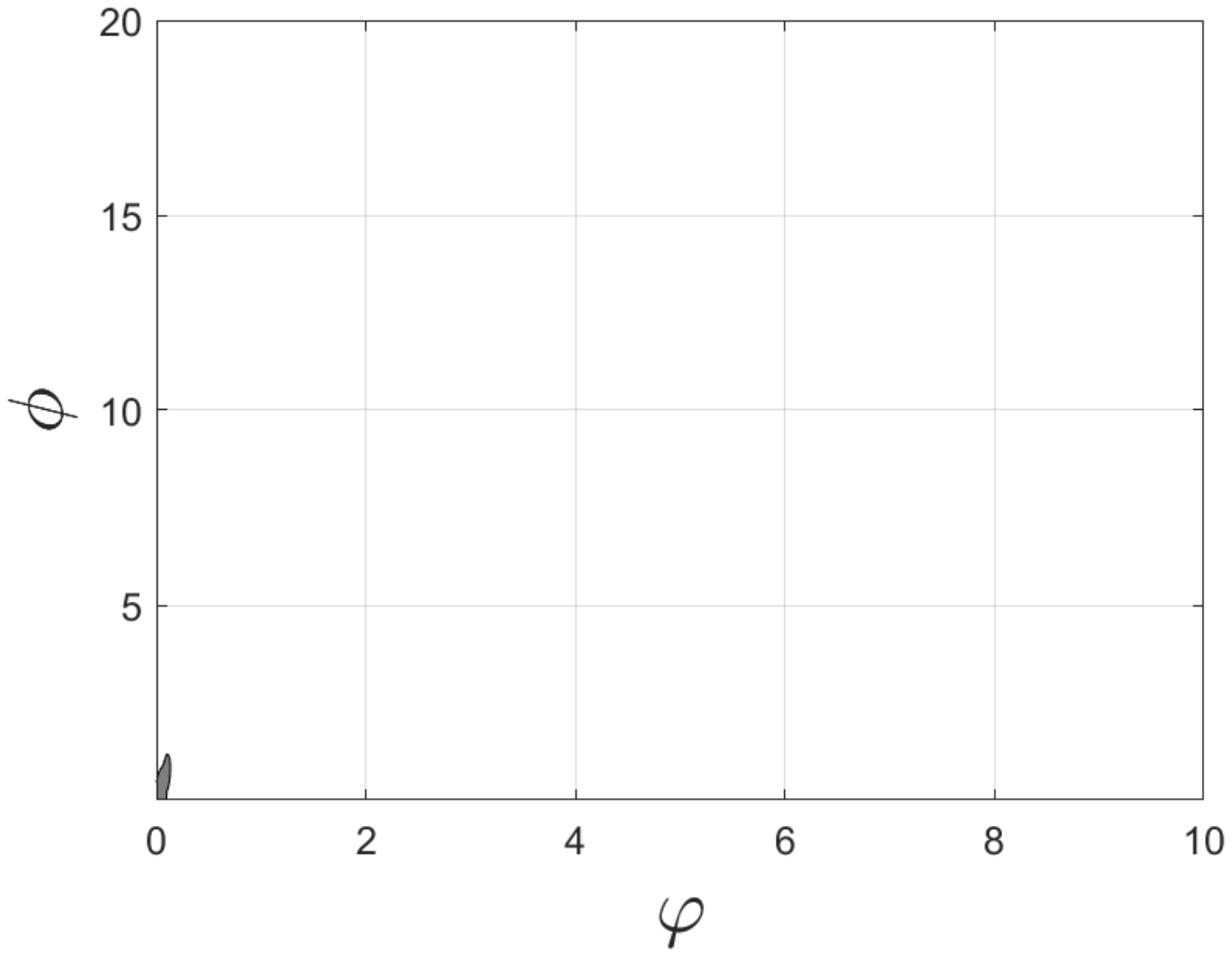}}
& \hspace{+.2cm}
{\includegraphics[angle=0,width=.23\textwidth,trim=120 280 140 230 ,
totalheight=.30\textwidth]{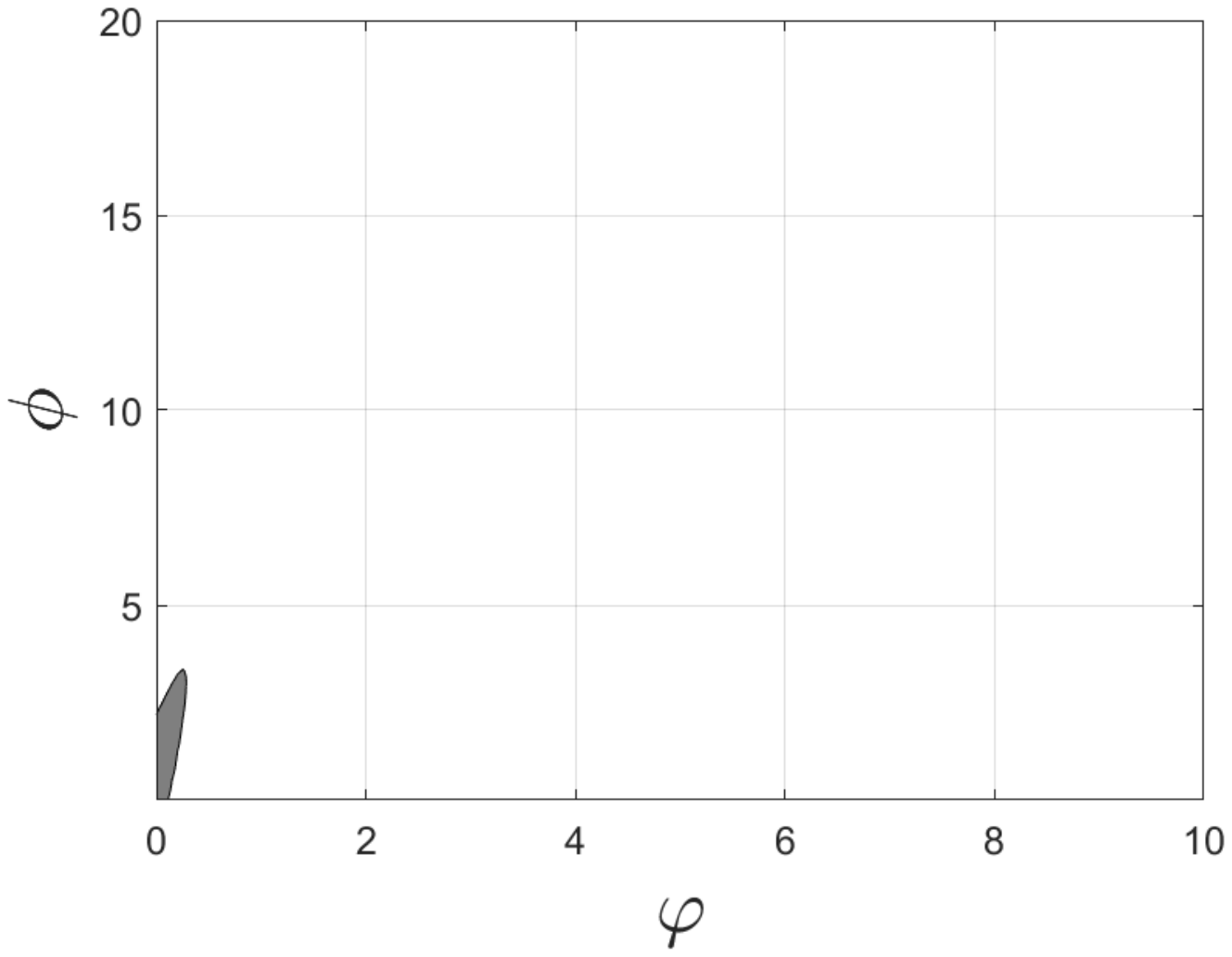}}
& \hspace{+.2cm}
{\includegraphics[angle=0,width=.23\textwidth, trim=120 280 140 230 ,
totalheight=.30\textwidth]{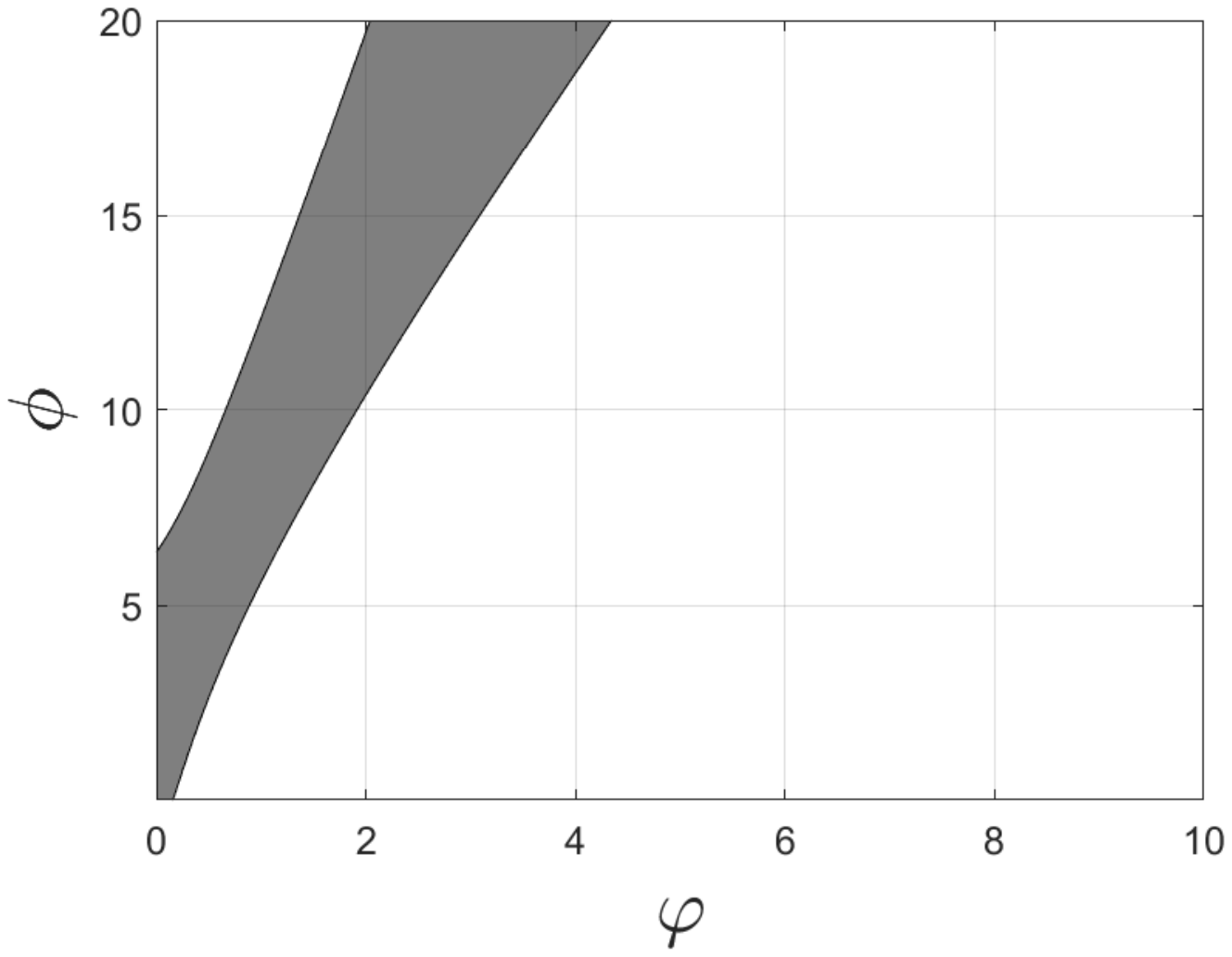}}
&\hspace{+.2cm}
{\includegraphics[angle=0,width=.23\textwidth, trim=120 280 140 230 ,
totalheight=.30\textwidth]{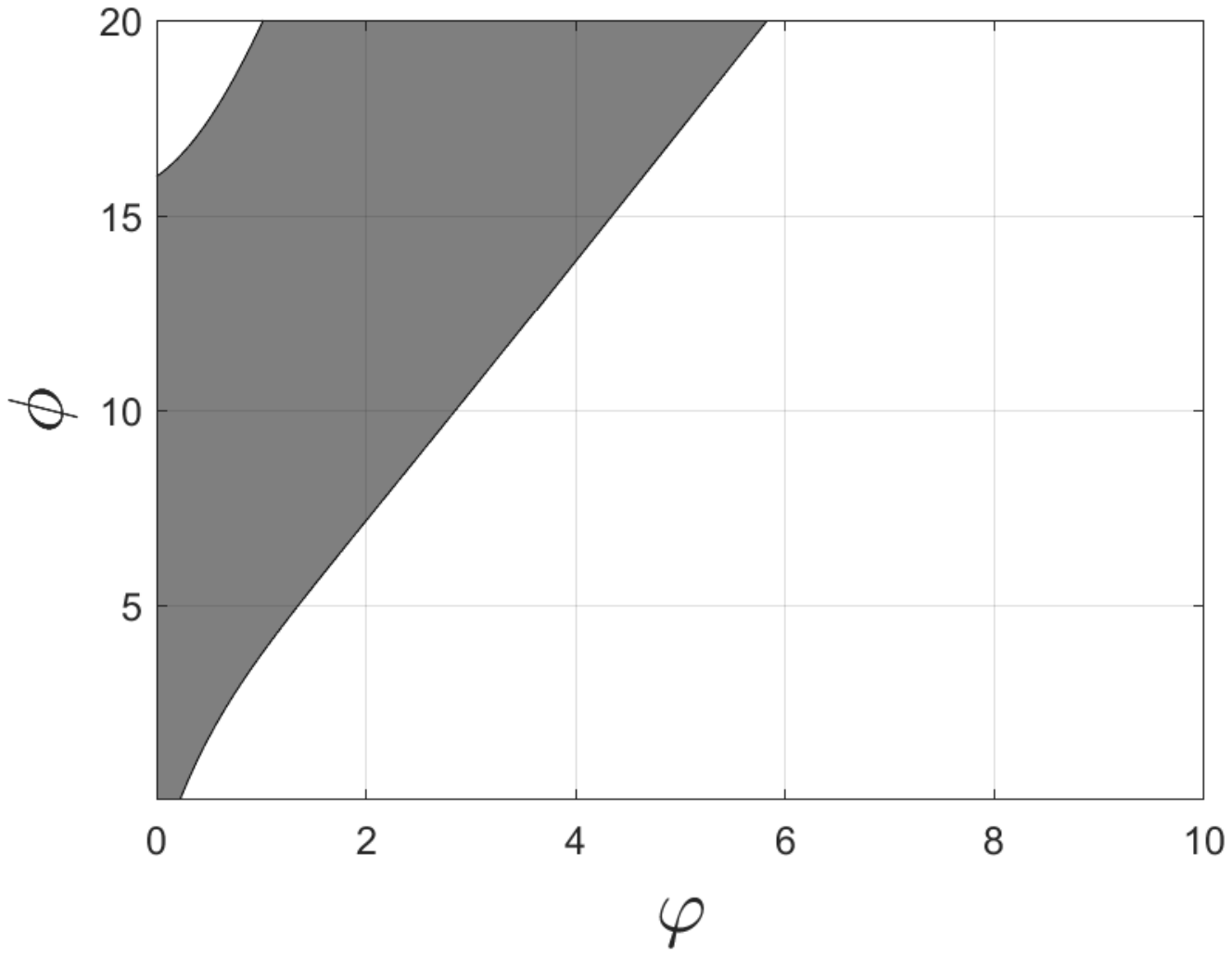}}\\[10pt]
&  {\small {(e)}} & {\small {(f)}}&{\small {(g)}}&{\small {(h)}}\vspace{-.4cm}\\
\raisebox{+5.7ex}{\rotatebox[origin=lt]{90}{qLL-S sets}}
\hspace{+.4cm}
&
{\includegraphics[angle=0,width=.23\textwidth,trim=120 280 140 230 ,
totalheight=.30\textwidth]{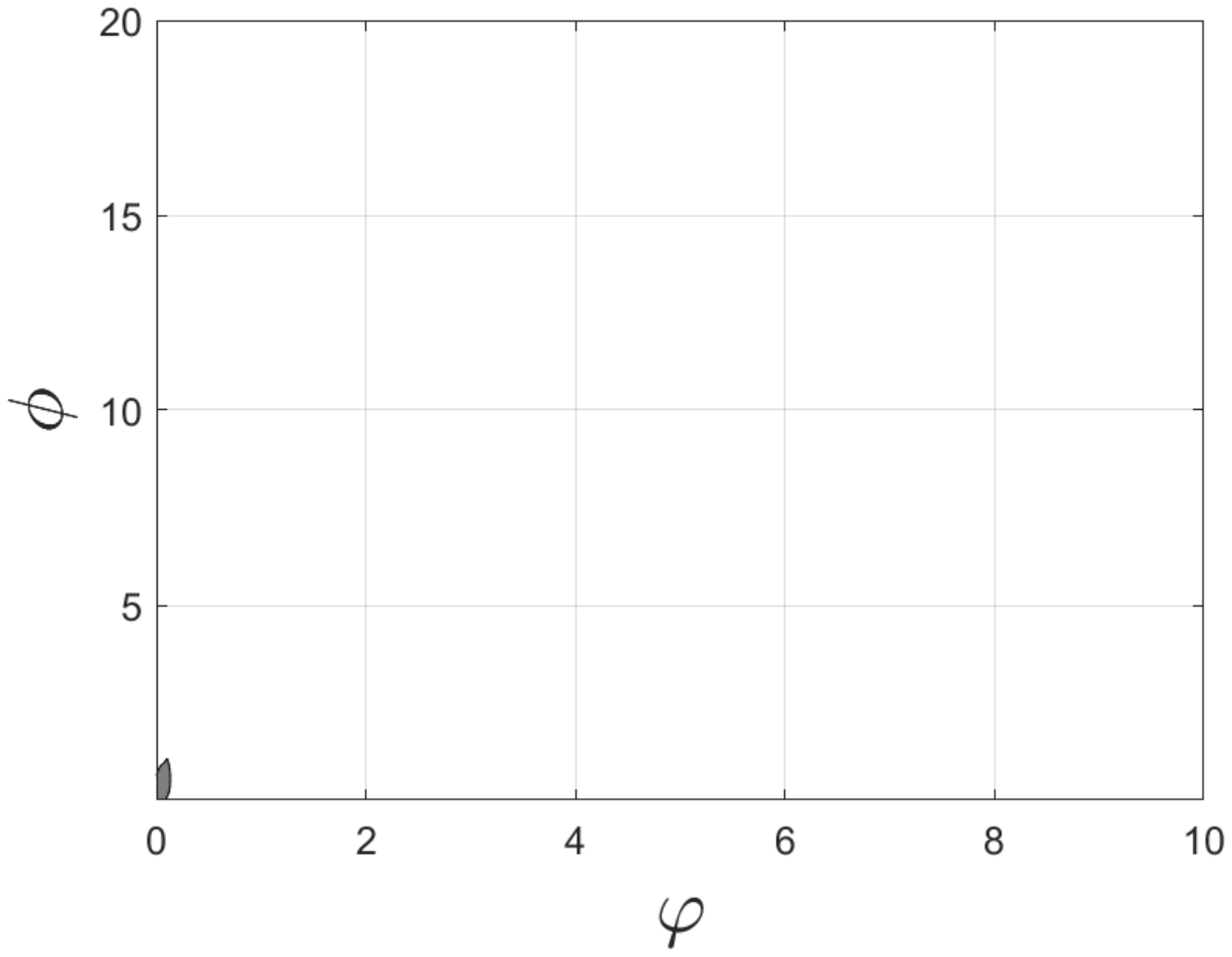}}
& \hspace{+.3cm}
{\includegraphics[angle=0,width=.23\textwidth,trim=120 280 140 230 ,
totalheight=.30\textwidth]{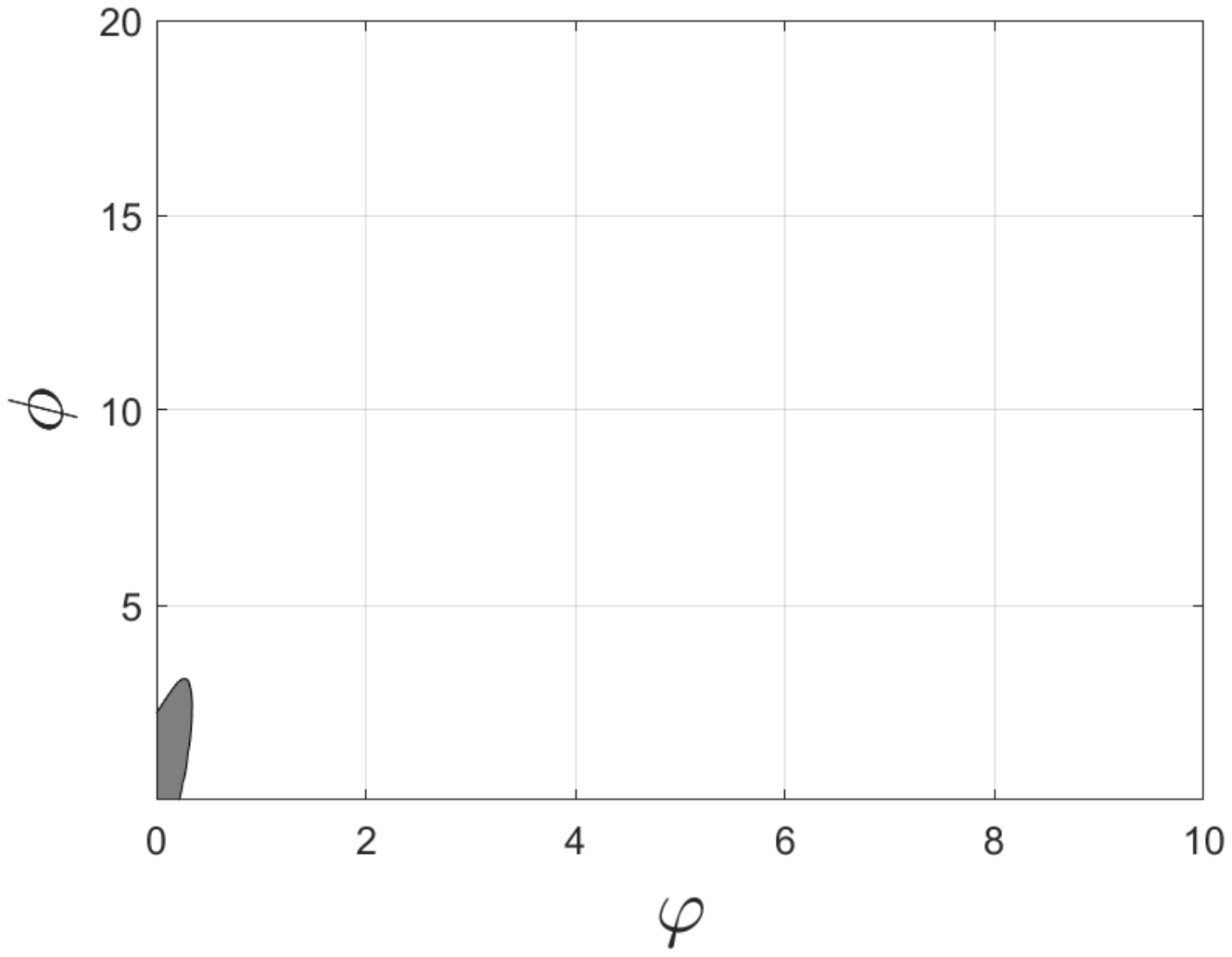}}
& \hspace{+.3cm}
{\includegraphics[angle=0,width=.23\textwidth, trim=120 280 140 230 ,
totalheight=.30\textwidth]{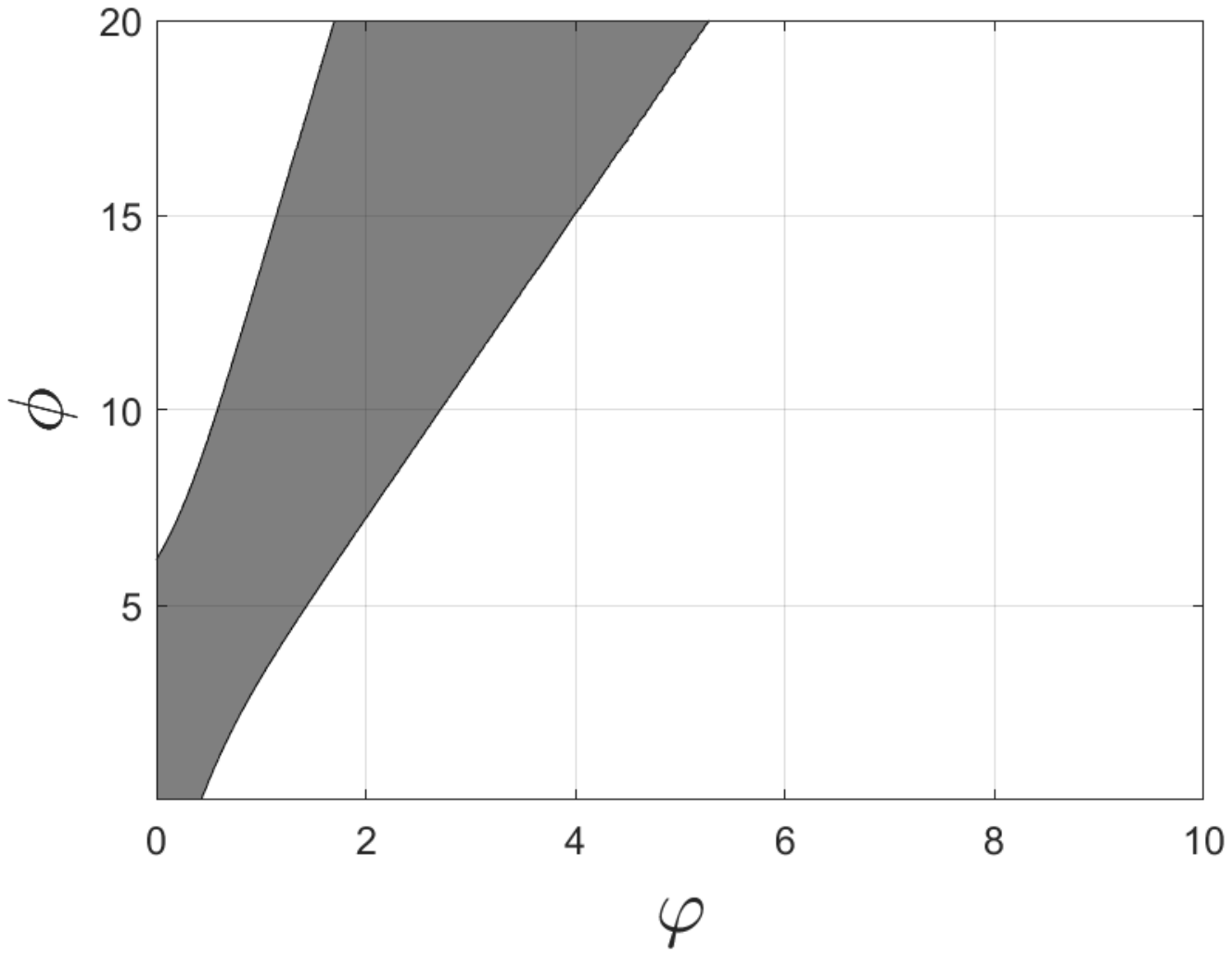}}
&
{\includegraphics[angle=0,width=.23\textwidth, trim=120 280 140 230 ,
totalheight=.30\textwidth]{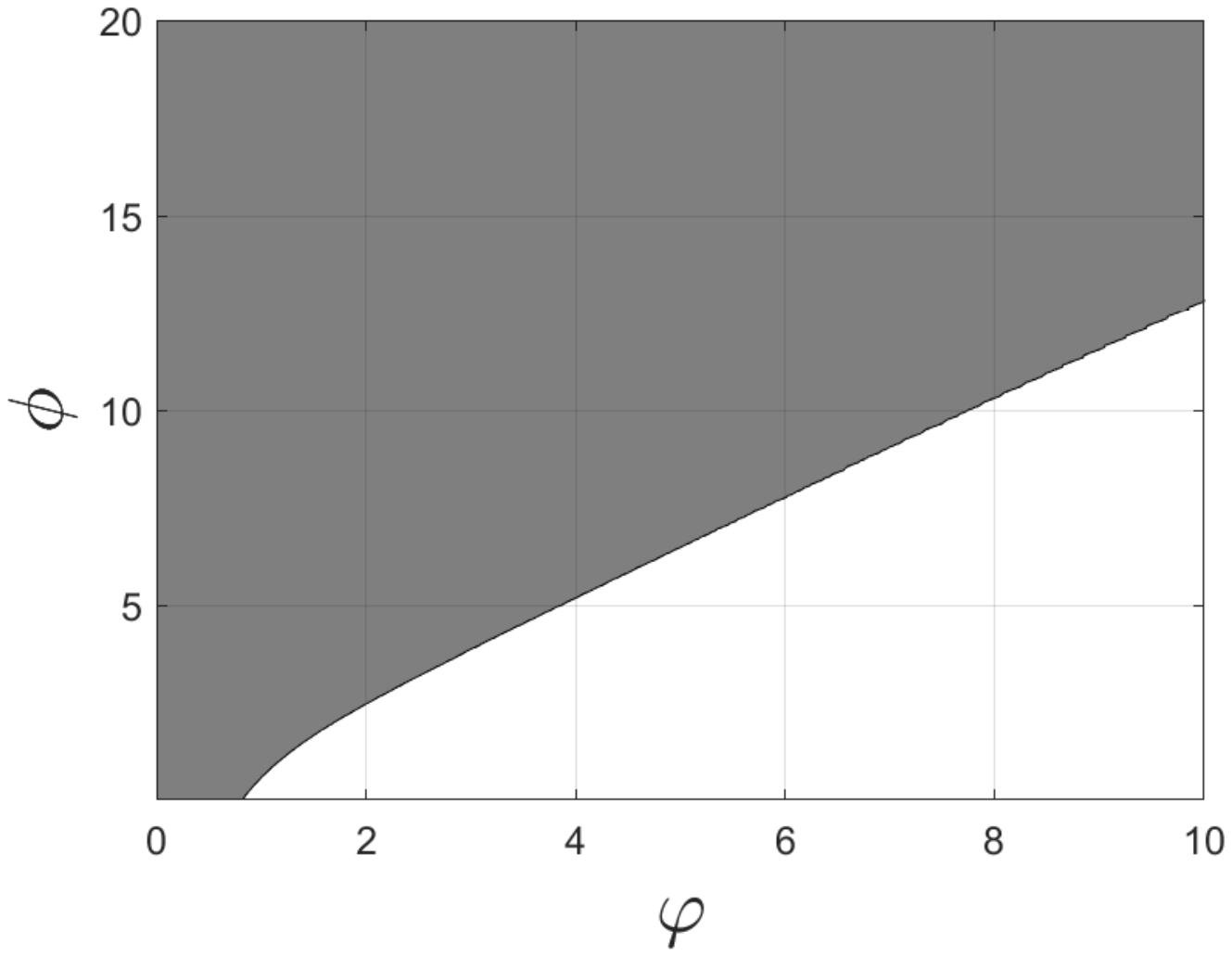}}\\[+10pt]
\cline{2-5}
& \multicolumn{4}{c}{Panel B: JPT Investment Proxy}\\[+5pt] \cline{2-5}
&   {\small {$\rho=0.0$}} & {\small {$\rho=0.6$}} & {\small {$\rho=0.8$}}& {\small {$\rho=0.9$}}\vspace{0.1cm}\\
& {\small {(a)}}&   {\small {(b)}} & {\small {(c)}} &  {\small {(d)}}\vspace{-.4cm}\\
\raisebox{+8.7ex}{\rotatebox[origin=lt]{90}{S sets }}
\hspace{+.4cm}
&
{\includegraphics[angle=0,width=.23\textwidth,trim=120 280 140 230 ,
totalheight=.30\textwidth]{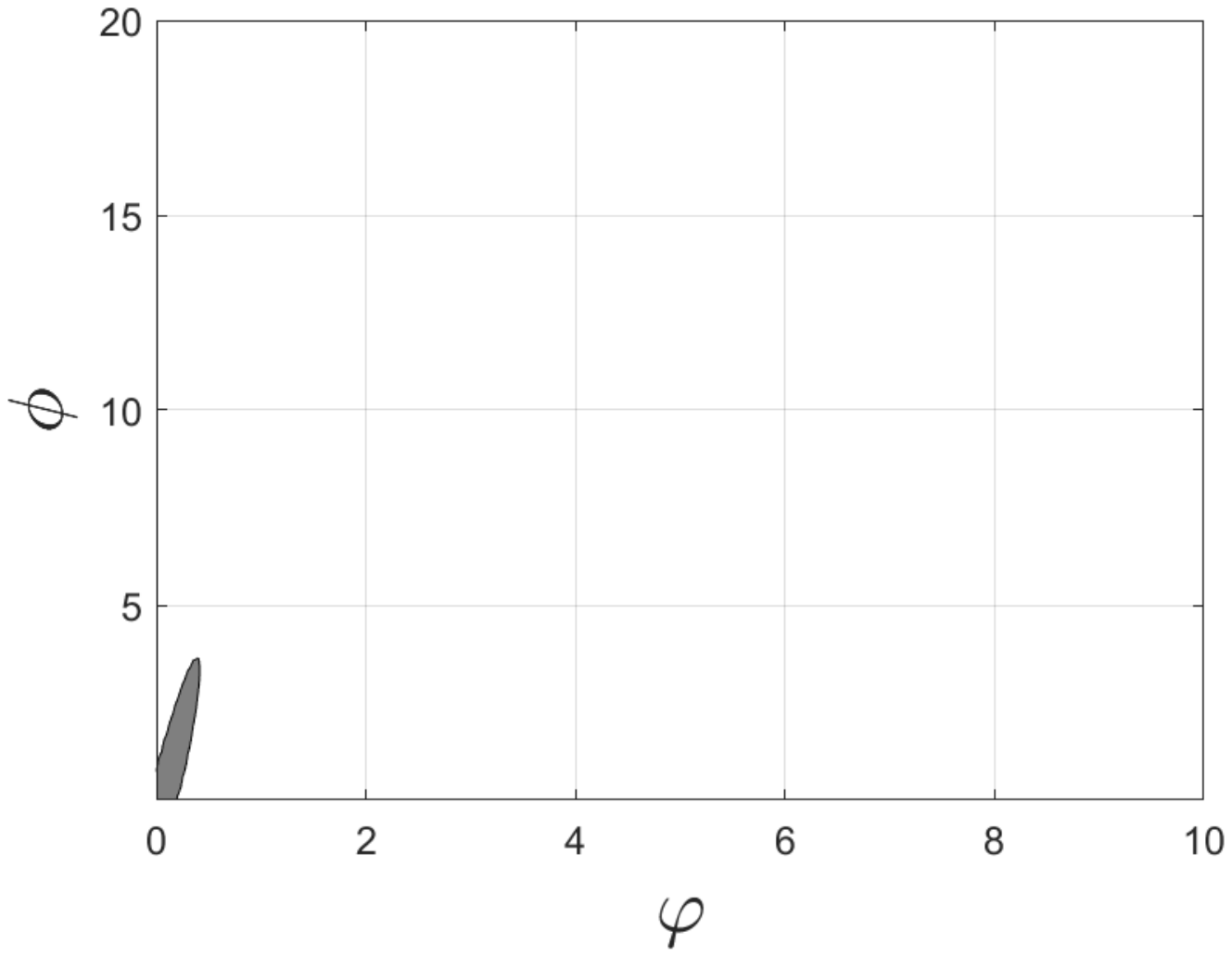}}
& \hspace{+.4cm}
{\includegraphics[angle=0,width=.23\textwidth,trim=120 280 140 230 ,
totalheight=.30\textwidth]{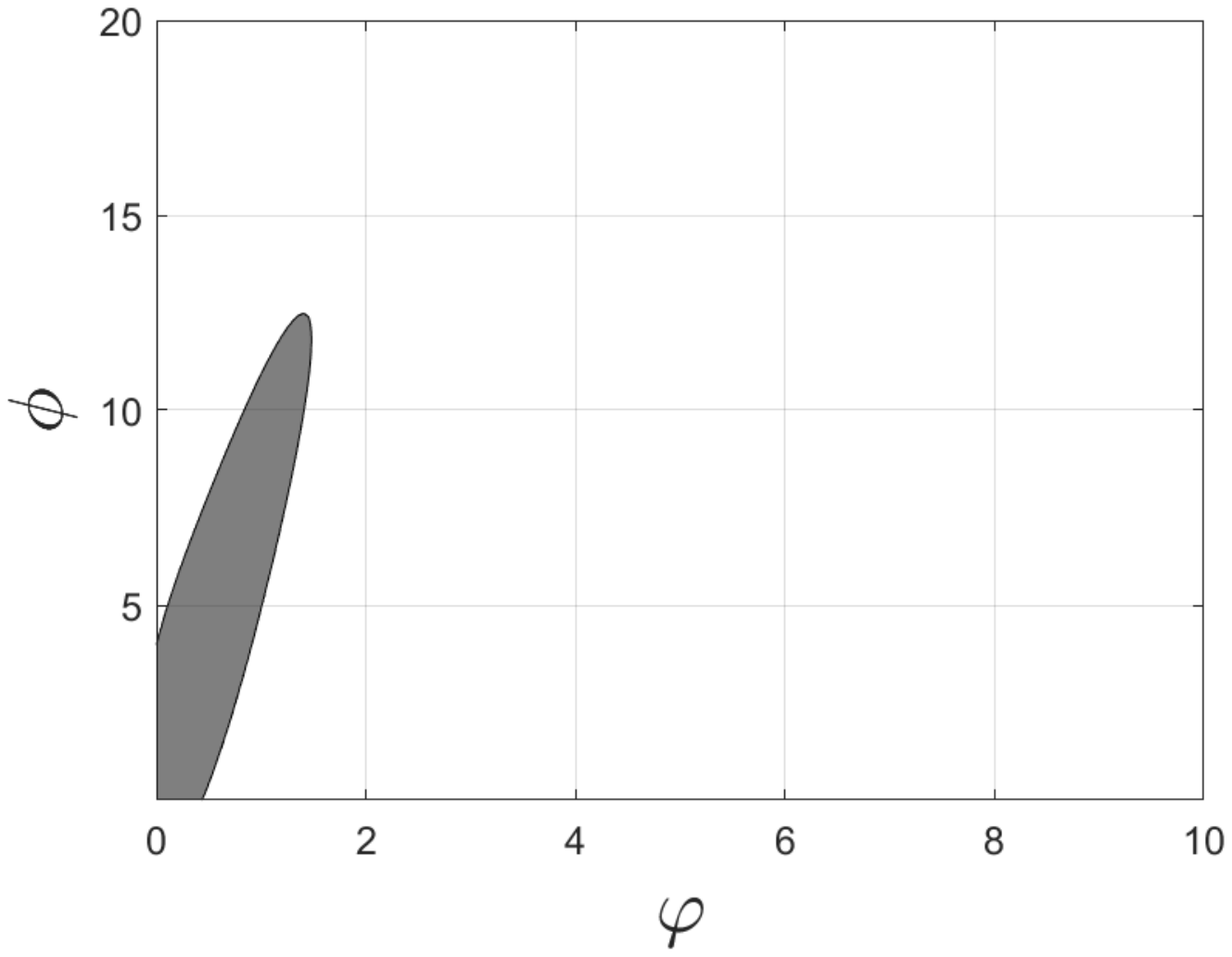}}
& \hspace{+.4cm}
{\includegraphics[angle=0,width=.23\textwidth, trim=120 280 140 230 ,
totalheight=.30\textwidth]{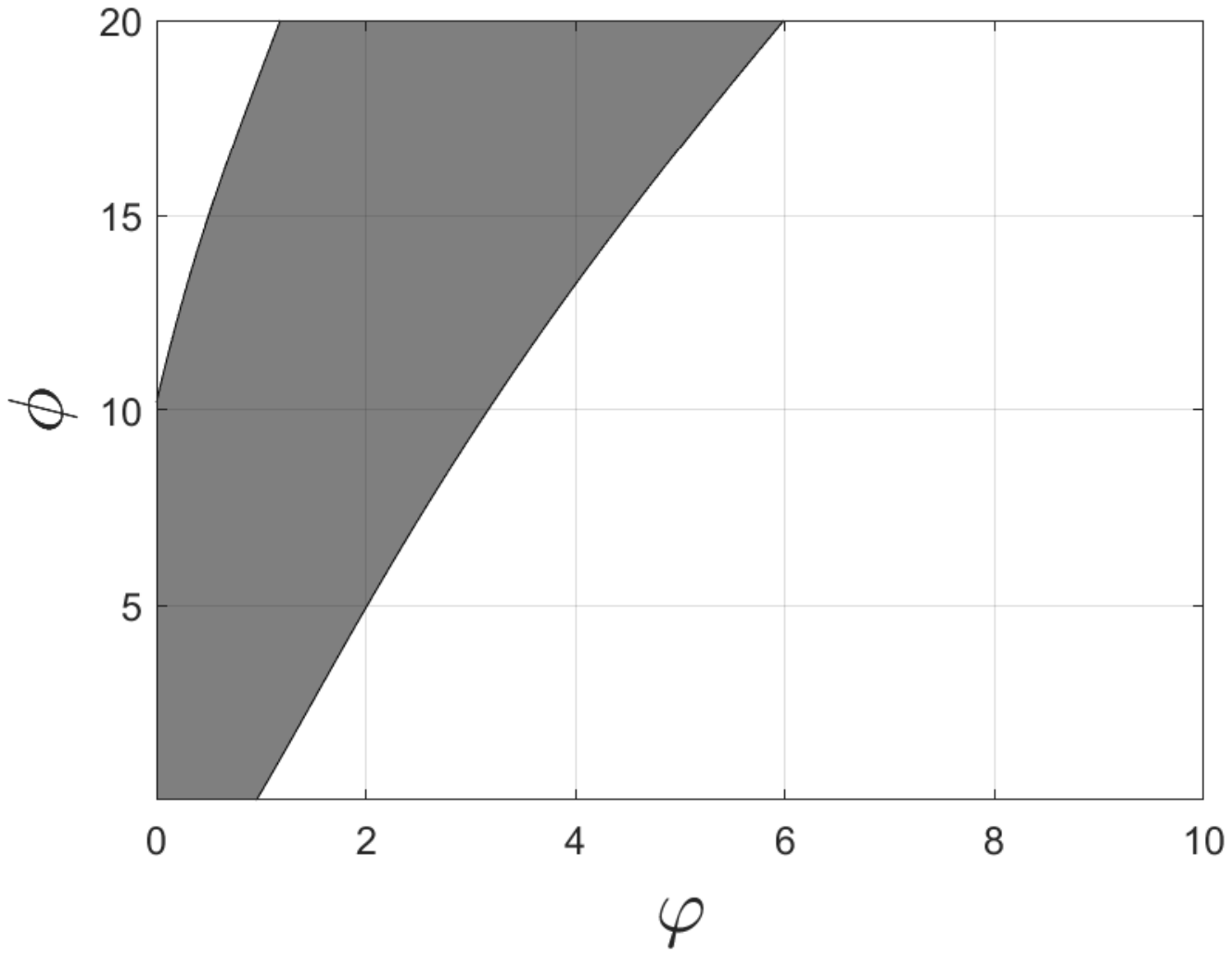}}
& \hspace{+.4cm}
{\includegraphics[angle=0,width=.23\textwidth, trim=120 280 140 230 ,
totalheight=.30\textwidth]{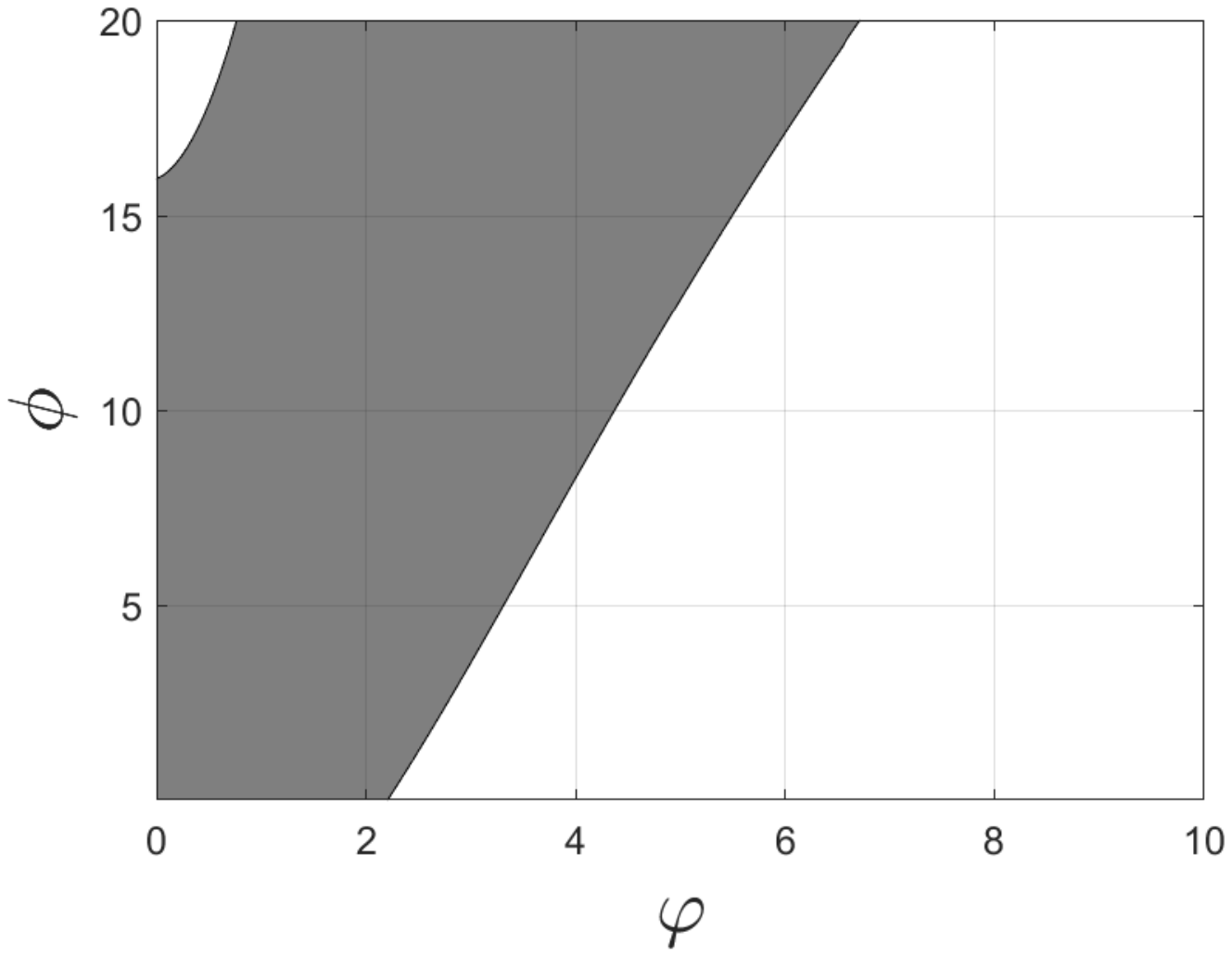}}\\[10pt]
&  {\small {(e)}} & {\small {(f)}}&{\small {(g)}}&{\small {(h)}}\vspace{-.4cm}\\
\raisebox{+5.7ex}{\rotatebox[origin=lt]{90}{qLL-S sets }}
\hspace{+.4cm}
&
{\includegraphics[angle=0,width=.23\textwidth,trim=120 280 140 230 ,
totalheight=.30\textwidth]{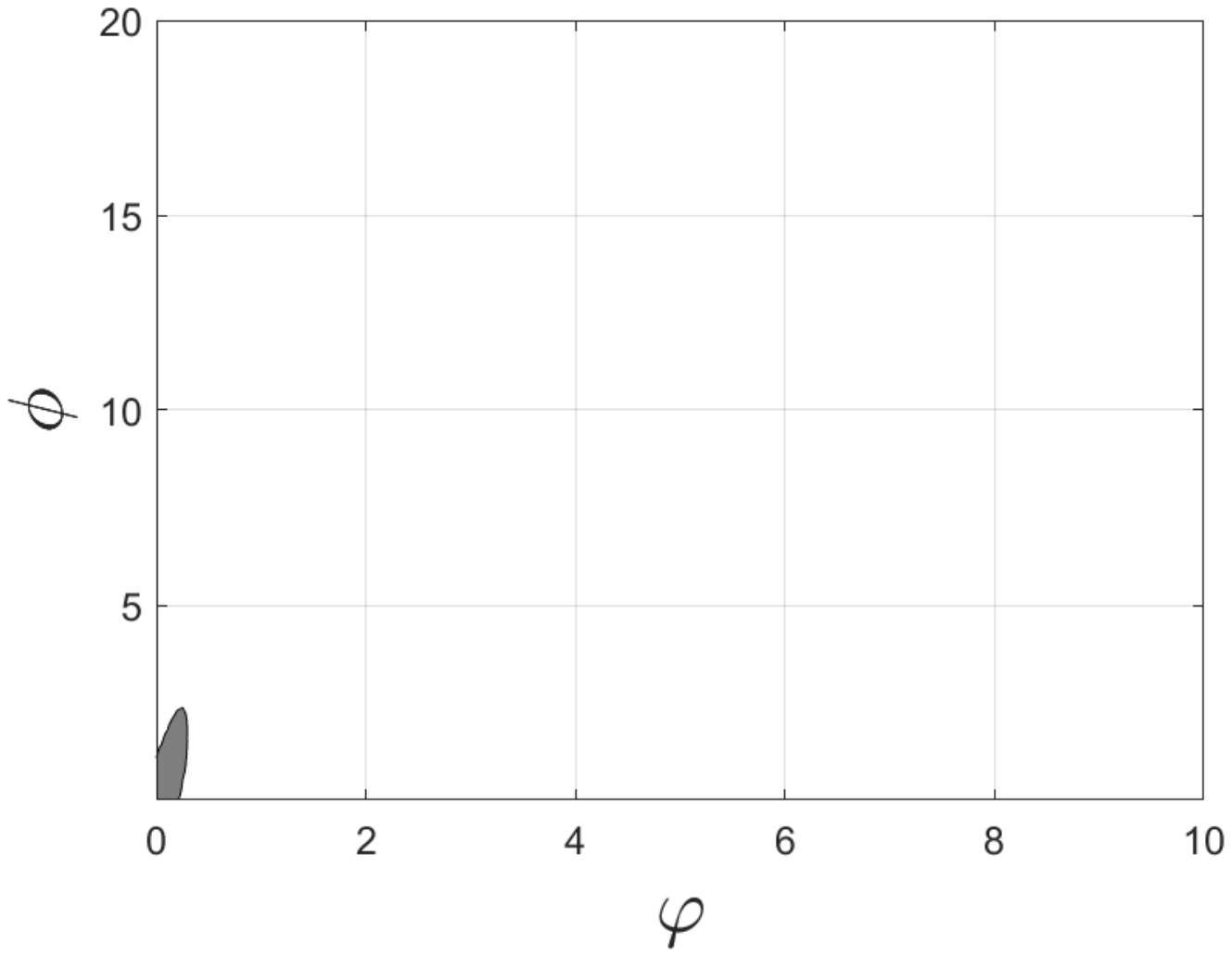}}
& \hspace{+.4cm}
{\includegraphics[angle=0,width=.23\textwidth,trim=120 280 140 230 ,
totalheight=.30\textwidth]{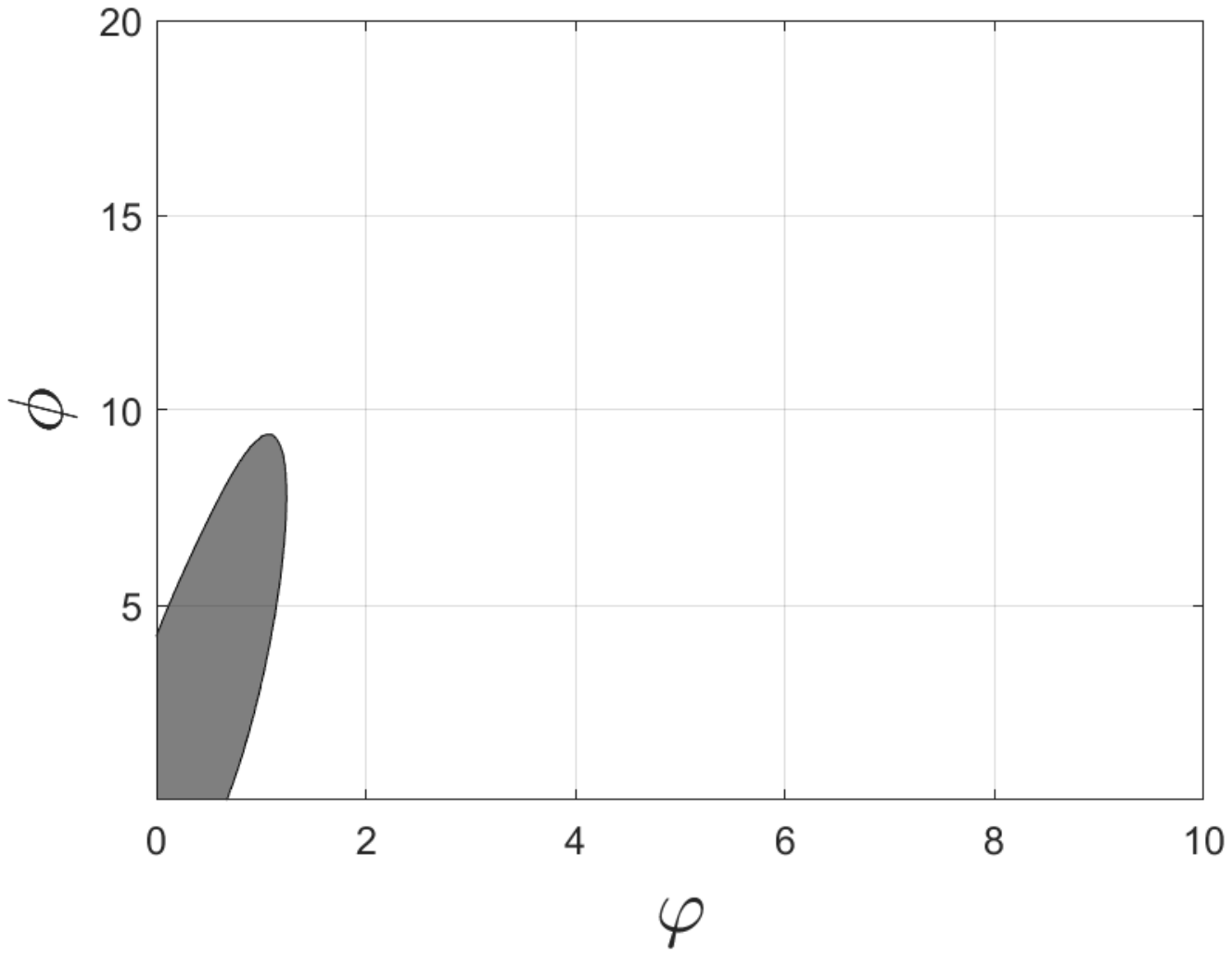}}
& \hspace{+.4cm}
{\includegraphics[angle=0,width=.23\textwidth, trim=120 280 140 230 ,
totalheight=.30\textwidth]{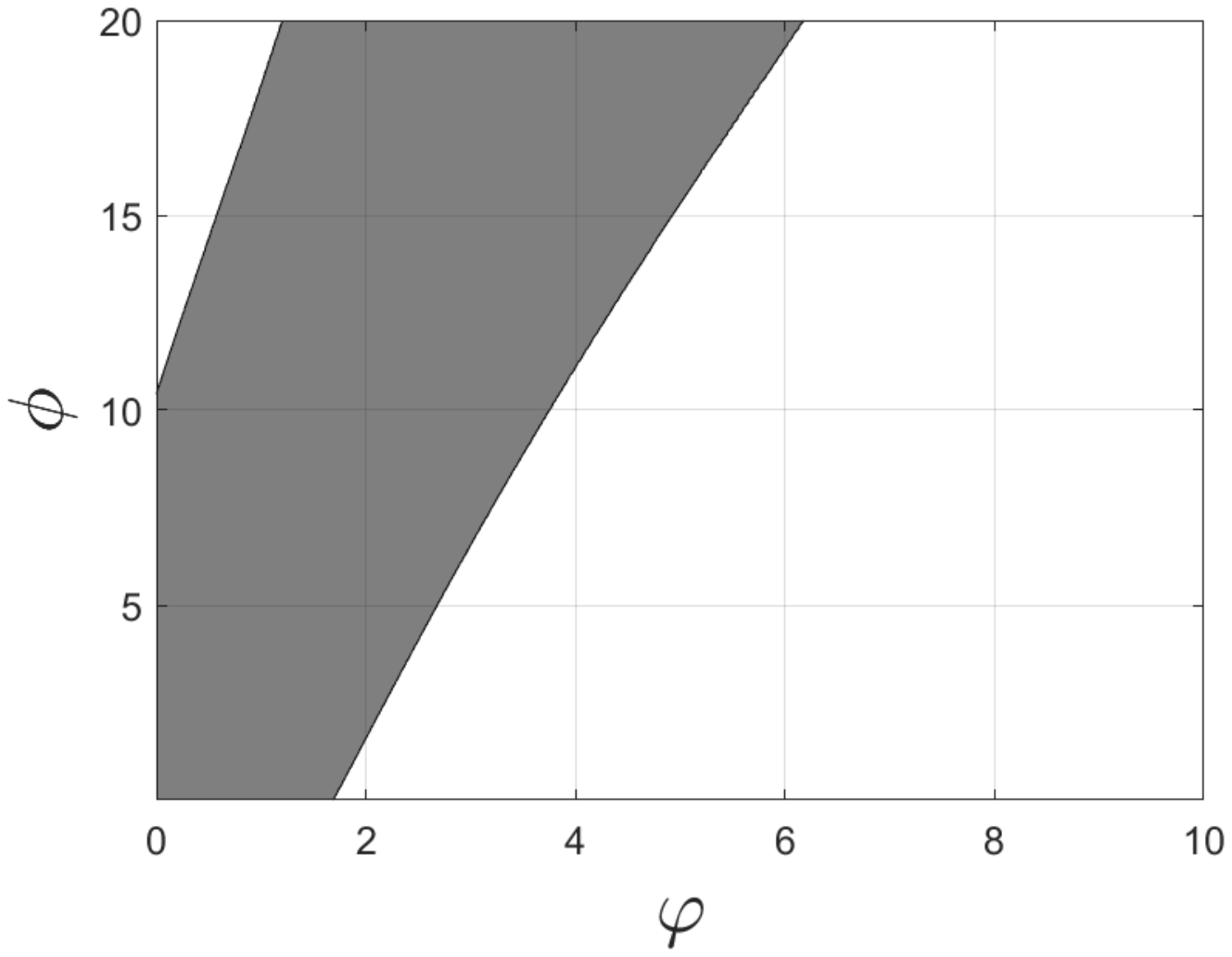}}
& \hspace{+.4cm}
{\includegraphics[angle=0,width=.23\textwidth, trim=120 280 140 230 ,
totalheight=.30\textwidth]{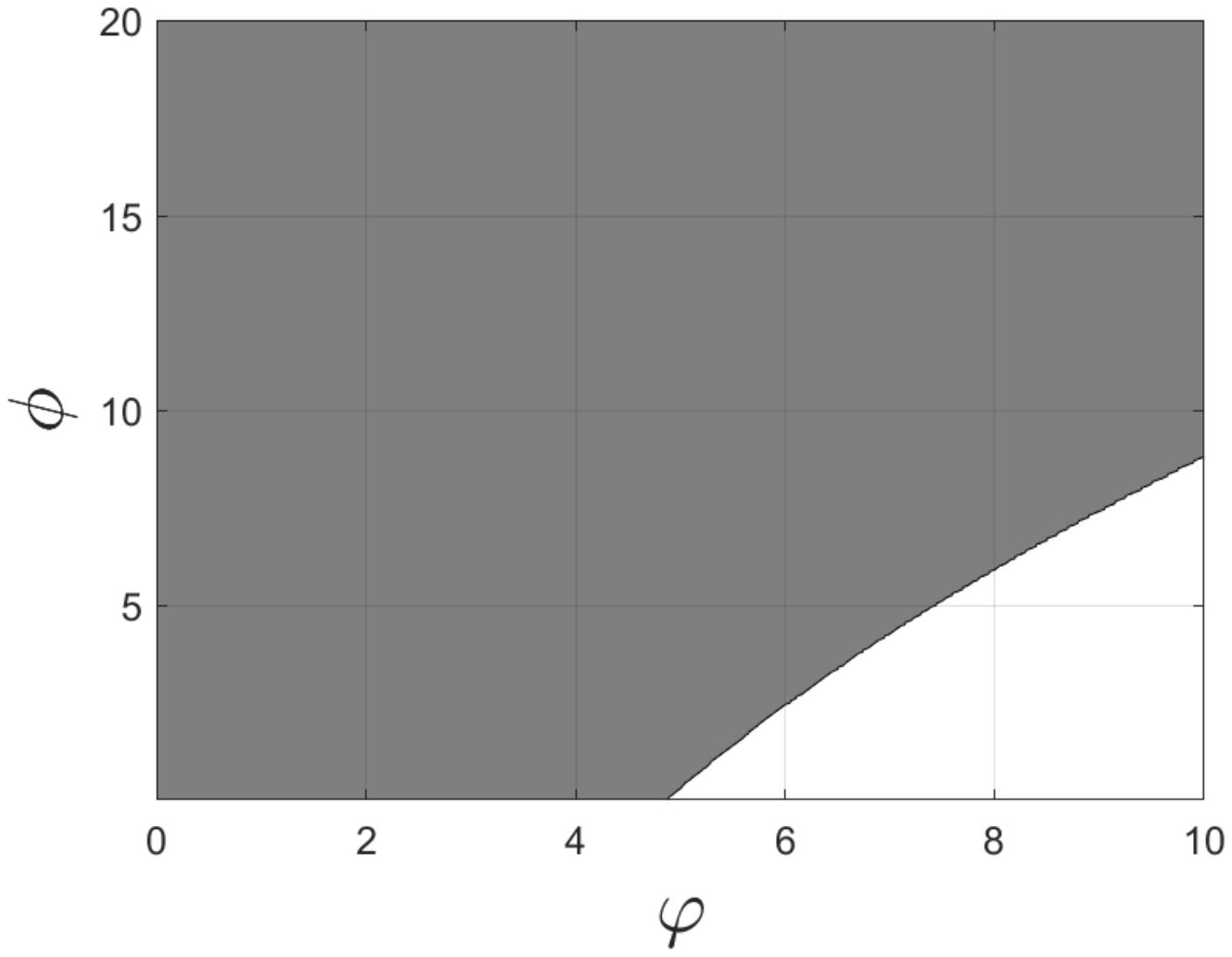}}\\[10pt]
\hline\hline
\end{tabular}}
\caption{90\% S and qLL-S confidence sets for
$(\varphi,\phi)$ derived from the investment Euler equation model
(\ref{eq: estimated_reduced}) with Fixed Private Investment and the sum of Gross Private
Domestic Investment and Personal Consumption Expenditure on Durable Goods as investment
proxies (Panels A and B, respectively). Instruments: a constant, $\Delta i_{t-1}$, $r^{p}_{t-2}$ and $u_{t-1}$.
\cite{Newey_West_1987} HAC. Period: 1967Q1-2019Q4.}%
\label{fig: reduced form}%
\end{figure}

For ease of exposition, we construct two-dimensional confidence sets for
$(\varphi,\phi)$ for fixed values of $\rho$ at 0.0, 0.6, 0.8, and 0.9. We estimate
(\ref{eq: estimated_reduced}) using the same set of instruments
and sample period as in Figure \ref{fig: baseline}. Figure \ref{fig: reduced form}
plots the resulting two-dimensional confidence sets for $(\varphi,\phi)$. 

%%%%%%%%%%%%%%%%%%%%%%%%%%%%%%%%%%%%%%%%%%%%%%%%%%
%Figure - Reduced Form parametrization
%%%%%%%%%%%%%%%%%%%%%%%%%%%%%%%%%%%%%%%%%%%%%%%%%%

The figure suggests that, when $\rho$ is between 0 and 0.6, $\varphi$ and $\phi$ are well-identified with their respective confidence intervals sitting tightly around 0. Looking at equation (\ref{eq: estimated_reduced}), this suggests limited responsiveness of investment growth to changes in capital utilization and the real interest rate. This implies that in addition to the weak identification issues discussed above, the mapping from reduced-form to structural parameters is also resulting in poorly identified confidence sets for the latter ones, even when the former ones are well-identified. For instance, despite $\phi$ being well identified when $\rho=0$ or $\rho=0.6$, its reciprocal $\kappa$ is not, since the confidence interval for $\phi$ includes 0.\footnote{So that, for example, values of $\phi$ between 0 and 0.2 imply possible values of $\kappa$ between five and infinity.} Similarly, even though both $\varphi$ and $\phi$ are well identified when the investment-specific shock is moderately persistent, the structural parameter $\zeta$ is not, since $\phi_{k}$ (which is calibrated in our earlier exercises) is very small, therefore resulting in poor identification of $\zeta$.\footnote{$\phi_{k}$, which depends on $\beta$ and $\delta$ as defined in (\ref{eqn:Basline_Euler_IAC_util_1}) in Appendix \ref{appsec: derivation}, is 0.0348 given the fairly standard calibration $\beta=0.99$ and $\delta=0.025.$}

 As for the sake of comparison, Table \ref{Table:range_phi_varphi} shows the estimates of the well-identified reduced-form parameters $(\varphi , \phi )$ implied by the estimates of the elasticity of capital utilization cost $(\zeta )$ and the investment adjustment cost $(\kappa)$ in Figure \ref{fig:FigureLit}. As explained above, because of the mapping from reduced-form to structural parameters, the large differences in the structural parameters does not translate into large differences in the reduced-form ones, whose implied values are consistent with our semi-structural estimation.

\begin{table}
\begin{center}
\captionsetup{justification=centering}
\caption{Values of the reduced-form parameters $\phi$ and $\varphi$ \\ implied by the estimates of $\kappa$ and $\zeta$ from Figure \ref{fig:FigureLit} }
%\resizebox{\textwidth}{!}{
\begin{tabularx}{0.55\textwidth}{Xcc}
 \hline \hline
\noalign{\vspace{0.5ex}}
%\resizebox{0.7\textwidth}{!}{%
& $\phi$ & $\varphi$ \\
\noalign{\vspace{0.5ex}}
\hline
\noalign{\vspace{0.5ex}}
CEE (2005) & 0.40 & 0.0001 \\
SW (2007) & [0.13-0.25] & [0.003-0.02] \\
JPT (2010) & [0.26-0.48] & [0.03-0.12] \\
ACEL (2011) & 0.67 & 0.26 \\
CTW (2011) & [0.05-0.10] & [0.0003-0.002] \\
CMR (2014) & 0.09 & 0.008 \\
AABC (2020) & [0.18-0.35] & [0.007-0.01]\\
IKR (2020) & [0.30-0.67] & [0.04-0.15] \\
\noalign{\vspace{0.5ex}}
\hline \hline
\end{tabularx}
 %}
\label{Table:range_phi_varphi}
\smallskip
		\begin{tablenotes}\setlength\labelsep{0pt} \footnotesize
			\item[] Notes: As in Figure \ref{fig:FigureLit}, the labels refer to the following papers: \citeauthor{Christiano_Eichenbaum_Evans_2005} - CEE(\citeyear{Christiano_Eichenbaum_Evans_2005}), \citeauthor{Smets_Wouters_2007} - SW(\citeyear{Smets_Wouters_2007}), \citeauthor{justiniano2010investment} - JPT(\citeyear{justiniano2010investment}), \citeauthor{Altig_Christiano_Eichenbaum_Linde_2011} - ACEL(\citeyear{Altig_Christiano_Eichenbaum_Linde_2011}), \citeauthor{Christiano_Trabandt_Walentin_2011} - CTW(\citeyear{Christiano_Trabandt_Walentin_2011}), \citeauthor{Christiano_Motto_Rostagno_2014} - CMR(\citeyear{Christiano_Motto_Rostagno_2014}), \citeauthor{aabc2020} - AABC(\citeyear{aabc2020}), \citeauthor{Inoue_Kuo_Rossi_2020} - IKR(\citeyear{Inoue_Kuo_Rossi_2020})
		\end{tablenotes}
 \end{center}
\end{table}

%\todo[inline, color=green]{LMM: move scatter plot here and adjust the text.}

%\sout{\todo[inline]{Point REF1(3): 'it would be great to show a first stage regression, a scatter plot, something that could explain the reader the relationship between instruments and instrumented variables. This way  could also motivate better the choice of instruments.' Sophocles' ANSWER: yes we can, for each rho, see my note cer.tex or pdf. \\Guido and Qazi: Below are the scatter plots for the baseline set of instruments. two questions: \\ 1) Shall we do them also for the external instruments below? I think there are already too many figures, at most we can put the plots for the external instruments in the ref response. \\2) is this the right place to put it? We are not sure. On the one hand, this is related to the discussion at the end of the semi-structural section about how low values of $\rho$ helps identification. Hence better to put in there. On the other hand, the ref talks about 'motivate better the choice of instruments'. }}

% \begin{figure}[hbt!]
% 	\centering
% 	{\includegraphics[width=\textwidth, height=12cm]{scatter.PNG}}%
% 	\caption*{Scatter plot}%
% 	\label{fig:Figure_scatter}%
% \end{figure}

\begin{figure}[htbhp]
\centering
\adjustbox{width=1.0\textwidth, height=.5\textwidth}{
\begin{tabular}{ccccc}
{\includegraphics[angle=0,width=.45\textwidth,trim=100 280 140 230 ,
totalheight=.4\textwidth]{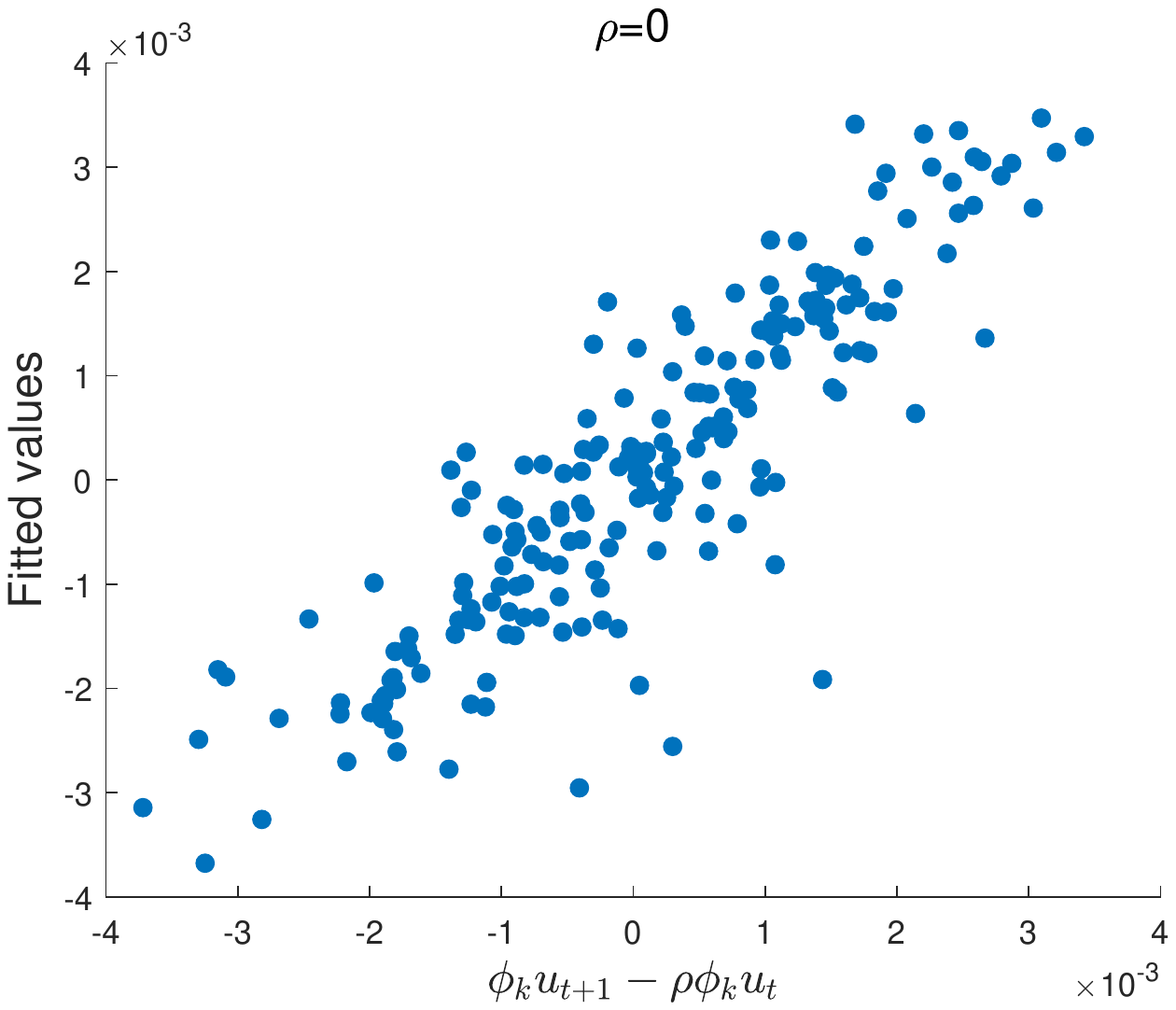}}
&& {\includegraphics[angle=0,width=.45\textwidth,trim=100 280 140 230 ,
totalheight=.4\textwidth]{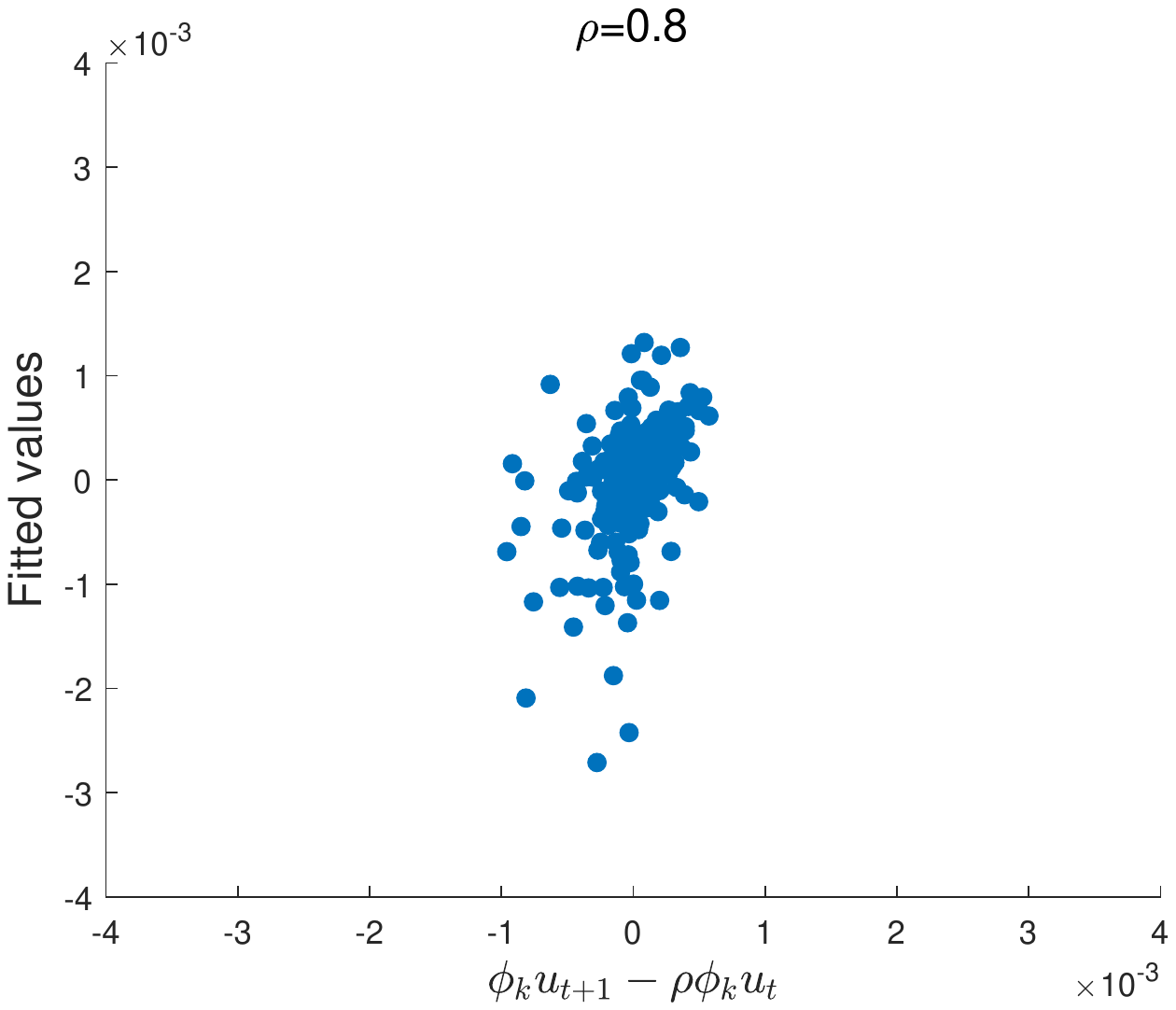}} 
\\[+25pt]
{\includegraphics[angle=0,width=.45\textwidth,trim=100 280 140 230 ,
totalheight=.4\textwidth]{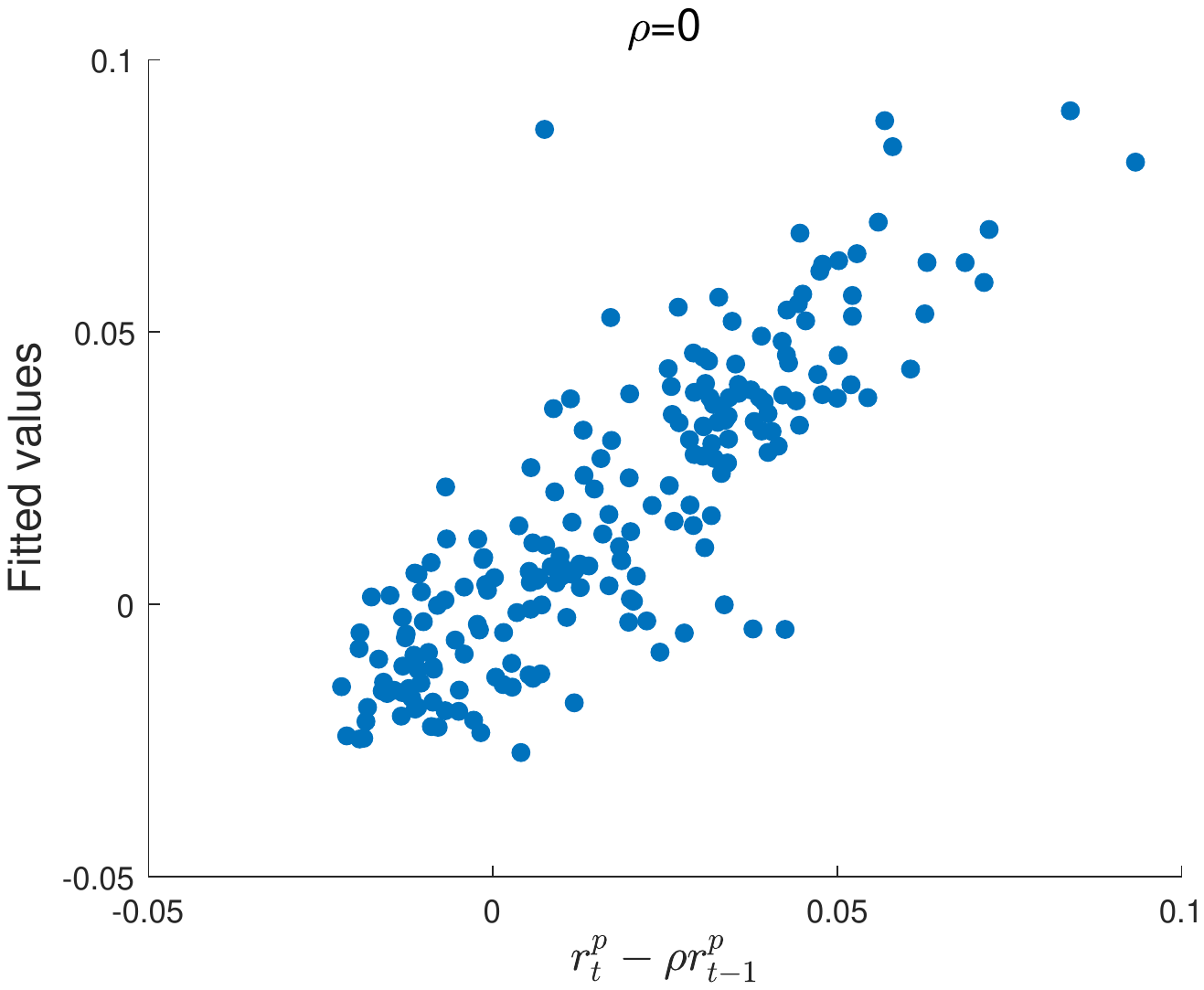}}
&& {\includegraphics[angle=0,width=.45\textwidth,trim=100 280 140 230 ,
totalheight=.4\textwidth]{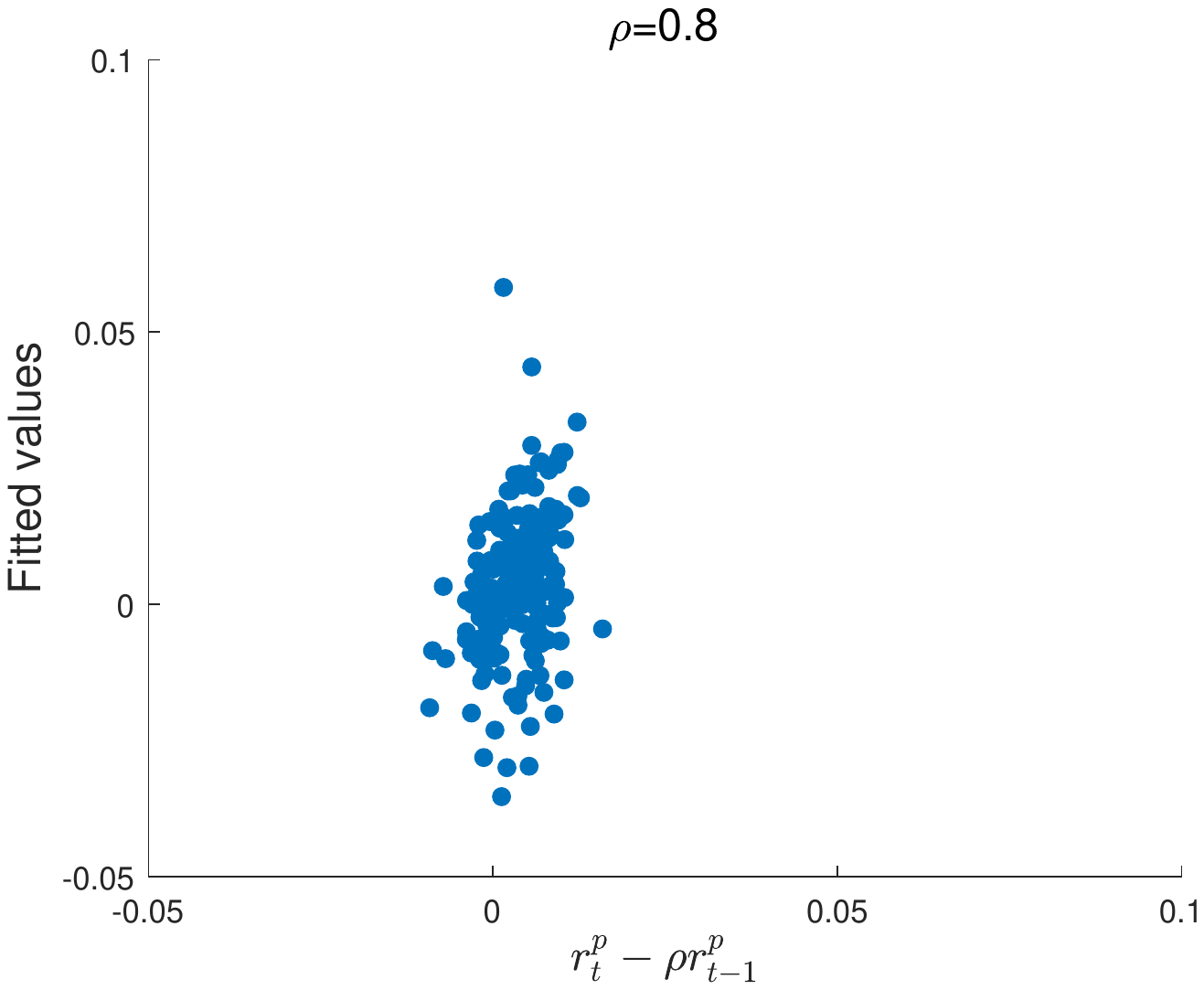}}

\\[+25pt]
\end{tabular}
}
\caption{Scatter plot of the fitted values of the endogenous regressors $(\phi_{k}u_{t+1}-\rho\phi_{k}u_{t})$ and $(r_{t}^{p}-\rho r_{t-1}^{p})$ from their respective first-stage regressions on the baseline instruments for two different values of $\rho$.}%
\label{fig:Figure_scatter}%
\end{figure}

To understand why identification of the semi-structural parameters $\varphi$ and $\phi$ worsens for higher values of $\rho$, we look at the autocorrelations of $r_{t}^{p}$ and $u_{t+1}$. The first  and second autocorrelations of $r_{t}^{p}$ in our baseline sample are 0.90 and 0.83 respectively, while the same autocorrelations for $u_{t+1}$ are 0.96 and 0.87, indicating that both series are highly persistent. Therefore, for small values of $\rho$, the instruments $r_{t-2}^{p}$ and $u_{t-1}$ are highly correlated with the endogenous variables of the system, resulting in better identification. In contrast, as the value of $\rho$ increases, $(u_{t+1}-\rho u_{t})$ and $(r_{t}^{p}-\rho r_{t-1}^{p})$ in equation (\ref{eq: estimated_reduced}) resemble white-noise processes, making lagged endogenous variables weak instruments. This is illustrated in Figure \ref{fig:Figure_scatter} which is a scatter plot of the fitted values of the endogenous regressors $(\phi_{k}u_{t+1}-\rho\phi_{k}u_{t})$ and $(r_{t}^{p}-\rho r_{t-1}^{p})$ from their respective first-stage regressions on the baseline set of instruments used to construct Figures \ref{fig: baseline} and \ref{fig: reduced form}. The left column, which plots the results for $\rho=0$, shows that the instruments are relatively strong as the fitted values mostly lie along the 45-degree line. In contrast, for high values of $\rho$, as in the right column which plots the results for $\rho=0.8$, the values of both $(\phi_{k}u_{t+1}-\rho\phi_{k}u_{t})$ and $(r_{t}^{p}-\rho r_{t-1}^{p})$ are tightly packed around zero and are uncorrelated with the fitted values from the first-stage regressions, suggesting that instruments are weak.

\section{Implications for DSGE Models \label{s: implications DSGE}}
In this Section, we discuss the main implications of our analysis. In the Introduction, we argue about the importance of using the GMM methodology to assess the empirical fit in aggregate data of the investment equation commonly used in DSGE models. Our results support the investment equation, in the sense that there is no evidence against it. Moreover, we ask whether the GMM methodology could shed some light on the values of the parameters of this equation given the wide range found in the literature. Unfortunately, the answer is negative, because these parameters are weakly identified. This raises two questions to which we turn in the next two subsections. The first relates to the sensitivity of the DSGE results to changes in the three main parameters of interest: are any of these parameters relatively more important for the dynamic properties of a standard DSGE model, and thus relatively more important to pin down? Second, how do DSGE models attain identification of these parameters?

\subsection{Sensitivity of DSGE Dynamics to the Parameters of the Investment Equation} \label{subsec: parameters}

\begin{sidewaysfigure}[htbp]
	\centering
	{\includegraphics[width=\textwidth, height=12cm]{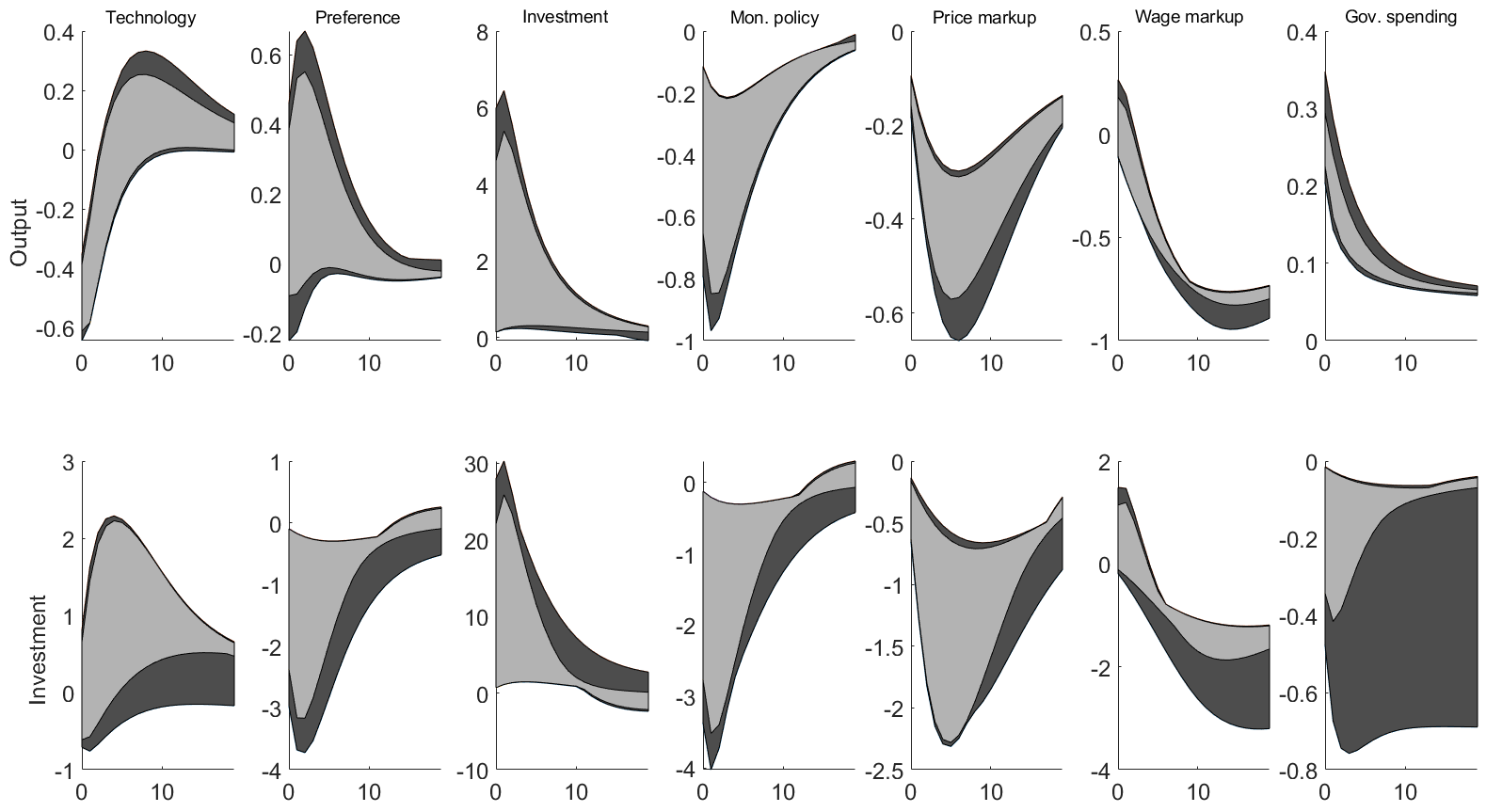}}%
	\caption{Impulse responses of output and investment to a one-standard deviation structural shock in JPT's model evaluated at the posterior median estimate of all parameters except $\kappa$ and $\zeta$. Dark shaded areas: both
	$\kappa$ and $\zeta$ vary in the 90\% S confidence set; light shaded areas: only $\kappa$ vary and $\zeta=5.30$ (posterior median estimate of JPT).}%
	\label{fig:Figure_irf_S_set_kappa}%
\end{sidewaysfigure}

\begin{sidewaysfigure}[htbp]
	\centering
	{\includegraphics[width=\textwidth, height=12cm]{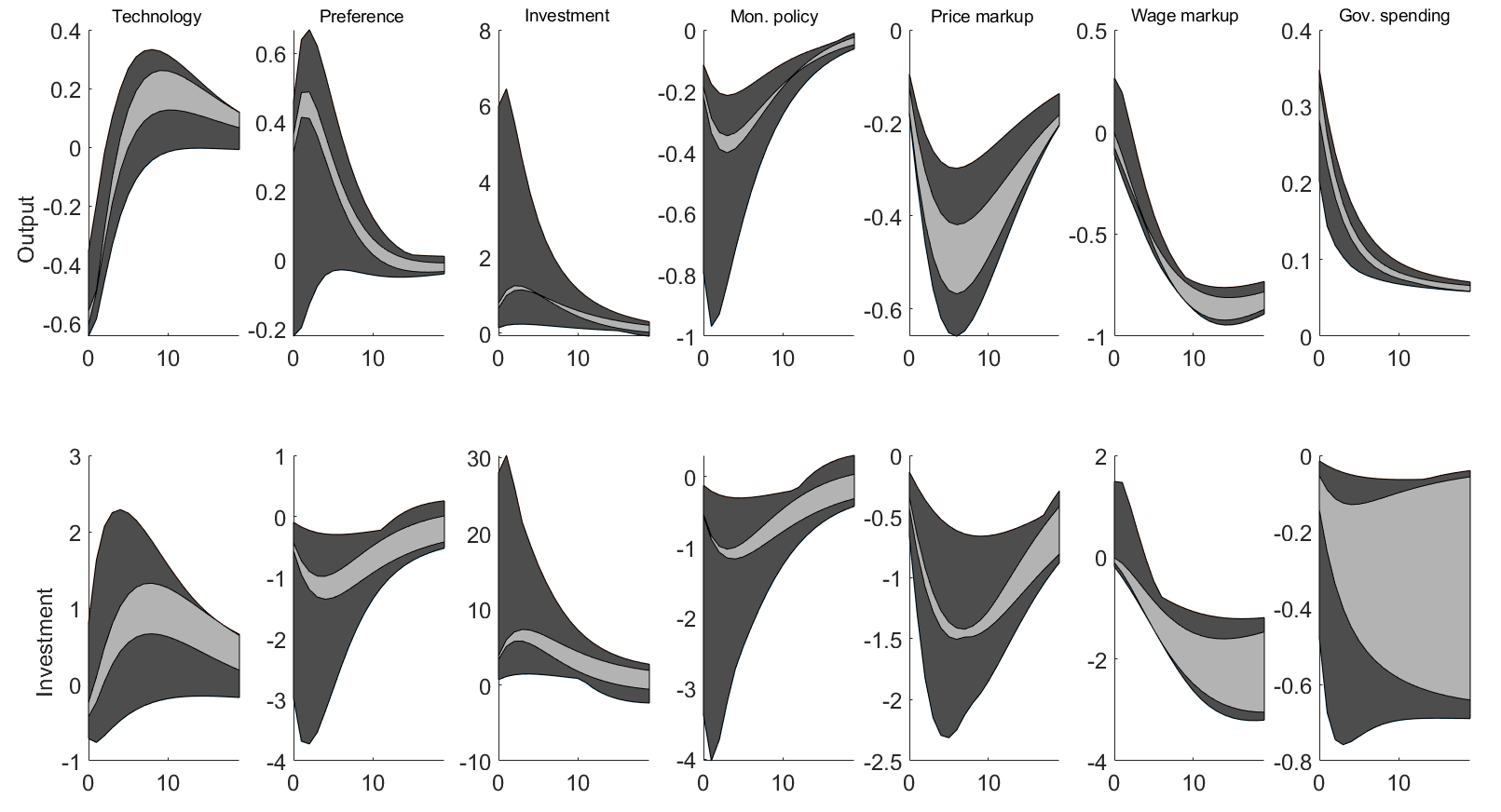}}%
	\caption{Impulse responses of output and investment to a one-standard deviation structural shock in JPT's model evaluated at the posterior median estimate of all parameters except $\kappa$ and $\zeta$. Dark shaded areas: both
	$\kappa$ and $\zeta$ vary in the 90\% S confidence set; light shaded areas: only $\zeta$ vary and $\kappa=2.85$ (posterior median estimate of JPT).}%
	\label{fig:Figure_irf_S_set_zeta}%
\end{sidewaysfigure}

Figures \ref{fig:Figure_irf_S_set_kappa}-\ref{fig:Figure_irf_S_set_zeta} mimic previous Figure \ref{fig:Figure_irf_out_inv}. The figures show how the impulse responses of output and investment to the seven structural shocks in JPT's model change when we allow the value of $\kappa$ and $\zeta$ to range in the 90\% baseline S confidence set, while the remaining parameters are fixed at the posterior median of JPT, including  $\rho$ which is set to 0.72. To highlight the relative sensitivity of the model dynamics to the values of $\kappa$ and $\zeta$, in both figures the dark shaded areas visualize all the possibilities when both $\kappa$ and $\zeta$  vary in the 90\% S confidence set. In Figure \ref{fig:Figure_irf_S_set_kappa}, the light shaded areas depict all the possibilities when $\kappa$ ranges in its 90\% S confidence set, while $\zeta$ is set to the JPT posterior median ($\zeta = 5.30$), while, in Figure \ref{fig:Figure_irf_S_set_zeta}, the light shaded areas depict all the possibilities when $\zeta$ ranges in its 90\% S confidence set, while $\kappa$ is set to the JPT posterior median ($\kappa =2.85$). Comparing the two figures, it is clear that the IRFs are very sensitive to the value of $\kappa,$ while they are little sensitive to the value of $\zeta.$ With the sole exception of the IRFs to a government spending shock (and marginally the wage markup shock), different values of $\zeta$ within the 90\% S confidence set change the IRFs only slightly, while most of the variation is due to the different values of $\kappa.$ This is particularly true for the shock that directly impacts the investment equation, that is, the investment-specific shock, as well as for the other two shocks that in these models are usually found to be the other main drivers of business cycle fluctuations, that is, the technology shock and the preference shock. 

\begin{figure}[hbt!]
	\centering
	{\includegraphics[width=\textwidth, height=12cm]{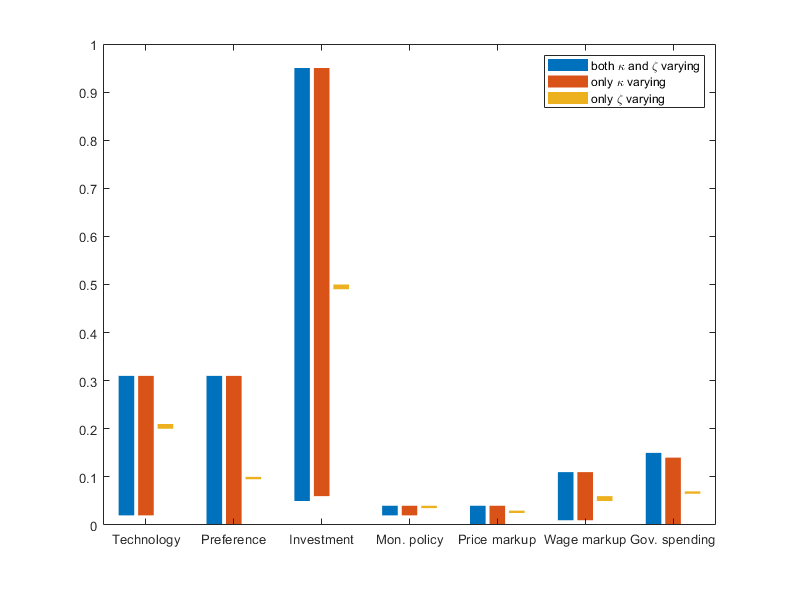}}%
	\caption{Forecast error variance decomposition of output growth in JPT's model evaluated at the posterior median estimate of all parameters except $\kappa$ and $\zeta$. Blue bars: both $\kappa$ and $\zeta$ vary in the 90\% S confidence set; orange bars: only $\kappa$ vary and $\zeta=5.30$ (posterior median estimate of JPT); yellow bars: only $\zeta$ vary and $\kappa=2.85$ (posterior median estimate of JPT).}%
	\label{fig:Figure_FEVD_JPT}%
\end{figure}

Figure \ref{fig:Figure_FEVD_JPT} uncovers the same message by looking at the variance decomposition, similarly to Table \ref{Table: VarDecom_GDPgrowth}. Figure \ref{fig:Figure_FEVD_JPT} shows how the variance decomposition of output growth changes when $\kappa$ and $\zeta$  vary in the 90\% baseline S confidence set, and, as before, the Figure displays the marginal effects of both $\kappa$ and $\zeta,$ obtained by varying each parameter at a time keeping the other fixed at the JPT posterior median.  Even more striking than for the IRFs, almost all the variation in the variance decomposition is due to changes in the values of $\kappa,$ while the role of $\zeta$ is negligible.

Together with Figure \ref{fig:Figure_irf_rho_grid}, these results suggest that $\kappa$ and $\rho$  are the key parameters that shape the dynamics of investment in a standard medium-scale DSGE model. The former defines the dynamic response of investment (and as a consequence of output) to the different shocks, and more prominently to the investment shock. A high value of the latter is fundamental to determine the persistent response of investment to the investment shock, and hence to replicate the persistent behaviour that characterizes the aggregate macroeconomic time series data. In contrast, $\zeta$ does not seem very relevant for determining the dynamic response of investment (and of output), so one could imagine that its value would be difficult to identify in the data. This is what we turn to next.

\subsection{Identification of the Parameters of the Investment Equation in DSGE Models} \label{subsec: identification DSGE}

\begin{figure}[ptbh]
\centering
\adjustbox{width=1.0\textwidth, max height=10cm}{
\begin{tabular}
{ccc}\hline\hline
\noalign{\vspace{2ex}}
\multicolumn{3}{c}{\LARGE{Panel A: JPT's dataset}}\\
\noalign{\vspace{2ex}} \hline 
%\\ 
\noalign{\vspace{1ex}}
%\cline{1-3} 
{\Large{Investment adjustment cost ($\kappa$)}} & {\Large{Elasticity capital utilization cost ($\zeta$)}} & {\Large{AR(1) coefficient investment shock ($\rho$)}}\\
\noalign{\vspace{1ex}}
\hline
\noalign{\vspace{2ex}}
{\Large (a)} & {\Large (b)} & {\Large (c)} \vspace{-1cm}\\
{\includegraphics[angle=0,width=0.6\textwidth, trim=120 270 140 230 ,
totalheight=.6\textwidth]{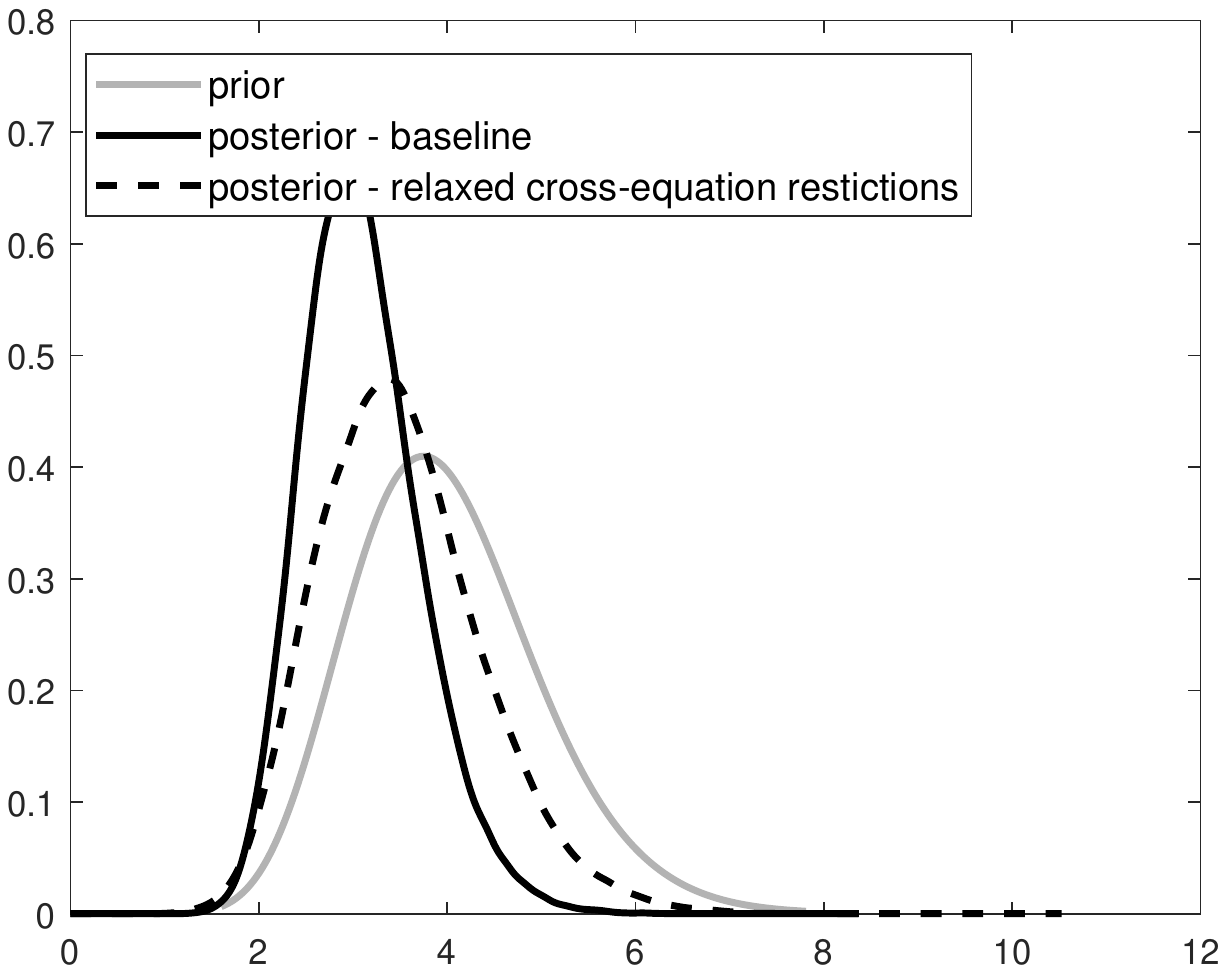}} &
\hspace{+.1cm}
{\includegraphics[angle=0,width=0.6\textwidth,trim=120 280 140 230 ,
totalheight=.6\textwidth]{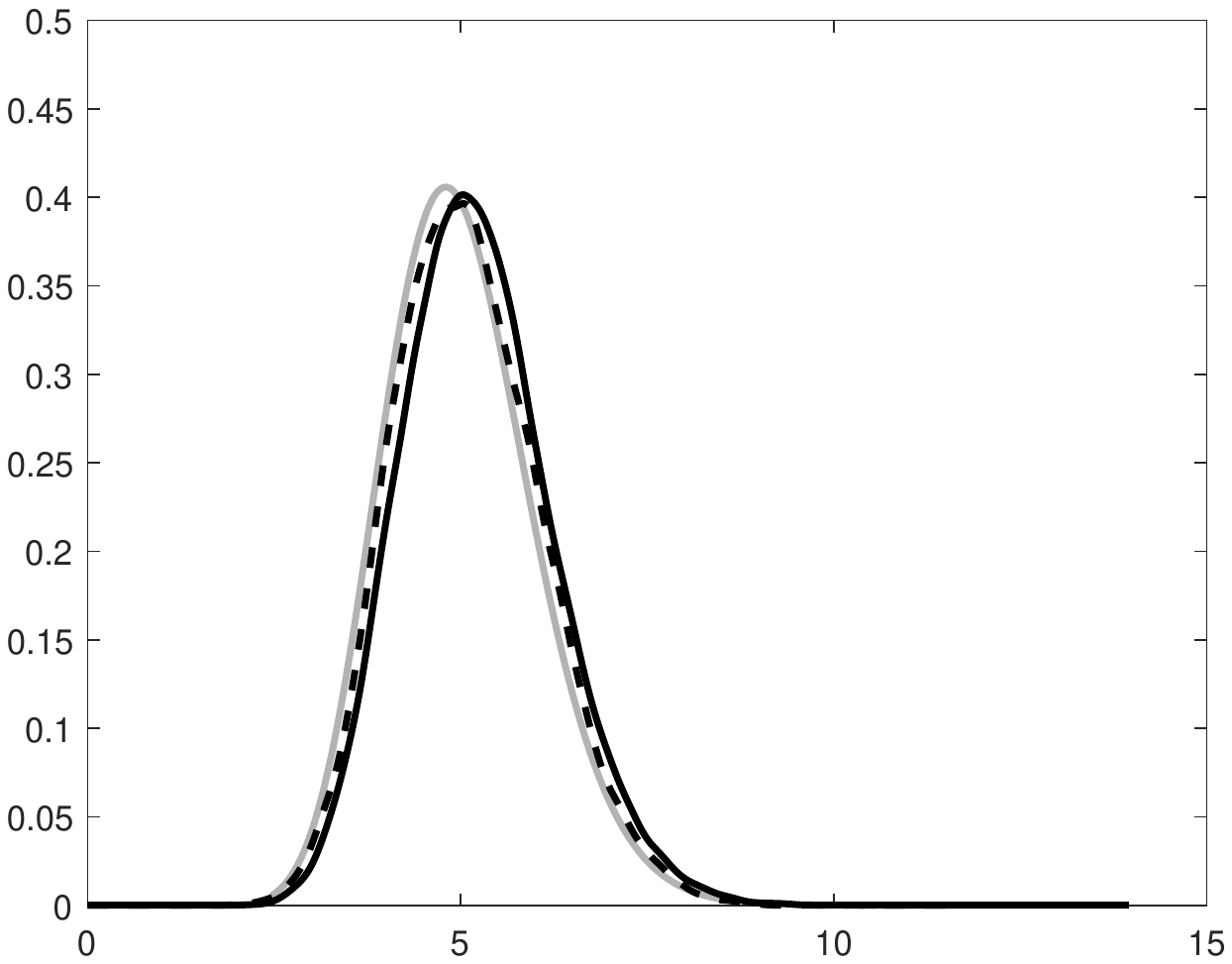}} & \hspace{+.1cm}
{\includegraphics[angle=0,width=0.6\textwidth,trim=120 280 140 230 ,
totalheight=.6\textwidth]{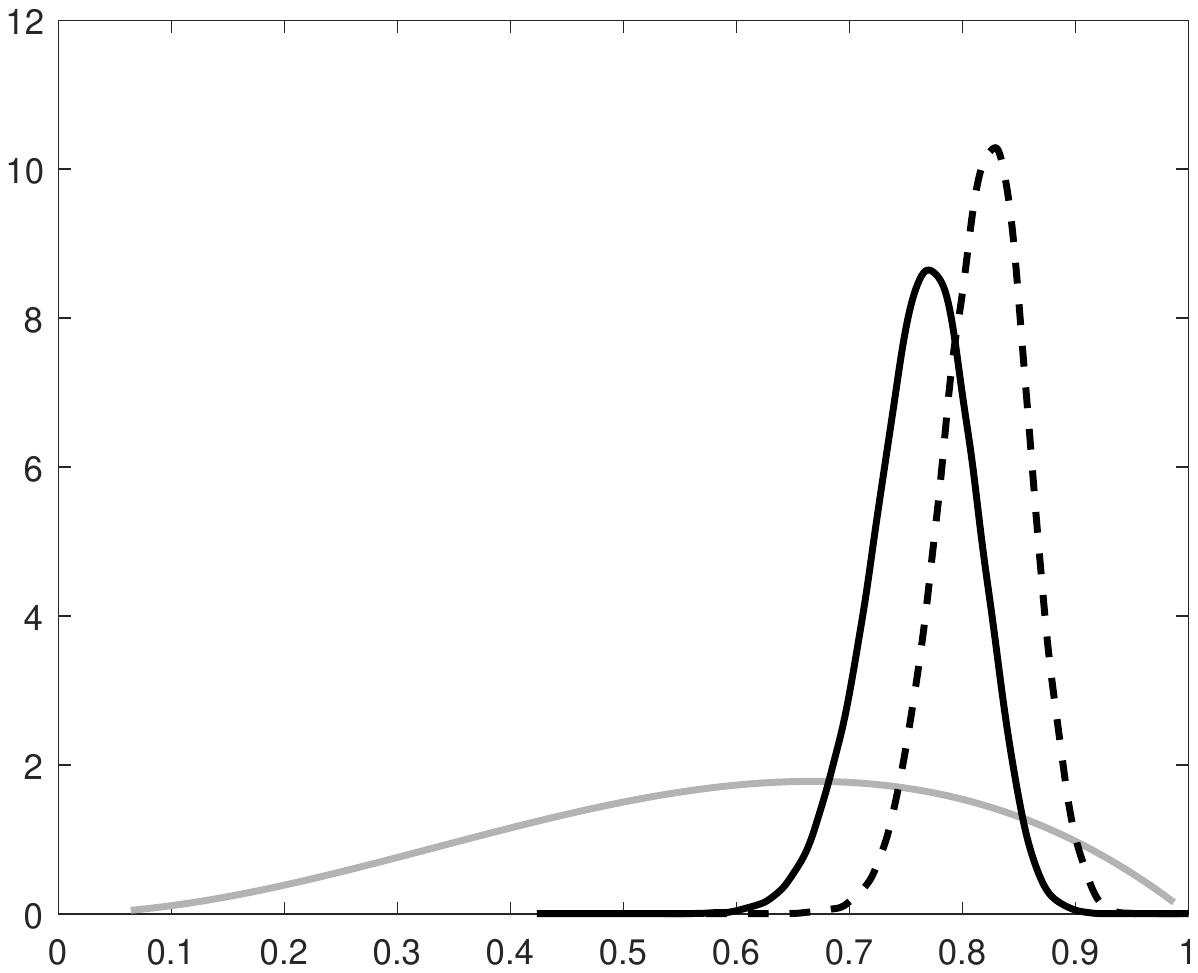}}\vspace{3ex}\\ 
{\Large (d)} & {\Large (e)} & {\Large (f)}\vspace{-.8cm}\\
{\includegraphics[angle=0,width=0.6\textwidth, trim=120 270 140 230 ,
totalheight=.6\textwidth]{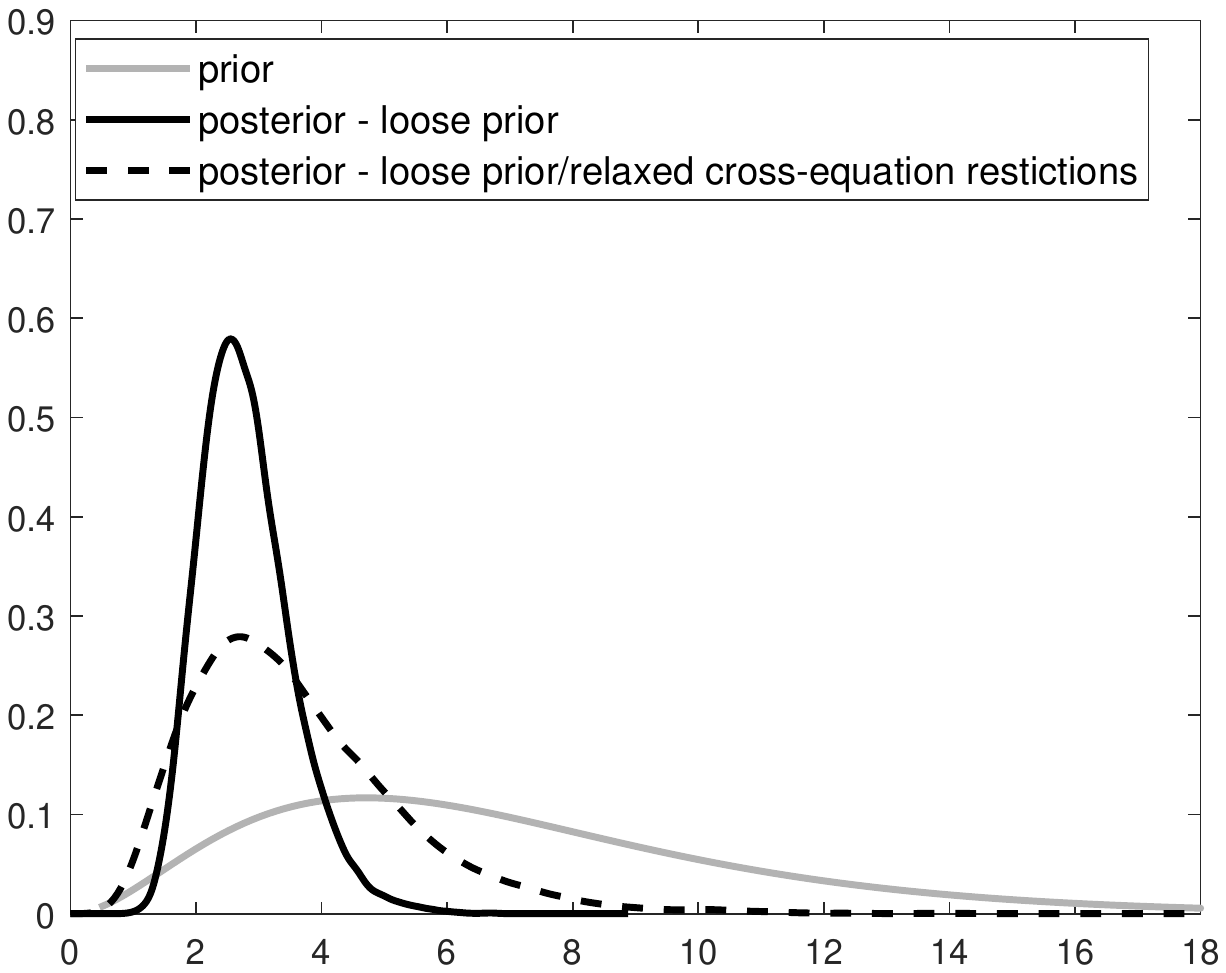}} &
\hspace{+.1cm}
{\includegraphics[angle=0,width=0.6\textwidth,trim=120 280 140 230 ,
totalheight=.6\textwidth]{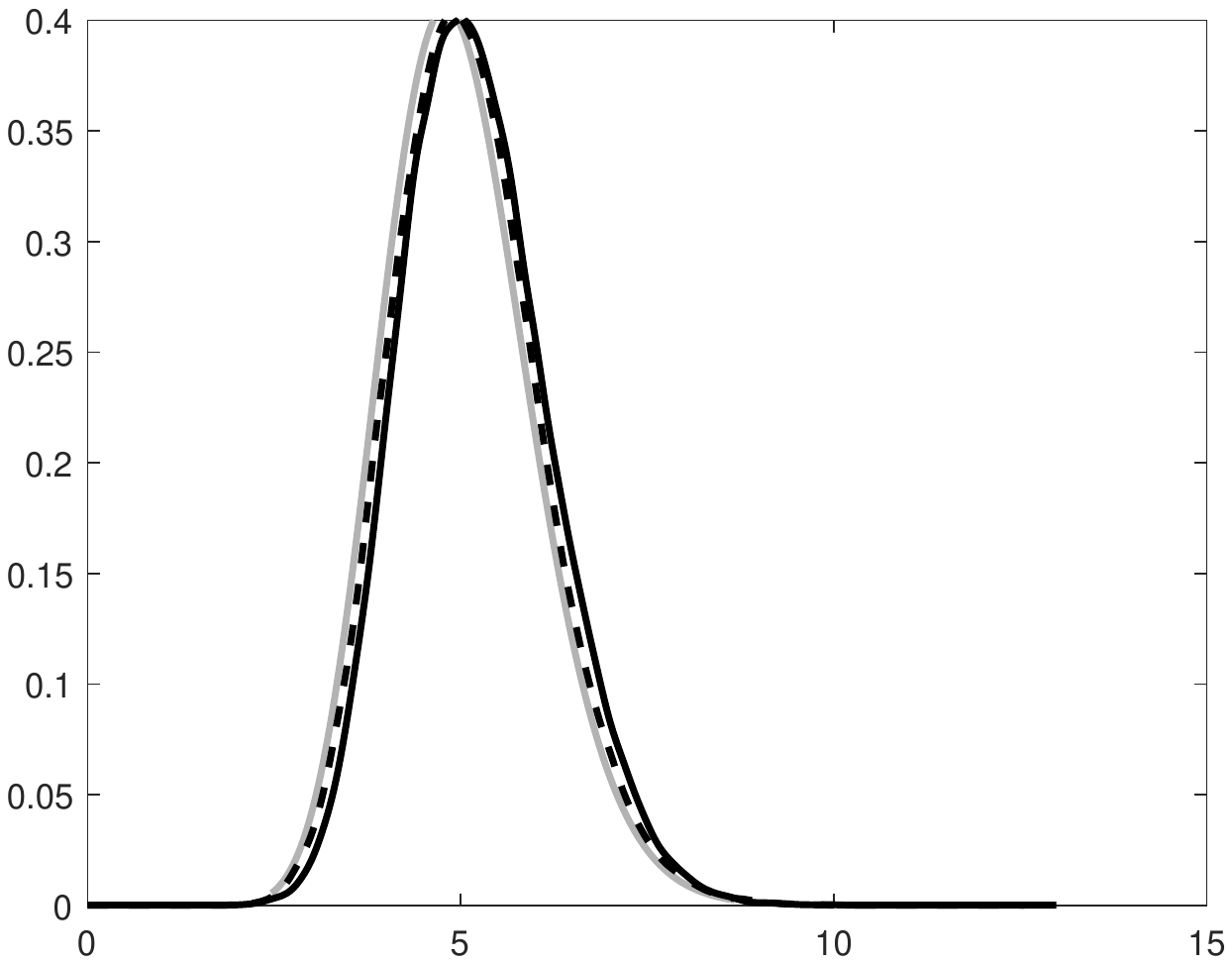}} & \hspace{+.1cm}
{\includegraphics[angle=0,width=0.6\textwidth,trim=120 280 140 230 ,
totalheight=.6\textwidth]{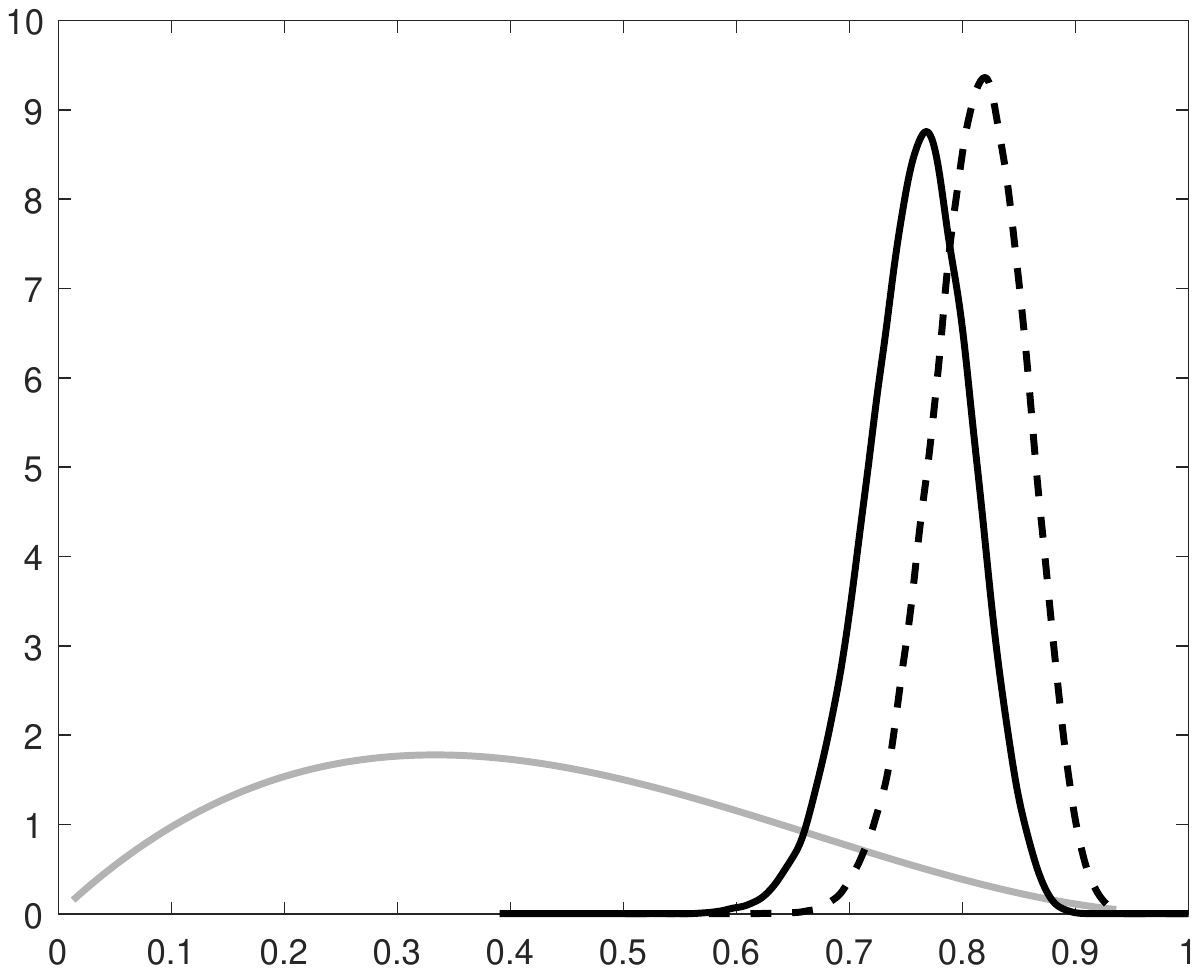}}\vspace{+10pt}\\
&  &  \\
\hline \hline
\noalign{\vspace{2ex}}
\multicolumn{3}{c}{\LARGE{Panel B: SW's dataset}}\\
\noalign{\vspace{2ex}} \hline 
%\\ 
\noalign{\vspace{1ex}}
%\cline{1-3} 
{\Large{Investment adjustment cost ($\kappa$)}} & {\Large{Elasticity capital utilization cost ($\zeta$)}} & {\Large{AR(1) coefficient investment shock ($\rho$)}}\\
\noalign{\vspace{1ex}}
\hline
\noalign{\vspace{2ex}}
{\Large (g)} & {\Large (h)} & {\Large (i)} \vspace{-1cm}\\
{\includegraphics[angle=0,width=0.6\textwidth, trim=120 270 140 230 ,
totalheight=.6\textwidth]{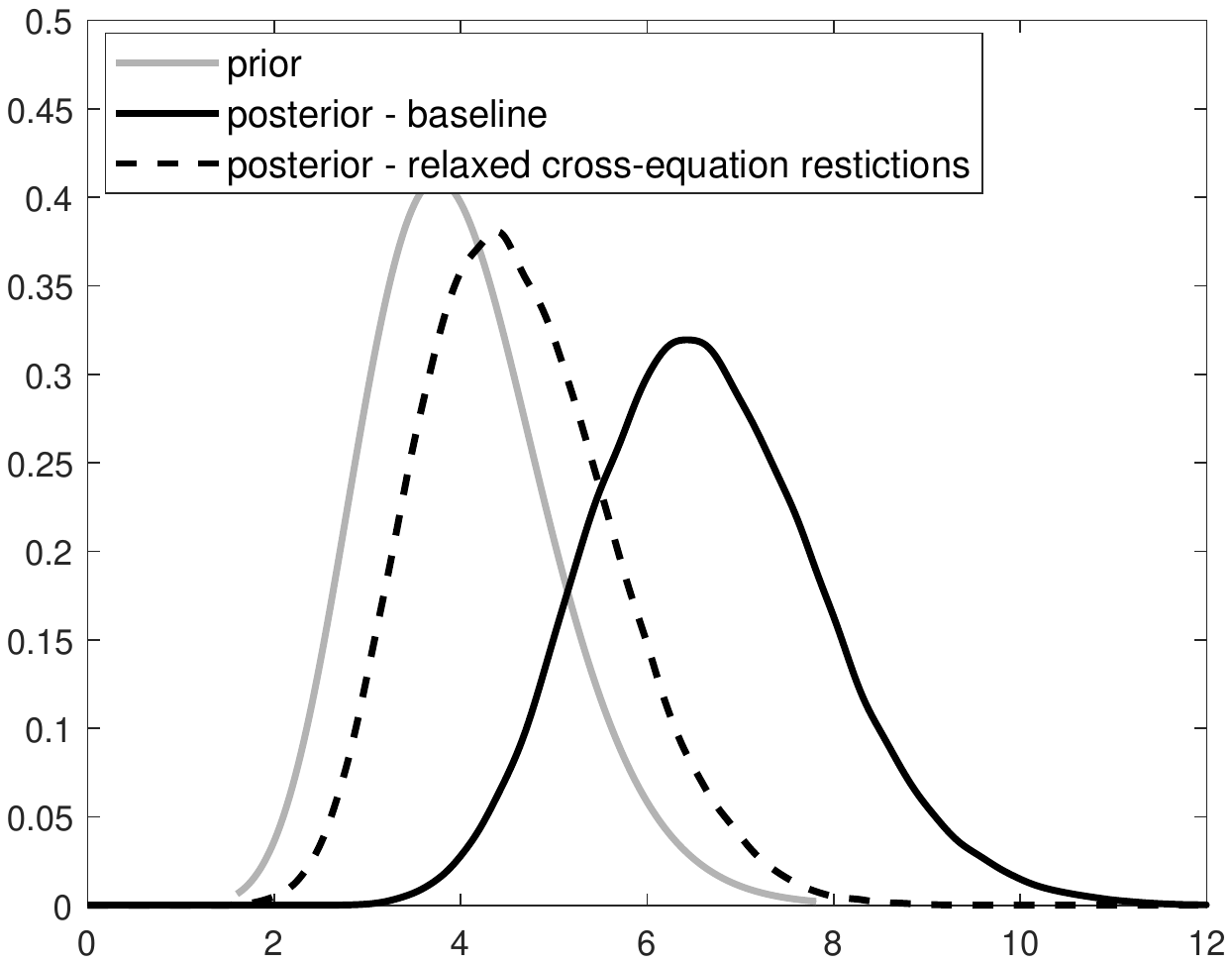}} &
\hspace{+.1cm}
{\includegraphics[angle=0,width=0.6\textwidth,trim=120 280 140 230 ,
totalheight=.6\textwidth]{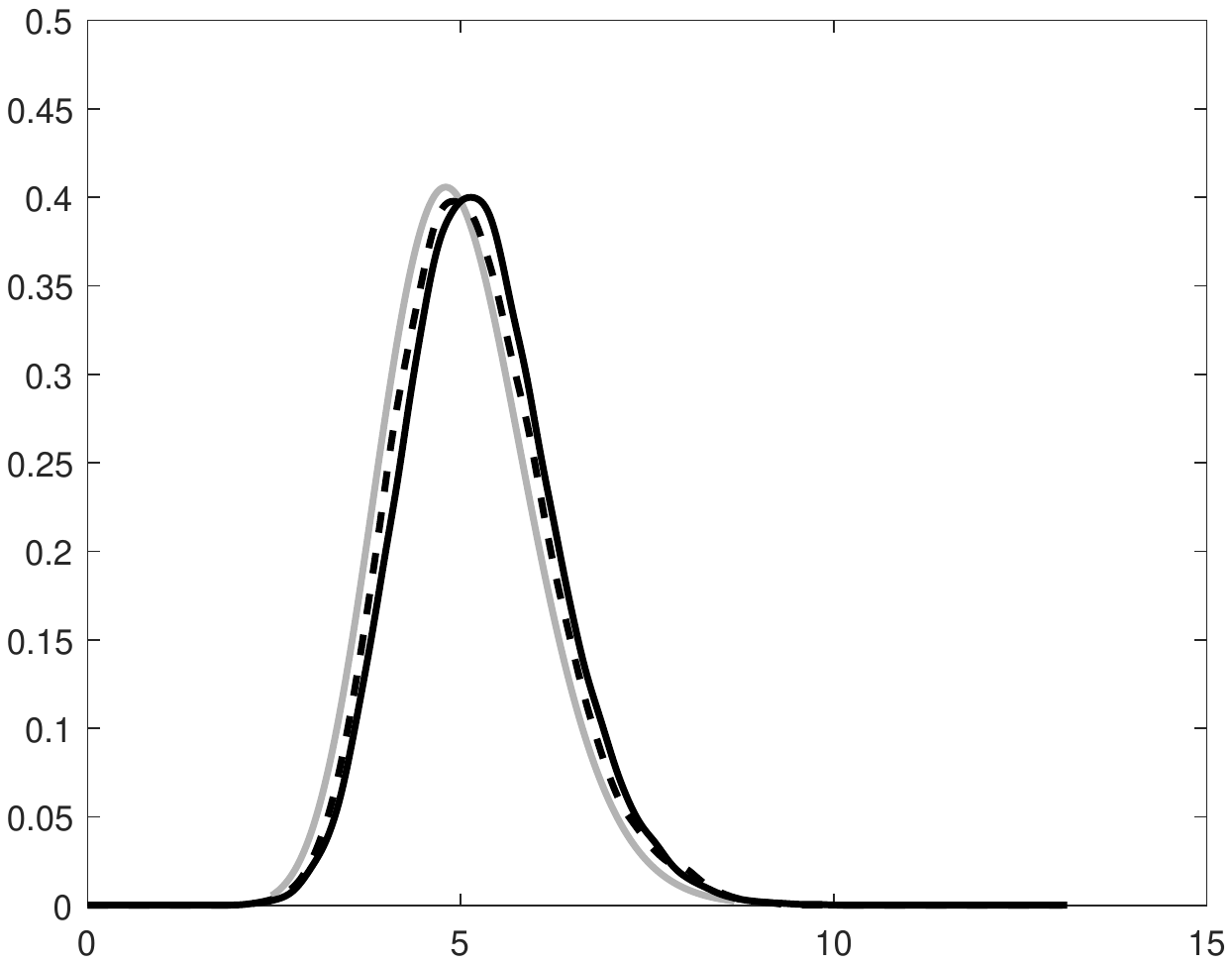}} & \hspace{+.1cm}
{\includegraphics[angle=0,width=0.6\textwidth,trim=120 280 140 230 ,
totalheight=.6\textwidth]{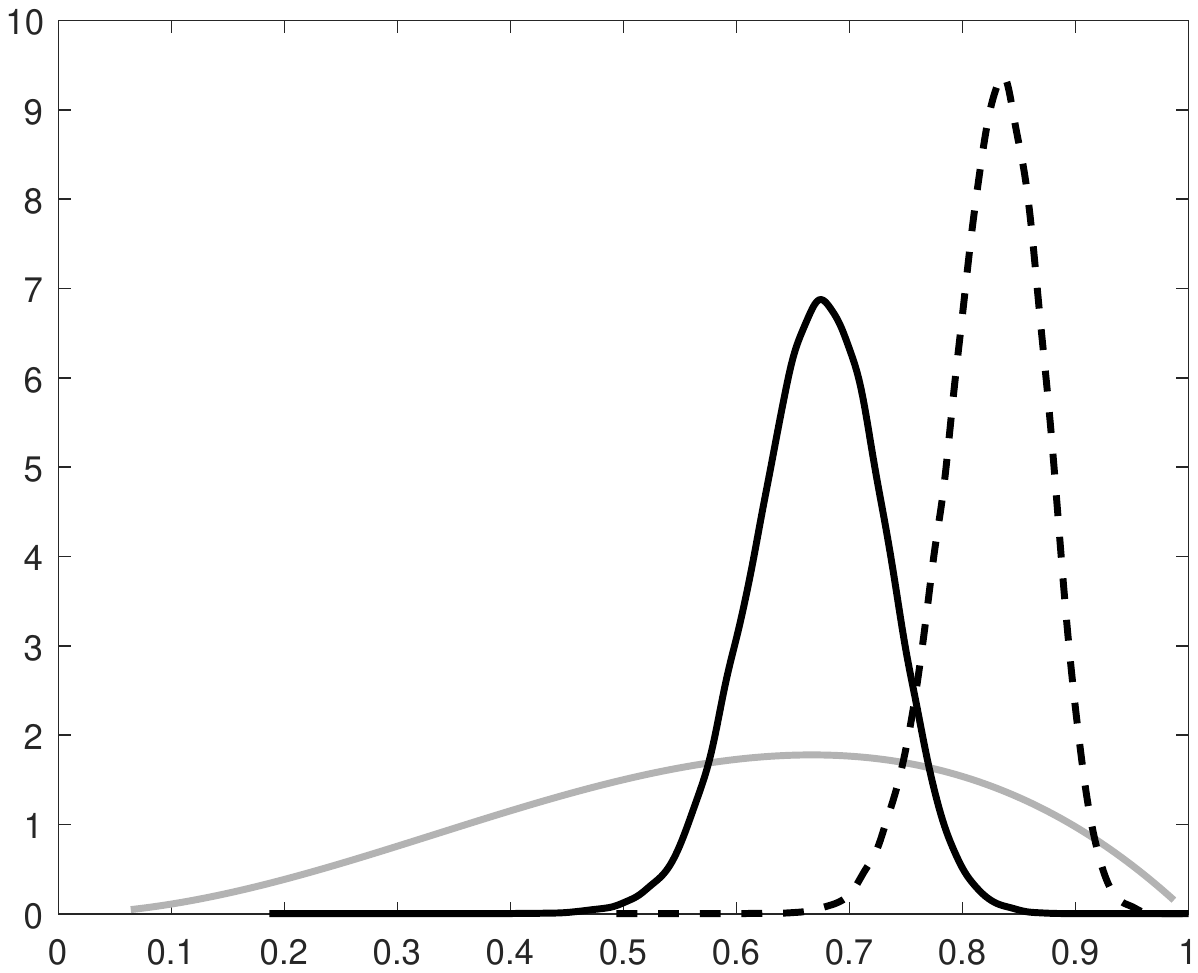}}\vspace{3ex}\\
{\Large (j)} & {\Large (k)} & {\Large (l)}\vspace{-.8cm}\\
{\includegraphics[angle=0,width=0.6\textwidth, trim=120 270 140 230 ,
totalheight=.6\textwidth]{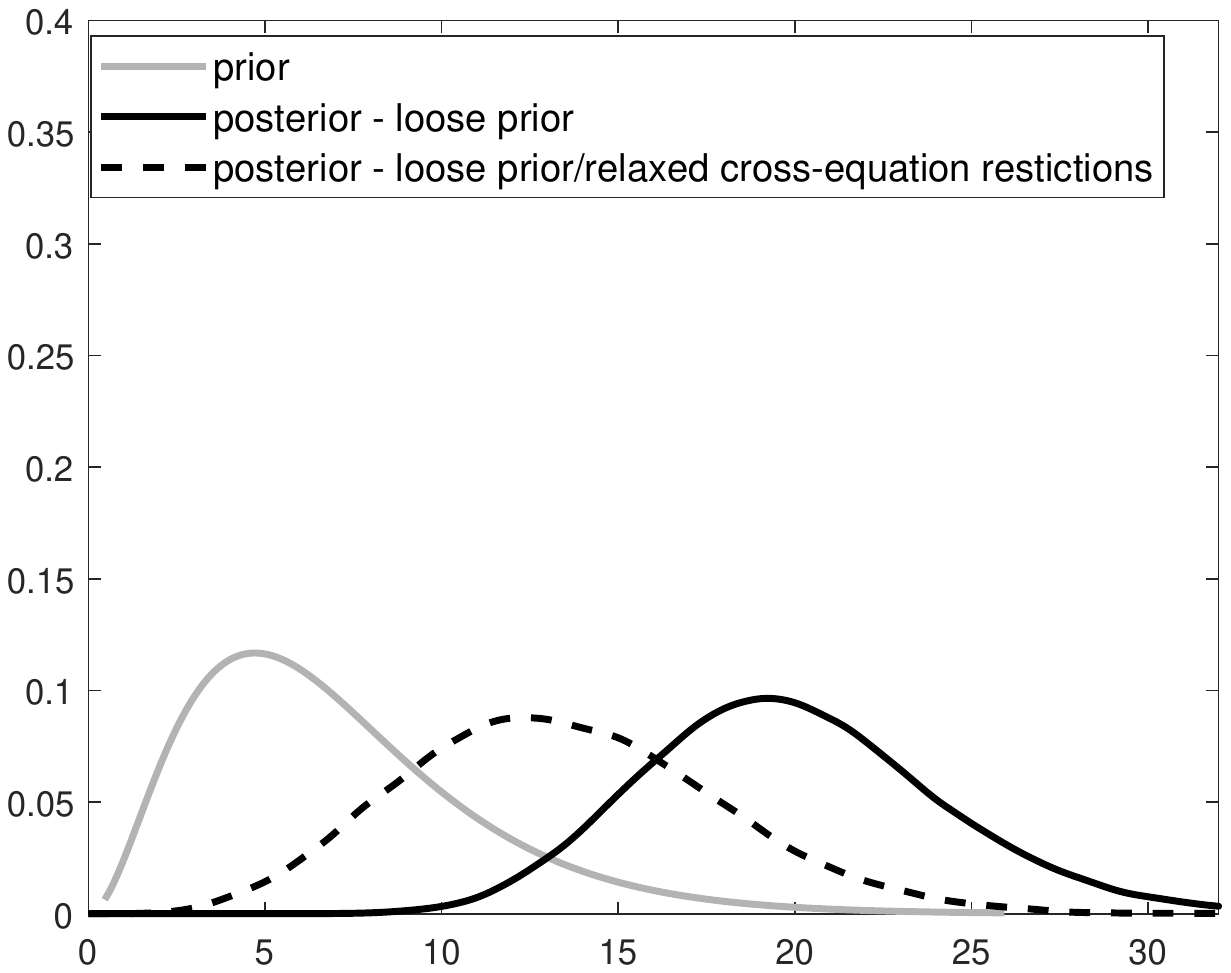}} &
\hspace{+.1cm}
{\includegraphics[angle=0,width=0.6\textwidth,trim=120 280 140 230 ,
totalheight=.6\textwidth]{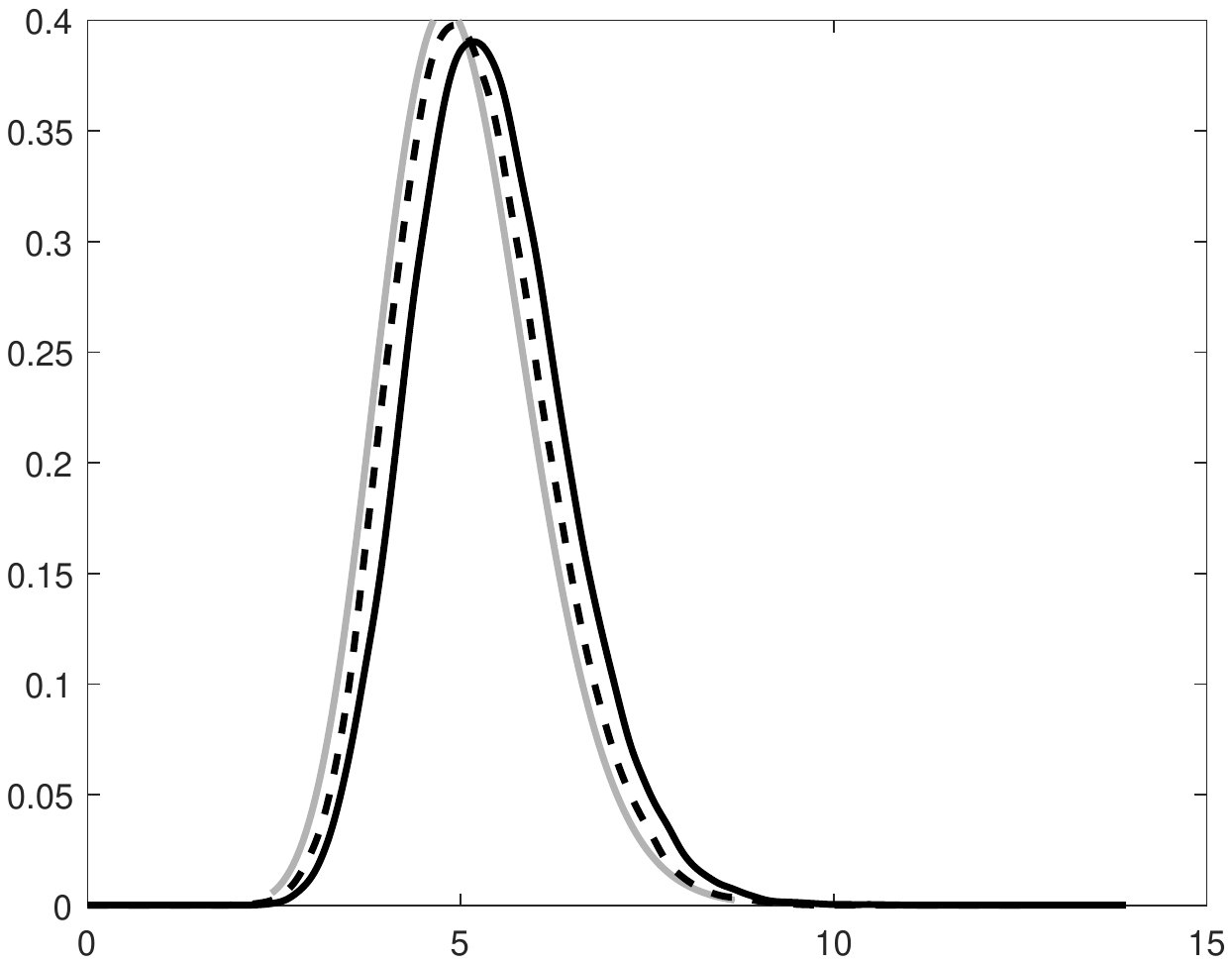}} & \hspace{+.1cm}
{\includegraphics[angle=0,width=0.6\textwidth,trim=120 280 140 230 ,
totalheight=.6\textwidth]{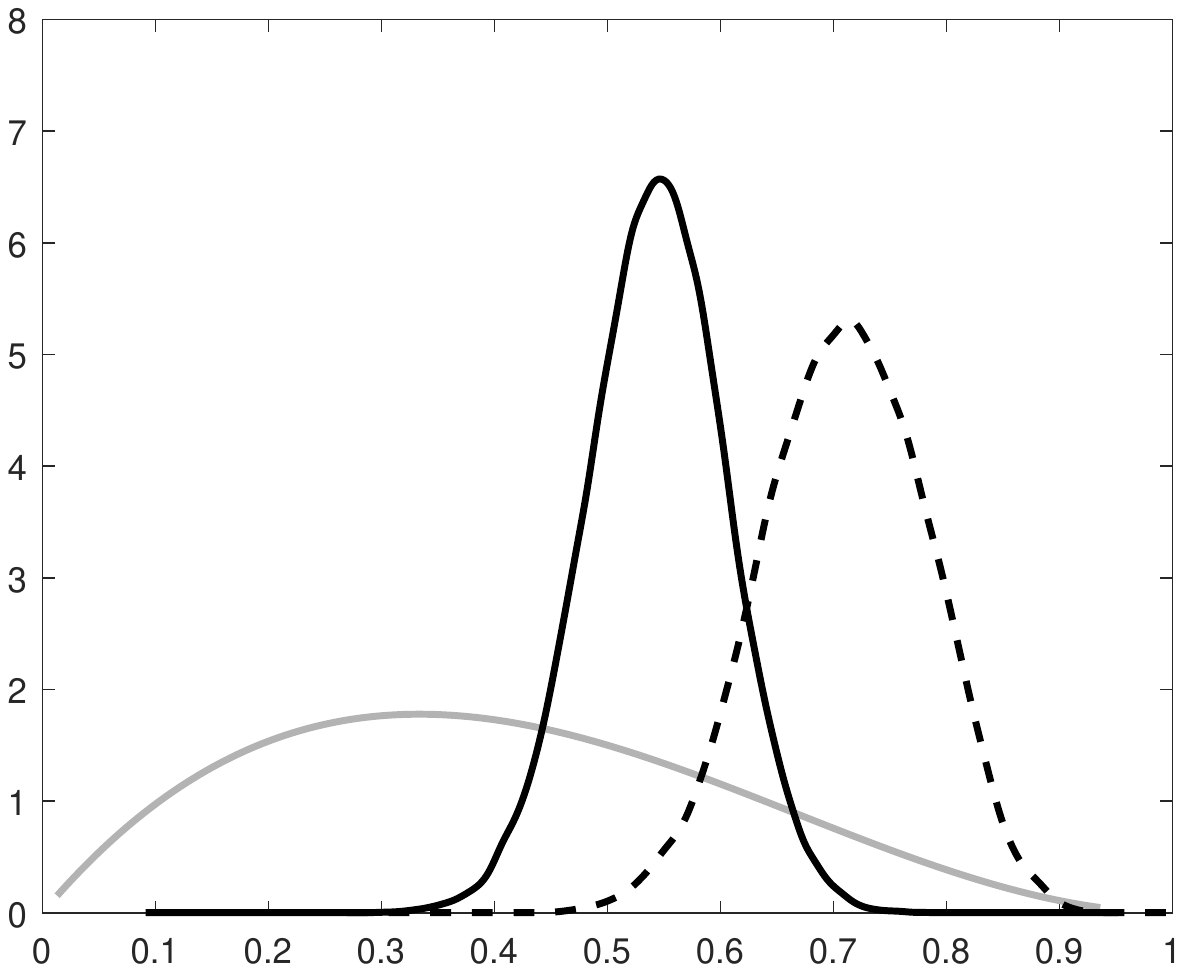}}\vspace{+10pt}\\
\hline \hline
\end{tabular}}
\caption{Prior-posterior plot: JPT's model}%
\label{fig:Prior_Posterior_JPTmodel}%
\end{figure}

In this Section we try to understand how DSGE models estimated with Bayesian methods obtain identification of the investment equation parameters, and compare this with our results. In comparison with our GMM single-equation approach, the DSGE system-based Bayesian estimation exhibits two main features that can help identification. The first is the use of priors in the Bayesian method. The second is the combination of the joint model dynamics of the variables in the system and the related cross-equation restrictions implied by rational expectations. We would like to distinguish the role of these two features. 
To this end, we perform the following exercise, displayed in Figure \ref{fig:Prior_Posterior_JPTmodel}: Panel A. Let us look first at the  top-left panel, regarding the parameter $\kappa$. As usually done in this type of models, the panel exhibits the difference between the prior (grey solid line) and the posterior (black solid line) to get a sense of how informative are the data. In addition, the panel features a third dashed line which is the posterior obtained by allowing the structural shocks of the model to be correlated in the estimation. The correlation among the shocks should `relax' the ties implied by the cross-equation restrictions in the model. Thus, a comparison between the posteriors with and without shock correlations - i.e., between the solid line and the dashed line - should give a sense of how much the structure of the model helps in achieving identification. The bottom-left panel displays the same lines as the top left-panel, but assuming a looser prior. Hence, a comparison between the top-left and the bottom-left panel should instead highlight the role of the prior in helping identification. Columns two and three in Figure \ref{fig:Prior_Posterior_JPTmodel}: Panel A show the same analysis for the parameters $\zeta$ and $\rho.$ Figure \ref{fig:Prior_Posterior_JPTmodel}: Panel A shows this exercise using JPT's model and data, while Figure \ref{fig:Prior_Posterior_JPTmodel}: Panel B does the same for JPT's model but with SW's data.\footnote{The results obtained when using the alternative combination, that is, SW model with SW's and JPT's data are similar and so we do not report them.}

Let us look at the implication of this analysis for the different parameters of interest. Regarding $\kappa$ (see the first column in Figure \ref{fig:Prior_Posterior_JPTmodel}), the top-left panel in Figure \ref{fig:Prior_Posterior_JPTmodel}: Panel A suggests that the cross-equation restrictions of the model do help in achieving identification because the dashed line is closer to the prior, while the solid line (where no correlations between the shocks are allowed) is instead much tighter. The bottom-left panel shows that the role of the prior in achieving identification is instead marginal because the posterior estimate is almost unchanged with a looser prior than in the top-left panel. However, the result is different if we estimate the same JPT model but with SW's data, as shown in Figure \ref{fig:Prior_Posterior_JPTmodel}: Panel B. While there is still an important role for cross-equation restrictions, the prior seems to be very important too, because the posterior in the bottom-left panel is much more dispersed than in the top-left one. Moreover, the SW's data set implies a larger value, i.e. posterior mode, for $\kappa.$

Regarding $\zeta$ (see the second column in Figure \ref{fig:Prior_Posterior_JPTmodel}), the results are very consistent with our GMM estimation: $\zeta$ is not identified by the data, and the identification is fully due to the prior for both data sets.\footnote{This is also consistent with JPT, who report a lack of identification of this parameter in their model. The Fischer information matrix analysis in DYNARE \citep[see][]{Iskrev_JME10} also points to the fact that $\zeta$ is not very well identified, contrary to $\kappa$ and $\rho.$} Moreover, this is consistent with our analysis in the previous Section \ref{subsec: parameters}, where we show that changes in the value of $\zeta$ are not affecting much the dynamics of the model, thus possibly impairing identification.

Regarding $\rho$ (see the third column in Figure \ref{fig:Prior_Posterior_JPTmodel}), the two Panels suggest that the data are quite informative and the parameter is very well identified. Both posteriors - with and without allowing for correlated shocks - move away distinctly from the prior and are quite peaked. The cross-equation restrictions call for a smaller value of this parameter, shifting the posterior towards the left, while SW's data set implies a lower value for the parameter and a relatively less peaked posterior.
The persistence parameter of the investment shock is thus well identified to be high, even if we relax the cross-equation restrictions and use a loose prior. We conjecture this to be due to the fact that the model wants to match the very persistent dynamics of the observable macroeconomic time series used in the Bayesian estimation. In contrast, our estimation procedure suggests that the investment equation is consistent with any value of $\rho,$ using fewer observables with respect to the system-based estimation. Interestingly, if anything, the GMM approach suggests a lower value of $\rho$, because, as noted earlier, with JPT's data Figures \ref{fig: mikusheva} (b) and (d) and Figure \ref{fig: Figure ext inst JPT} (c) shrinks the confidence set of $\rho$ to values roughly lower than 0.5. Moreover, in Section \ref{s: semistructural} we analyze the implications of different calibrations of $\rho$ for the identification of the reduced-form parameters in a semi-structural estimated equation, and we find that lower values of $\rho$ lead to sharper identification of the semi-structural parameters. If one wants to estimate $\rho$ more precisely from the data (e.g., from IRFs in DSGEs), one would need to make additional assumptions about the specification of the other variables that would allow us to identify the shock $\widetilde{\nu}_{t}$ in equation \eqref{eq: Basline_Euler_IAC_util_2} and hence estimate its autocorrelation directly. We cannot do that in a limited information approach.

Our results point out that identification of the structural parameters in DSGE models could be achieved through cross-equation restrictions implied by the joint dynamics of the full system. On the other hand, these assumptions could potentially lead to biased estimates whenever some other parts of the system are misspecified, as we show next.

\subsection{On Cross-equation Restrictions and Identification}\label{s: cross-eq-restrictions}
In this Subsection, we use a simple example to demonstrate how cross-equation restrictions from a system method can achieve identification of a model that is not identified using a single-equation GMM approach at the cost of losing robustness to misspecification.

Recall that GMM estimates the single equation
(\ref{eq: Basline_Euler_IAC_util_2}), where we have also assumed that $\beta$
and $\delta$ are known. For the purpose of this discussion it suffices to
simplify the exposition to the case of a single unknown parameter. Hence,
assume $\rho=0$ and $\kappa$ is known, so there is only one unknown parameter,
$\zeta$, and the model in equation (\ref{eq: Basline_Euler_IAC_util_2}) can be written as
\begin{equation}
E_{t}y_{t+2}=\zeta E_{t}x_{t+1}+\widetilde{\nu}_{t}\label{eq: very simple}%
\end{equation}
where $y_{t+2}:=\kappa\left(  \Delta\widetilde{i}_{t}-\left(  \beta+\phi
_{q}\right)  \Delta\widetilde{i}_{t+1}+\beta\phi_{q}\Delta\widetilde{i}%
_{t+2}\right)  +\widetilde{r}_{t}^{p}$ and $x_{t}:=\phi_{k}\widetilde{u}_{t}$,
with $\phi_{k}=1-\left(  1-\delta\right)  \beta.$

Now, $\zeta$ is identified in (\ref{eq: very simple}) if and only if
$var\left(  E_{t}x_{t+1}\right)  >0$. But in a limited-information setting, we do not observe $E_{t}x_{t+1},$ so we have to instrument for it using
predetermined variables $Z_{t}$ that belong to the information set at time
$t-1$. Specifically, the corresponding single-equation GMM regression for
(\ref{eq: very simple}) is%
\begin{equation}
y_{t+2}=\zeta x_{t+1}+\underbrace{\left[  \widetilde{\nu}_{t}+y_{t+2}%
-E_{t}y_{t+2}-\zeta\left(  x_{t+1}-E_{t}x_{t+1}\right)  \right]  }_{\xi_{t}%
},\label{eq: very simple IV}%
\end{equation}
with $E_{t-1}\xi_{t}=0,$ and any predetermined variable is a valid instrument for $x_{t+1}$. So, for the (single-equation) GMM approach to identify $\zeta$ it is necessary that $var\left(  E_{t-1}x_{t+1}\right)  >0$.

It is possible to come up with examples where a system method will identify
$\zeta$ while the single-equation GMM approach will not. Suppose
\begin{equation}
x_{t}=\omega_{t}+\theta\omega_{t-1},\quad\left\vert \theta\right\vert
<1,\label{eq: x MA}%
\end{equation}
(an invertible first-order moving average process) and $\omega_{t}$ (the
structural shock driving capacity utilization $x_{t}=\phi_{k}\widetilde{u}_{t}$)
is orthogonal to the investment-specific technology shock $\widetilde{\nu}_{t}$. Equation (\ref{eq: x MA}) implies
that $E_{t}x_{t+1}=\theta\omega_{t}$, while $E_{t-1}x_{t+1}=0$. So, $\zeta$
can be identified if we use the additional equation (\ref{eq: x MA}), but it
is not identified from a single-equation approach that does not make enough
assumptions to pin down $E_{t}x_{t+1}$.

In terms of implementation, because of the triangular nature of this simple
example, i.e., because (\ref{eq: x MA}) does not involve $\widetilde{\nu}_{t}$
or $y_{t}$, we can demonstrate how identification works as
follows. First estimate (\ref{eq: x MA}) to obtain $\omega_{t}$ and $\theta$,
next, compute $z_{t}:=\theta\omega_{t}=E_{t}x_{t+1}$, and finally, estimate
$\zeta$ from the regression%
\begin{equation}
y_{t+2}=\zeta z_{t}+\underbrace{\left(  \widetilde{\nu}_{t}+y_{t+2}%
-E_{t}y_{t+2}\right)  }_{e_{t}}\text{.}\label{eq: FI regression}%
\end{equation}
Unless $\theta=0$ (rank condition), which would imply $z_{t}=0$ for all $t$,
the above regression identifies $\zeta$, so a system\ analysis will produce
bounded confidence sets, while the single-equation GMM analysis based on
(\ref{eq: very simple}) that uses only predetermined variables to instrument
for $x_{t+1}$ will produce unbounded confidence sets.

\paragraph{Misspecification}

The increased precision of the system approach\ comes at the cost of lower
robustness to misspecification. Suppose the true law of motion for $x_{t}$
were given by%
\begin{equation}
x_{t}=\gamma x_{t-1}+\omega_{t},\label{eq: x AR}%
\end{equation}
i.e., an AR(1) instead of an MA(1). Using (\ref{eq: x MA}) instead of
(\ref{eq: x AR}), one would get an inconsistent estimate of $E_{t}x_{t+1},$
say $z_{t}^{\ast}=\theta^{\ast}\omega_{t}^{\ast}$, where $\theta^{\ast}%
,\omega_{t}^{\ast}$ are the pseudo-true values of $\theta,\omega_{t}$ in the
MA(1) specification (\ref{eq: x MA}), when the data is generated according to
(\ref{eq: x AR}), instead of the true $z_{t}=\gamma x_{t}$ from
(\ref{eq: x AR}). So, instead of using (\ref{eq: FI regression}), a
misspecified system approach would be estimating $\zeta$ from the incorrect
regression%
\begin{equation}\label{eq: misspecified regression}
y_{t+2}=\zeta z_{t}^{\ast}+\underbrace{\left[  \zeta\left(  z_{t}-z_{t}^{\ast
}\right)  +e_{t}\right]  }_{e_{t}^{\ast}}, \qquad cov(z_t^*,e_t^*)\neq 0,
\end{equation}
which suffers from omitted variable bias because $z_{t}^{\ast}$ correlates with
$e_{t}^{\ast},$ see Appendix \ref{app: misspecified derivation}. Thus, the system estimate of $\zeta$ will be biased.

A DSGE model allows us to use cross-equation restrictions to determine
$E_{t}x_{t+1}$ under rational expectations, see Footnote
\ref{footnote: Sargent}. In this present simple example, one may think that
we are not actually using any cross-equation restrictions because
$E_{t}x_{t+1}=\theta\omega_{t}$ does not involve the structural parameter
$\zeta$ of the original target equation (\ref{eq: very simple}). However, in
more general (e.g., non-triangular) settings where $x_{t}$ is allowed to be
simultaneously determined with $y_{t}$, $E_{t}x_{t+1}$ will depend also on
$\zeta$, and system estimation will indeed impose cross-equation restrictions.

\section{Conclusions} \label{s: concl}

We assess the empirical performance of the most commonly employed specification of investment behavior in modern operational DSGE models employed for policy analysis. The specification of the investment block of these models is based on the investment adjustment cost specification proposed by \cite{Christiano_Eichenbaum_Evans_2005} together with variable capital utilization. We employ the same limited-information methodology that was used in the extant literature on the empirical performance of other key parts of DSGE models, such as the New Keynesian Phillips curve, the monetary policy rule, and the consumption Euler equation. 
%Thus, this paper fills a gap in the literature by carrying out the same exercise for another fundamental part of the DSGE models, that is, the investment block, which is the other main component of aggregate demand.
%From an econometric point of view, this work falls in the same research agenda of earlier works by the same co-authors on the NKPC \citep{KleibergenMavroeidis09jbes,MM14},  the Taylor rule \citep{Mavr_AER10} and the Euler equation of consumption \citep{AMM2021}, by employing state-of-the-art robust weak-IV methods. 

Our results are mixed. On the one hand, the investment equation is not rejected by the data. On the other hand, there is little information that aggregate data could provide to identify the key structural parameters of the investment block of current medium-scale DSGE models. Hence, this identification has to come from the cross-equation restrictions that other parts of the model imply. However, semi-structural estimation shows that investment is insensitive to changes in capital utilization and the real interest rate. In fact, the semi-structural parameters - the elasticity of investment to the real interest rate and capital utilization - are quite tightly estimated to be near zero when the persistence of the investment-specific shock is assumed to be low. Finally, similar to the results in \cite{AMM2021}, structural change is not as informative for the identification as it was found to be for the NKPC, and there is no evidence of parameter instability.

Finally, we investigate how DSGE models estimated with Bayesian methods obtain identification of the investment Euler equation parameters, in contrast to our methodology. Using the JPT model, we find that the investment adjustment cost parameter is mainly identified by the model cross-equation restrictions, while the role of the prior in achieving identification is marginal. The elasticity of capital utilization is not identified by the data. This is consistent with the evidence in JPT and with our analysis, which shows that changes in the value of $\zeta$ are not affecting much the dynamics of the model, thus possibly impairing identification. The data are quite informative, instead, on the persistence parameter of the investment-specific shock which is thus well identified to be high, even if we relax the cross-equation restrictions and use a loose prior. We conjecture this to be due to the fact that the model wants to match the very persistent dynamics of the observable macroeconomic time series used in the Bayesian estimation. 
In conclusion, our results suggest that identification of the structural parameters in DSGE models could be achieved through cross-equation restrictions implied by the joint dynamics of the full system. 
However, these assumptions could potentially lead to biased estimates whenever some other parts of the system are misspecified.

% \newpage
\section*{Appendix}
\appendix

\section{Derivation of equation (\ref{eq: estimated}) \label{appsec: derivation}}

The representative household chooses ${I_{t}}${, }${\hat{K}_{t+1}}${, }${B_{t+1}}$,
and ${u_{t}}$ to maximise (\ref{eq: UF}) under the period-by-period budget
constraint (\ref{eq: bc}) and capital accumulation equation (\ref{eq: CAE}).
The first-order conditions are
\begin{align*}
I_{t}:\hspace{12pt}1= &  \nu_{t}Q_{t}\left[  1-\mathcal{S}\left(  \dfrac
{I_{t}}{I_{t-1}}\right)  -\mathcal{S}^{\prime}\left(  \dfrac{I_{t}}{I_{t-1}%
}\right)  \left(  \dfrac{I_{t}}{I_{t-1}}\right)  \right]  \\
&  +\beta E_{t}\left\{  \dfrac{\lambda_{t+1}}{\lambda_{t}}\mathcal{S}^{\prime
}\left(  \dfrac{I_{t+1}}{I_{t}}\right)  \left(  \dfrac{I_{t+1}}{I_{t}}\right)
^{2}\nu_{t+1}Q_{t+1}\right\}  ,\\
\hat{K}_{t+1}:\hspace{12pt}Q_{t}= &  \beta E_{t}\left\{  \dfrac{\lambda_{t+1}%
}{\lambda_{t}}\left[  r_{t+1}^{k}u_{t+1}-a(u_{t+1})+Q_{t+1}\left(
1-\delta\right)  \right]  \right\}  ,\\
B_{t+1}:\hspace{12pt}1= &  \beta E_{t}\left\{  \dfrac{\lambda_{t+1}}%
{\lambda_{t}}\dfrac{R_{t}}{\pi_{t+1}}\right\}  ,\ \text{and}\\
u_{t}:\hspace{12pt}r_{t}^{k}= &  a^{\prime}\left(  u_{t}\right),  
\end{align*}
where $Q_{t}$ denotes the marginal $\mathcal{Q}$, defined as the ratio of the
Lagrange multipliers associated with the capital accumulation equation and the
budget constraint $\left(  \lambda_{t}\right)  $, and $\pi_{t+1}$ is the
inflation rate in period $t+1$.

Log-linearizing the above first-order conditions around the non-stochastic
steady state yields
\begin{align}
\widetilde{q}_{t} &  =\kappa\left(  \widetilde{i}_{t}-\widetilde{i}%
_{t-1}\right)  -\beta\kappa\left(  E_{t}\widetilde{i}_{t+1}-\widetilde{i}%
_{t}\right)  -\widetilde{\nu}_{t},\label{eqn:Euler_IAC_util_Investment}\\
\widetilde{q}_{t} &  =E_{t}\widetilde{\lambda}_{t+1}-\widetilde{\lambda}%
_{t}+\beta\bar{r}^{k}E_{t}\widetilde{r}_{t+1}^{k}+\beta\left(  1-\delta
\right)  E_{t}\widetilde{q}_{t+1},\label{eqn:Euler_IAC_util_Capital}\\
E_{t}\widetilde{\lambda}_{t+1}-\widetilde{\lambda}_{t} &  =-(\widetilde{r}%
_{t}-E_{t}\widetilde{\pi}_{t+1}), \ \text{and}\label{eqn:Euler_IAC_util_Bond}\\
\widetilde{r}_{t}^{k} &  =\zeta\widetilde{u}_{t},\label{eq: rk}%
\end{align}
where lowercase letters with a tilde denote the respective log deviations of
the variables from their steady state. Although
(\ref{eqn:Euler_IAC_util_Investment})-(\ref{eq: rk}) can be estimated, the empirical literature
has struggled to find an appropriate proxy for $\tilde{q}_{t}$, the marginal $\mathcal{Q}$, which is
unobservable. \cite{Hayashi_1982} showed that under some regularity conditions the
average $\mathcal{Q}$ is equivalent to marginal $\mathcal{Q}$. However,
subsequent empirical studies have confirmed such regularity conditions to be
unsatisfactory, finding insignificant coefficients on average $\mathcal{Q}$.
Thus, we follow the treatment in \cite{Groth_Khan_2010} and get rid of $\tilde{q}_{t}$ from the log-linearized conditions. 
Substituting (\ref{eqn:Euler_IAC_util_Investment}) and
(\ref{eqn:Euler_IAC_util_Bond}) into (\ref{eqn:Euler_IAC_util_Capital}) yields
our preferred baseline investment Euler equation with IAC
\begin{align}
\widetilde{i}_{t}= &  \dfrac{1/\kappa}{1+\beta+\phi_{q}}\left[  \phi_{k}%
E_{t}\widetilde{r}_{t+1}^{k}-E_{t}\widetilde{r}_{t}^{p}+\widetilde{\nu}%
_{t}\right]  \nonumber\\
&  +\dfrac{1}{1+\beta+\phi_{q}}\widetilde{i}_{t-1}+\dfrac{\beta+\phi
_{q}(1+\beta)}{1+\beta+\phi_{q}}E_{t}\widetilde{i}_{t+1}-\dfrac{\beta\phi_{q}%
}{1+\beta+\phi_{q}}E_{t}\widetilde{i}_{t+2}-\dfrac{\phi_{q}/\kappa}%
{1+\beta+\phi_{q}}E_{t}\widetilde{\nu}_{t+1}%
,\label{eqn:Basline_Euler_IAC_util_1}%
\end{align}
where $\widetilde{r}_{t}^{p}$ denotes the log-deviation of ex-ante real
interest rate from steady state, i.e., $\widetilde{r}_{t}^{p}=\widetilde{r}%
_{t}-\widetilde{\pi}_{t+1}$, and $\phi_{q}=(1-\delta)/(1-\delta+\bar{r}^{k})$;
$\phi_{k}=\bar{r}^{k}/(1-\delta+\bar{r}^{k})$; and $\bar{r}^{k}=1/\beta
-1+\delta$.\footnote{Replacing $\bar{r}^{k}$ into the equations of $\phi_{q}$
and $\phi_{k}$ results in $\phi_{q}=\beta\left(  1-\delta\right)  $ and
$\phi_{k}=1-\beta\left(  1-\delta\right)  $ respectively.}

We can use (\ref{eq: rk}) to substitute out the rental rate of capital,
$\widetilde{r}_{t}^{k}$, which is an unobservable variable, with the capacity
utilization, $\widetilde{u}_{t}$, for which a time series is available. So,
(\ref{eqn:Basline_Euler_IAC_util_1}) becomes%
\begin{align*}
\widetilde{i}_{t}= &  \dfrac{1/\kappa}{1+\beta+\phi_{q}}\left[  \phi_{k}\zeta
E_{t}\widetilde{u}_{t+1}-E_{t}\widetilde{r}_{t}^{p}+\widetilde{\nu}%
_{t}\right]  %\nonumber
\\
&  +\dfrac{1}{1+\beta+\phi_{q}}\widetilde{i}_{t-1}+\dfrac{\beta+\phi
_{q}(1+\beta)}{1+\beta+\phi_{q}}E_{t}\widetilde{i}_{t+1}-\dfrac{\beta\phi_{q}%
}{1+\beta+\phi_{q}}E_{t}\widetilde{i}_{t+2}-\dfrac{\phi_{q}/\kappa}%
{1+\beta+\phi_{q}}E_{t}\widetilde{\nu}_{t+1}%
.%\label{eqn:Basline_Euler_IAC_util_2}%
\end{align*}
Note that this equation is the same as equation (\ref{eq: Basline_Euler_IAC_util_2}) in the main text, that is simply rewritten in first differences of the investment terms.
For any variable $x_{t}$ the rational expectations (RE) forecast error is
$\eta_{t|t-1}^{x}=x_{t}-E_{t-1}(x_{t})$, which implies that $E_{t}(x_{t+1}%
)=x_{t+1}-\eta_{t+1|t}^{x}.$ Moreover, $E_{t}\widetilde{\nu}_{t+1}%
=\rho\widetilde{\nu}_{t}$, since $\widetilde{\nu}_{t}\sim AR(1):\widetilde{\nu
}_{t}=\rho\widetilde{\nu}_{t-1}+\varepsilon_{t}^{v}.$ Finally, define
$E_{t}(\widetilde{i}_{t+2})=\widetilde{i}_{t+2}-\eta_{t+2|t}^{i}$, then
\begin{align*}
\widetilde{i}_{t}= &  \dfrac{1/\kappa}{1+\beta+\phi_{q}}\left[  \phi_{k}%
\zeta\widetilde{u}_{t+1}-\phi_{k}\zeta\eta_{t+1|t}^{u}-\widetilde{r}_{t}%
^{p}+\eta_{t+1|t}^{\pi}\right]  +\dfrac{1}{1+\beta+\phi_{q}}\widetilde{i}%
_{t-1}\\
&  +\dfrac{\beta+\phi_{q}(1+\beta)}{1+\beta+\phi_{q}}\left(  \widetilde{i}%
_{t+1}-\eta_{t+1|t}^{i}\right)  -\dfrac{\beta\phi_{q}}{1+\beta+\phi_{q}}\left(
\widetilde{i}_{t+2}-\eta_{t+2|t}^{i}\right)  +\dfrac{1/\kappa\left(  1-\phi
_{q}\rho\right)  }{1+\beta+\phi_{q}}\widetilde{\nu}_{t}\text{,}%
\end{align*}
or%
\begin{align}
\widetilde{i}_{t}=&\dfrac{1/\kappa}{1+\beta+\phi_{q}}\left[  \phi_{k}%
\zeta\widetilde{u}_{t+1}-\widetilde{r}_{t}^{p}\right]  +\dfrac{1}{1+\beta
+\phi_{q}}\widetilde{i}_{t-1}+\dfrac{\beta+\phi_{q}(1+\beta)}{1+\beta+\phi
_{q}}\widetilde{i}_{t+1}\nonumber\\
&-\dfrac{\beta\phi_{q}}{1+\beta+\phi_{q}}%
\widetilde{i}_{t+2}+\varepsilon_{t}, \label{app eq: euler eq}%
\end{align}
where $
\varepsilon_{t}=  \tfrac{1/\kappa}{1+\beta+\phi_{q}}\left[  -\phi_{k}%
\zeta\eta_{t+1|t}^{u}+\eta_{t+1|t}^{\pi}\right]  -\tfrac{\beta+\phi_{q}(1+\beta
)}{1+\beta+\phi_{q}}\eta_{t+1|t}^{i}+\tfrac{\beta\phi_{q}}{1+\beta+\phi_{q}}%
\eta_{t+2|t}^{i} 
+\tfrac{1/\kappa\left(  1-\phi_{q}\rho\right)  }{1+\beta+\phi_{q}%
}\widetilde{\nu}_{t}\text{.}$

Using the facts that
$
\widetilde{i}_{t} =\tfrac{1}{1+\beta+\phi_{q}}\widetilde{i}_{t}%
+\tfrac{\beta+\phi_{q}}{1+\beta+\phi_{q}}\widetilde{i}_{t}$, and $
\tfrac{\beta+\phi_{q}(1+\beta)}{1+\beta+\phi_{q}}\widetilde{i}_{t+1} 
=\tfrac{\beta+\phi_{q}}{1+\beta+\phi_{q}}\widetilde{i}_{t+1}+\tfrac{\beta
\phi_{q}}{1+\beta+\phi_{q}}\widetilde{i}_{t+1},%
$
the terms in $\widetilde{i}$ could be written as first difference, so
equation (\ref{app eq: euler eq}) becomes
\[
\dfrac{1}{1+\beta+\phi_{q}}\Delta\widetilde{i}_{t}=\dfrac{1/\kappa}%
{1+\beta+\phi_{q}}\left[  \phi_{k}\zeta\widetilde{u}_{t+1}-\widetilde{r}%
_{t}^{p}\right]  +\dfrac{\beta+\phi_{q}}{1+\beta+\phi_{q}}\Delta
\widetilde{i}_{t+1}-\dfrac{\beta\phi_{q}}{1+\beta+\phi_{q}}\Delta
\widetilde{i}_{t+2}+\varepsilon_{t}\text{,}%
\]
or%
\begin{equation}
\Delta\widetilde{i}_{t}=\dfrac{\phi_{k}}{\kappa}\zeta\widetilde{u}%
_{t+1}-\dfrac{1}{\kappa}\widetilde{r}_{t}^{p}+\left(  \beta+\phi_{q}\right)
\Delta\widetilde{i}_{t+1}-\beta\phi_{q}\Delta\widetilde{i}_{t+2}+\left(
1+\beta+\phi_{q}\right)  \varepsilon_{t}\text{.}\label{basedeltaiIAC}%
\end{equation}
We then just eliminate $\widetilde{\nu}_{t}$ in the error term $\varepsilon
_{t}$, again by lagging (\ref{basedeltaiIAC}), multiplying it by $\rho$, which
results in
\[
\rho\Delta\widetilde{i}_{t-1}=\dfrac{\rho\phi_{k}}{\kappa}\zeta\widetilde{u}%
_{t}-\dfrac{\rho}{\kappa}\widetilde{r}_{t-1}^{p}+\rho\left(  \beta+\phi
_{q}\right)  \Delta\widetilde{i}_{t}-\rho\beta\phi_{q}\Delta\widetilde{i}%
_{t+1}+\rho\left(  1+\beta+\phi_{q}\right)  \varepsilon_{t-1}\text{,}%
\]
and subtracting the result from (\ref{basedeltaiIAC}), such that%
\begin{align*}
\Delta\widetilde{i}_{t}-\rho\Delta\widetilde{i}_{t-1} = &\  \dfrac{\phi_{k}%
}{\kappa}\zeta\left(  \widetilde{u}_{t+1}-\rho\widetilde{u}_{t}\right)
-\dfrac{1}{\kappa}\left(  \widetilde{r}_{t}^{p}-\rho\widetilde{r}_{t-1}%
^{p}\right)  +\left(  \beta+\phi_{q}\right)  \left(  \Delta\widetilde{i}%
_{t+1}-\rho\Delta\widetilde{i}_{t}\right)& \nonumber
\\
&  -\beta\phi_{q}\left(  \Delta\widetilde{i}_{t+2}-\rho\Delta\widetilde{i}%
_{t+1}\right)  +\left(  1+\beta+\phi_{q}\right)  \left(  \varepsilon_{t}%
-\rho\varepsilon_{t-1}\right)  \text{.}%\label{app eq: base_IAC_final}%
\end{align*}
Rearranging terms we obtain the baseline specification (\ref{eq: estimated})%
\begin{align*}
\left[  1+\rho\left(  \beta+\phi_{q}\right)  \right]  \Delta\widetilde{i}_{t}
=&\  \rho\Delta\widetilde{i}_{t-1}+\left(  \beta+\phi_{q}+\rho\beta\phi
_{q}\right)  \Delta\widetilde{i}_{t+1}-\beta\phi_{q}\Delta\widetilde{i}%
_{t+2}\\
&  +\dfrac{\phi_{k}}{\kappa}\zeta\widetilde{u}_{t+1}-\dfrac{\rho\phi_{k}%
}{\kappa}\zeta\widetilde{u}_{t}-\dfrac{1}{\kappa}\widetilde{r}_{t}^{p}%
+\dfrac{\rho}{\kappa}\widetilde{r}_{t-1}^{p}+\epsilon_{t}\text{,}%
\end{align*}
where%
\begin{equation}
\epsilon_{t}:=\left(  1+\beta+\phi_{q}\right)  \left(  \varepsilon_{t}%
-\rho\varepsilon_{t-1}\right)  \text{,}\label{app eq: error}%
\end{equation}
$\left(  \varepsilon_{t}-\rho\varepsilon_{t-1}\right) =\tfrac{1/\kappa}{1+\beta+\phi_{q}}\left[  -\phi_{k}\eta_{t+1|t}^{u}%
+\eta_{t+1|t}^{\pi}-\rho\left(  -\phi_{k}\eta_{t|t-1}^{u}+\eta_{t|t-1}^{\pi}\right)
\right]  -\tfrac{\beta+\phi_{q}(1+\beta)}{1+\beta+\phi_{q}}\left(  \eta
_{t+1|t}^{i}-\rho\eta_{t|t-1}^{i}\right)$\newline $  
+\tfrac{\beta\phi_{q}}{1+\beta+\phi_{q}}\left(  \eta_{t+2|t}^{i}%
-\rho\eta_{t+1|t}^{i}\right)  +\tfrac{1/\kappa\left(  1-\phi_{q}\rho\right)
}{1+\beta+\phi_{q}}\varepsilon_{t}^{v}%
$, and $\varepsilon_{t}^{v}=\widetilde{\nu}_{t}-\rho\widetilde{\nu
}_{t-1}$.
% \begin{align*}
% &  \dfrac{1/\kappa}{1+\beta+\phi_{q}}\left[  -\phi_{k}\eta_{t+1}^{r^{k}}%
% +\eta_{t+1}^{\pi}-\rho\left(  -\phi_{k}\eta_{t}^{r^{k}}+\eta_{t}^{\pi}\right)
% \right]  -\dfrac{\beta+\phi_{q}(1+\beta)}{1+\beta+\phi_{q}}\left(  \eta
% _{t+1}^{i}-\rho\eta_{t}^{i}\right)  \\
% &  \quad+\dfrac{\beta\phi_{q}}{1+\beta+\phi_{q}}\left(  \eta_{t+2}^{i}%
% -\rho\eta_{t+1}^{i}\right)  +\dfrac{1/\kappa\left(  1-\phi_{q}\rho\right)
% }{1+\beta+\phi_{q}}\underbrace{\left(  \widetilde{\nu}_{t}-\rho\widetilde{\nu
% }_{t-1}\right)  }_{\varepsilon_{t}^{v}}.%
% \end{align*}

\section{Computational Details\label{appsec: comp. details}}

The empirical moments of the linear model can be represented by
$\frac{1}{T}\sum_{t=1}^{T}f_{t}\left(  \theta,d\right)$, where   
% =\frac{1}{T}\sum_{t=1}^{T}Z_{t}^{\prime}\epsilon_{t}\left(  \theta,d\right)  
$f_{t}\left(  \theta,d\right)=Z_t' \left(Y_{t}b\left(  \theta\right) -X_{t}d\right)$, $Y_{t} = \left[
\begin{array}
[c]{cccccccc}%
\Delta\widetilde{i}_{t} & \Delta\widetilde{i}_{t-1} & \Delta\widetilde{i}%
_{t+1} & \Delta\widetilde{i}_{t+2} & \widetilde{r}_{t}^{p} & \widetilde{r}%
_{t-1}^{p} & \widetilde{u}_{t} & \widetilde{u}_{t+1}%
\end{array}
\right]$,
$Z_{t}=(X_{t},Z_{2,t})$
is the set of instrumental variables partitioned into included ($X_{t}$) and
excluded ($Z_{2,t}$) instruments, $
b\left(  \theta\right)' = \left[
1+\rho\left(  \beta+\phi_{q}\right),  
-\rho,
-\left(  \beta+\phi_{q}+\rho\beta\phi_{q}\right), 
\beta\phi_{q},
\dfrac{1}{\kappa},
-\dfrac{\rho}{\kappa},
\phi_{k}\frac{\rho\zeta}{\kappa},
-\phi_{k}\frac{\zeta}{\kappa}
\right]'
$
is a vector which
contains the structural parameters and $d$ are the strongly identified
parameters, which are estimated before the computation of the statistical
tests. The variable $X_{t}$ is $1$ corresponding to the constant in the
estimated regression specification, which captures all the steady-state terms.
We use $Z_{2,t}=\left\{  \Delta i_{t-1},r_{t-2}^{p},u_{t-1}\right\}  $ in our
baseline results. The sample size is $T$.

\subsection{S and qLL-S tests}

Under $H_{0}:\theta=\theta_{0}$, $b\left(  \theta_{0}\right)  $ is fixed. The
S statistic is 
\begin{equation}
S\left(  \theta_{0}\right)  =\;\underset{d}{\min}\frac{1}%
{T}\sum_{t=1}^{T}f_{t}\left(  \theta_0,d\right)'\widehat{V}\left(
\theta_{0},d\right)  ^{-1}\sum_{t=1}^{T}f_{t}\left(  \theta_0,d\right). \label{eq: objec fun}%
\end{equation}
The minimand in the above expression is the so-called continuously updated GMM objective function, evaluated at the continuously updated estimator for the untested parameter $d$ 
under $H_{0}$, see \cite{SW00}. The variance estimator $\widehat{V}%
\left(  \theta_{0},d\right)  $ is a heteroskedasticity and autocorrelation
consistent (HAC) estimator of $\operatorname*{Var}\left(  \frac{1}{\sqrt{T}%
}\sum_{t=1}^{T}f_{t}\left(  \theta_0,d\right)  \right)  $%
\begin{equation*}
\widehat{V}=\widehat{\Gamma}_{0}+\sum_{j=1}^{T}\omega_{j}\left(
\widehat{\Gamma}_{j}+\widehat{\Gamma}_{j}^{\prime}\right)  \text{,}
\label{eq: Sigma_hat}%
\end{equation*}
where $\widehat{\Gamma}_{j}=\left[  \frac{1}{T}\sum_{t=j+1}^{T}\widehat{w}%
_{t}\widehat{w}_{t}^{\prime}\right]  $, $\widehat{w}_{t}$ is \ $f_{t}\left(
\theta_{0},\hat{d}\right)  -\bar{f}_{T}\left(  \theta_{0},\hat{d}\right)
$,\ $\bar{f}_{T}\left(  \theta_{0},\hat{d}\right)  =\frac{1}{T}\sum_{t=1}%
^{T}f_{t}\left(  \theta_{0},\hat{d}\right)  $. The parameter $\omega_{j}%
$\ represents the Barlett kernel.

The S statistic is obtained by plugging $\hat{d}\left(  \theta_{0}\right)  $
into the objective function (\ref{eq: objec fun}). The S test at level $\alpha$ rejects $H_0:\theta=\theta_0$ when the S statistic exceeds the $1-\alpha$ quantile of the 
$\chi^2$ distribution with $k_{z}-k_{x}$ degrees of freedom, where $k_{z}$
and $k_{x}$ are the number of elements in vectors $Z_{t}$\ and $X_{t}$, respectively.

The qLL-S test rejects for large values of the statistic
\[
\text{qLL--S}(\theta_{0})=\frac{10}{11}\text{S}\left(  \theta_{0}\right)
+\text{qLL--S}^{B}(\theta_{0})
\]
where S$\left(  \theta_{0}\right)  $ is the S statistic evaluated at
$\theta=\theta_{0}$, and qLL--S$^{B}(\theta_{0})$ is the statistic that
detects violations of the moment conditions in subsamples. The algorithm for
computing the qLL--S$^{B}$ is detailed in \cite{MM14}, where one can also find
tables of critical values.

The confidence sets derived from the tests are obtained by performing a grid
search over the parameter space. The 90\% confidence sets are formed by the
collection of points that do not reject $H_{0}:\theta=\theta_{0}$ at 10\%
significance level.

\subsection{Split-sample S test}\label{s: mikusheva}
We derive a GMM version of the split-sample Anderson-Rubin test proposed by \cite{Mikusheva2021} for linear models.
Let $\bar{Y}_{t}$ be the demeaned values of $Y_{t}$.  Define $\bar{W}_{t}\left(  \theta\right)  =\bar{Y}_{t}\frac{\partial b\left(
\theta\right)  }{\partial\theta^{\prime}}$, which is of dimension $1\times3$,
and let $\mathbf{W}\left(  \theta\right)  $ be the $T\times3$\ matrix with
stacked terms $\bar{W}_{t}\left(  \theta\right)  $, $t=1,\ldots,T$. \ Define also
the matrices $\mathbf{Y}$ and $\mathbf{Z}$ of dimensions $T\times8$ and
$T\times k$ ($k=3$ in the baseline case) of stacked elements of $\bar{Y}_{t}$ and demeaned excluded instruments  $\bar{Z}_{2,t}$. Partition $\mathbf{W}\left(
\theta\right)  =\mathbf{Y}\frac{\partial b\left(  \theta\right)  }%
{\partial\theta^{\prime}}$, $\mathbf{Y}$ and $\mathbf{Z}$ as
$
\mathbf{W}\left(  \theta\right)  =\left[
\mathbf{W}_{1}\left(  \theta\right) :
\mathbf{W}_{p}\left(  \theta\right) :
\mathbf{W}_{2}\left(  \theta\right)
\right]
$, $\mathbf{Y}=\left[
\mathbf{Y}_{1}:
\mathbf{Y}_{p}:
\mathbf{Y}_{2}
\right]$, and $\mathbf{Z}=\left[
\mathbf{Z}_{1}:
\mathbf{Z}_{p}:
\mathbf{Z}_{2}
\right]
$.

% \[
% \mathbf{W}\left(  \theta\right)  =\left[
% \begin{array}
% [c]{c}%
% \mathbf{W}_{1}\left(  \theta\right) \\
% \mathbf{W}_{p}\left(  \theta\right) \\
% \mathbf{W}_{2}\left(  \theta\right)
% \end{array}
% \right]  \text{, }\mathbf{Y}=\left[
% \begin{array}
% [c]{c}%
% \mathbf{Y}_{1}\\
% \mathbf{Y}_{p}\\
% \mathbf{Y}_{2}%
% \end{array}
% \right]  \text{\ and }\mathbf{Z}=\left[
% \begin{array}
% [c]{c}%
% \mathbf{Z}_{1}\\
% \mathbf{Z}_{p}\\
% \mathbf{Z}_{2}%
% \end{array}
% \right]
% \]

In our case, the first subsample corresponds to 45\% of 
the initial observations. The terms $\mathbf{W}_{p}\left(  \theta\right)  $ and $\mathbf{Z}_{p}$\ are
not used in the procedure in order to keep the exogeneity assumption valid. Following \citet[p. 30]{Mikusheva2021}, we set $p=3$ because the error $\epsilon_t$ in (\ref{app eq: error}) is adapted to the $t+2$ information set and the instruments include variables dated $t-1$.
Then, estimate the fitted value of $\mathbf{W}_{2}\left(  \theta\right)  $ as 
$
\mathbf{\hat{W}}_{2}\left(  \theta\right)  =\mathbf{Z}_{2}\hat{\pi}_{1}\left(
\theta\right)$, where $
\hat{\pi}_{1}\left(  \theta\right)  =\left(  \mathbf{Z}_{1}^{\prime}%
\mathbf{Z}_{1}\right)  ^{-1}\mathbf{Z}_{1}^{\prime} \mathbf{W}%
_{1}\left(  \theta\right)$ and $\mathbf{W}%
_{1}\left(  \theta\right) =\mathbf{Y}_{1}\tfrac{\partial b\left(
\theta\right)  }{\partial\theta^{\prime}}.%
$

% \[
% \mathbf{\hat{W}}_{2}\left(  \theta\right)  =\mathbf{Z}_{2}\hat{\pi}_{1}\left(
% \theta\right),\quad 
% \text{where}\quad
% \hat{\pi}_{1}\left(  \theta\right)  =\left(  \mathbf{Z}_{1}^{\prime}%
% \mathbf{Z}_{1}\right)  ^{-1}\mathbf{Z}_{1}^{\prime}\underbrace{\mathbf{W}%
% _{1}\left(  \theta\right)  }_{\mathbf{Y}_{1}\frac{\partial b\left(
% \theta\right)  }{\partial\theta^{\prime}}}.%
% \]
Finally, we compute the split-sample S statistic as
\[
S_{M}(\theta)=\frac{1}{T_{2}}b\left(  \theta\right)  ^{\prime}\mathbf{Y}_{2}^{\prime
}\mathbf{\hat{W}}_{2}\left(  \theta\right)  \left[ \widehat{\Omega}\left(
\theta\right)  \right]  ^{-1}\mathbf{\hat{W}}_{2}\left(  \theta\right)
^{\prime}\mathbf{Y}_{2}b\left(  \theta\right),
\]
where $\widehat{\Omega}\left(  \theta\right)  $ is the HAC estimator of the variance of $\frac{1}%
{\sqrt{T_{2}}}\sum_{t=t_{2}}^{T}\hat{\bar{W}}_{t}\left(  \theta\right)  ^{\prime
}\bar{Y}_{t}b\left(  \theta\right)  $ and $T_{2}$\ corresponds to the number of
observations of the last subsample.

The split-sample S test at level $\alpha$ rejects $H_0: \theta = \theta_0$ when $S_M(\theta_0)$ exceeds the $1-\alpha$ quantile of a $\chi^2$ distribution with 3 degrees of freedom.

\section{Data\label{appsec: data}}

\subsection{Data Sources for baseline analysis}

\begin{itemize}
\item {\textbf{Gross Private Domestic Investment [\texttt{GPDI}]:} %\newline
Billions of Dollars, Seasonally Adjusted Annual Rate; Source: U.S. Bureau of
Economic Analysis; FRED - \url{https://fred.stlouisfed.org/series/GPDI}.}

\item {\textbf{Fixed Private Investment [\texttt{FPI}]:} %\newline 
Billions of
Dollars, Seasonally Adjusted Annual Rate; Source: U.S. Bureau of Economic
Analysis; FRED - \url{https://fred.stlouisfed.org/series/FPI}.}

\item {\textbf{Personal Consumption Expenditures: Durable Goods [\texttt{PCDG}%
]:} %\newline 
Billions of Dollars, Seasonally Adjusted Annual Rate; Source: U.S.
Bureau of Economic Analysis; FRED -
\url{https://fred.stlouisfed.org/series/PCDG}.}

\item {\textbf{Gross Domestic Product (implicit price deflator)
[\texttt{GDPDEF}]:} %\newline 
Index 2012=100, Seasonally Adjusted; Source: U.S.
Bureau of Economic Analysis; FRED -
\url{https://fred.stlouisfed.org/series/GDPDEF}.}

\item {\textbf{Gross Private Domestic Investment (implicit price deflator)
[\texttt{A006RD3Q086SBEA}]:} %\newline 
Index 2012=100, Seasonally Adjusted;
Source: U.S. Bureau of Economic Analysis; FRED -
\url{https://fred.stlouisfed.org/series/A006RD3Q086SBEA}.}

\item {\textbf{Gross Private Domestic Investment: Fixed Investment (implicit
price deflator) [\texttt{A007RD3Q086SBEA}]:} %\newline 
Index 2012=100, Seasonally
Adjusted; Source: U.S. Bureau of Economic Analysis; FRED -\newline
\url{https://fred.stlouisfed.org/series/A007RD3Q086SBEA#0}.}

\item {\textbf{Personal Consumption Expenditures: Durable goods (implicit
price deflator) [\texttt{DDURRD3Q086SBEA}]:} %\newline 
Index 2012=100, Seasonally
Adjusted; Source: U.S. Bureau of Economic Analysis; FRED -\newline
\url{https://fred.stlouisfed.org/series/DDURRD3Q086SBEA}.}

\item {\textbf{Effective Federal Funds Rate [\texttt{FEDFUNDS}]:} %\newline
Percent, Not Seasonally Adjusted; Source: Board of Governors of the Federal
Reserve System; FRED - \url{https://fred.stlouisfed.org/series/FEDFUNDS}.}

\item {\textbf{Capacity Utilization: Total Index [\texttt{TCU}]:} %\newline
Percent of Capacity, Seasonally Adjusted; Source: Board of Governors of the
Federal Reserve System; FRED - \url{https://fred.stlouisfed.org/series/TCU#0}.}
\end{itemize}

\subsection{Additional exogenous instruments}

\label{appsubsec: external instruments}

\begin{itemize}
\item {\textbf{\cite{romer2004new}'s narrative-based monetary policy shock}
(1969m3-2007m12); retrieved from Valerie A. Ramey's website under \textit{Data
and Programs for \enquote{Macroeconomic Shocks and Their Propagation}, 2016
Handbook of Macroeconomics.} -
\url{https://econweb.ucsd.edu/~vramey/research.html#data}.}

\item {\textbf{\cite{ramey2018government}'s military news shock}
(1967q1-2015q4); retrieved from Valerie A. Ramey's website under
\textit{Programs and Data for
\enquote{Government Spending Multipliers in Good Times and in Bad} with Sarah
Zubairy, April 2018 Journal of Political Economy.} - 
\url{https://econweb.ucsd.edu/~vramey/research.html#data}.}

\item {\textbf{Spot Crude Oil Price: West Texas Intermediate (WTI)
[\texttt{WTISPLC}]} (1967q1-2019q4); Deflated using CPI, Not Seasonally
Adjusted; Source: Federal Reserve Bank of St. Louis; FRED - 
\url{https://fred.stlouisfed.org/series/WTISPLC}.}

\item {\textbf{VXO} (1967q1-2019q4); Source: Chicago Board of Options
Exchange (CBOE) and retrieved from FRED - 
\url{https://fred.stlouisfed.org/series/VXOCLS}. \newline Note: This index is
unavailable before 1986. Following \cite{bloom2009impact}, pre-1986 monthly
return volatilities are computed as the monthly standard deviation of the
daily S\&P500 index normalized to the same mean and variance as the VXO index
when they overlap from 1986 onward.}

\end{itemize}

\subsection{Data Transformation}

\label{appsubsec: data transf.}

\textbf{Investment:} Investment series is first divided by the civilian
non-institutional population (16 years or over) to convert into per capita
terms and the resulting per capita series is then deflated using the
respective implicit price deflators. Two per capita measures of investment are
used in the analysis. They are:

\begin{enumerate}
\item SW - Real Fixed Private Investment (FPI). 

\item JPT - sum of Real Gross Private Domestic Investment (GPDI) and Real Personal
Consumption Expenditure: Durables Goods (PCDG).  
\end{enumerate}
\noindent The investment measures are computed, respectively, as
$\tfrac{FPI}{P_{fpi}} \quad \text{and} \quad
\tfrac{GDPI}{P_{gpdi}} + \frac{PCDG}{P_{pcdg}}$,
where $P_{gpdi}$, $P_{pcdg}$ and $P_{fpi}$ are the respective implicit price
deflators. Growth rates of investment are then computed as the log difference
of the resulting series.\vspace{+4pt}

\noindent\textbf{Inflation:} Log difference of the quarterly implicit GDP
price deflator. \vspace{+4pt}

\noindent\textbf{Real (ex-post) interest rate:} Difference between the Federal
Funds Rate and the GDP deflator inflation rate. \vspace{+4pt}

\noindent\textbf{Capacity utilization:} Log of the capacity utilization index.\vspace{+4pt}

\noindent\textbf{Narrative-based monetary policy shock:} Quarterly average of
the monthly series from \cite{romer2004new}. \vspace{+4pt}

\noindent\textbf{Narrative-based military news shock:} Defense news variable
of \cite{ramey2016defense} scaled by trend GDP following
\cite{ramey2018government}. \vspace{+4pt}

\noindent\textbf{VXO:} Quarterly average of the monthly series, demeaned and
standardized. \vspace{+4pt}

\noindent\textbf{Oil:} Log difference of the real oil price series.

\section{Robustness checks\label{appsec: further results}}

\begin{figure}[ptbh]
\adjustbox{max width=\textwidth}{\centering
\begin{tabular}
[c]{cccccc}\hline\hline
& \multicolumn{2}{c}{SW Investment} && \multicolumn{2}{c}{JPT Investment}\\[+2pt]
\cline{2-3}\cline{5-6}\\
& {\small {One Lag}} & {\small {Two Lags}} && {\small {One Lag}}
& {\small {Two Lags}}\\
& {\small (a)} & {\small (b)} && {\small (c)} & {\small (d)}\vspace{-.5cm}\\
\raisebox{+8.7ex}{\rotatebox[origin=lt]{90}{S sets }}\hspace{+.1cm} &
{\includegraphics[angle=0,width=.18\textwidth, trim=120 270 140 230 ,
totalheight=.25\textwidth]{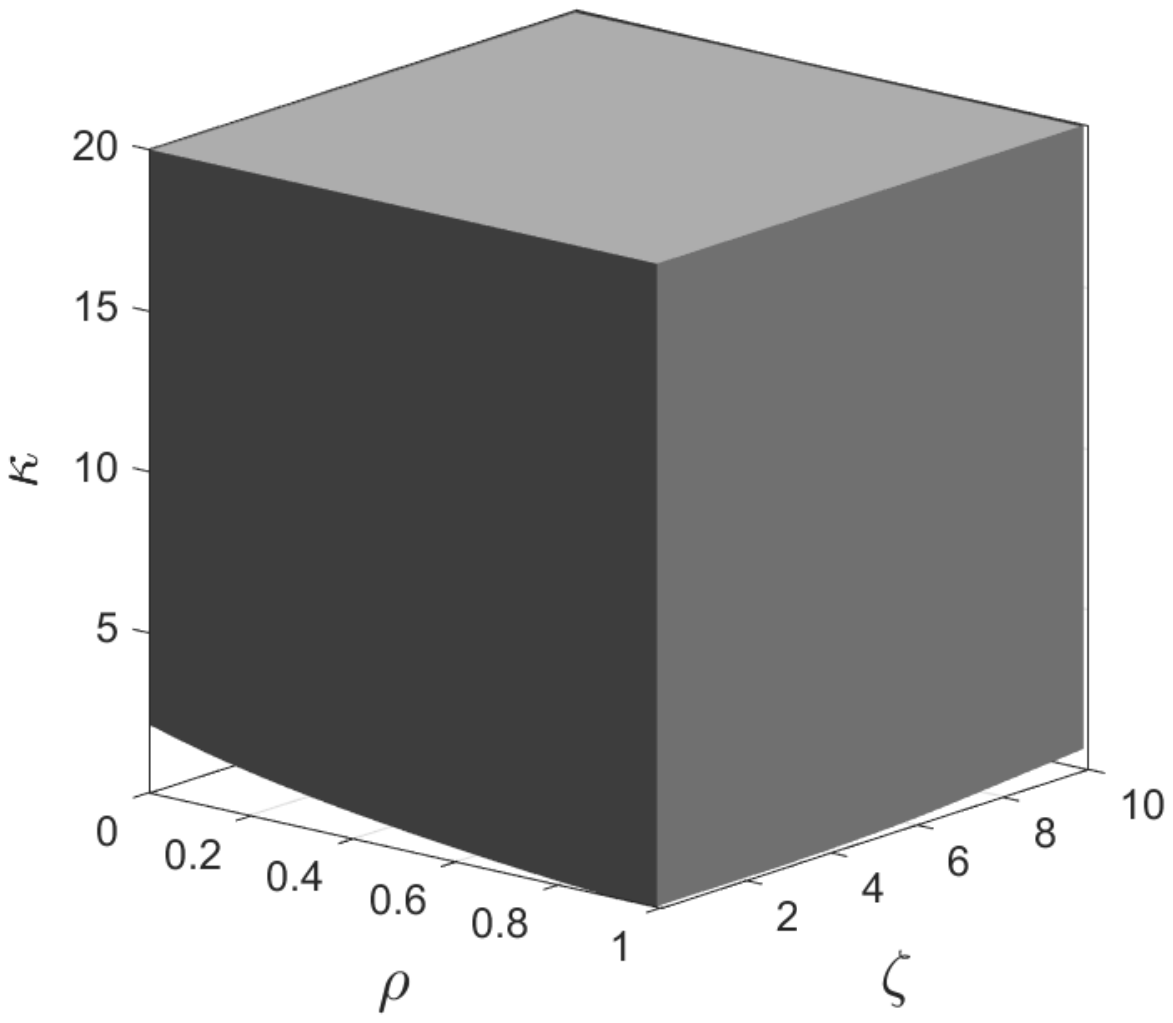}} &
\hspace{+.1cm}
{\includegraphics[angle=0,width=.18\textwidth, trim=120 270 140 230 ,
totalheight=.25\textwidth]{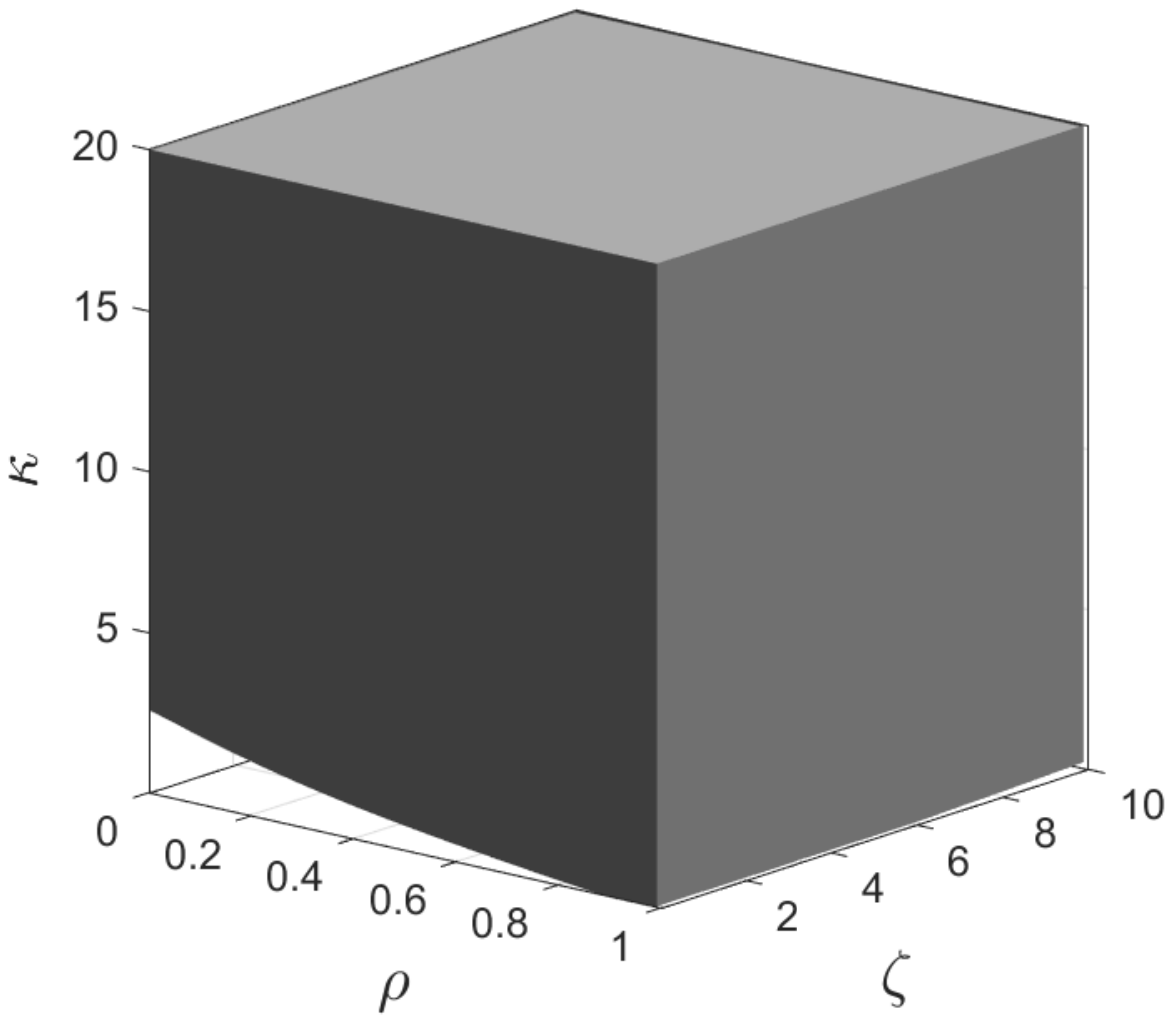}} && \hspace{+.1cm}
{\includegraphics[angle=0,width=.18\textwidth, trim=120 270 140 230 ,
totalheight=.25\textwidth]{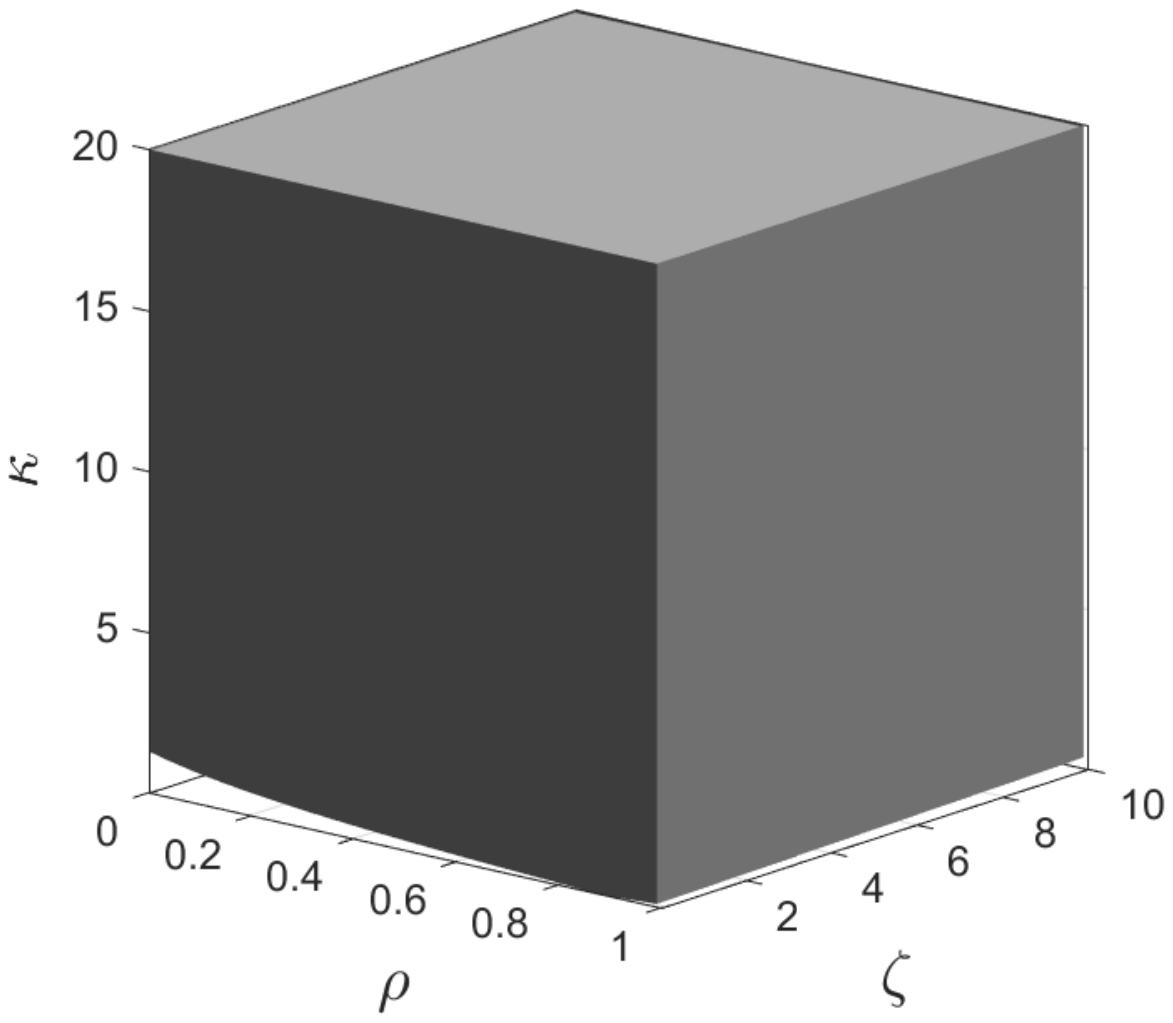}} &
\hspace{+.1cm}
{\includegraphics[angle=0,width=.20\textwidth,trim=120 280 140 230 ,
totalheight=.25\textwidth]{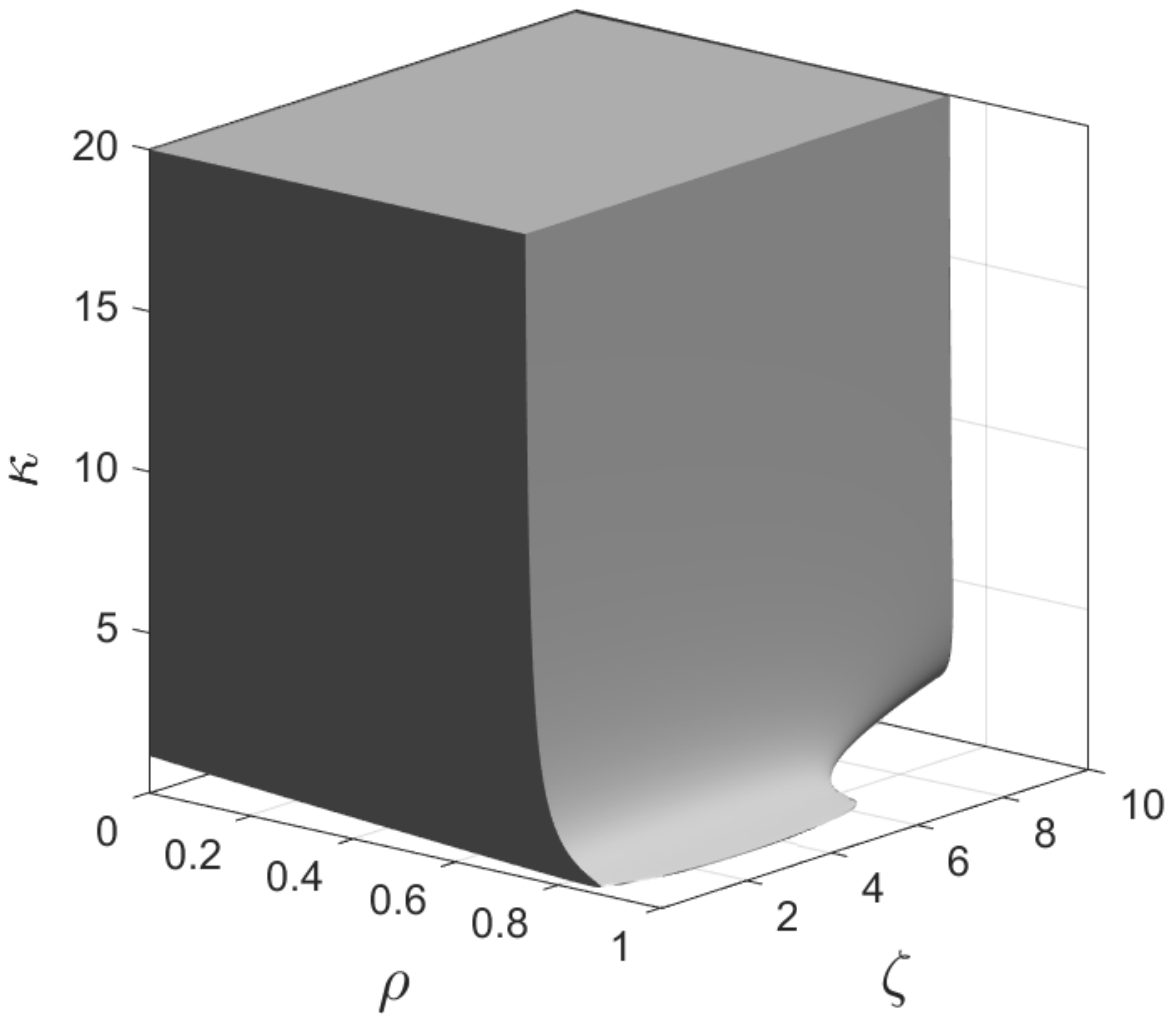}}\\[+3pt]
& {\small (e)} & {\small (f)} && {\small (g)} & {\small (h)}\vspace{-.4cm}\\
\raisebox{+6.7ex}{\rotatebox[origin=lt]{90}{qLL-S set }}\hspace{+.1cm} &
{\includegraphics[angle=0,width=.20\textwidth,trim=120 280 140 230 ,
totalheight=.25\textwidth]{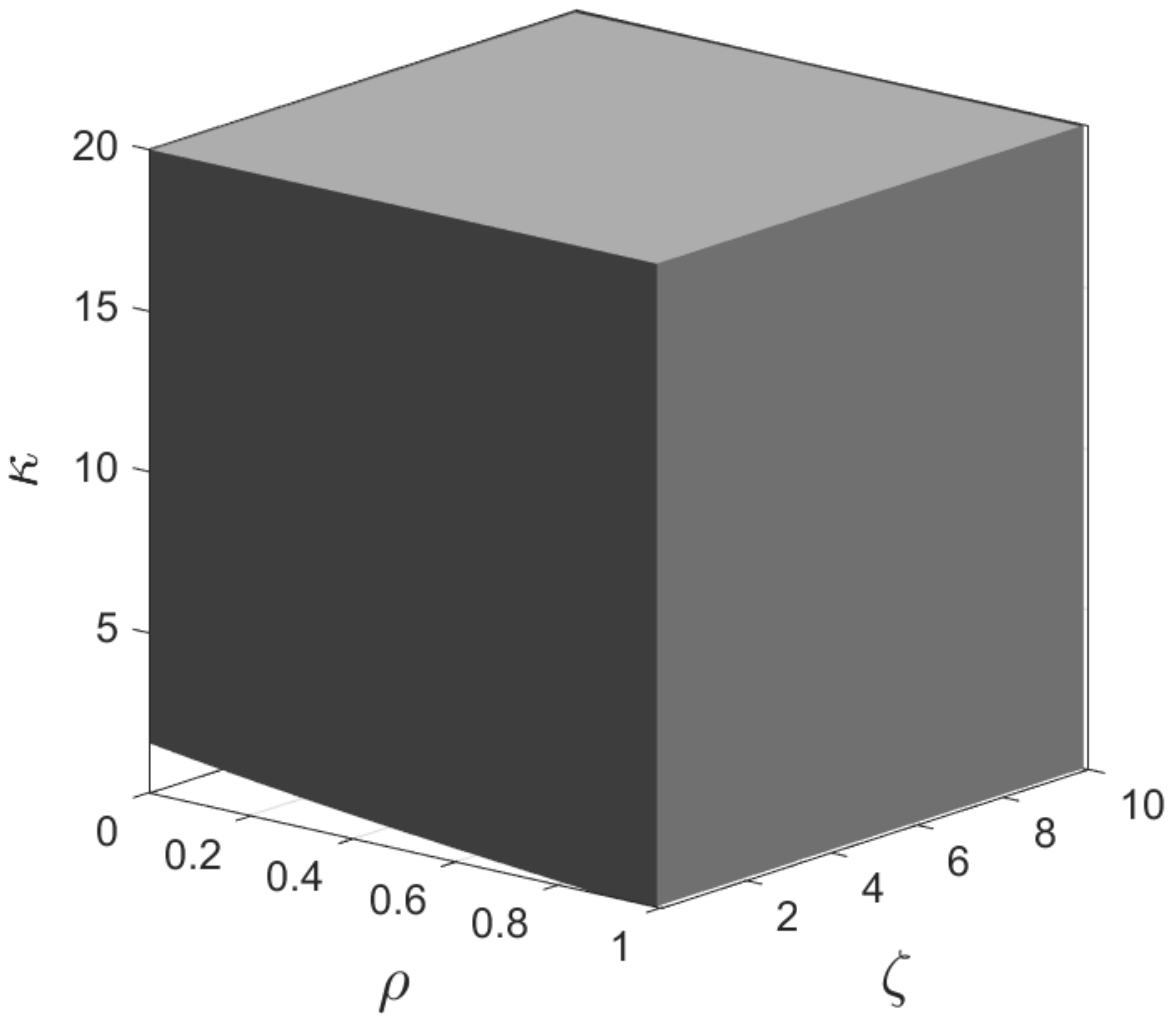}} & \hspace{+.1cm}
{\includegraphics[angle=0,width=.20\textwidth,trim=120 280 140 230 ,
totalheight=.25\textwidth]{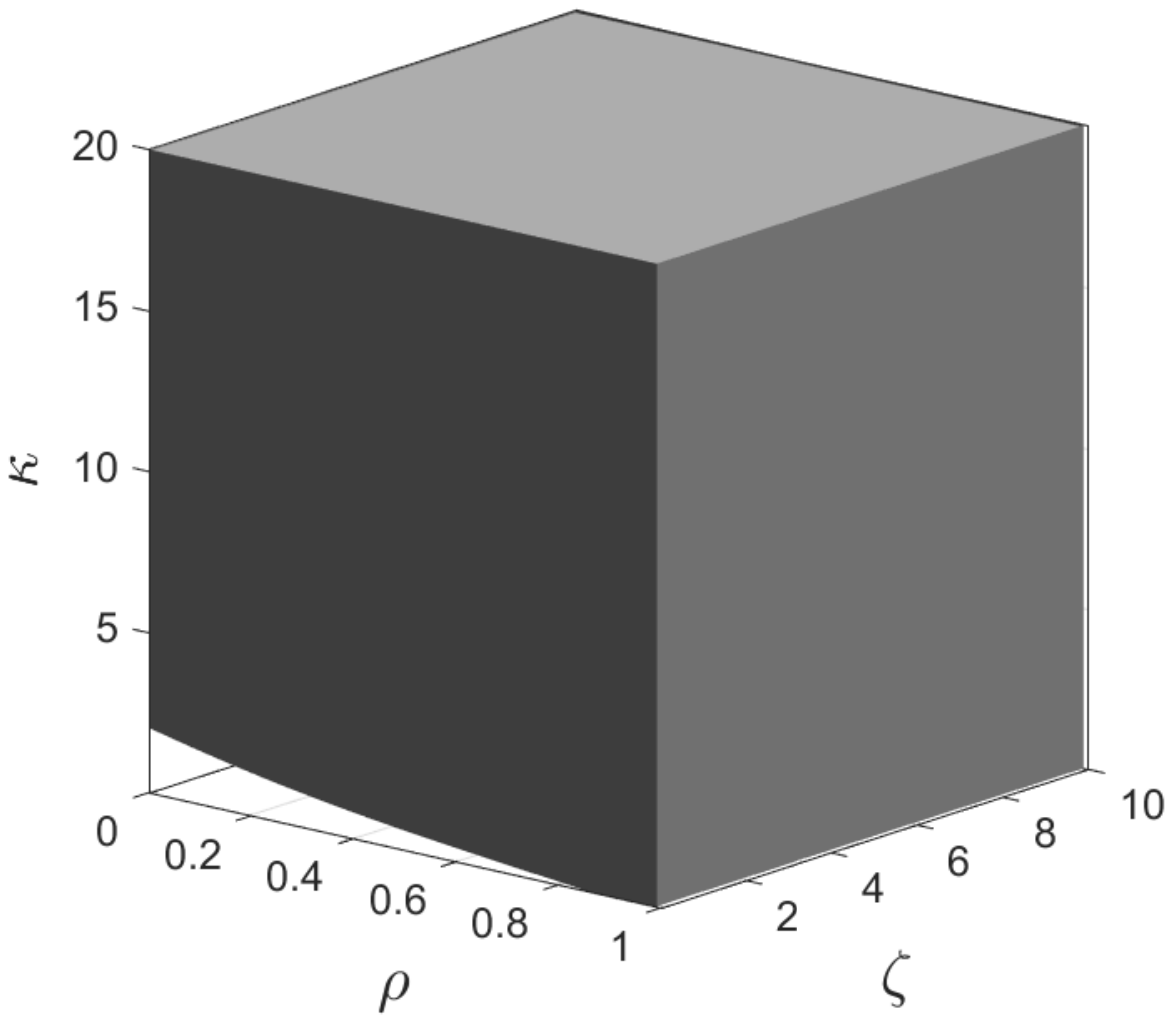}} &&
\hspace{+.1cm}
{\includegraphics[angle=0,width=.20\textwidth,trim=120 280 140 230 ,
totalheight=.25\textwidth]{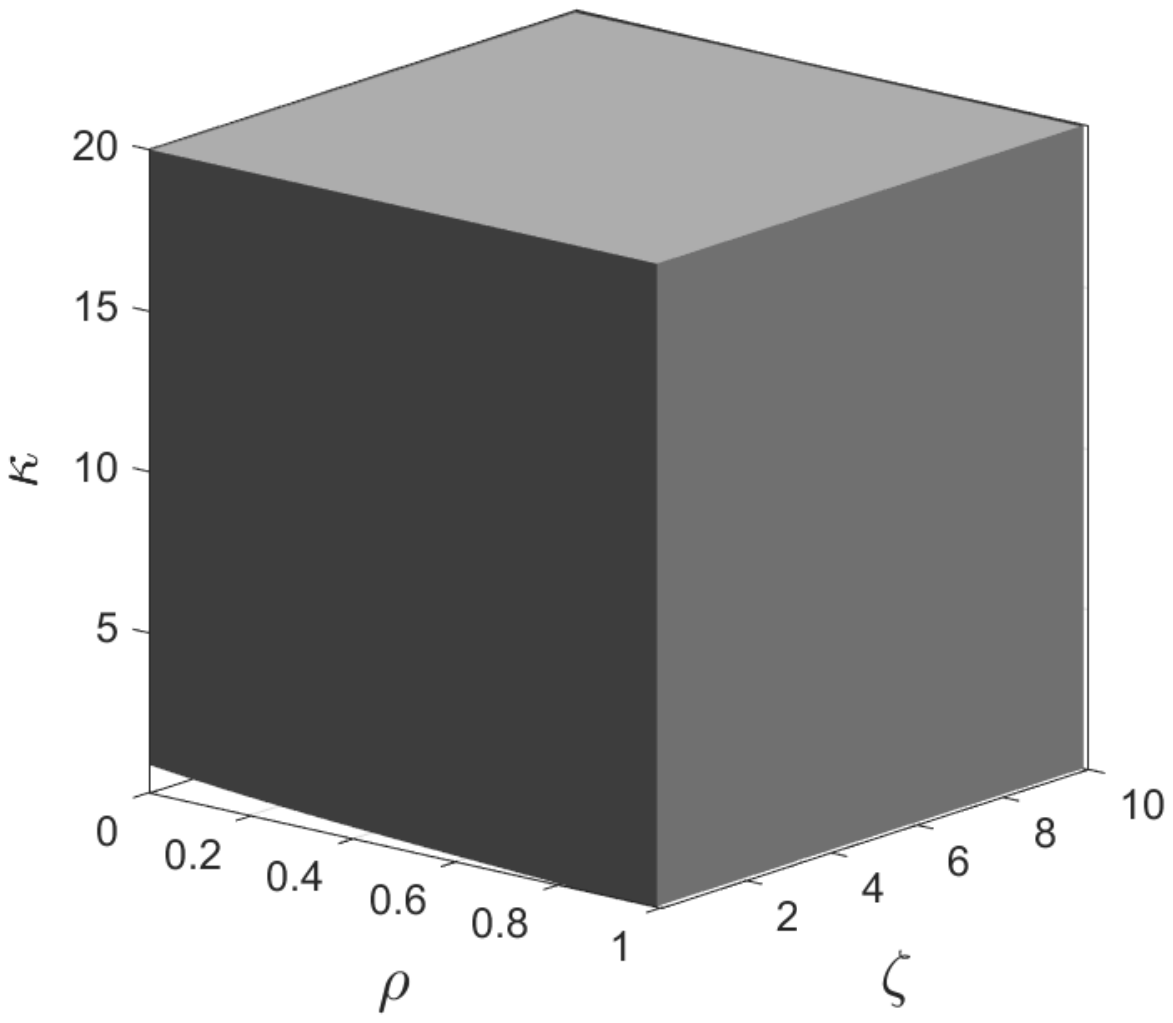}} & \hspace{+.1cm}
{\includegraphics[angle=0,width=.20\textwidth,trim=120 280 140 230 ,
totalheight=.25\textwidth]{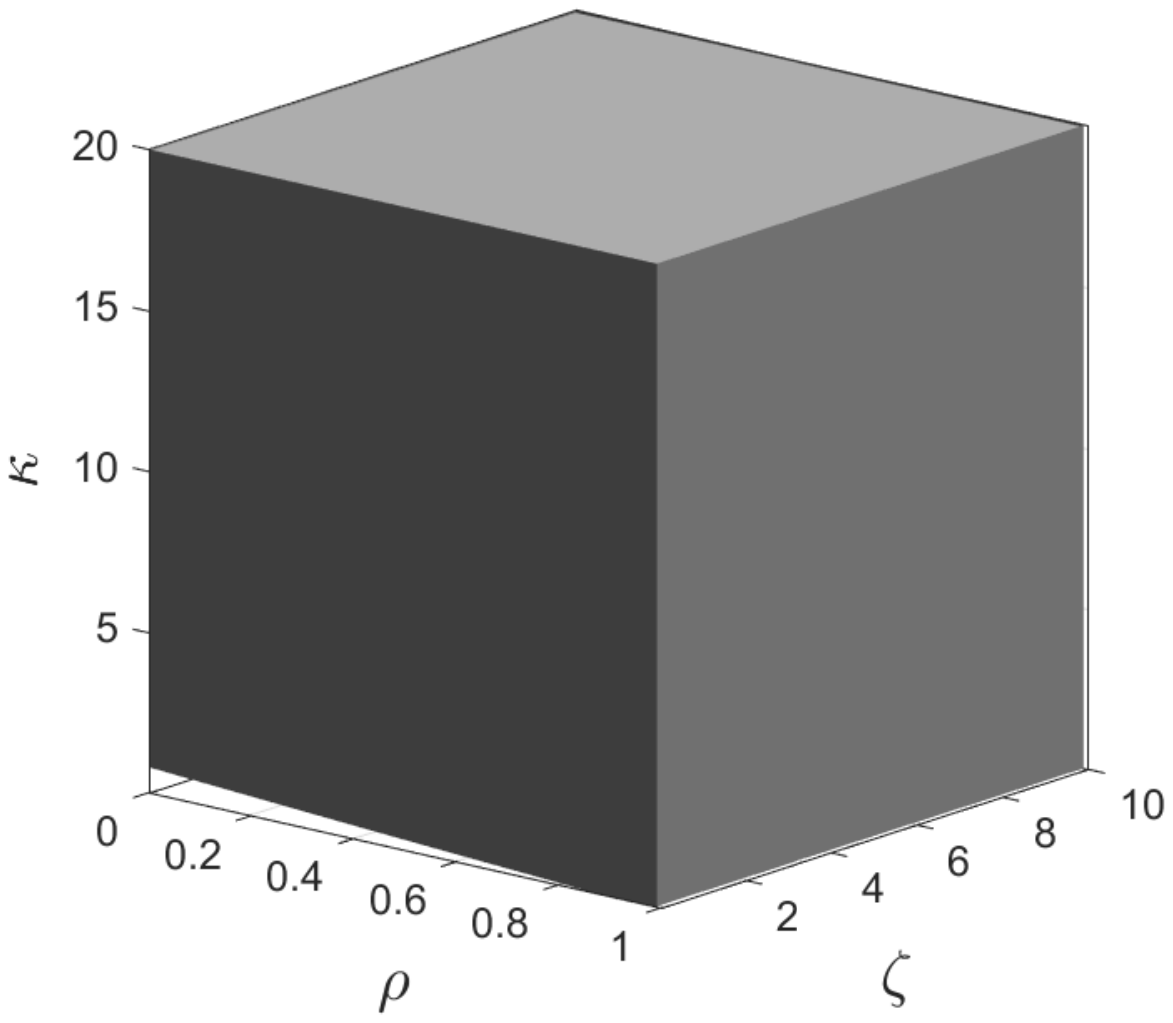}}\vspace{+2pt}\\\hline\hline
\end{tabular}} \vspace{-.4cm} \caption{ 90\% S and qLL-S confidence sets for
$\theta=(\rho,\kappa,\zeta)$ in the investment euler equation model
(\ref{eq: estimated}). Instruments: \protect\underline{One lag} - constant,
$\Delta i_{t-1}$, $r^{p}_{t-2}$, $u_{t-1}$; \protect\underline{Two lags} -
constant, $\Delta i_{t-1}$, $\Delta i_{t-2}$, $r^{p}_{t-2}$, $r^{p}_{t-3}$,
$u_{t-1}$, $u_{t-2}$. Left two columns show the results based on using Fixed Private Investment as investment proxy, while right two columns use
the sum of Gross Private
Domestic Investment and Personal Consumption Expenditure on Durable Goods as investment proxy. Period:
1967Q1-2019Q4. \cite{Newey_West_1987} HAC. }%
\label{appfig: Two lags}%
\end{figure}

\begin{figure}[ptbh]
\adjustbox{max width=\textwidth}{\centering
\begin{tabular}
[c]{cccccc}\hline\hline
& \multicolumn{2}{c}{SW Investment} && \multicolumn{2}{c}{JPT Investment}\\[+3pt]
\cline{2-3}\cline{5-6}\\
& {\small {1967Q1-2004Q4}} & {\small {1967Q1-2019Q4}} && {\small {1967Q1-2004Q4}}
& {\small {1967Q1-2019Q4}} \\
& {\small (a)} & {\small (b)} && {\small (c)} & {\small (d)}\vspace{-.5cm}\\
\raisebox{+8.7ex}{\rotatebox[origin=lt]{90}{S sets }}\hspace{+.1cm} &
{\includegraphics[angle=0,width=.18\textwidth, trim=120 270 140 230 ,
totalheight=.25\textwidth]{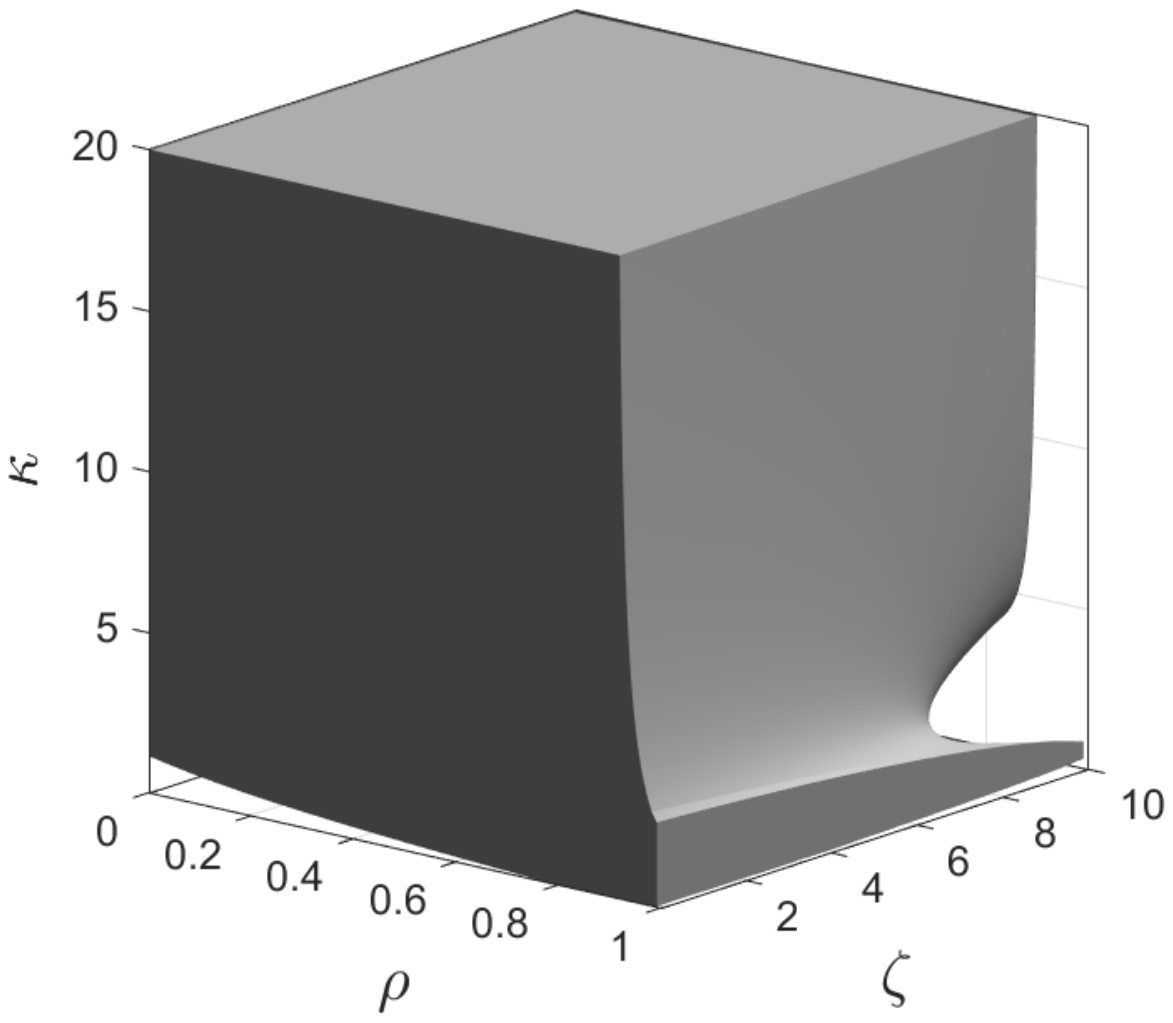}}  &
\hspace{+.1cm}
{\includegraphics[angle=0,width=.18\textwidth, trim=120 270 140 230 ,
totalheight=.25\textwidth]{Figure Appendix/Figure two lags/SW_onelag/cr_Z1_GMM_S.pdf}}
&&
\hspace{+.1cm}
{\includegraphics[angle=0,width=.20\textwidth,trim=120 280 140 230 ,
totalheight=.25\textwidth]{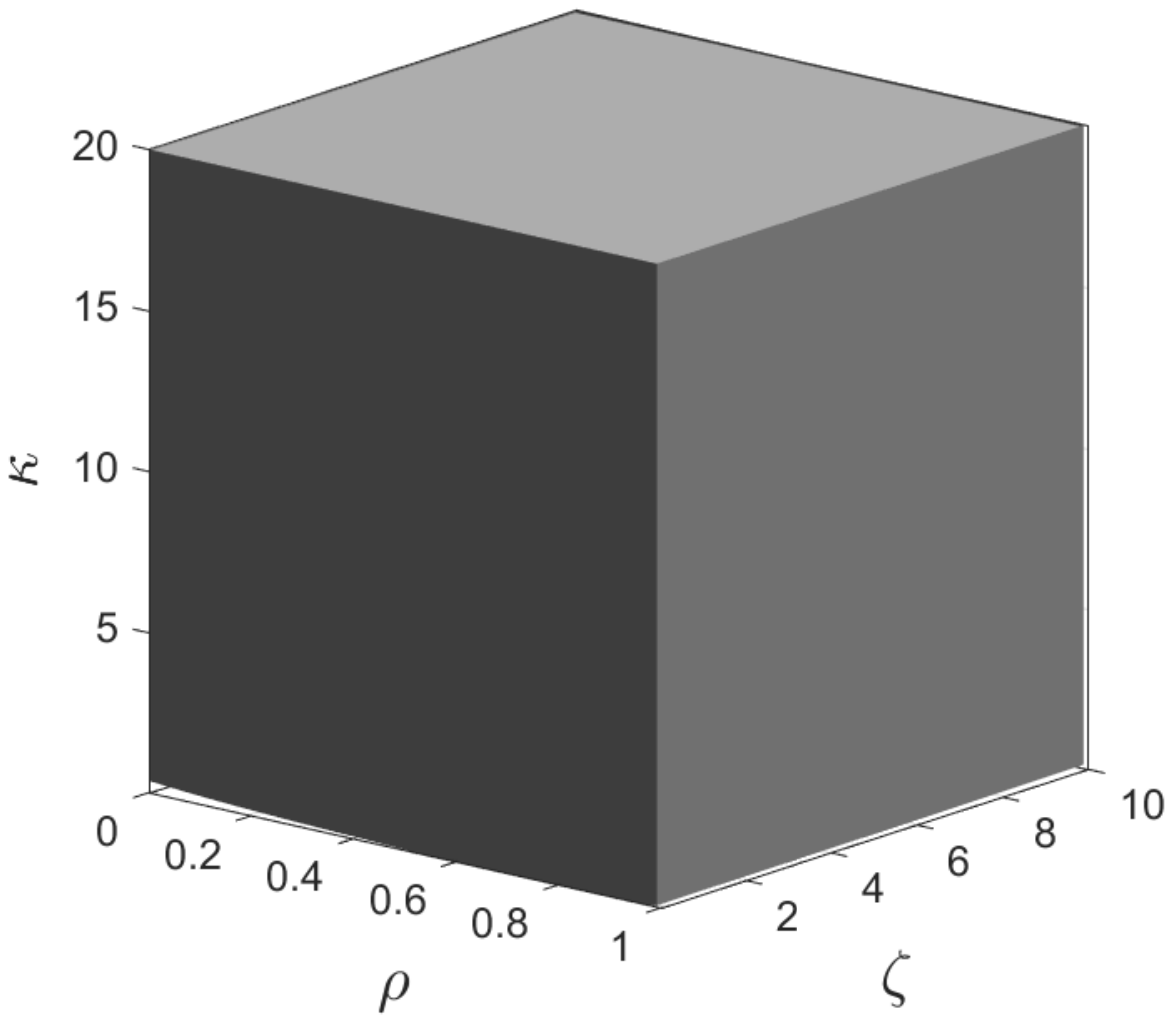}} &
\hspace{+.1cm}
{\includegraphics[angle=0,width=.18\textwidth, trim=120 270 140 230 ,
totalheight=.25\textwidth]{Figure Appendix/Figure two lags/JPT_onelag/cr_Z1_GMM_S.pdf}}\\[+3pt]
& {\small (e)} & {\small (f)} && {\small (g)} & {\small (h)}\vspace{-.4cm}\\
\raisebox{+6.7ex}{\rotatebox[origin=lt]{90}{qLL-S set }}\hspace{+.1cm} &
{\includegraphics[angle=0,width=.20\textwidth,trim=120 280 140 230 ,
totalheight=.25\textwidth]{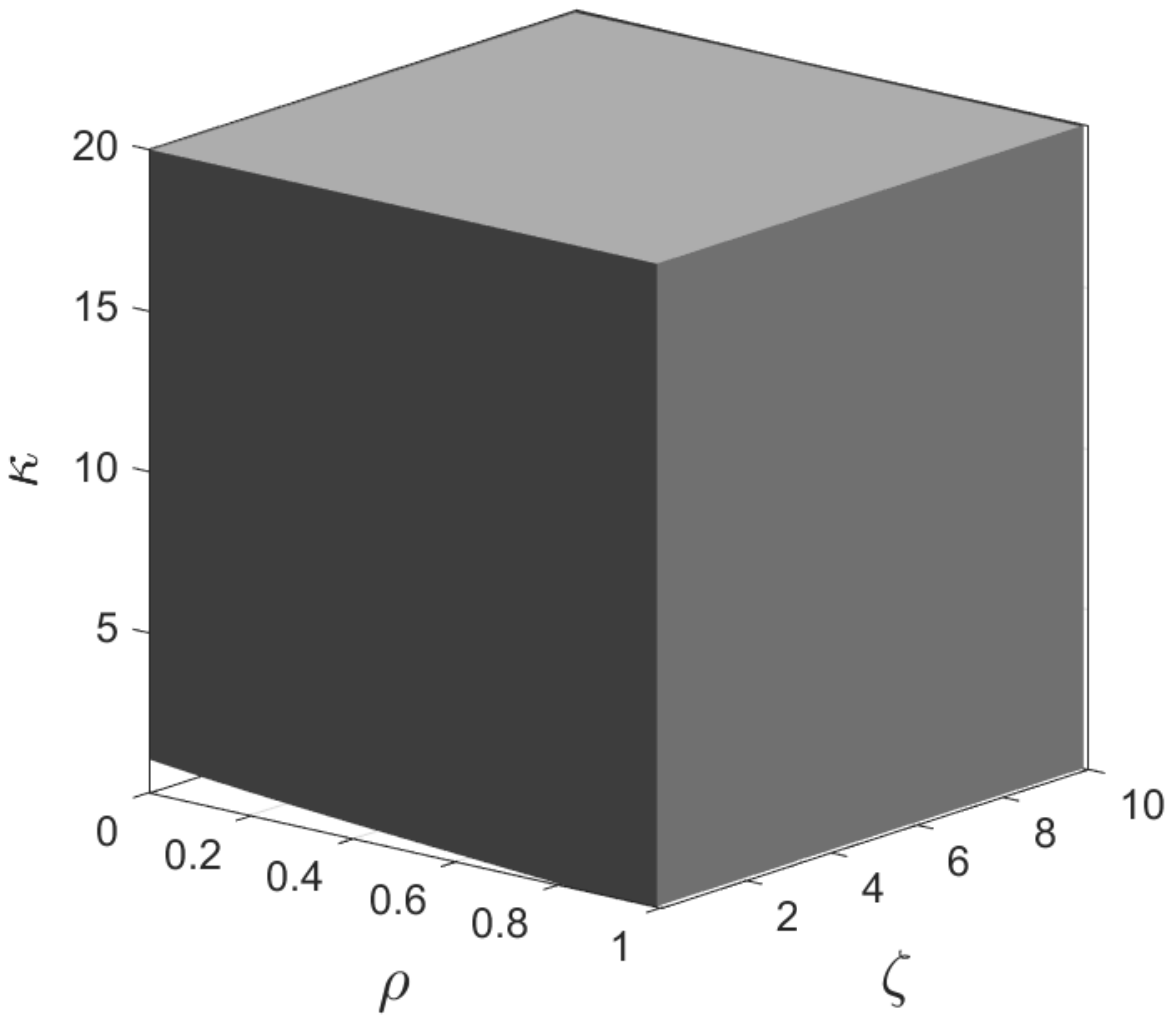}}
& \hspace{+.1cm}
{\includegraphics[angle=0,width=.20\textwidth,trim=120 280 140 230 ,
totalheight=.25\textwidth]{Figure Appendix/Figure two lags/SW_onelag/cr_Z1_GenS_qLL.pdf}}
&&
\hspace{+.1cm}
{\includegraphics[angle=0,width=.20\textwidth,trim=120 280 140 230 ,
totalheight=.25\textwidth]{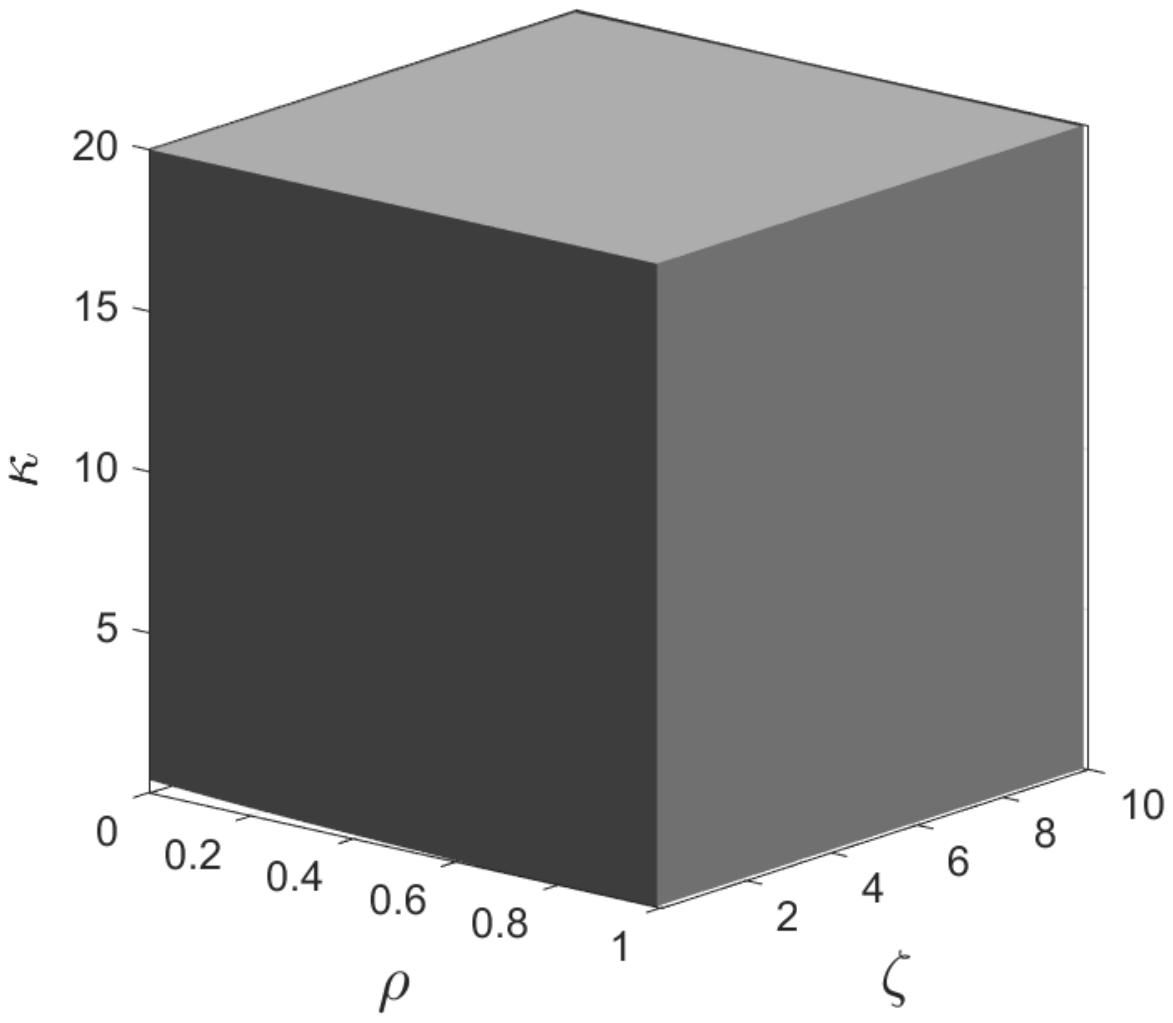}}
& \hspace{+.1cm}
{\includegraphics[angle=0,width=.20\textwidth,trim=120 280 140 230 ,
totalheight=.25\textwidth]{Figure Appendix/Figure two lags/JPT_onelag/cr_Z1_GenS_qLL.pdf}}
\vspace{+2pt}\\\hline\hline
\end{tabular}} \vspace{-.4cm} \caption{ 90\% S and qLL-S confidence sets for
$\theta=(\rho,\kappa,\zeta)$ in the investment euler equation model
(\ref{eq: estimated}). Instruments: constant, $\Delta i_{t-1}$, $r^{p}_{t-2}$,
$u_{t-1}$. Left two columns show the results based on using Fixed Private Investment as investment proxy, while right two columns use
the sum of Gross Private
Domestic Investment and Personal Consumption Expenditure on Durable Goods as investment proxy.  \cite{Newey_West_1987} HAC. }%
\label{appfig: 67_04}%
\end{figure}

\begin{figure}[ptbh]
\centering
\adjustbox{min width=\textwidth,max width=\textwidth, max height=9.5cm}{
\begin{tabular}
[c]{ccccc}\hline\hline
& \multicolumn{4}{c}{Exogenous Instruments with SW Investment Proxy }\\ \cline{2-5}
& {\small {Baseline}} & {\small {Mon. pol. shock}} & {\small {Military news}}
& {\small {Oil}}\\
& {\small {1967Q1-2019Q4}} & {\small {1969Q2-2007Q4}} &
{\small {1967Q1-2015Q4}} & {\small {1967Q1-2019Q4}}\\
& {\small (a)} & {\small (b)} & {\small (c)} & {\small (d)}\vspace{-.5cm}\\
\raisebox{+8.7ex}{\rotatebox[origin=lt]{90}{S sets }}\hspace{+.4cm} &
{\includegraphics[angle=0,width=.18\textwidth, trim=120 270 140 230 ,
totalheight=.25\textwidth]{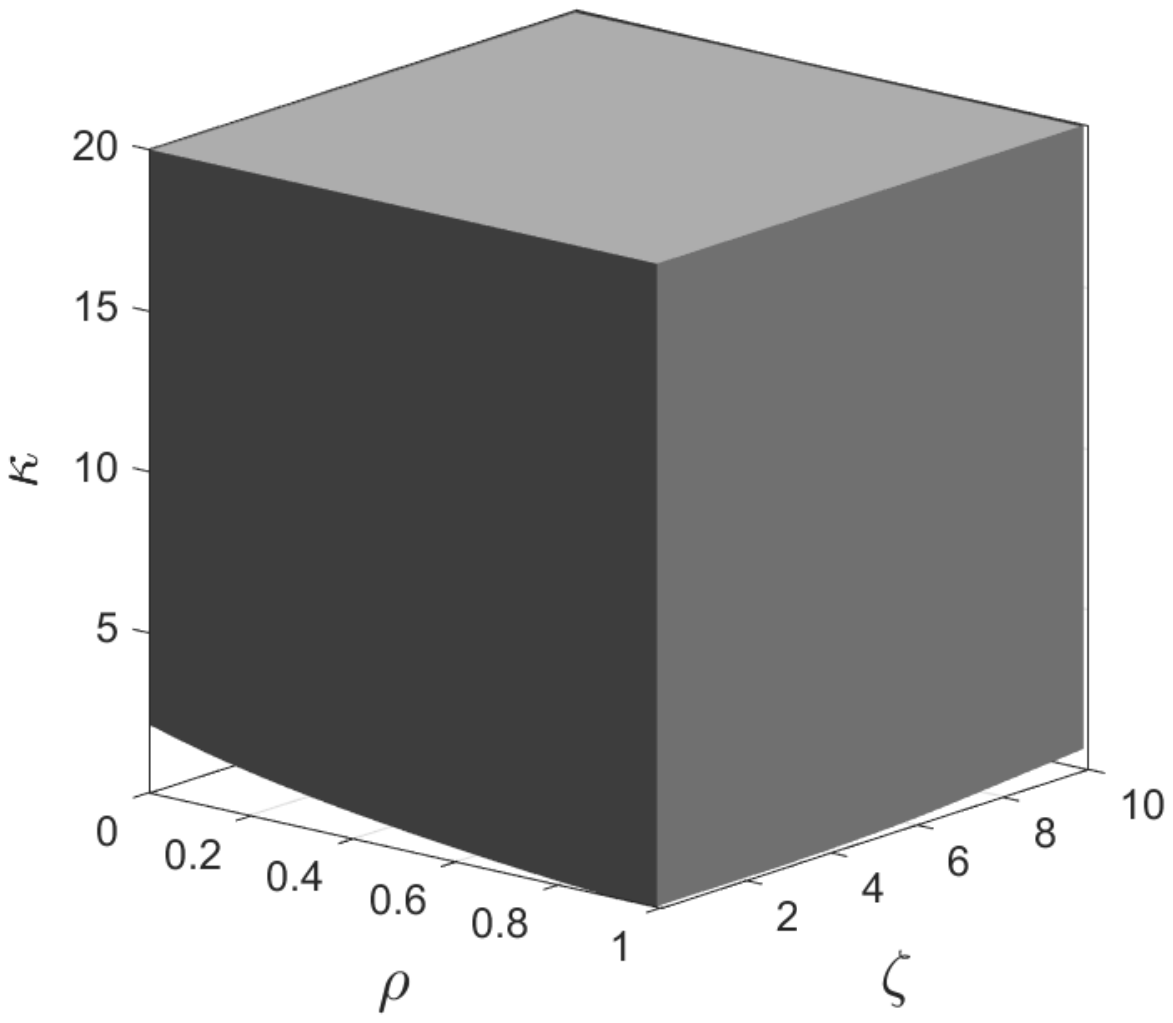}} &
\hspace{+.1cm}
{\includegraphics[angle=0,width=.20\textwidth,trim=120 280 140 230 ,
totalheight=.25\textwidth]{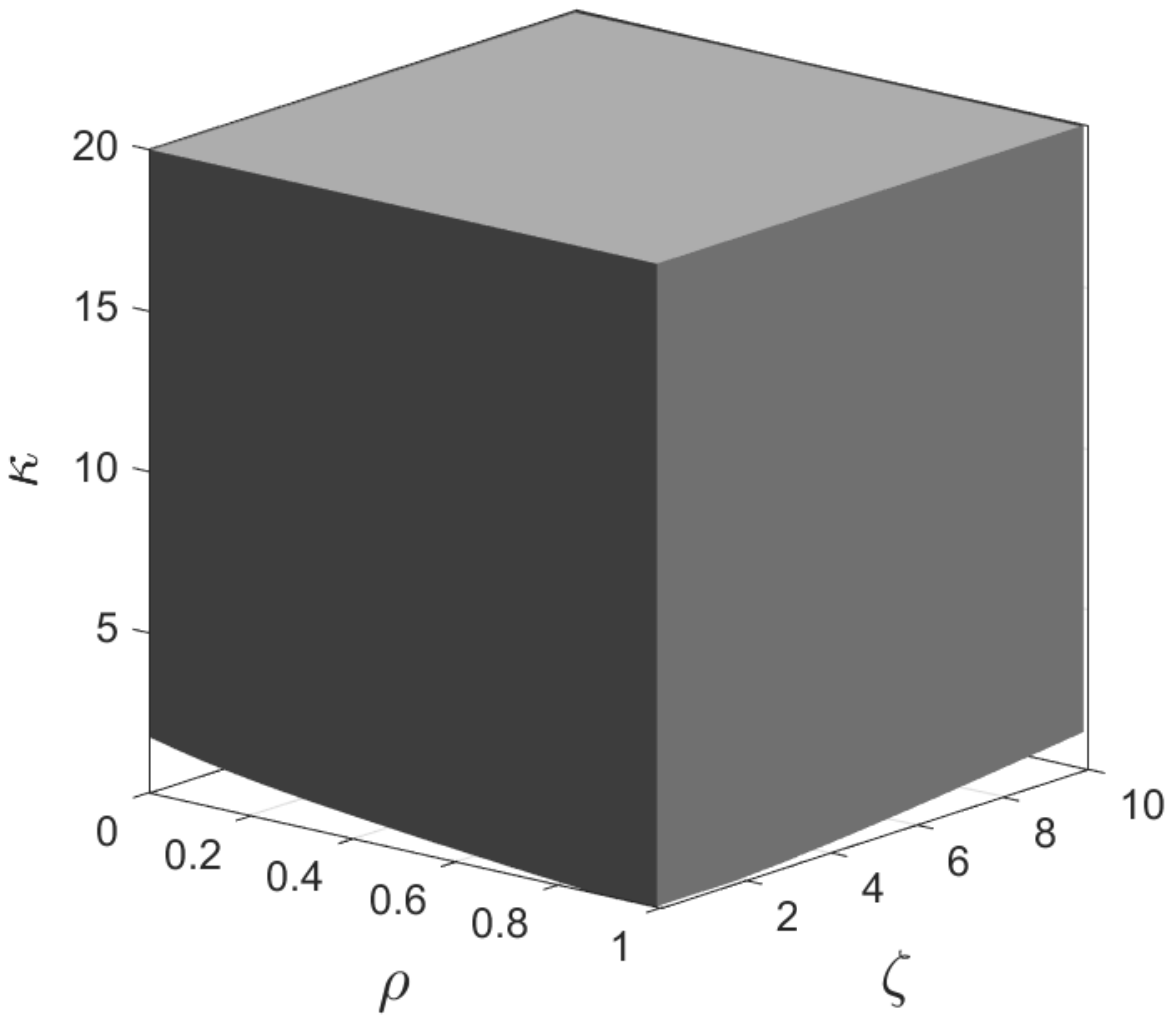}} & \hspace{+.1cm}
{\includegraphics[angle=0,width=.20\textwidth,trim=120 280 140 230 ,
totalheight=.25\textwidth]{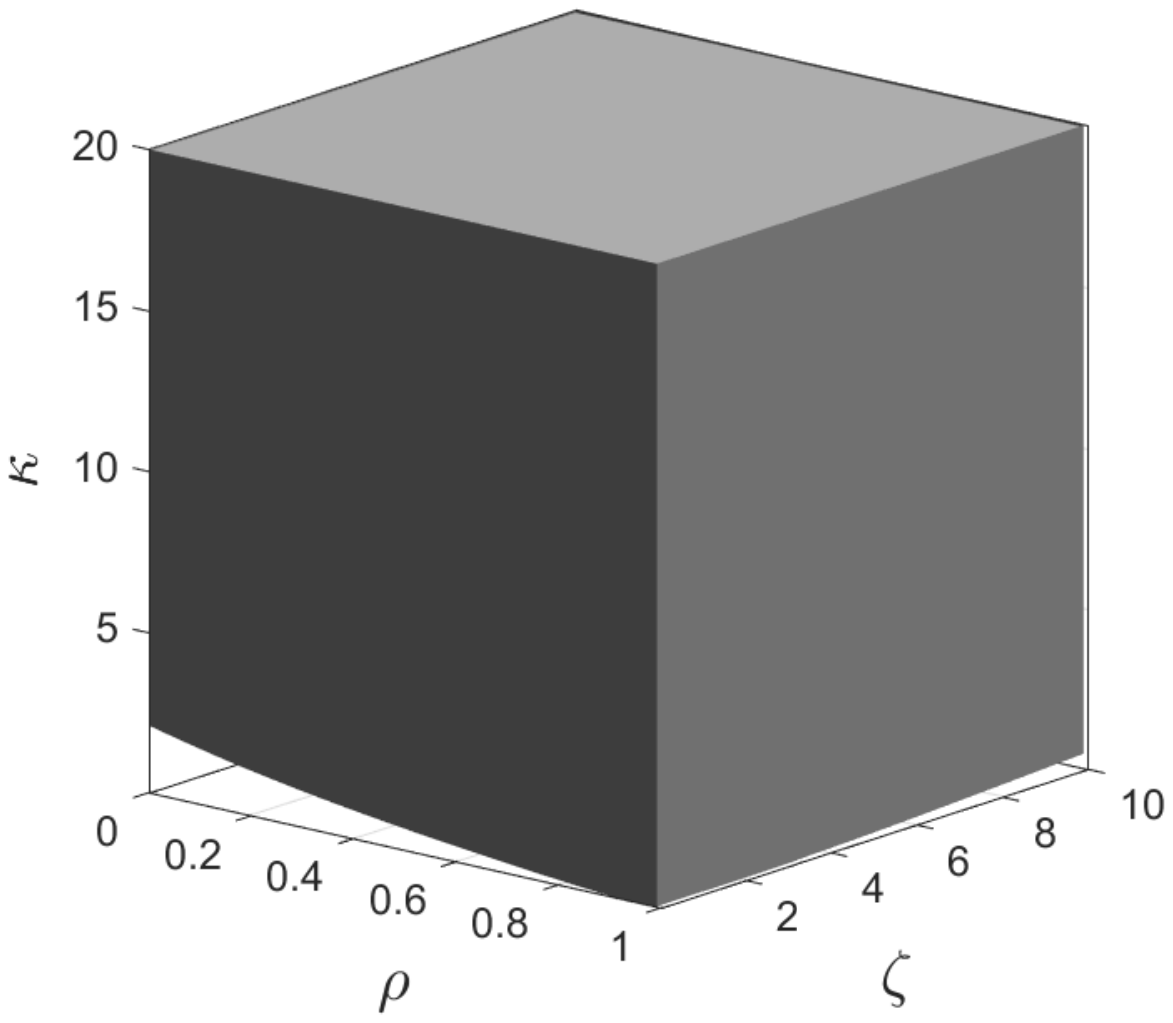}} &
\hspace{+.1cm}
{\includegraphics[angle=0,width=.20\textwidth,trim=120 280 140 230 ,
totalheight=.25\textwidth]{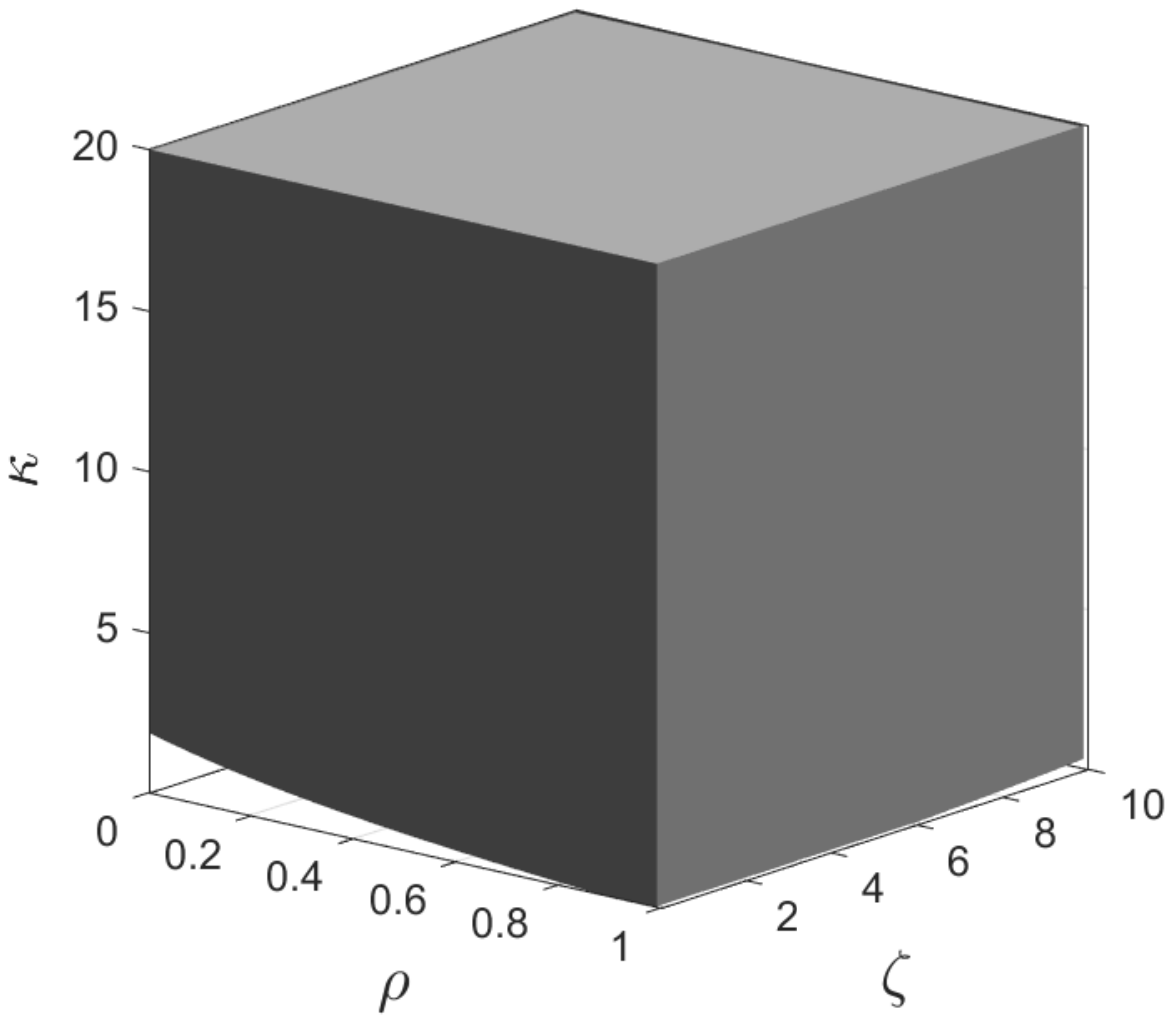}}\\[+3pt]
& {\small {VXO}} & {\small (b)+(c)} & {\small (d)+(e)} &
{\small {(b)+(c)+(d)+(e)}}\\
& {\small {1967Q1-2019Q4}} & {\small {1969Q2-2007Q4}} &
{\small {1967Q1-2019Q4}} & {\small {1969Q2-2007Q4}}\\
& {\small (e)} & {\small (f)} & {\small (g)} & {\small (h)}\vspace{-.4cm}\\
\raisebox{+8.7ex}{\rotatebox[origin=lt]{90}{S sets }}\hspace{+.4cm}
&
{\includegraphics[angle=0,width=.18\textwidth, trim=120 280 140 230 ,
totalheight=.25\textwidth]{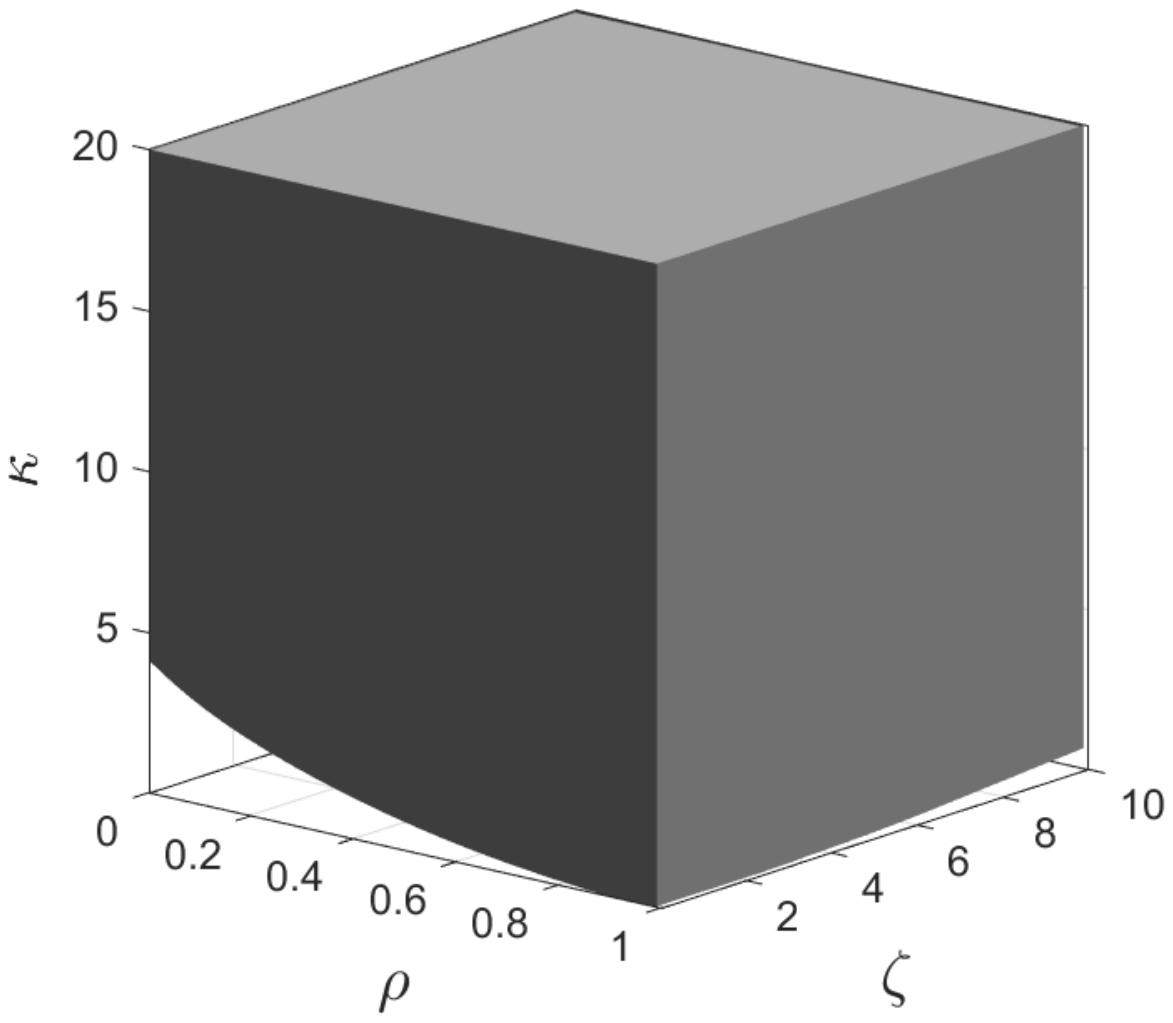}}
& \hspace{+.1cm}
{\includegraphics[angle=0,width=.20\textwidth,trim=120 280 140 230 ,
totalheight=.25\textwidth]{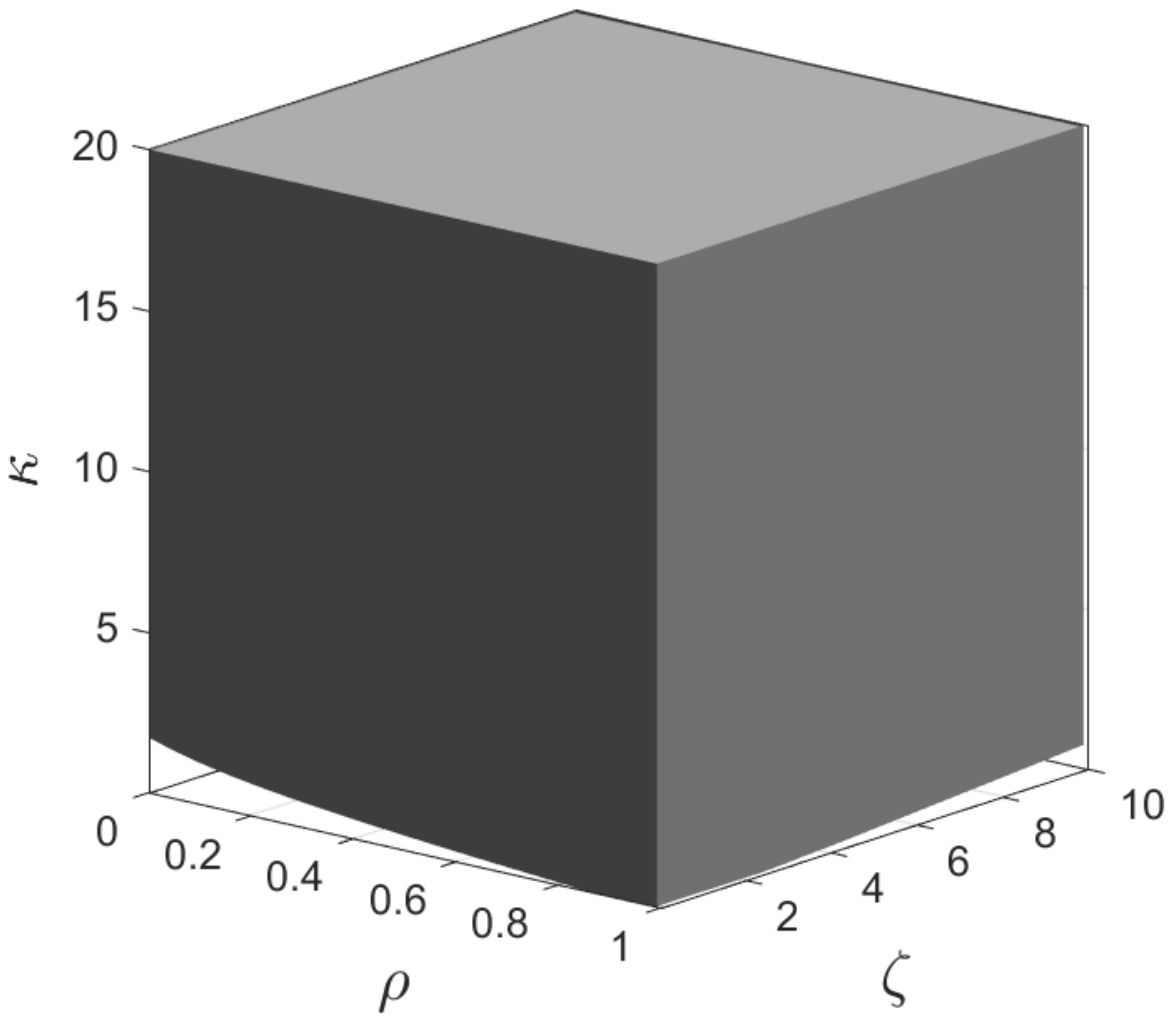}}
& \hspace{+.1cm}
{\includegraphics[angle=0,width=.20\textwidth,trim=120 280 140 230 ,
totalheight=.25\textwidth]{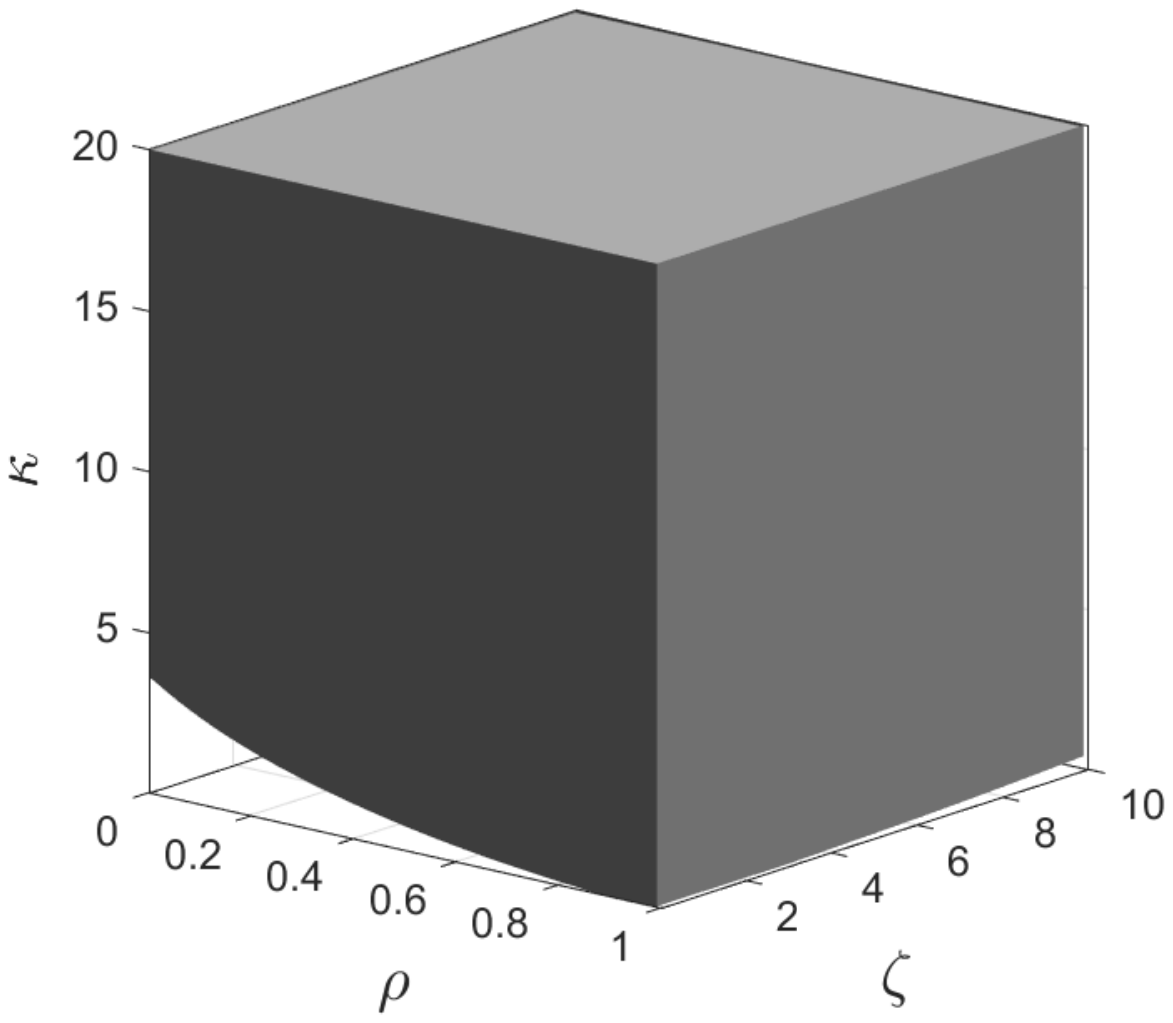}}
& \hspace{+.1cm}
{\includegraphics[angle=0,width=.20\textwidth,trim=120 280 140 230 ,
totalheight=.25\textwidth]{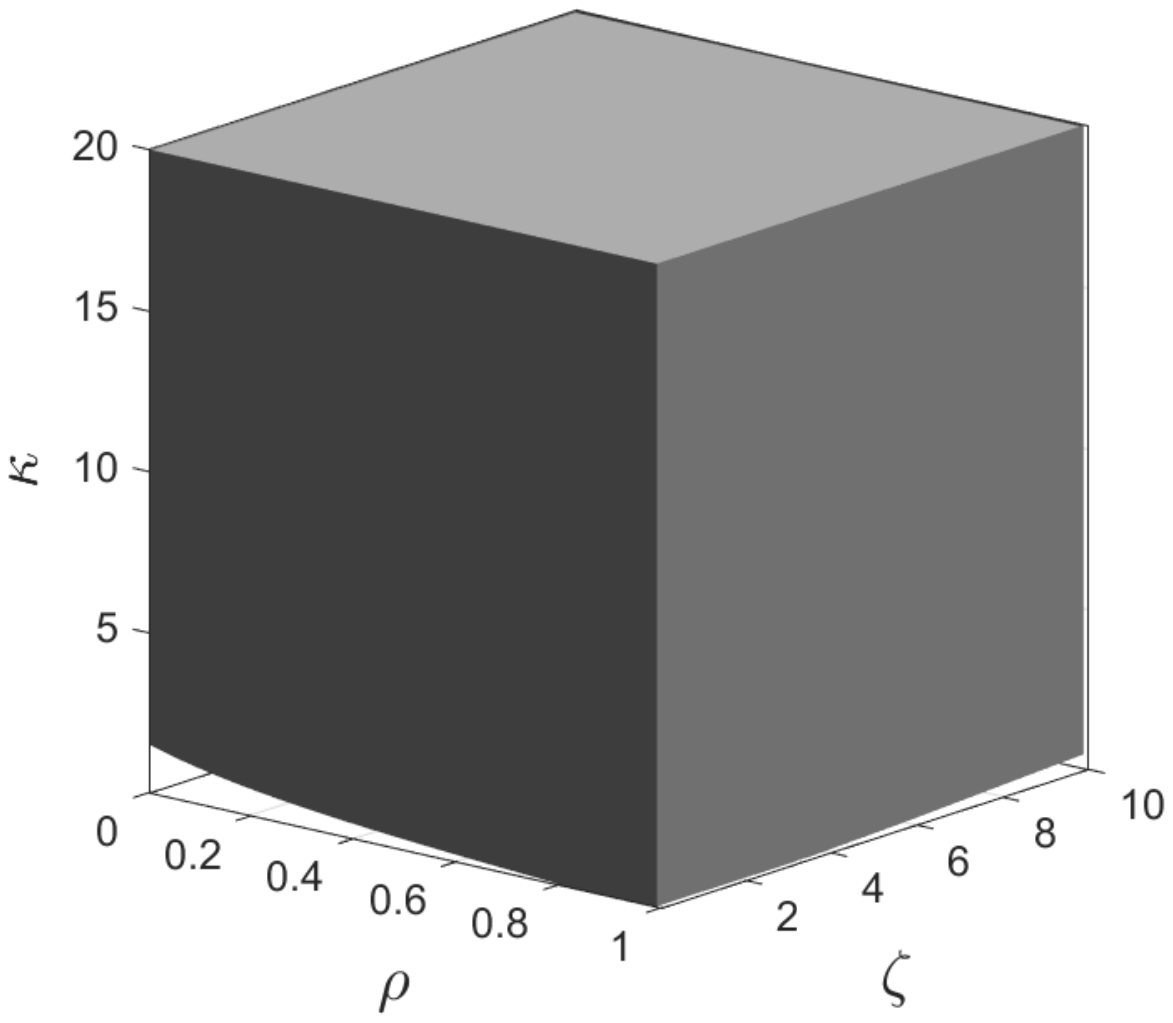}}\vspace{+10pt}\\
&  &  &  & \\
& {\small {Baseline}} & {\small {Mon. pol. shock}} & {\small {Military news}}
& {\small {Oil}}\\
& {\small {1967Q1-2019Q4}} & {\small {1969Q2-2007Q4}} &
{\small {1967Q1-2015Q4}} & {\small {1967Q1-2019Q4}}\\
& {\small (i)} & {\small (j)} & {\small (k)} & {\small (l)}\vspace{-.4cm}\\
\raisebox{+5.7ex}{\rotatebox[origin=lt]{90}{qLL-S sets }}\hspace{+.4cm} &
{\includegraphics[angle=0,width=.18\textwidth, trim=120 270 140 230 ,
totalheight=.25\textwidth]{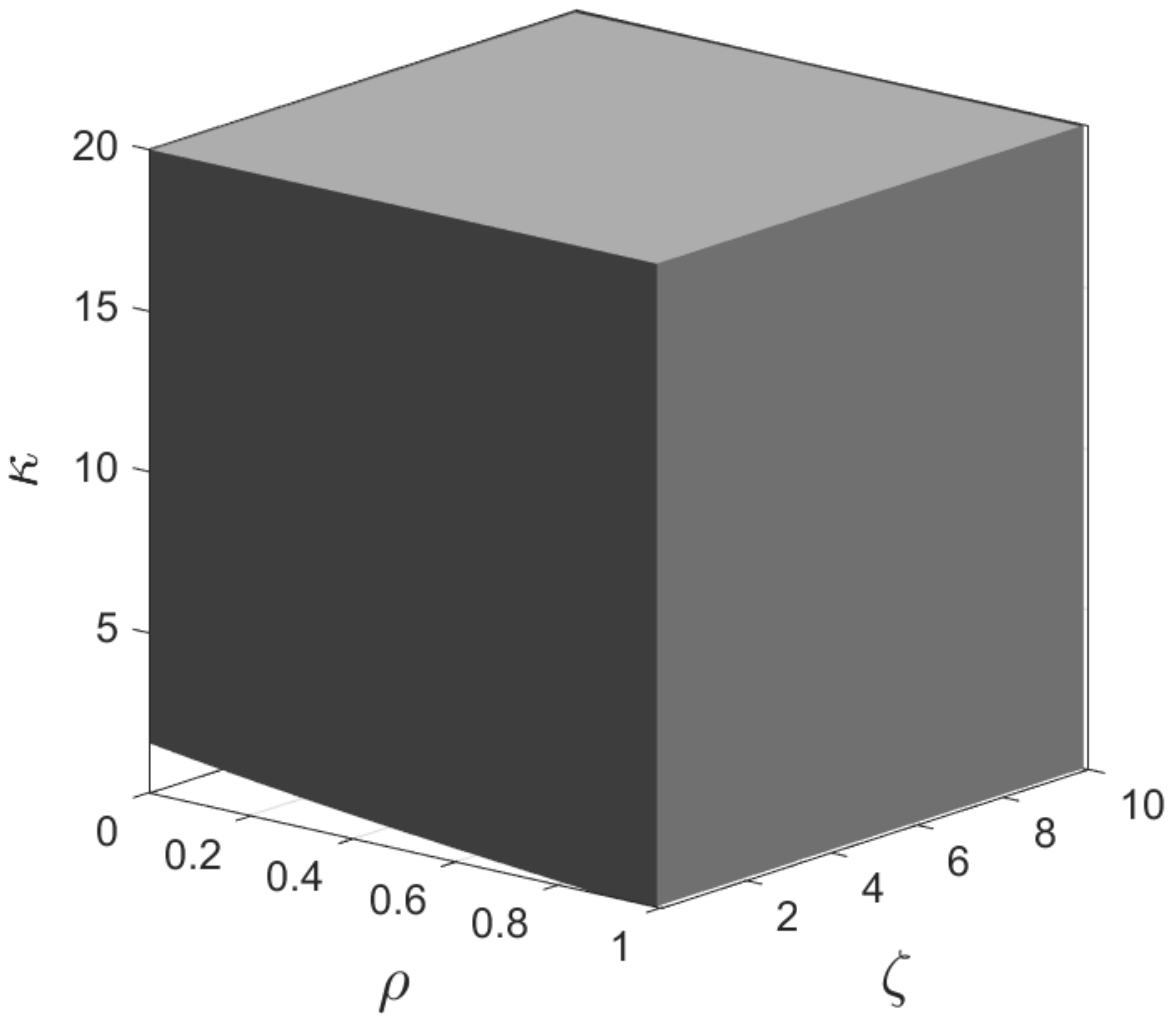}}
& \hspace{+.1cm}
{\includegraphics[angle=0,width=.20\textwidth,trim=120 280 140 230 ,
totalheight=.25\textwidth]{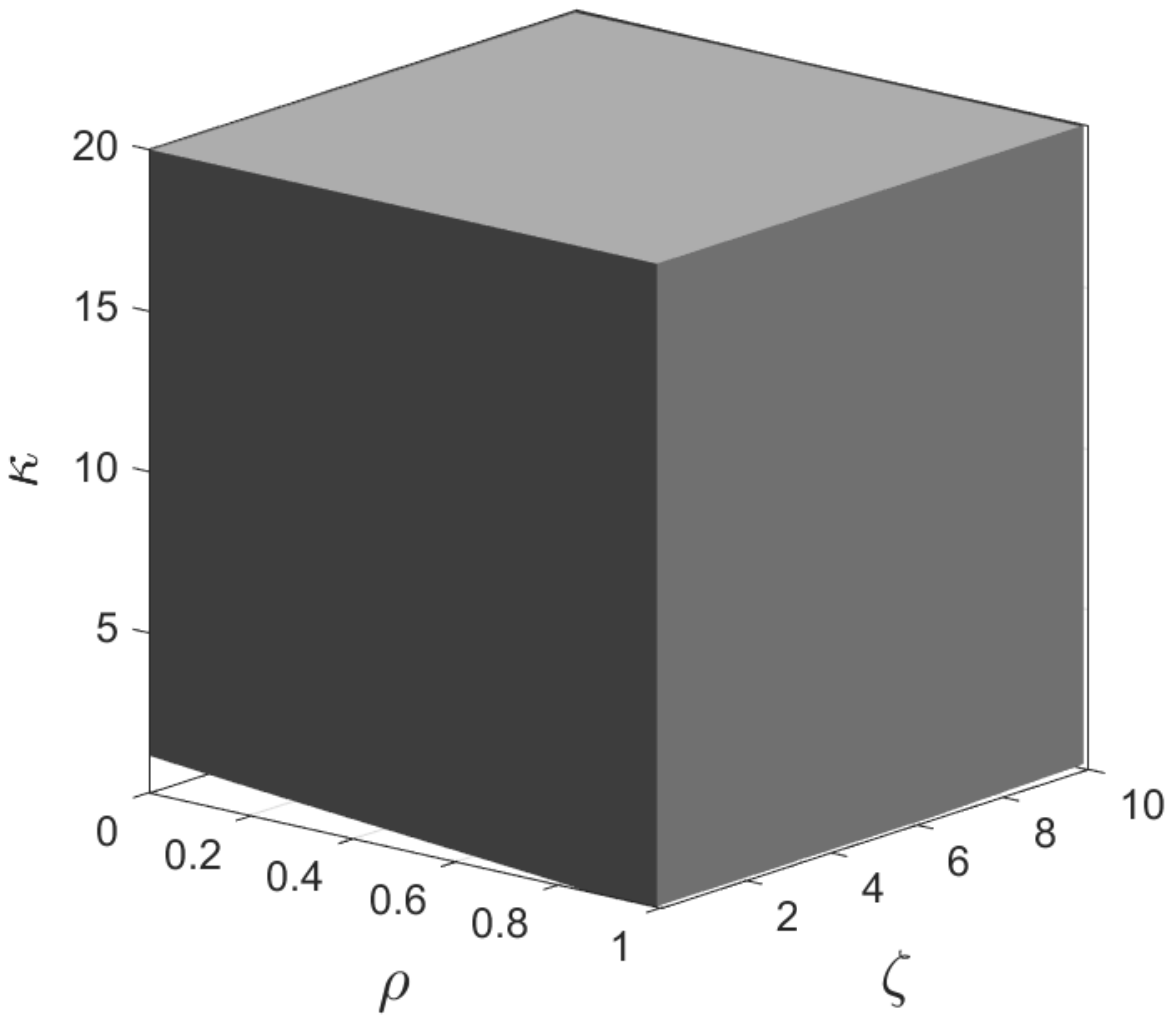}}
& \hspace{+.1cm}
{\includegraphics[angle=0,width=.20\textwidth,trim=120 280 140 230 ,
totalheight=.25\textwidth]{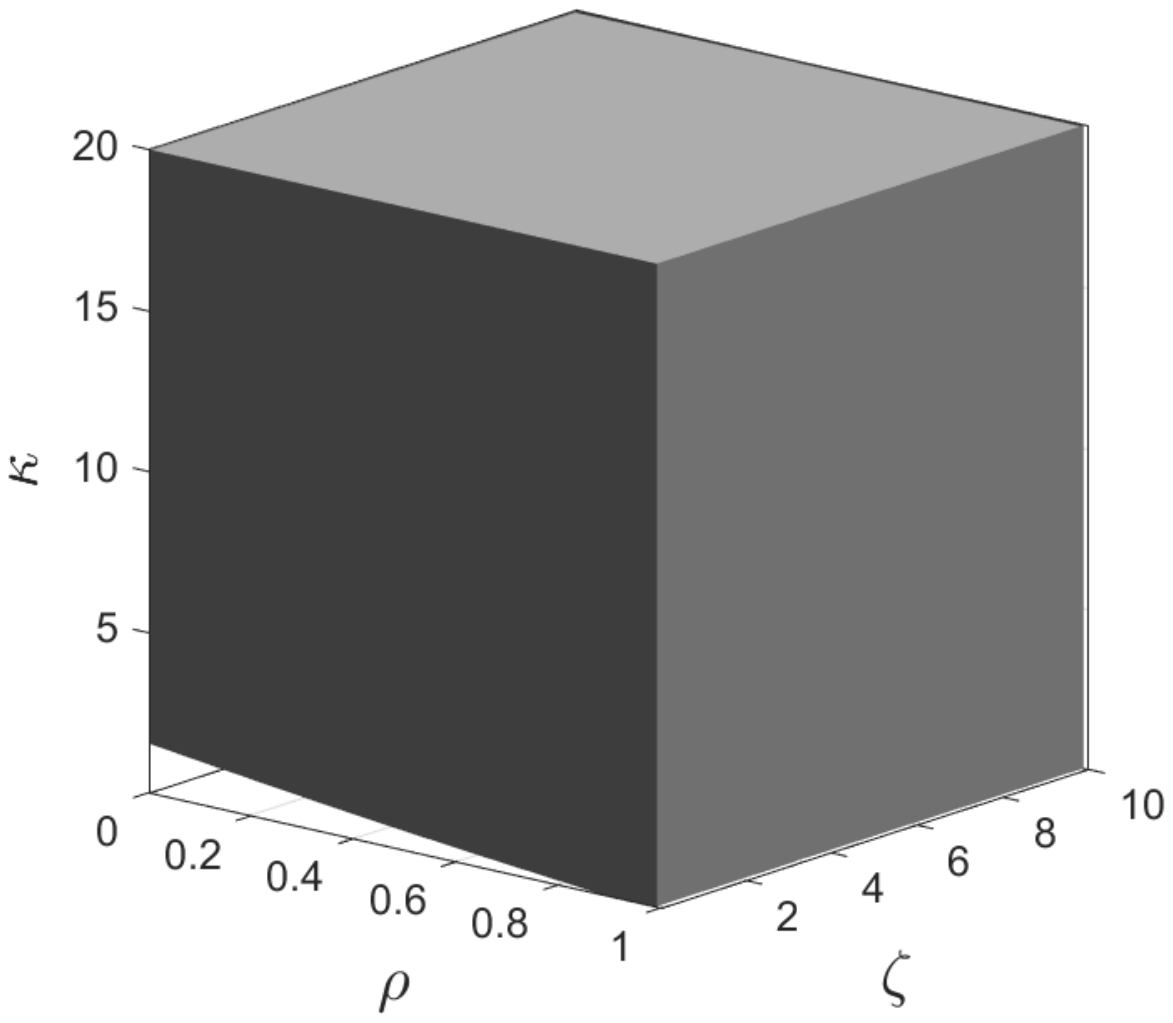}}
& \hspace{+.1cm}
{\includegraphics[angle=0,width=.20\textwidth,trim=120 280 140 230 ,
totalheight=.25\textwidth]{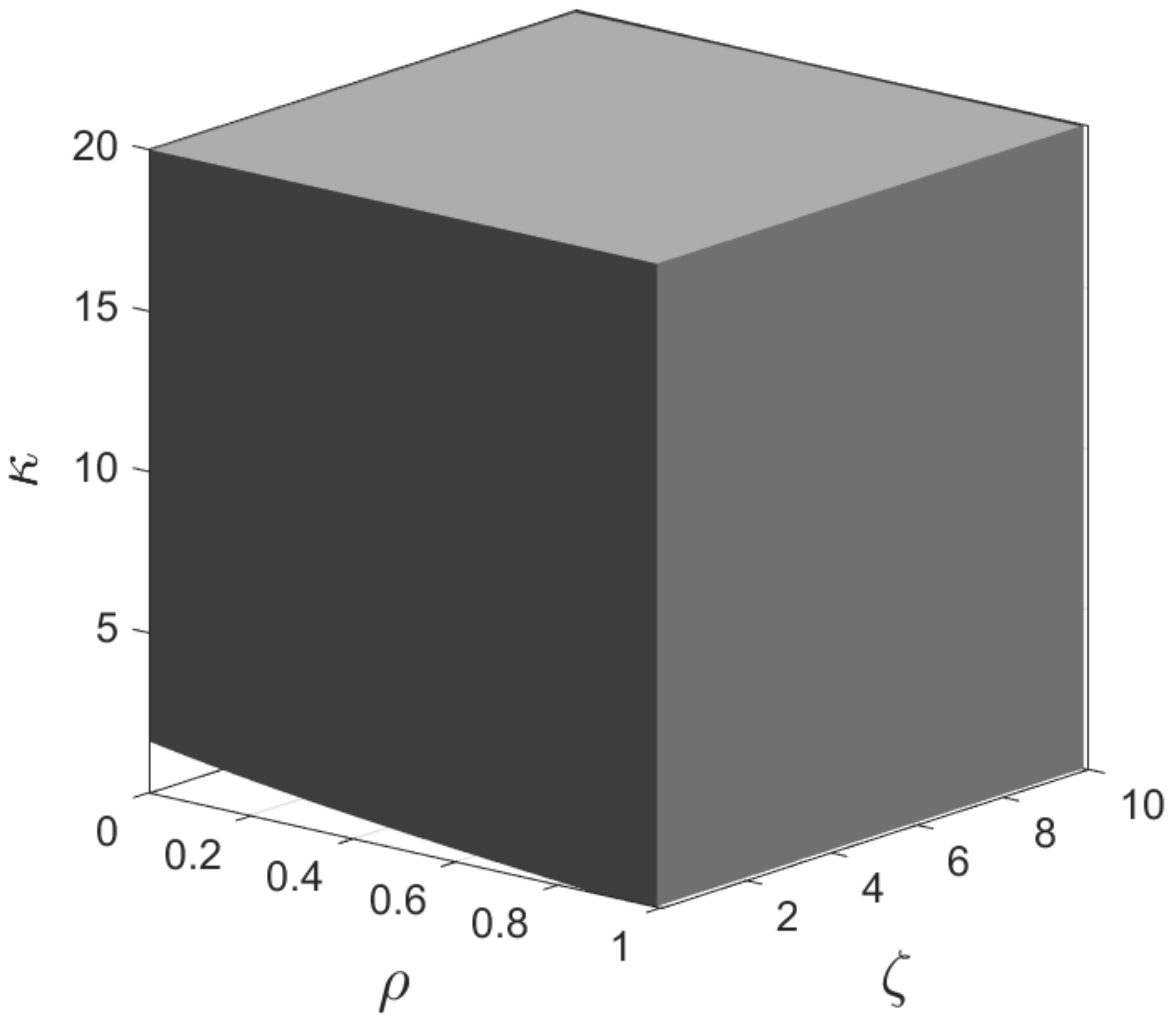}}\\[+3pt]
& {\small {VXO}} & {\small (j)+(k)} & {\small (l)+(m)} &
{\small {(j)+(k)+(l)+(m)}}\\
& {\small {1967Q1-2019Q4}} & {\small {1969Q2-2007Q4}} &
{\small {1967Q1-2019Q4}} & {\small {1969Q2-2007Q4}}\\
& {\small (m)} & {\small (n)} & {\small (o)} & {\small (p)}\vspace{-.4cm}\\
\raisebox{+5.7ex}{\rotatebox[origin=lt]{90}{qLL-S sets }}\hspace{+.4cm}
&
{\includegraphics[angle=0,width=.18\textwidth, trim=120 280 140 230 ,
totalheight=.25\textwidth]{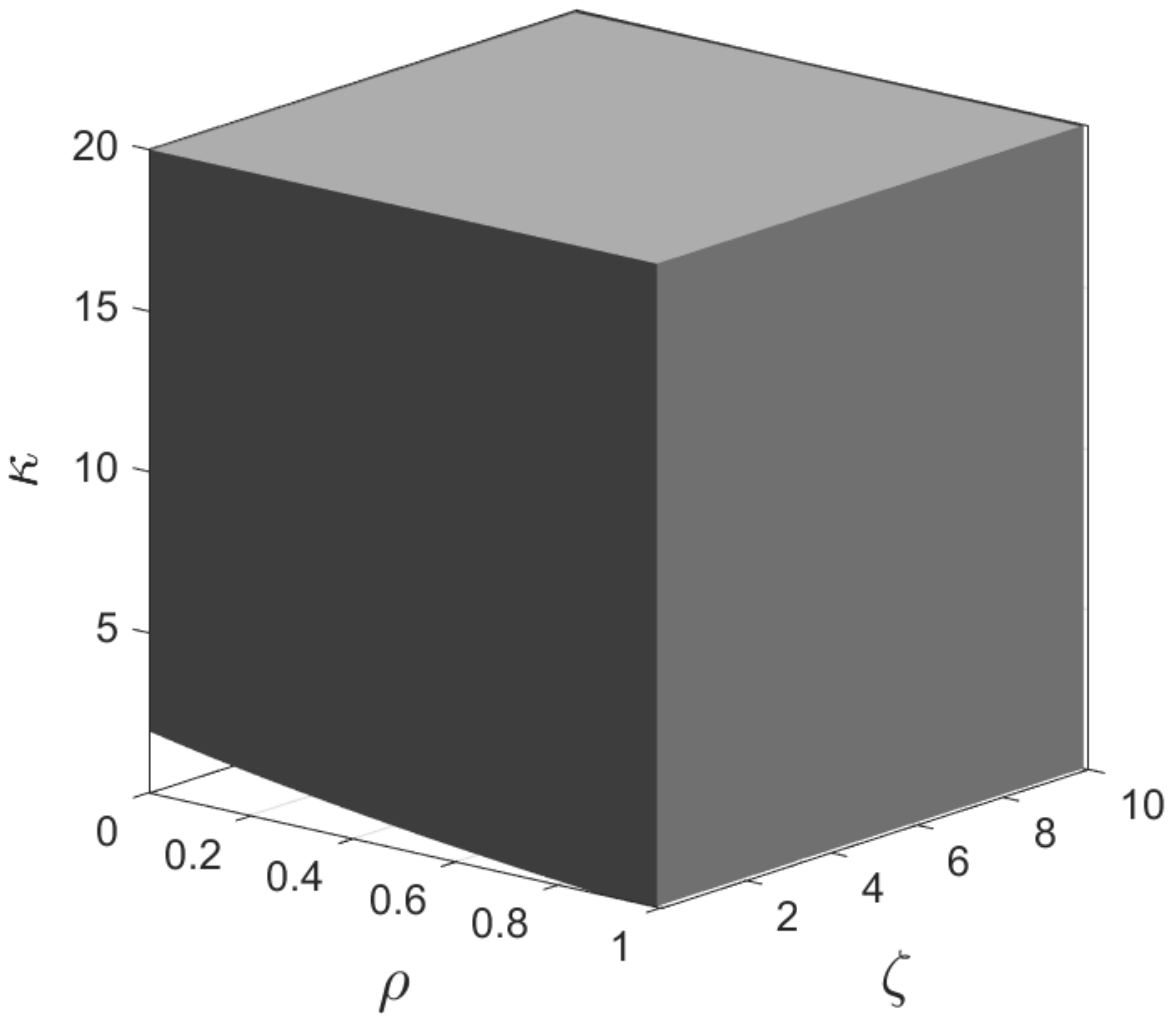}}
& \hspace{+.1cm}
{\includegraphics[angle=0,width=.20\textwidth,trim=120 280 140 230 ,
totalheight=.25\textwidth]{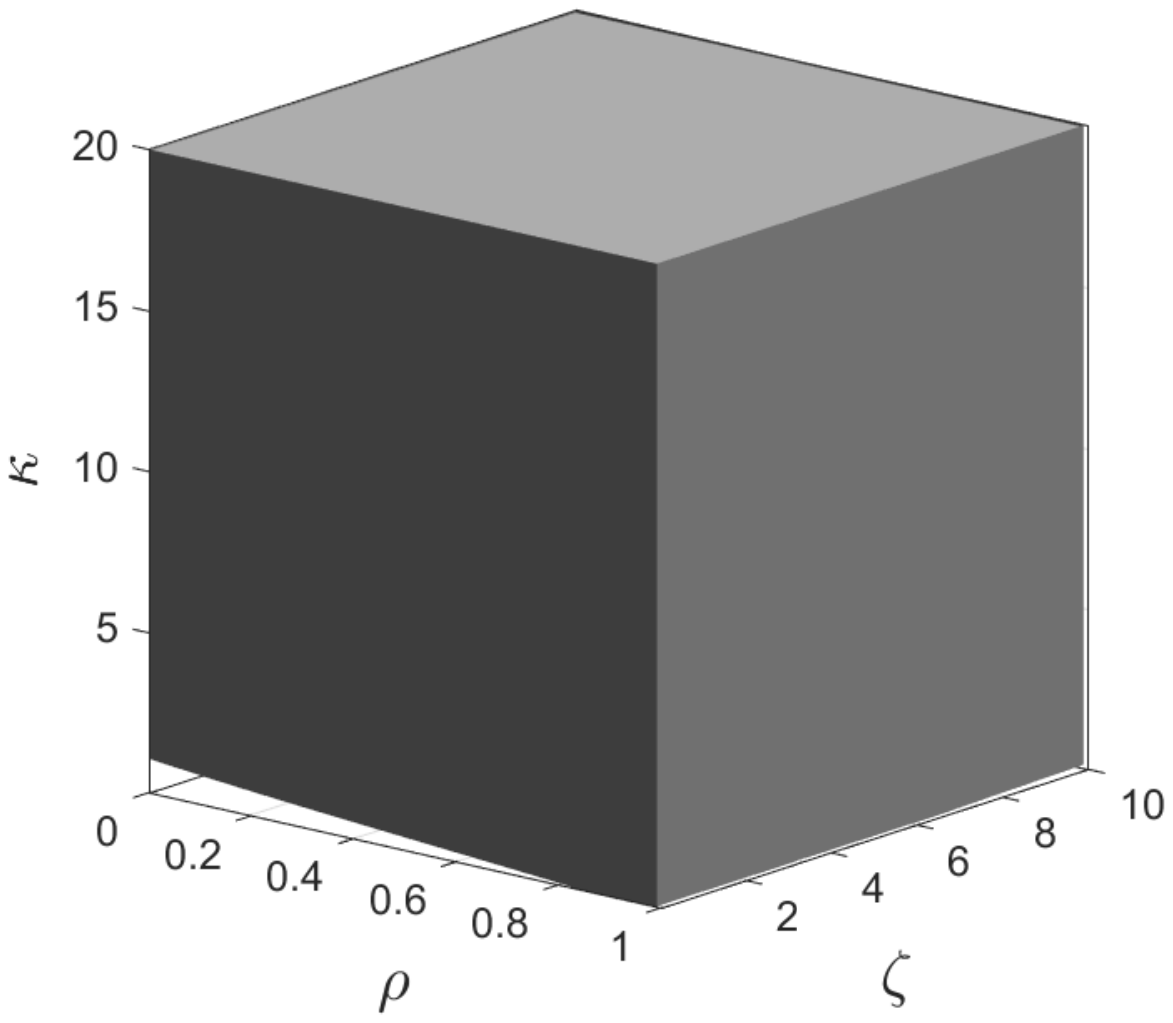}}
& \hspace{+.1cm}
{\includegraphics[angle=0,width=.20\textwidth,trim=120 280 140 230 ,
totalheight=.25\textwidth]{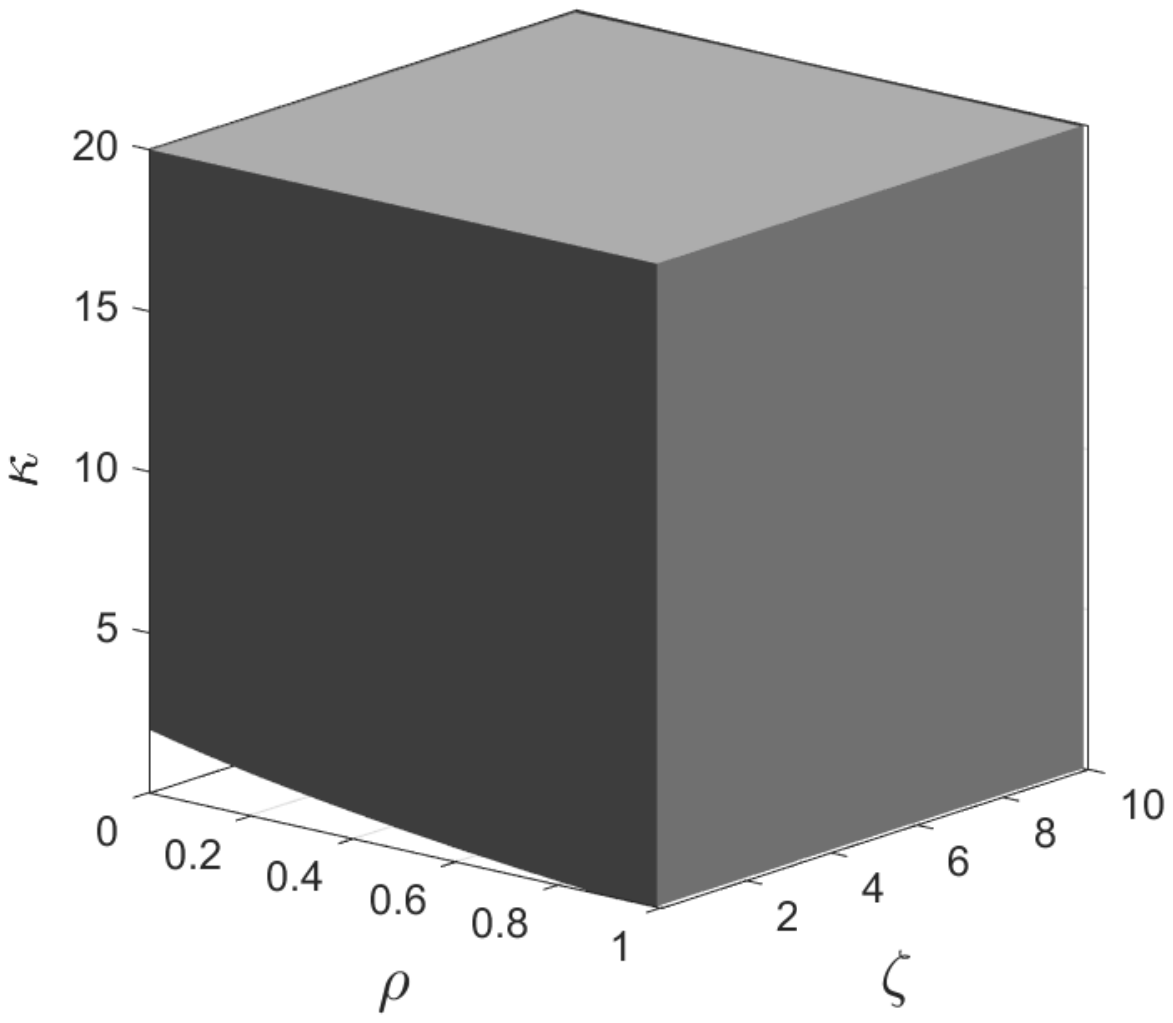}}
& \hspace{+.1cm}
{\includegraphics[angle=0,width=.20\textwidth,trim=120 280 140 230 ,
totalheight=.25\textwidth]{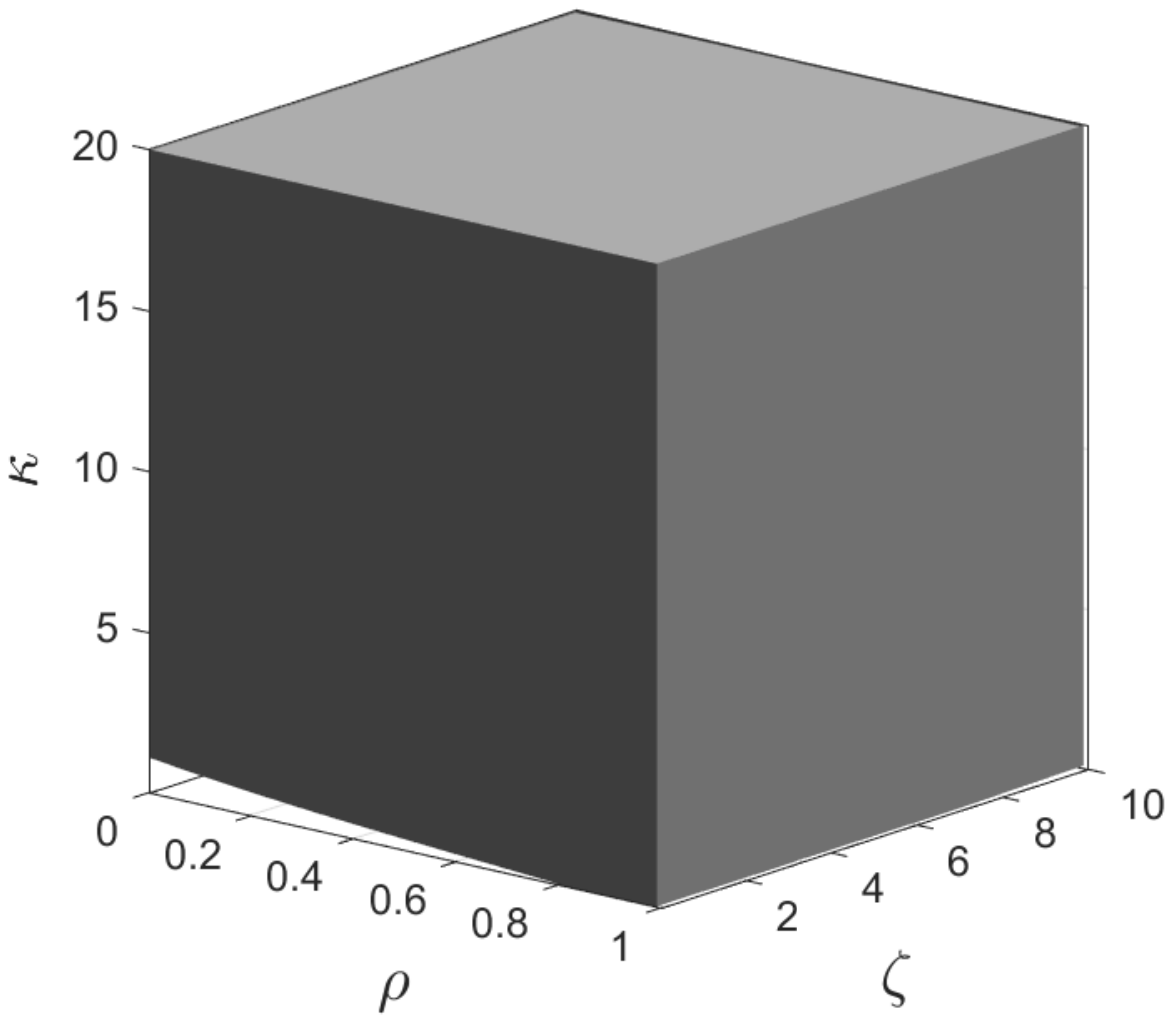}}\\\hline\hline
\end{tabular}} \caption{ 90\% S and qLL-S confidence sets for $\theta
=(\rho,\kappa,\zeta)$ derived from the investment Euler equation model
(\ref{eq: estimated}) using Fixed Private Investment as investment
proxy. A constant, $\Delta i_{t-1}$, $r_{t-2}^{p}$, and $u_{t-1}$ are common
instruments in all specifications. The additional instrument(s) by
specification is (are): \protect\underline{Mon. pol. shock}:
\citeauthor{romer2004new}'s \citeyearpar{romer2004new} monetary policy shock;
\protect\underline{Military news}: \citeauthor{ramey2018government}'s
\citeyearpar{ramey2018government} military news shock; \protect\underline{Oil}%
: growth rate of real oil price; \protect\underline{VXO}: financial
uncertainty. }%
\label{appfig: SW rob inst}%
\end{figure}

\begin{figure}[ptbh]
\centering
\adjustbox{min width=\textwidth,max width=\textwidth, max height=9.5cm}{
\begin{tabular}
[c]{ccccc}\hline\hline
& \multicolumn{4}{c}{Exogenous Instruments with JPT Investment Proxy}\\ \cline{2-5}
& {\small {Baseline}} & {\small {Mon. pol. shock}} & {\small {Military news}}
& {\small {Oil}}\\
& {\small {1967Q1-2019Q4}} & {\small {1969Q2-2007Q4}} &
{\small {1967Q1-2015Q4}} & {\small {1967Q1-2019Q4}}\\
& {\small (a)} & {\small (b)} & {\small (c)} & {\small (d)}\vspace{-.5cm}\\
\raisebox{+8.7ex}{\rotatebox[origin=lt]{90}{S sets }}\hspace{+.4cm} &
{\includegraphics[angle=0,width=.18\textwidth, trim=120 270 140 230 ,
totalheight=.25\textwidth]{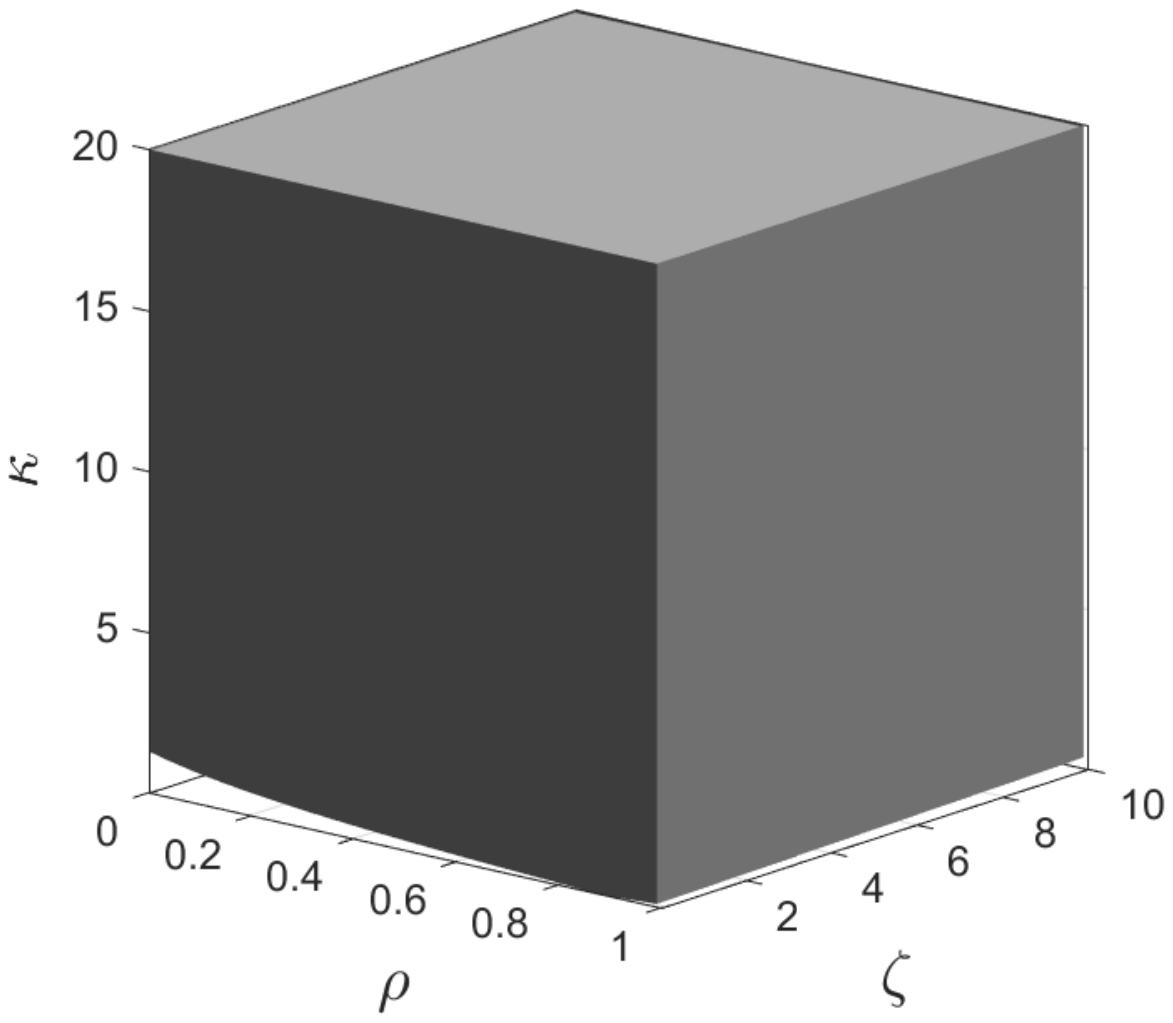}} &
\hspace{+.1cm}
{\includegraphics[angle=0,width=.20\textwidth,trim=120 280 140 230 ,
totalheight=.25\textwidth]{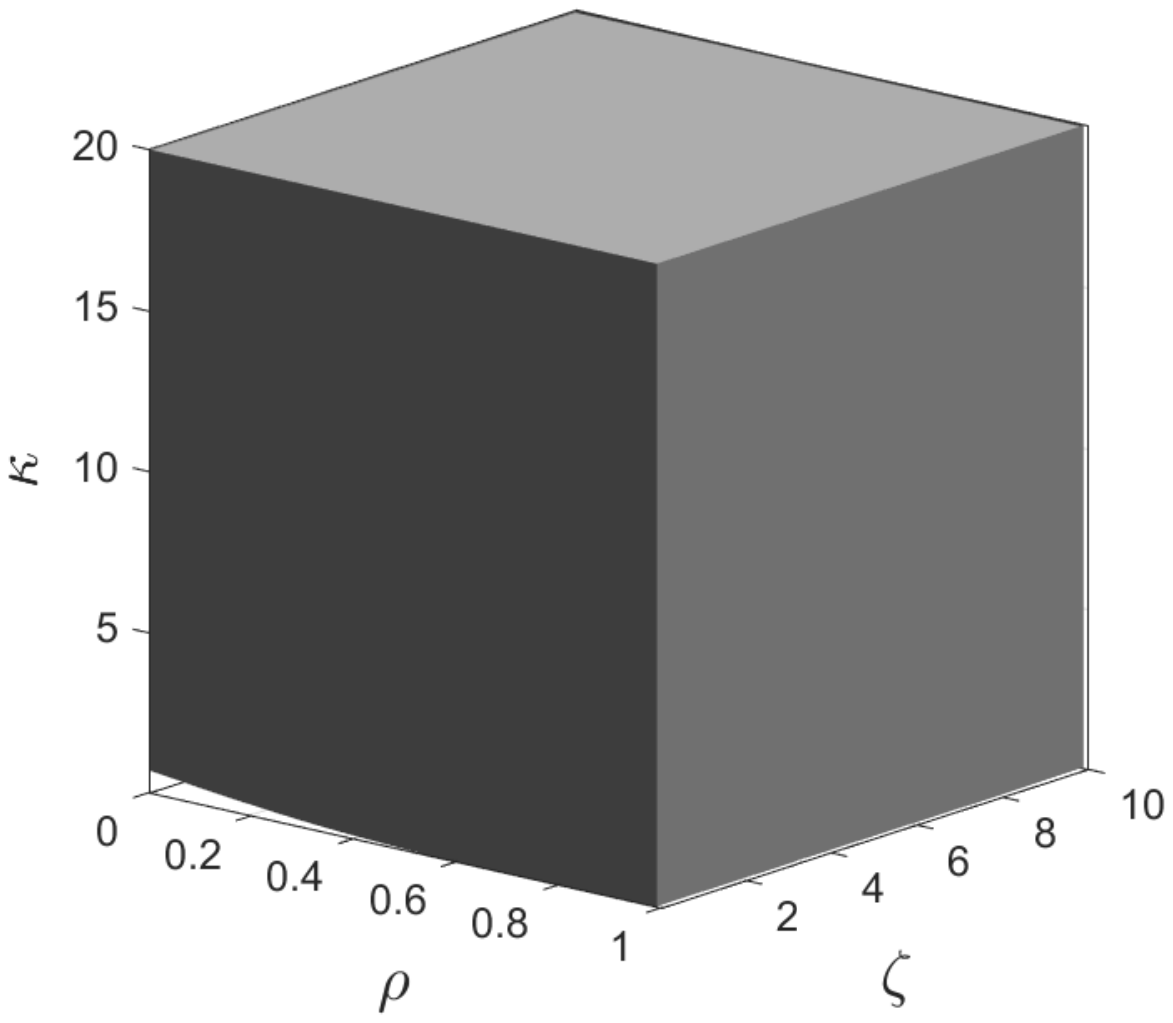}} & \hspace{+.1cm}
{\includegraphics[angle=0,width=.20\textwidth,trim=120 280 140 230 ,
totalheight=.25\textwidth]{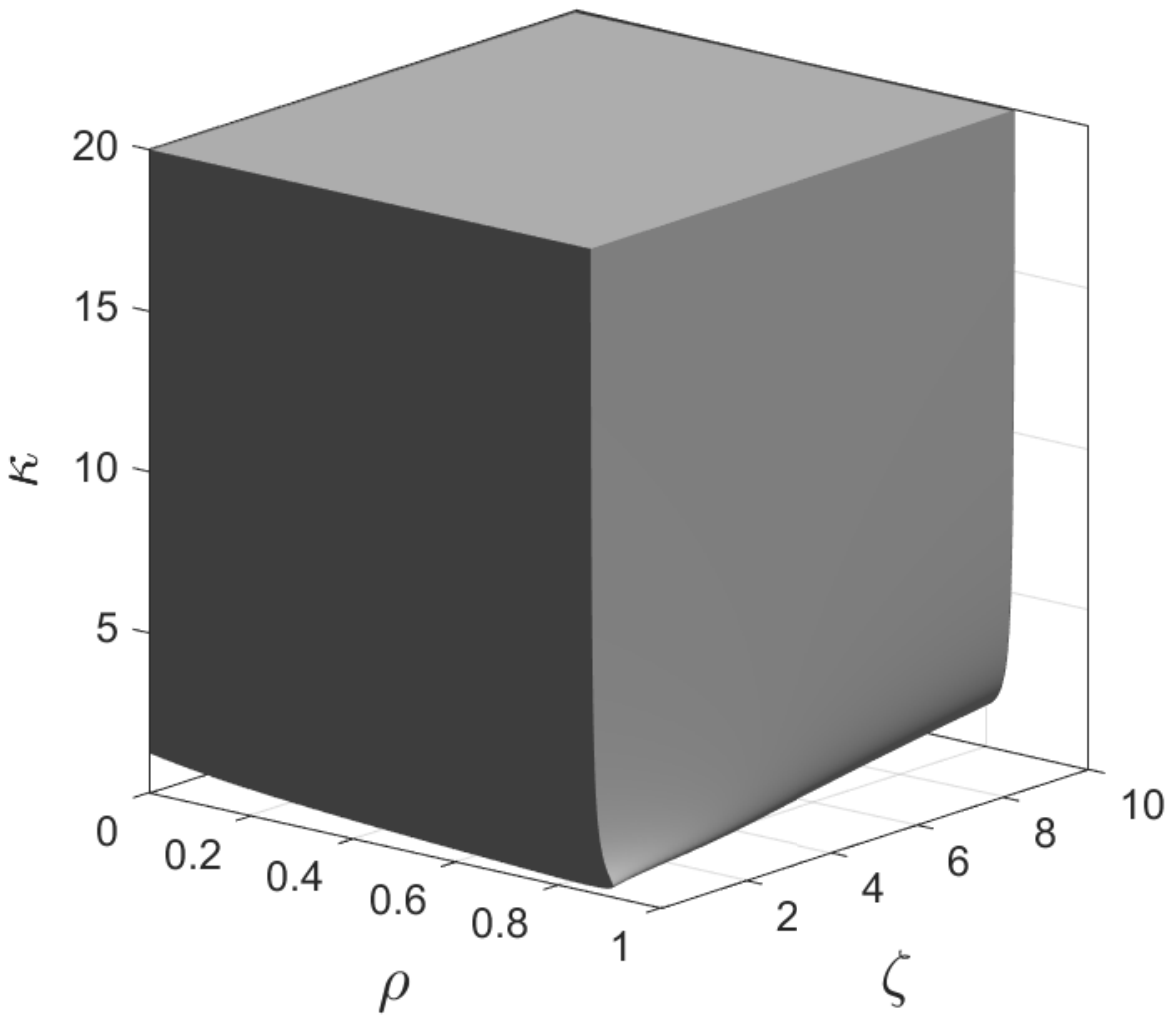}} &
\hspace{+.1cm}
{\includegraphics[angle=0,width=.20\textwidth,trim=120 280 140 230 ,
totalheight=.25\textwidth]{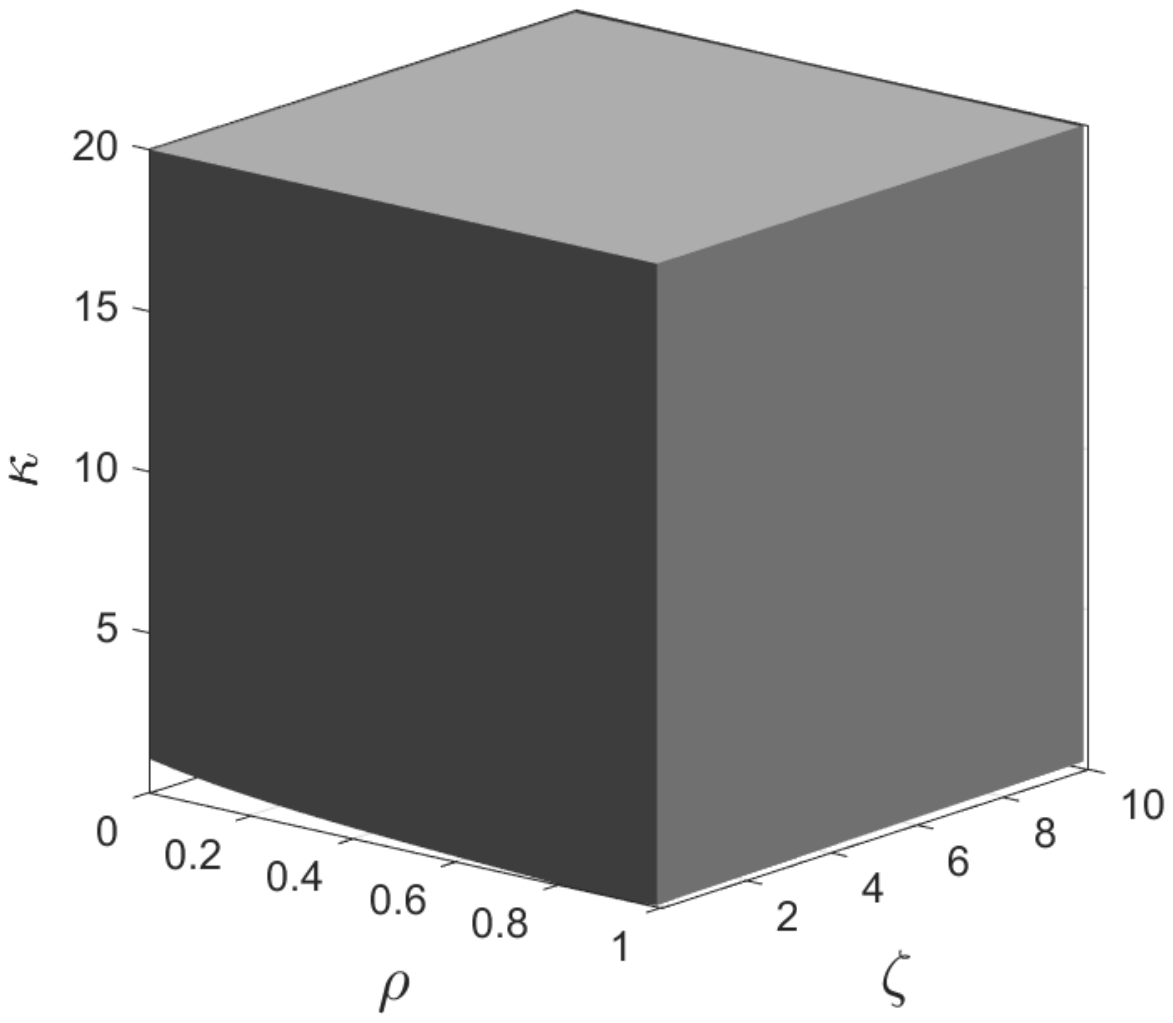}}\\[+3pt]
& {\small {VXO}} & {\small (b)+(c)} & {\small (d)+(e)} &
{\small {(b)+(c)+(d)+(e)}}\\
& {\small {1967Q1-2019Q4}} & {\small {1969Q2-2007Q4}} &
{\small {1967Q1-2019Q4}} & {\small {1969Q2-2007Q4}}\\
& {\small (e)} & {\small (f)} & {\small (g)} & {\small (h)}\vspace{-.4cm}\\
\raisebox{+8.7ex}{\rotatebox[origin=lt]{90}{S sets }}\hspace{+.4cm} &
{\includegraphics[angle=0,width=.18\textwidth, trim=120 280 140 230 ,
totalheight=.25\textwidth]{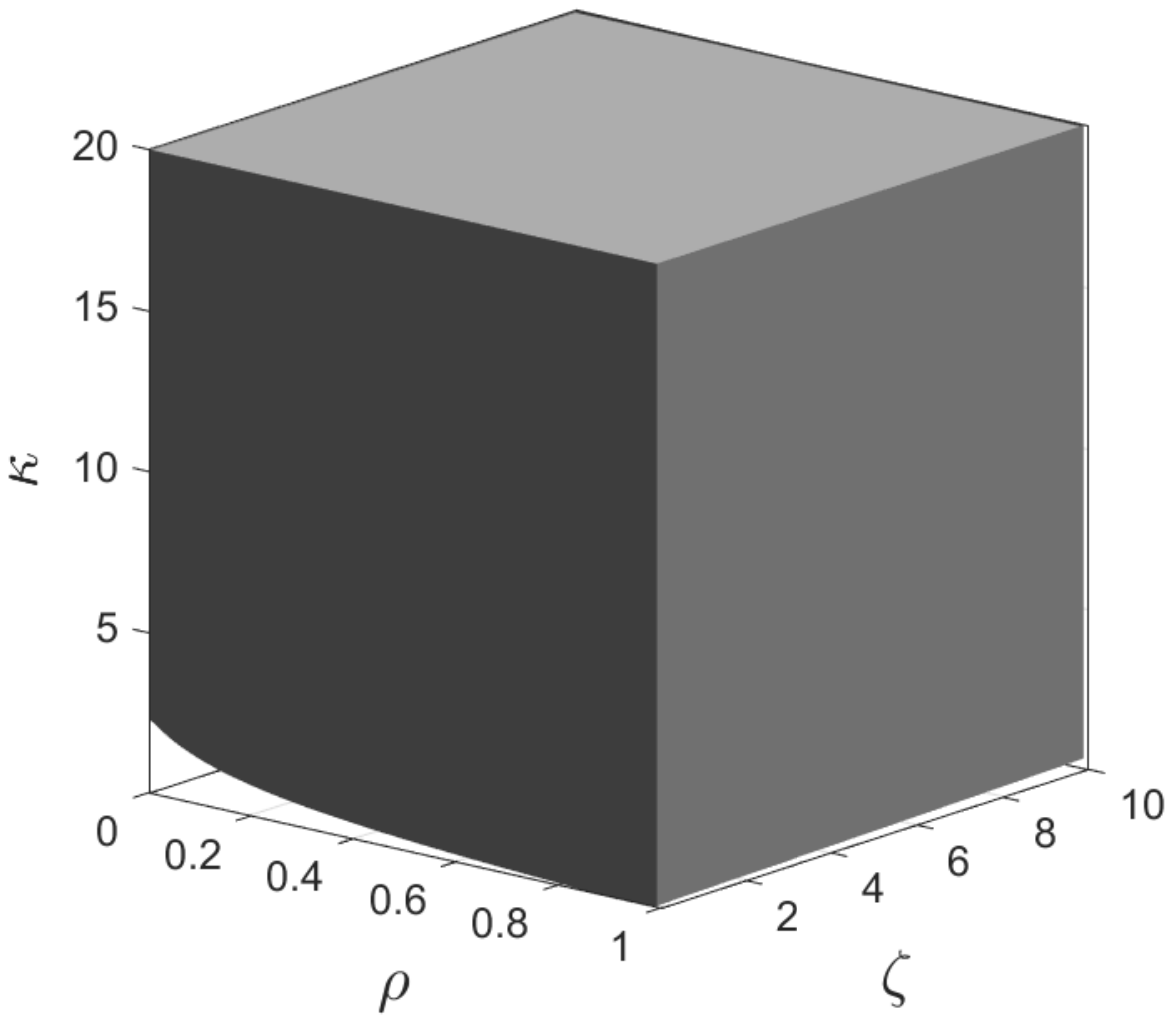}} & \hspace{+.1cm}
{\includegraphics[angle=0,width=.20\textwidth,trim=120 280 140 230 ,
totalheight=.25\textwidth]{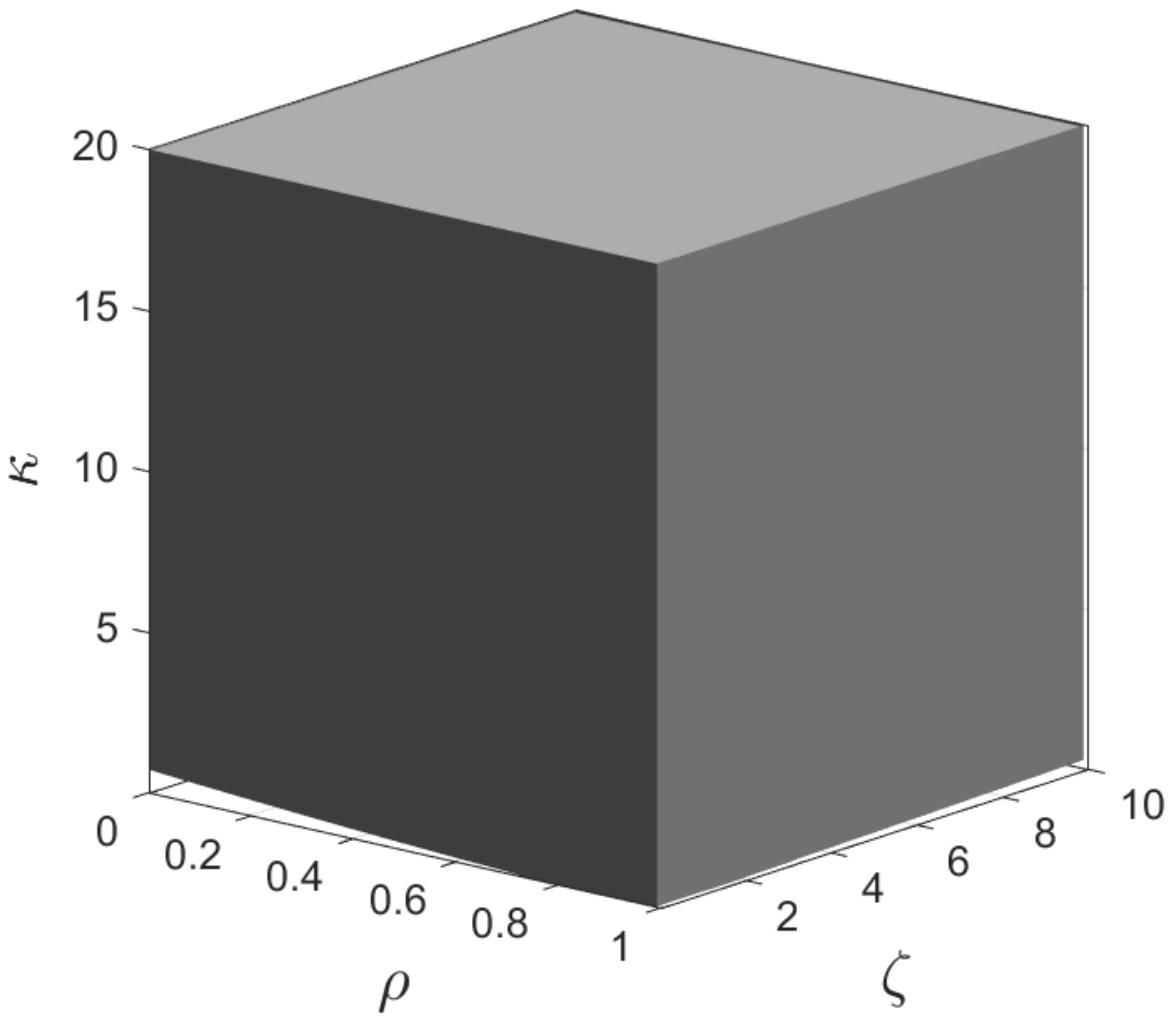}} &
\hspace{+.1cm}
{\includegraphics[angle=0,width=.20\textwidth,trim=120 280 140 230 ,
totalheight=.25\textwidth]{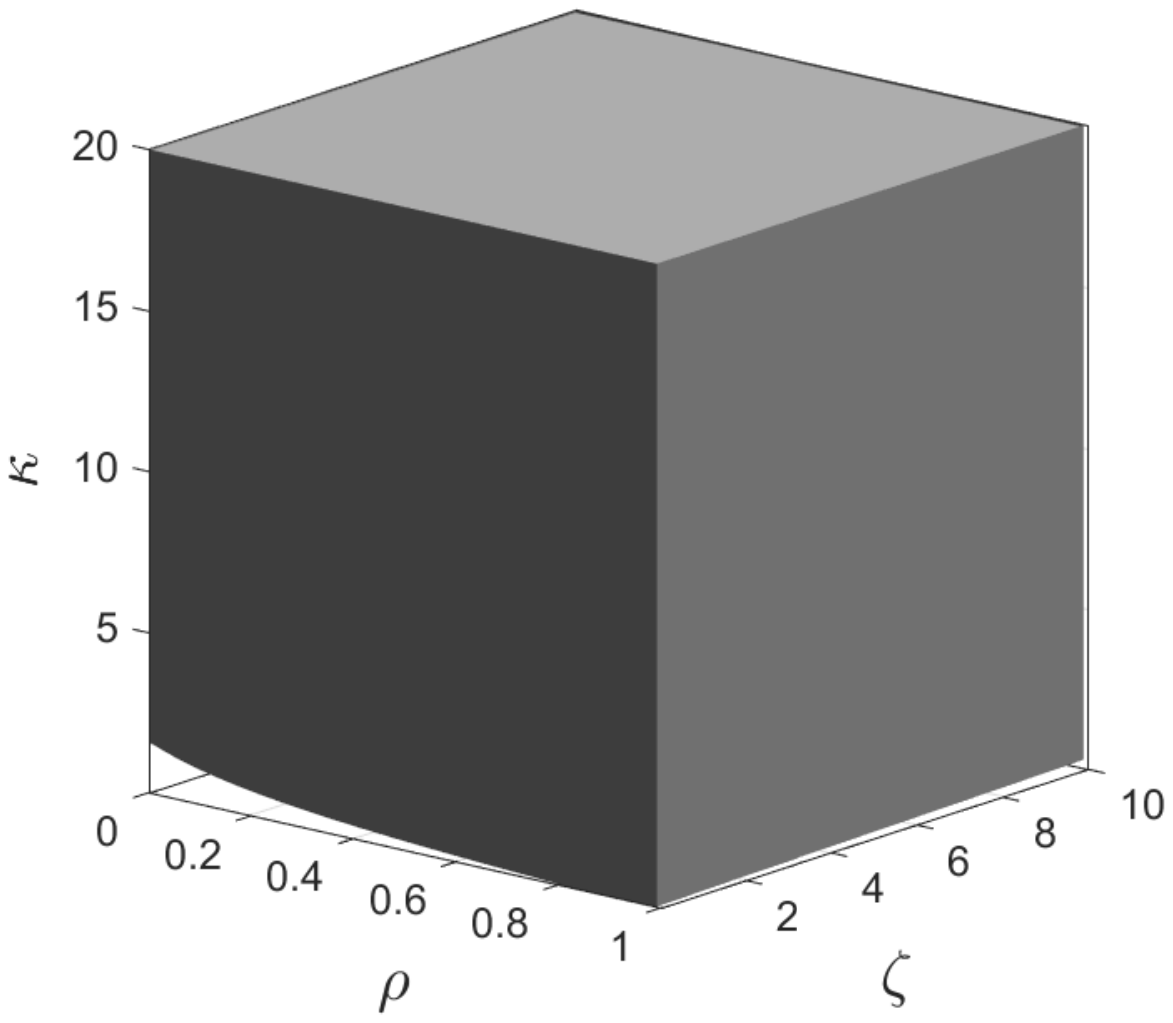}} & \hspace{+.1cm}
{\includegraphics[angle=0,width=.20\textwidth,trim=120 280 140 230 ,
totalheight=.25\textwidth]{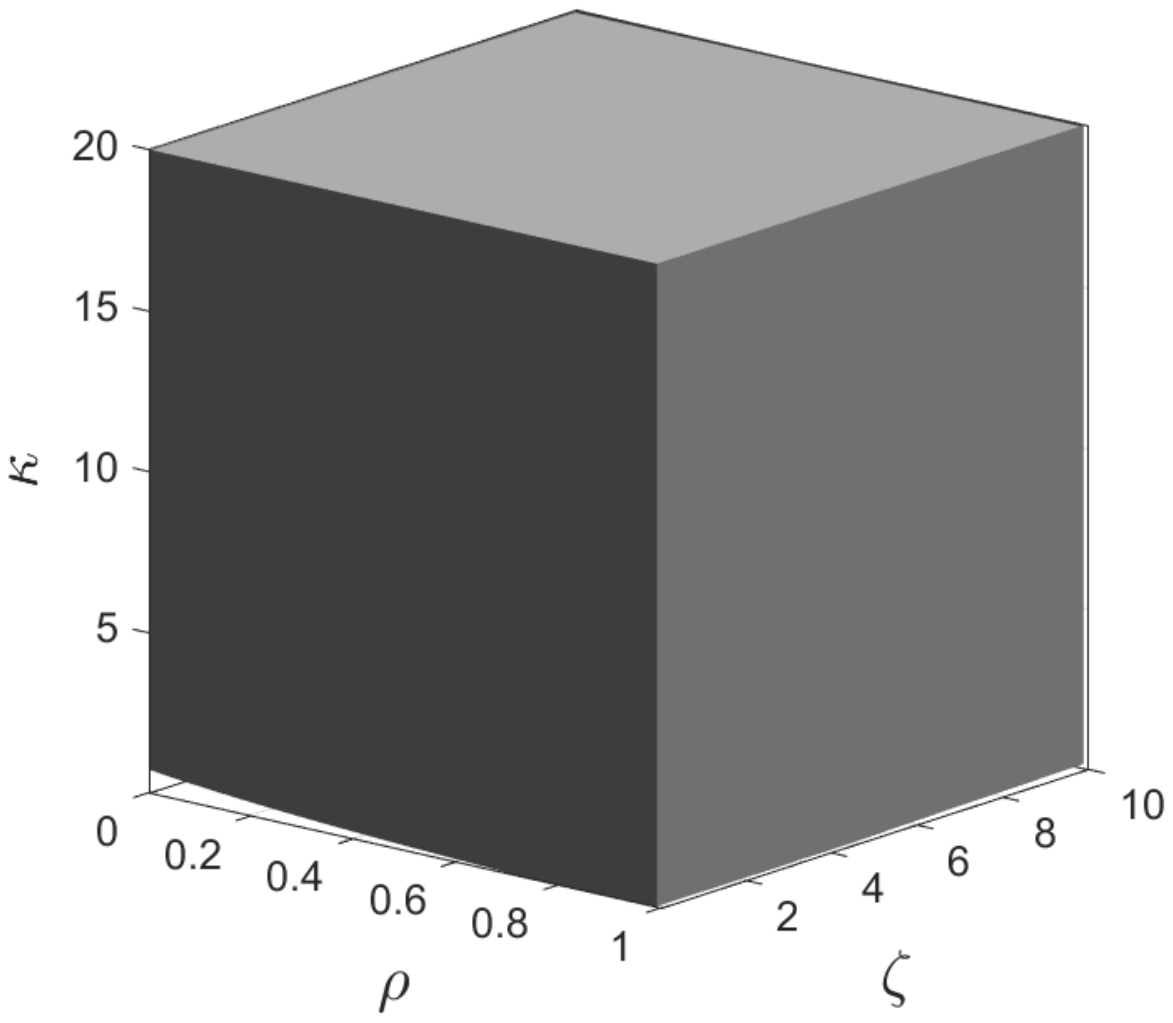}}\vspace{+10pt}\\
&  &  &  & \\
& {\small {Baseline}} & {\small {Mon. pol. shock}} & {\small {Military news}}
& {\small {Oil}}\\
& {\small {1967Q1-2019Q4}} & {\small {1969Q2-2007Q4}} &
{\small {1967Q1-2015Q4}} & {\small {1967Q1-2019Q4}}\\
& {\small (i)} & {\small (j)} & {\small (k)} & {\small (l)}\vspace{-.4cm}\\
\raisebox{+5.7ex}{\rotatebox[origin=lt]{90}{qLL-S sets }}\hspace{+.4cm} &
{\includegraphics[angle=0,width=.18\textwidth, trim=120 270 140 230 ,
totalheight=.25\textwidth]{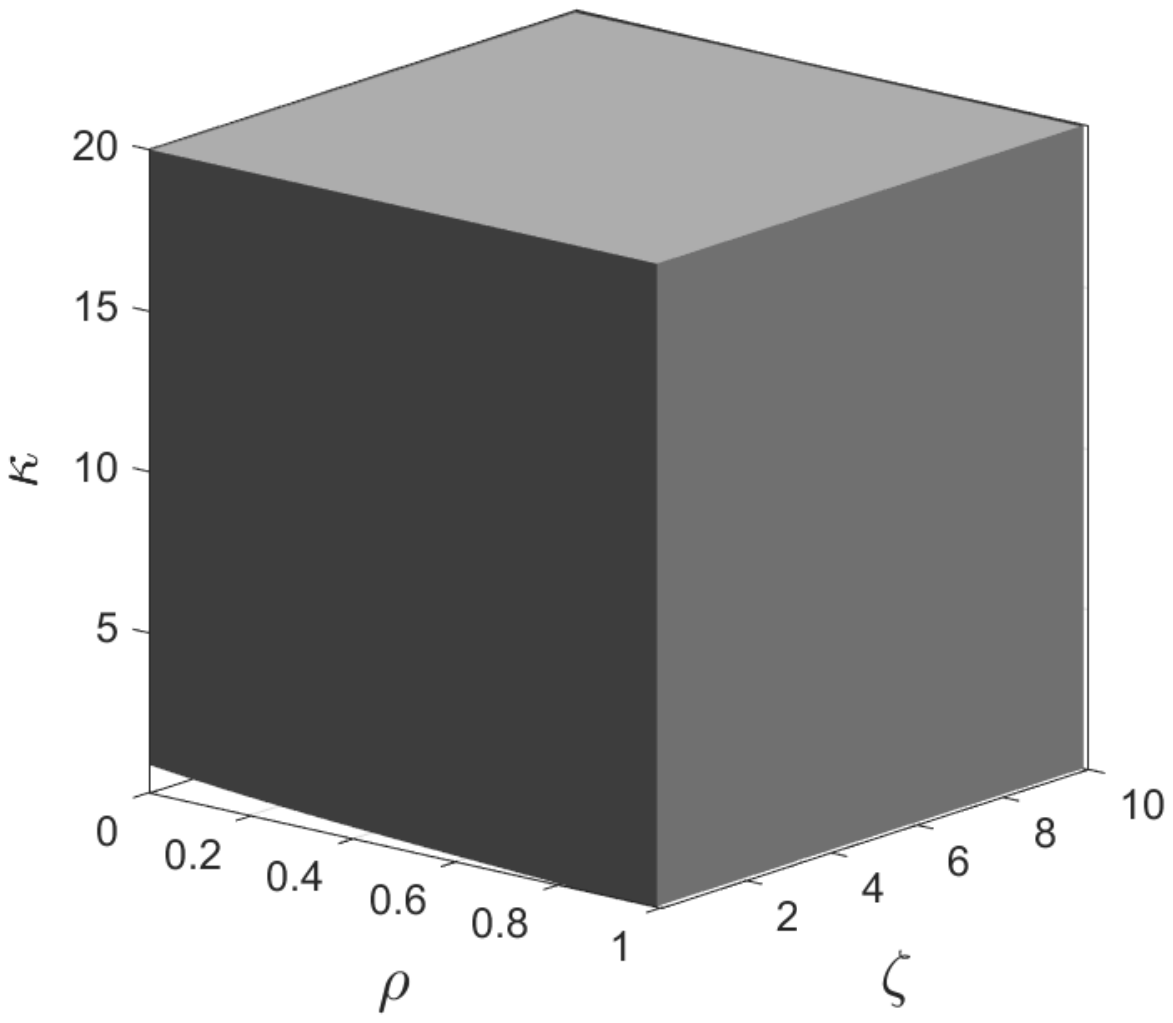}} &
\hspace{+.1cm}
{\includegraphics[angle=0,width=.20\textwidth,trim=120 280 140 230 ,
totalheight=.25\textwidth]{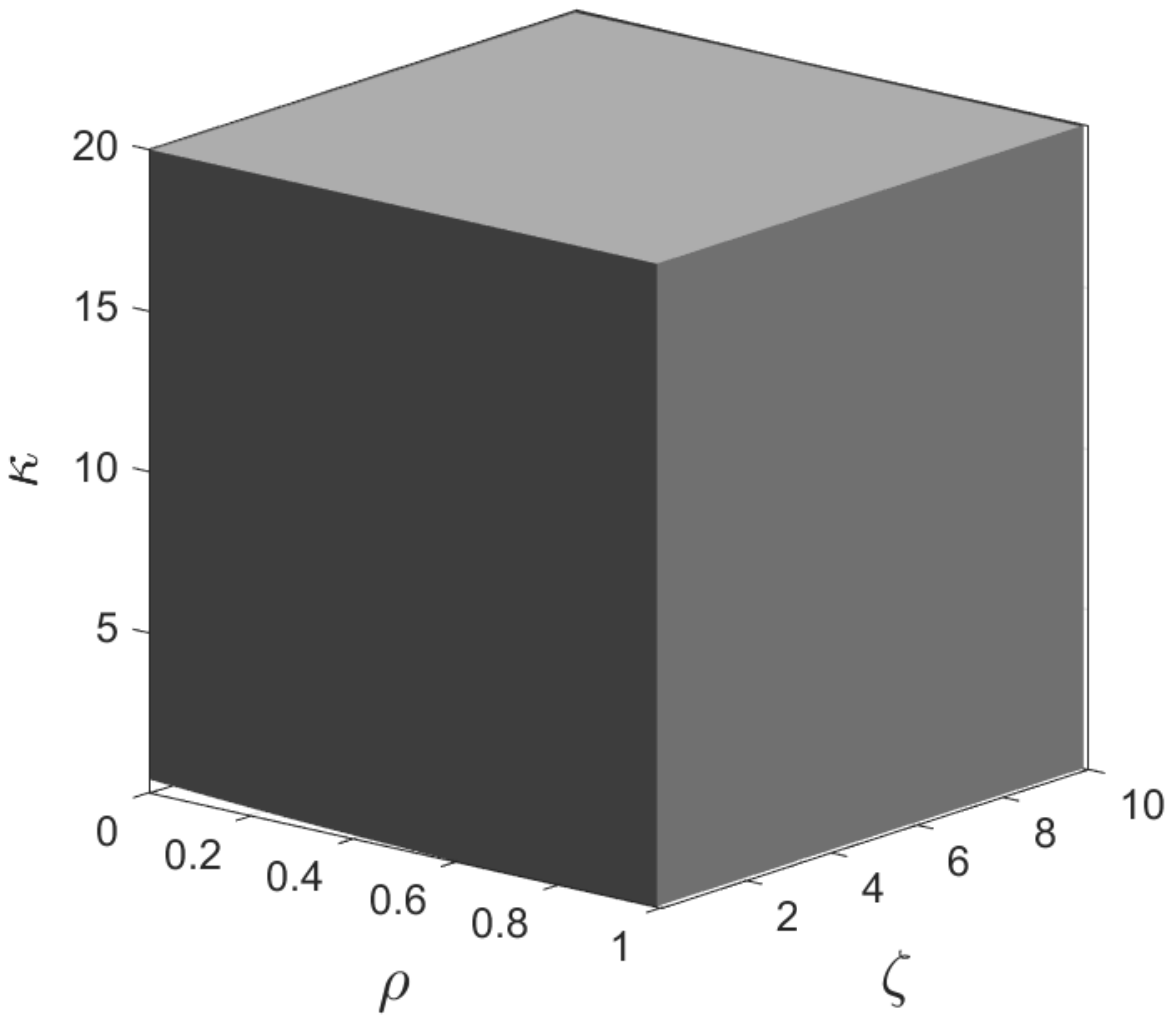}} &
\hspace{+.1cm}
{\includegraphics[angle=0,width=.20\textwidth,trim=120 280 140 230 ,
totalheight=.25\textwidth]{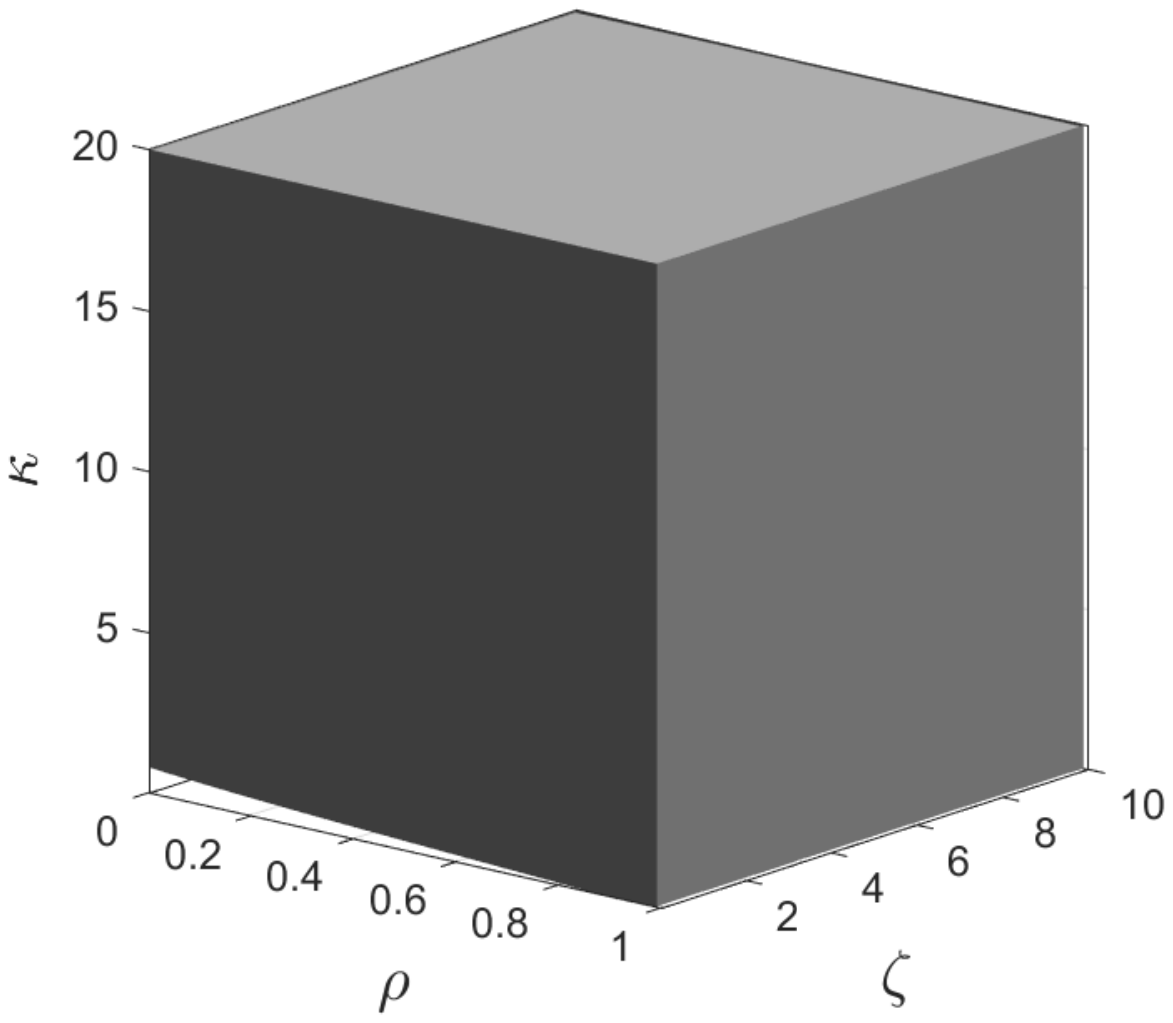}} &
\hspace{+.1cm}
{\includegraphics[angle=0,width=.20\textwidth,trim=120 280 140 230 ,
totalheight=.25\textwidth]{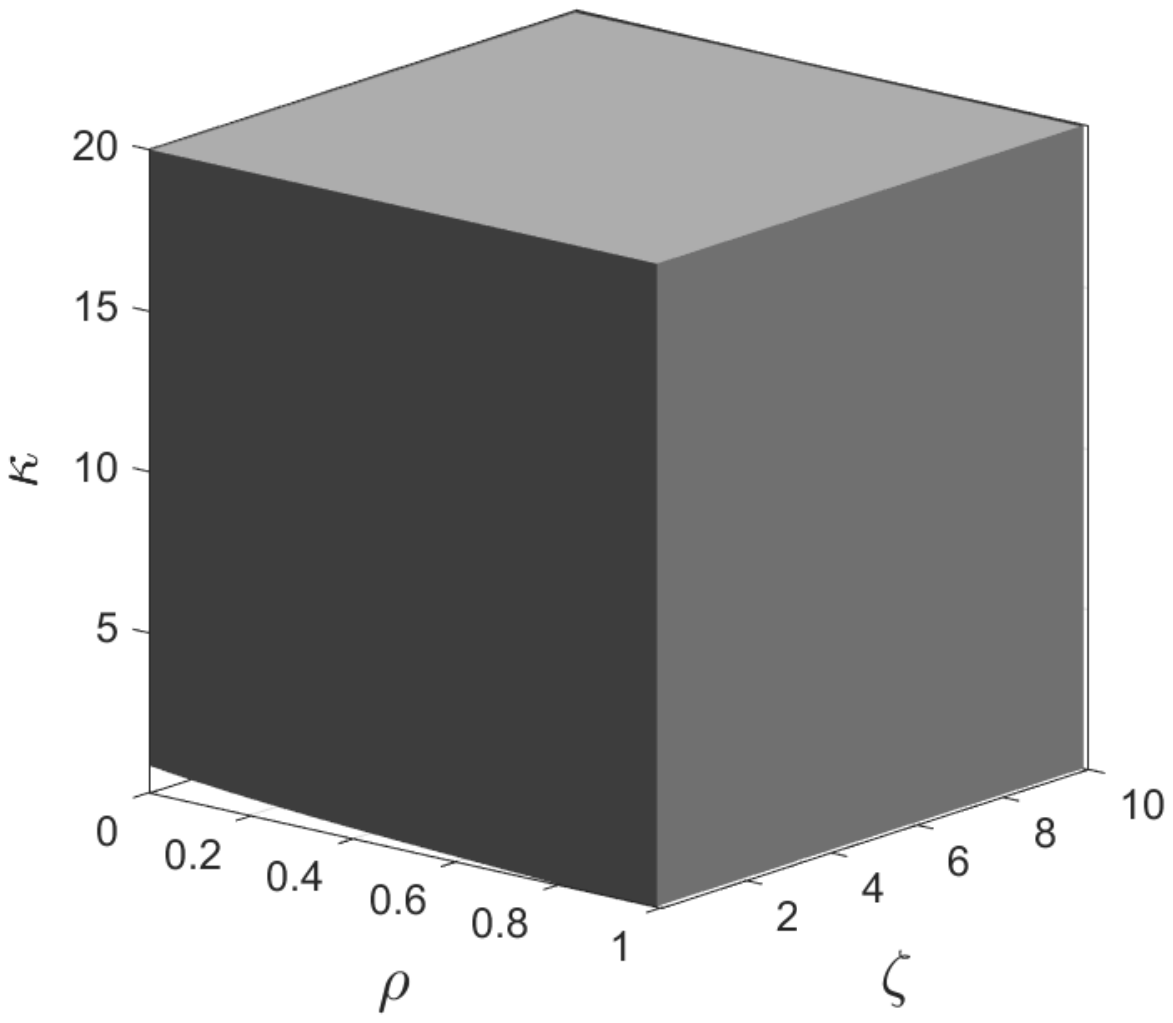}}\\[+3pt]
& {\small {VXO}} & {\small (j)+(k)} & {\small (l)+(m)} &
{\small {(j)+(k)+(l)+(m)}}\\
& {\small {1967Q1-2019Q4}} & {\small {1969Q2-2007Q4}} &
{\small {1967Q1-2019Q4}} & {\small {1969Q2-2007Q4}}\\
& {\small (m)} & {\small (n)} & {\small (o)} & {\small (p)}\vspace{-.4cm}\\
\raisebox{+5.7ex}{\rotatebox[origin=lt]{90}{qLL-S sets }}\hspace{+.4cm} &
{\includegraphics[angle=0,width=.18\textwidth, trim=120 280 140 230 ,
totalheight=.25\textwidth]{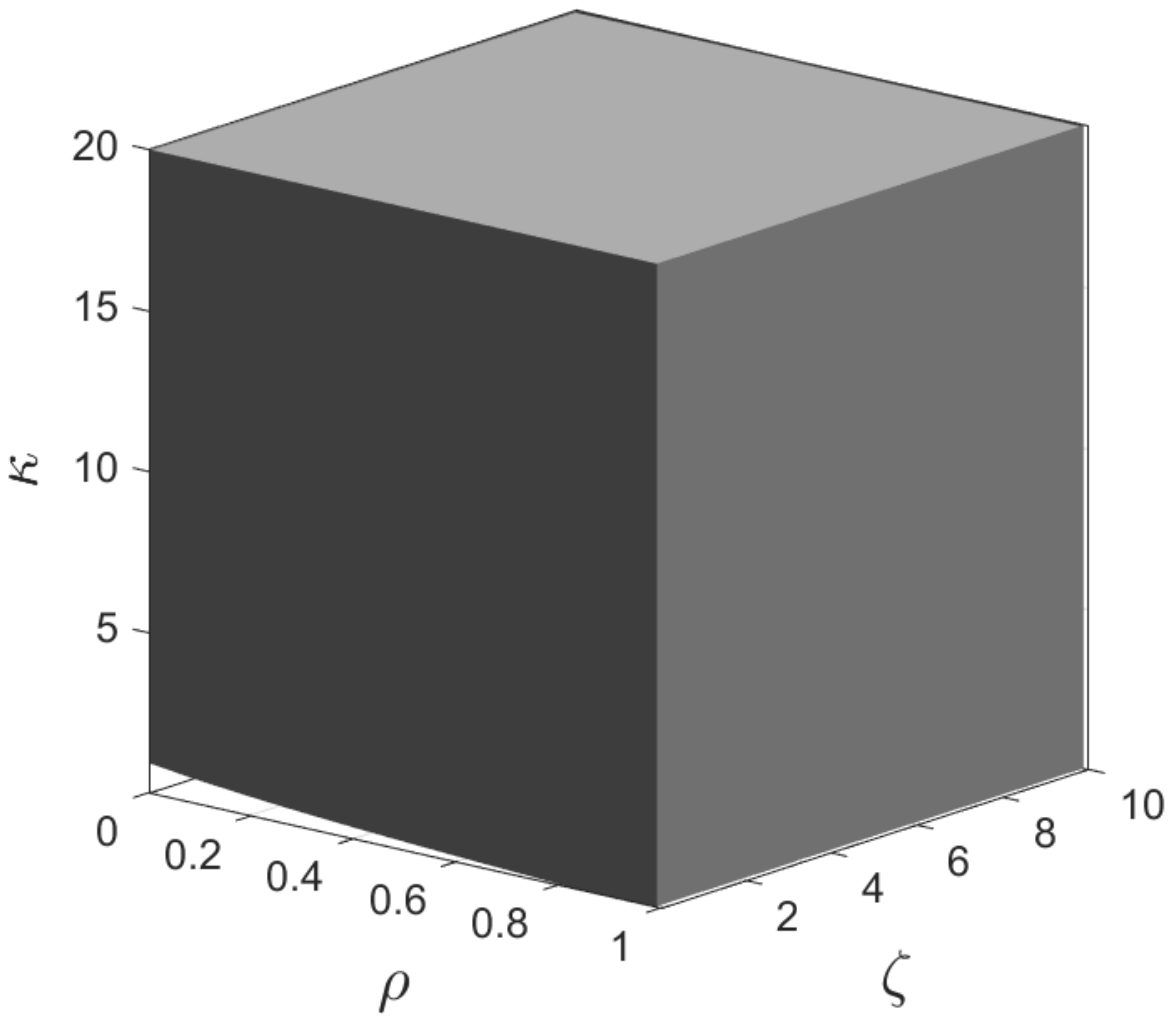}} & \hspace{+.1cm}
{\includegraphics[angle=0,width=.20\textwidth,trim=120 280 140 230 ,
totalheight=.25\textwidth]{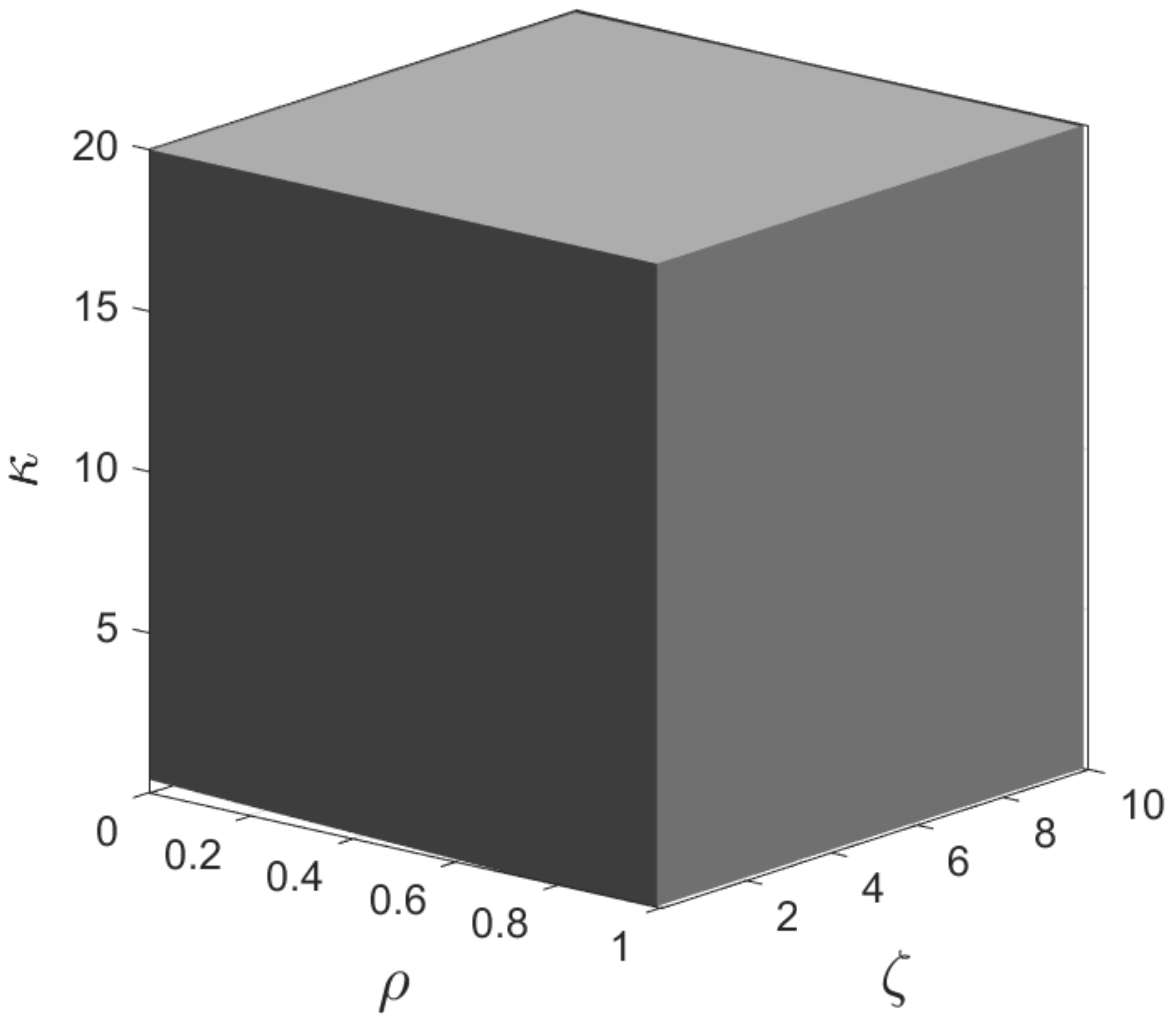}} &
\hspace{+.1cm}
{\includegraphics[angle=0,width=.20\textwidth,trim=120 280 140 230 ,
totalheight=.25\textwidth]{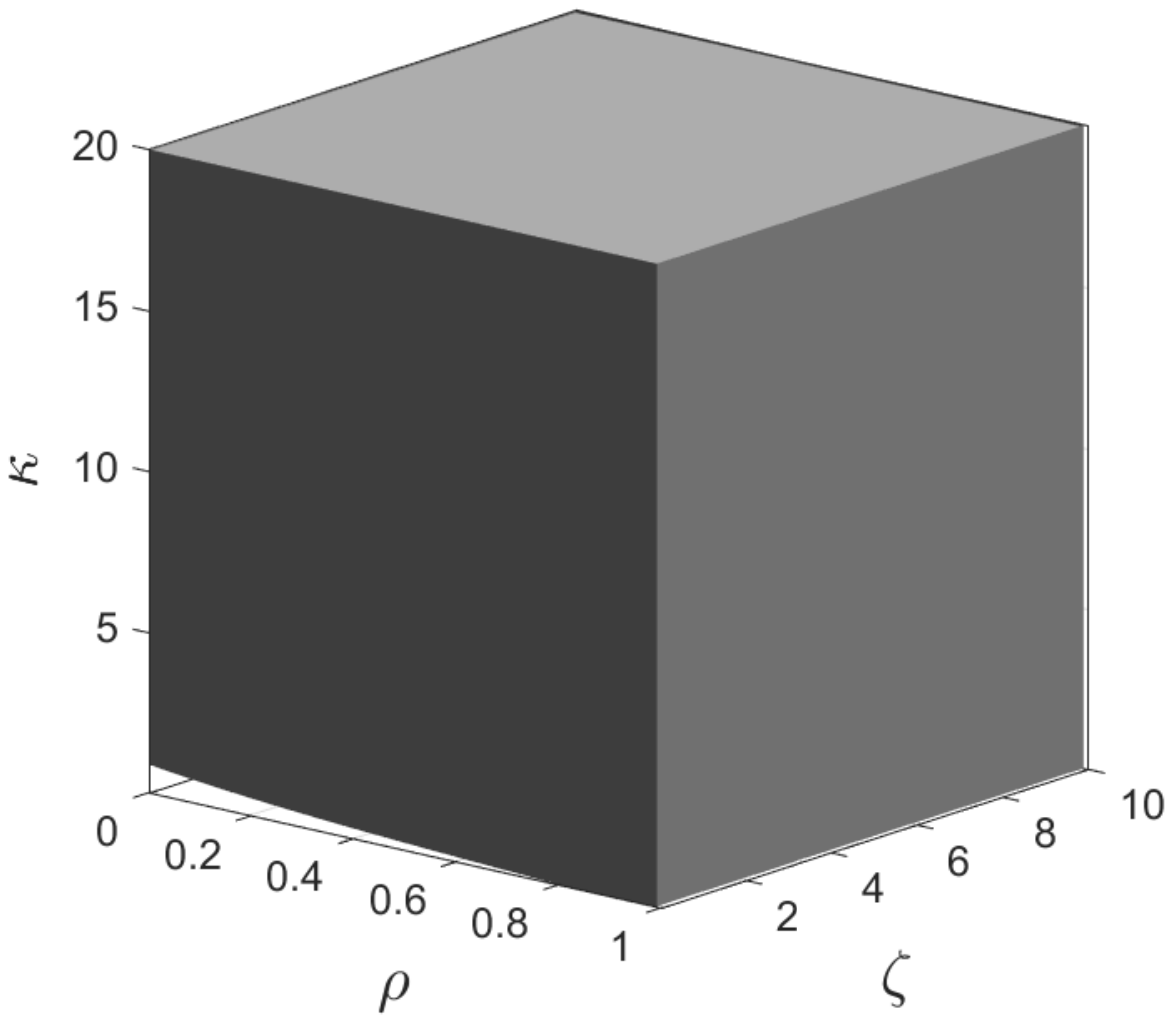}} &
\hspace{+.1cm}
{\includegraphics[angle=0,width=.20\textwidth,trim=120 280 140 230 ,
totalheight=.25\textwidth]{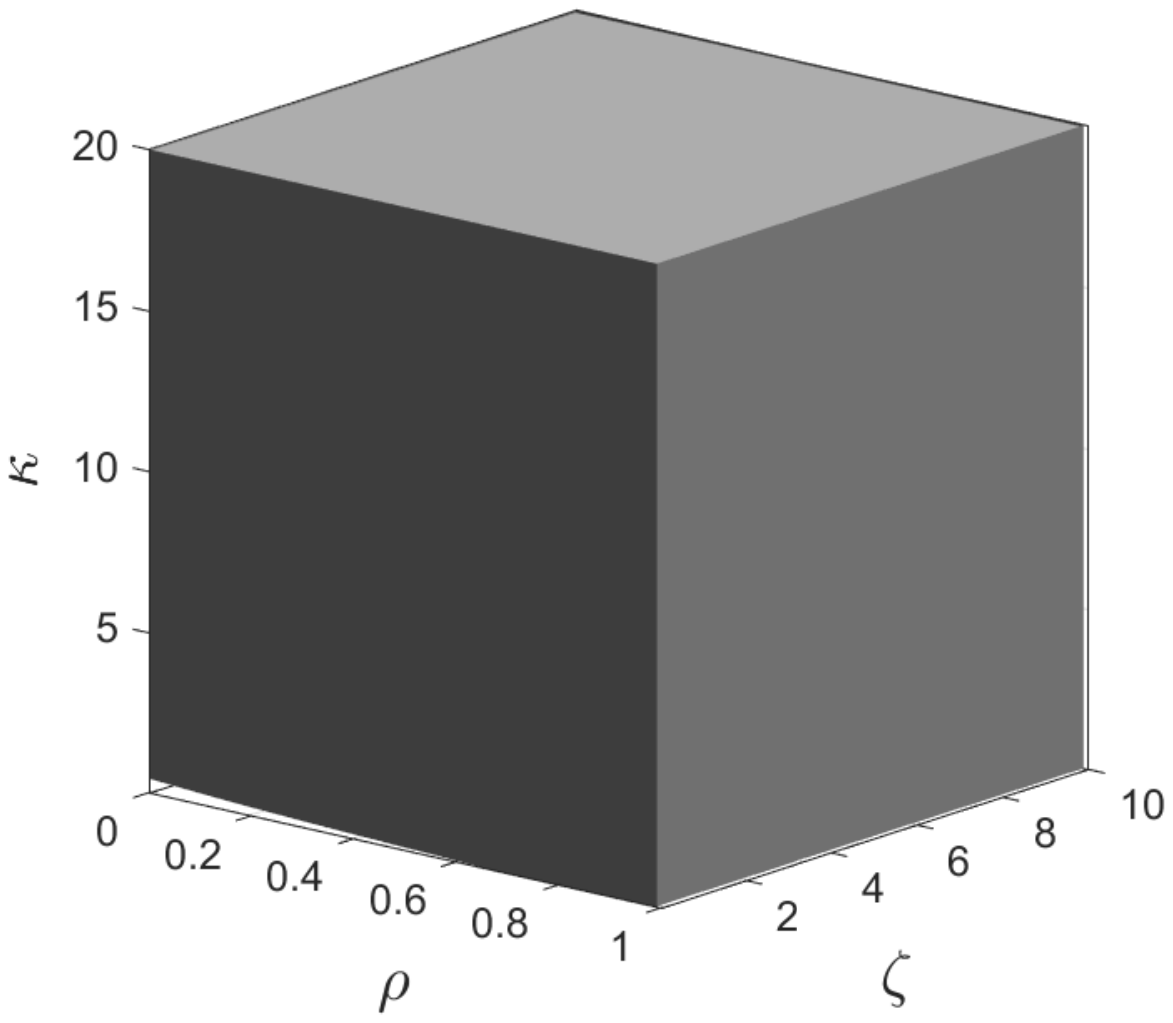}}\\\hline\hline
\end{tabular}} \caption{ 90\% S and qLL-S confidence sets for $\theta
=(\rho,\kappa,\zeta)$ derived from the investment Euler equation model
(\ref{eq: estimated}) using the sum of Gross Private
Domestic Investment and Personal Consumption Expenditure on Durable Goods as
investment proxy. A constant, $\Delta i_{t-1}$, $r_{t-2}^{p}$, and $u_{t-1}$
are common instruments in all specifications. The additional instrument(s) by
specification is (are): \protect\underline{Mon. pol. shock}:
\citeauthor{romer2004new}'s \citeyearpar{romer2004new} monetary policy shock;
\protect\underline{Military news}: \citeauthor{ramey2018government}'s
\citeyearpar{ramey2018government} military news shock; \protect\underline{Oil}%
: growth rate of real oil price; \protect\underline{VXO}: financial
uncertainty. }%
\label{appfig: JPT rob inst}%
\end{figure}

In this section we report further results to investigate the robustness of the empirical results reported in the main text. Figure
\ref{appfig: Two lags} shows the results when we use two lags of endogenous variables as instruments and compares them with our baseline results using one lag. Figure \ref{appfig: 67_04} shows the results when we restrict our estimation sample to 2004Q4 (as in SW and JPT) and compares it with our baseline sample ending in 2019Q4. Figures \ref{appfig: SW rob inst} and \ref{appfig: JPT rob inst} correspond to Figures \ref{fig: Figure ext inst SW} and \ref{fig: Figure ext inst JPT} in the main text, respectively, but the external instruments are now added together with $r_{t-2}^{p}$ and ${u}_{t-1}$ in the set of instruments. The results are unchanged when using external instruments both for the SW and JPT investment measures, respectively. 
%\sout{Likewise, Figures \ref{appfig: rob shocks SW} and \ref{appfig: rob shocks JPT} correspond to Figures \ref{fig: Figure shocks SW} and \ref{fig: Figure shocks JPT} in the main text, respectively. Again, $r_{t-2}^{p}$ and ${u}_{t-1}$ are now added as instruments together with $\Delta i_{t-1}$, when using smoothed estimates of shocks from the DSGE model of SW and JPT, respectively. Again, the  Figures show the robustness of our results to alternative combinations of instruments.} 
%Figure \ref{appfig: rob rho shocks} plots the results for the semi-structural model when using lagged investment growth, technology and price markup shocks as instruments.
The main conclusion from these sensitivity analyses is that the results reported in the paper remain largely robust.

%%%%%%%%%%%%%%%%%%%%%%%%%%%%%%%%%%%%%%%%%%%%%%%%%%%%%%%%
%Figure Exogenous Instruments with SW Investment Proxy
%%%%%%%%%%%%%%%%%%%%%%%%%%%%%%%%%%%%%%%%%%%%%%%%%%%%%%%%

%%%%%%%%%%%%%%%%%%%%%%%%%%%%%%%%%%%%%%%%%%%%%%%%%%%%%%%%
%Figure Exogenous Instruments with JPT Investment Proxy
%%%%%%%%%%%%%%%%%%%%%%%%%%%%%%%%%%%%%%%%%%%%%%%%%%%%%%%%

\section{The Capital Adjustment Cost Model}\label{app: s: CAC}

In this section, we derive the investment Euler equation with capital adjustment cost.
Similar to the capital accumulation equation (\ref{eq: CAE}), the representative household accumulates
end-of-period $t$ capital%
\begin{equation}
\hat{K}_{t+1}=\nu_{t}I_{t}+(1-\delta)\hat{K}_{t}-D(\hat{K}_{t},I_{t})\text{.}%
\label{eq: Euler_CAE_CAC}%
\end{equation}
The function $D(\hat{K}_{t},I_{t})$ is the capital adjustment cost (CAC) which can
be defined as
\[
D(\hat{K}_{t},I_{t})=\frac{\sigma}{2}\left(  \frac{I_{t}}{\hat{K}_{t}}-\delta\right)
^{2}\hat{K}_{t}\text{,}%
\]
where $\sigma>0$ governs the magnitude of adjustment costs to capital
accumulation and $\delta$ is the depreciation rate. This functional form is a
variant of the one considered in \cite{Lucas_1967} and \cite{Lucas_Prescott_1971}, and
has reappeared more recently in the DSGE literature, see, for example,
\cite{Christiano_Eichenbaum_Rebelo_2011} and \cite{Basu_Bundick_2017}.

The representative household still chooses ${I_{t}}${, }${\hat{K}_{t+1}}${,
}${B_{t+1}}$, and ${u_{t}}${ }to maximise (\ref{eq: UF}) under the
period-by-period budget constraint (\ref{eq: bc}) and capital accumulation
equation (\ref{eq: Euler_CAE_CAC}).The relevant first order conditions
required to derive the log-linearized investment Euler equations are%
\begin{align*}
I_{t}:\hspace{12pt}\nu_{t}Q_{t}= &  \left[  1+\sigma\left(  \dfrac{I_{t}%
}{\hat{K}_{t}}-\delta\right)  Q_{t}\right]  ,\\
\hat{K}_{t+1}:\hspace{12pt}Q_{t}= &  \beta E_{t}\left\{  \dfrac{\lambda_{t+1}%
}{\lambda_{t}}\left[  r_{t+1}^{k}u_{t+1}-a\left(  u_{t+1}\right)
-\dfrac{\sigma}{2}\left(  \dfrac{I_{t+1}}{\hat{K}_{t+1}}-\delta\right)  ^{2}\right]
\right\}  \\
&  +\beta E_{t}\left\{  \dfrac{\lambda_{t+1}}{\lambda_{t}}\left[
\sigma\left(  \dfrac{I_{t+1}}{\hat{K}_{t+1}}-\delta\right)  \dfrac{I_{t+1}}{\hat{K}_{t+1}%
}+Q_{t+1}\left(  1-\delta\right)  \right]  \right\} \text{,} \\
B_{t+1}:\hspace{12pt}1= &  \beta E_{t}\left\{  \dfrac{\lambda_{t+1}}%
{\lambda_{t}}\dfrac{R_{t}}{\pi_{t+1}}\right\}  , \text{ and}\\
u_{t}:\hspace{12pt}r_{t}^{k}= &  a^{\prime}\left(  u_{t}\right), 
\end{align*}
where $Q_{t}$ denotes the marginal $\mathcal{Q}$, defined as the ratio of the
Lagrange multipliers associated with the capital accumulation equation and the
budget constraint $\left(  \lambda_{t}\right)  $, and $\pi_{t+1}$ is the
inflation rate in period $t+1$.

Log-linearization of the FOCs around the steady state yields
\begin{align}
\widetilde{q}_{t} &  =\sigma\delta\left(  \widetilde{i}_{t}-\widetilde{k}%
_{t}\right)  -\widetilde{\nu}_{t},\label{eqn:Euler_CAC_Investment}\\
\widetilde{q}_{t} &  =E_{t}\widetilde{\lambda}_{t+1}-\widetilde{\lambda}%
_{t}+\beta\sigma\delta^{2}\left( E_{t} \widetilde{i}_{t+1}-\widetilde{k}%
_{t+1}\right)  +\beta(1-\delta)E_{t}\widetilde{q}_{t+1}+\beta\bar{r}^{k}%
E_{t}\widetilde{r}_{t+1}^{k},\label{eqn:Euler_CAC_capital}\\
E_{t}\widetilde{\lambda}_{t+1}-\widetilde{\lambda}_{t} &  =-(\widetilde{r}%
_{t}-E_{t}\widetilde{\pi}_{t+1}), \text{ and}\label{eqn:Euler_CAC_Bond}\\
\widetilde{r}_{t}^{k} &  =\zeta\widetilde{u}_{t}.\label{eqn: rk_CAC}%
\end{align}
\bigskip Similar to the IAC\ model, combining (\ref{eqn:Euler_CAC_Investment})-(\ref{eqn:Euler_CAC_Bond}) yields our baseline investment Euler equation with CAC%
\begin{equation}
\widetilde{i}_{t}=\widetilde{k}_{t}+\dfrac{1}{\sigma\delta}\left[  \beta
\bar{r}^{k}E_{t}\widetilde{r}_{t+1}^{k}-E_{t}\widetilde{r}_{t}^{p}\right]
+\beta\left(  E_{t}\widetilde{i}_{t+1}-\widetilde{k}_{t+1}\right)
+\dfrac{1}{\sigma\delta}\widetilde{\nu}_{t}-\dfrac{\beta\left(  1-\delta
\right)  }{\sigma\delta}E_{t}\widetilde{\nu}_{t+1}%
.\label{eqn:Basline_Euler_CAC}%
\end{equation}
Using equation (\ref{eqn: rk_CAC}), we replace $\widetilde{r}_{t+1}^{k}$ by
$\zeta\widetilde{u}_{t+1}$ to obtain%
\[
\widetilde{i}_{t}=\widetilde{k}_{t}+\dfrac{1}{\sigma\delta}\left[  \beta
\bar{r}^{k}\zeta E_{t}\widetilde{u}_{t+1}-E_{t}\widetilde{r}_{t}^{p}\right]
+\beta\left(  E_{t}\widetilde{i}_{t+1}-\widetilde{k}_{t+1}\right)
+\dfrac{1}{\sigma\delta}\widetilde{\nu}_{t}-\dfrac{\beta\left(  1-\delta
\right)  }{\sigma\delta}E_{t}\widetilde{\nu}_{t+1}.
\]

As before, we replace the variables in expectations with the observed values and
its respective rational expectation forecast error and also replace
$E_{t}\widetilde{\nu}_{t+1}$ by $\rho\widetilde{\nu}_{t}$, which results in%
\[
\widetilde{i}_{t}=\widetilde{k}_{t}+\dfrac{1}{\sigma\delta}\left[  \beta
\bar{r}^{k}\zeta\left(  \widetilde{u}_{t+1}-\eta_{t+1|t}^{u}\right)
-(\widetilde{r}_{t}^{p}+\eta_{t+1|t}^{\pi})\right]  +\beta\left(  \widetilde{i}%
_{t+1}-\eta_{t+1|t}^{i}-\widetilde{k}_{t+1}\right)
+\dfrac{1-\beta\left(  1-\delta\right)  \rho}{\sigma\delta}\widetilde{\nu}%
_{t}\text{,}%
\]
or%
\begin{equation}
\widetilde{i}_{t}=\widetilde{k}_{t}+\dfrac{1}{\sigma\delta}\left[  \beta
\bar{r}^{k}\zeta\widetilde{u}_{t+1}-\widetilde{r}_{t}^{p}\right]
+\beta\left(  \widetilde{i}_{t+1}-\widetilde{k}_{t+1}\right)  +\varepsilon
_{t}\text{,}\label{eq: Basline_Euler_CAC_util_1}%
\end{equation}
where $\varepsilon_{t}=-\tfrac{1}{\sigma\delta}\left[  \beta\bar{r}^{k}\zeta
\eta_{t+1|t}^{u}+\eta_{t+1|t}^{\pi}\right]  -\beta  \eta_{t+1|t}^{i}
+\tfrac{1-\beta\left(  1-\delta\right)  \rho}%
{\sigma\delta}\widetilde{\nu}_{t}%
$. 
Since in steady state $\overline{I}=\delta \overline{K}$ and $\overline{\nu}=1$, log-linearizing the
capital accumulation equation (\ref{eq: Euler_CAE_CAC}) results in
\[
\widetilde{k}_{t+1}=\delta(\widetilde{\nu}_{t}+\widetilde{i}_{t}%
)+(1-\delta)\widetilde{k}_{t}.%
\]

Therefore, multiplying (\ref{eq: Basline_Euler_CAC_util_1}) by $(1-\delta)$,
lagging it, and subtracting from the original equation results in%
\begin{align*}
\widetilde{i}_{t}-(1-\delta)\widetilde{i}_{t-1} =&\  \delta(\widetilde{\nu
}_{t-1}+\widetilde{i}_{t-1})+\dfrac{1}{\sigma\delta}\left[  \beta\bar{r}%
^{k}\zeta\widetilde{u}_{t+1}-\tilde{r}_{t}^{p}-(1-\delta)\left(  \beta\bar
{r}^{k}\zeta\widetilde{u}_{t}-\tilde{r}_{t-1}^{p}\right)  \right]  +\\
&  +\beta\left(  \widetilde{i}_{t+1}-(1-\delta)\widetilde{i}_{t}\right)
-\beta\delta\left(  \widetilde{\nu}_{t}+\widetilde{i}_{t}\right)
+\varepsilon_{t}-(1-\delta)\varepsilon_{t-1},%
\end{align*}
where $\widetilde{k}_{t}-(1-\delta)\widetilde{k}_{t-1}$ is replaced by
$\delta(\widetilde{\nu}_{t-1}+\widetilde{i}_{t-1})$. Further simplification
leads us to
\begin{equation}
\Delta\widetilde{i}_{t}=\beta\Delta\widetilde{i}_{t+1}+\dfrac{1}{\sigma\delta
}\left[  \beta\bar{r}^{k}\zeta\widetilde{u}_{t+1}-\tilde{r}_{t}^{p}%
-(1-\delta)\left(  \beta\bar{r}^{k}\zeta\widetilde{u}_{t}-\tilde{r}_{t-1}%
^{p}\right)  \right]  +\tilde{\varepsilon}_{t-1},\label{basedi}%
\end{equation}
where $\tilde{\varepsilon}_{t-1}= -\beta\delta\widetilde{\nu}_{t}+\delta
\widetilde{\nu}_{t-1}+\varepsilon_{t}-(1-\delta)\varepsilon_{t-1}$, or%
\begin{equation*}
\begin{aligned}
\tilde{\varepsilon}_{t-1}= &  -\left(  \beta\delta-\dfrac{1-\beta\left(  1-\delta\right)  \rho}%
{\sigma\delta}\right)  \widetilde{\nu}_{t}+\left[  \delta-(1-\delta
)\dfrac{1-\beta\left(  1-\delta\right)  \rho}{\sigma\delta}\right]
\widetilde{\nu}_{t-1}\\
&  -\dfrac{1}{\sigma\delta}\left[  \beta\bar{r}^{k}\eta_{t+1|t}^{u}%
+\eta_{t+1|t}^{\pi}-(1-\delta)\left[  \beta\bar{r}^{k}\eta_{t|t-1}^{u}+\eta
_{t|t-1}^{\pi}\right]  \right] -\beta\left[    \eta_{t+1|t}^{i}  -(1-\delta
)  \eta_{t|t-1}^{i}  \right].
\end{aligned}
\end{equation*}
We need to get rid of the $\tilde{\nu}$\ in the error term. Therefore, lagging
(\ref{basedi}), multiplying it by $\rho,$ and subtracting from (\ref{basedi})
results in%
\begin{align*}
\Delta\widetilde{i}_{t}-\rho\Delta\widetilde{i}_{t-1}  = &\ \beta\left(
\Delta\widetilde{i}_{t+1}-\rho\Delta\widetilde{i}_{t}\right)   +\dfrac{1}{\sigma\delta}\left[  \beta\bar{r}^{k}\zeta\left(  \widetilde{u}%
_{t+1}-\rho\widetilde{u}_{t}\right)  -\left(  \tilde{r}_{t}^{p}-\rho\tilde
{r}_{t-1}^{p}\right)  \right]  \\
& -\dfrac{(1-\delta)}{\sigma\delta}\left[  \left(  \beta\bar{r}^{k}%
\zeta\left(  \widetilde{u}_{t}-\rho\widetilde{u}_{t-1}\right)  -\left(
\tilde{r}_{t-1}^{p}-\rho\tilde{r}_{t-2}^{p}\right)  \right)  \right]  +\tilde{\varepsilon}_{t-1}-\rho\tilde{\varepsilon}_{t-2},
\end{align*}
or
\begin{align*}
\Delta\widetilde{i}_{t}\left(  1+\rho\beta\right)   &  =\rho\Delta
\widetilde{i}_{t-1}+\beta\Delta\widetilde{i}_{t+1}+\dfrac{\beta\bar{r}^{k}%
}{\sigma\delta}\zeta\widetilde{u}_{t+1}-\dfrac{1}{\sigma\delta}\tilde{r}%
_{t}^{p}-\dfrac{(1-\delta)\beta\bar{r}^{k}}{\sigma\delta}\zeta\widetilde{u}_{t}%
+\dfrac{1-\delta}{\sigma\delta}\tilde{r}_{t-1}^{p}\\
&  -\dfrac{\rho\beta\bar{r}^{k}}{\sigma\delta}\zeta\widetilde{u}_{t}%
+\dfrac{\rho}{\sigma\delta}\tilde{r}_{t-1}^{p}+\dfrac{\rho(1-\delta)\beta
\bar{r}^{k}}{\sigma\delta}\zeta\widetilde{u}_{t-1}-\dfrac{\rho(1-\delta
)}{\sigma\delta}\tilde{r}_{t-2}^{p}+\tilde{\varepsilon}_{t-1}-\rho
\tilde{\varepsilon}_{t-2}.%
\end{align*}
The above equation can be rewritten as
\begin{align}
\Delta\widetilde{i}_{t}  &  =\frac{\rho}{1+\rho\beta}\Delta\widetilde{i}%
_{t-1}+\frac{\beta}{1+\rho\beta}\Delta\widetilde{i}_{t+1}+\dfrac{\beta\bar
{r}^{k}}{\sigma\delta\left(  1+\rho\beta\right)  }\zeta\widetilde{u}_{t+1}%
-\dfrac{1}{\sigma\delta\left(  1+\rho\beta\right)  }\tilde{r}_{t}%
^{p}\nonumber\\
& -\dfrac{\left(  1-\delta\right)  \beta\bar{r}^{k}}{\sigma\delta\left(
1+\rho\beta\right)  }\zeta\widetilde{u}_{t} +\dfrac{1-\delta}{\sigma\delta\left(  1+\rho\beta\right)  }\tilde{r}%
_{t-1}^{p}-\dfrac{\rho\beta\bar{r}^{k}}{\sigma\delta\left(  1+\rho
\beta\right)  }\zeta\widetilde{u}_{t}+\dfrac{\rho}{\sigma\delta\left(
1+\rho\beta\right)  }\tilde{r}_{t-1}^{p}\nonumber\\
& +\dfrac{\rho(1-\delta)\beta\bar{r}%
^{k}}{\sigma\delta\left(  1+\rho\beta\right)  }\zeta\widetilde{u}_{t-1}%
-\dfrac{\rho(1-\delta)}{\sigma\delta\left(  1+\rho\beta\right)  }\tilde
{r}_{t-2}^{p}+\overline{\varepsilon}_{t-1},\label{basedi2}
\end{align}
where $
\overline{\varepsilon}_{t-1}\left(  1+\rho\beta\right)  :=\left(  \tilde
{\varepsilon}_{t-1}-\rho\tilde{\varepsilon}_{t-2}\right)
$ is 
\begin{align*}
-& \beta\delta\widetilde{\nu}_{t}+\delta\widetilde{\nu}_{t-1}-\dfrac
{1}{\sigma\delta}\left[  \beta\bar{r}^{k}\eta_{t+1|t}^{u}+\eta_{t+1|t}^{\pi
}\right]  -\beta \eta_{t+1|t}^{i}  +\dfrac
{1-\beta\left(  1-\delta\right)  \rho}{\sigma\delta}\widetilde{\nu}_{t}\\
&  -(1-\delta)\left[  -\dfrac{1}{\sigma\delta}\left[  \beta\bar{r}^{k}\eta
_{t|t-1}^{u}+\eta_{t|t-1}^{\pi}\right]  -\beta \eta_{t|t-1}^{i}%
  +\dfrac{1-\beta\left(  1-\delta\right)  \rho}{\sigma\delta
}\widetilde{\nu}_{t-1}\right] \\
&  +\rho\beta\delta\widetilde{\nu}_{t-1}-\rho\delta\widetilde{\nu}%
_{t-2}+\dfrac{\rho}{\sigma\delta}\left[  \beta\bar{r}^{k}\eta_{t|t-1}^{u}%
+\eta_{t|t-1}^{\pi}\right]  +\rho\beta \eta_{t|t-1}^{i}
-\rho\dfrac{1-\beta\left(  1-\delta\right)  \rho}{\sigma\delta}\widetilde{\nu
}_{t-1}\\
&  +\rho(1-\delta)\left[  -\dfrac{1}{\sigma\delta}\left[  \beta\bar{r}^{k}%
\eta_{t-1|t-2}^{u}+\eta_{t-1|t-2}^{\pi}\right]  +\rho\beta\eta_{t-1|t-2}^{i}%
  -\rho\dfrac{1-\beta\left(  1-\delta\right)  \rho
}{\sigma\delta}\widetilde{\nu}_{t-2}\right].
\end{align*}
Further simplifying the above equation becomes
\begin{align}
\overline{\varepsilon}_{t-1}\left(  1+\rho\beta\right)   = & -\beta
\delta\varepsilon_{t}^{v}+\delta\varepsilon_{t-1}^{v}-\dfrac{1}{\sigma\delta
}\left[  \beta\bar{r}^{k}\eta_{t+1|t}^{u}+\eta_{t+1|t}^{\pi}\right]
-\beta\eta_{t+1|t}^{i}  +\dfrac{1-\beta\left(
1-\delta\right)  \rho}{\sigma\delta}\varepsilon_{t}^{v}\nonumber\\
&  -(1-\delta)\left[  -\dfrac{1}{\sigma\delta}\left[  \beta\bar{r}^{k}\eta
_{t|t-1}^{u}+\eta_{t|t-1}^{\pi}\right]  -\beta\eta_{t|t-1}^{i}%
  +\dfrac{1-\beta\left(  1-\delta\right)  \rho}{\sigma\delta
}\varepsilon_{t-1}^{v}\right] \nonumber \\
&  +\dfrac{\rho}{\sigma\delta}\left[  \beta\bar{r}^{k}\eta_{t|t-1}^{u}%
+\eta_{t|t-1}^{\pi}\right]  +\rho\beta\eta_{t|t-1}^{i} \nonumber \\
&  +\rho(1-\delta)\left[  -\dfrac{1}{\sigma\delta}\left[  \beta\bar{r}^{k}%
\eta_{t-1|t-2}^{u}+\eta_{t-1|t-2}^{\pi}\right]  +\rho\beta\eta_{t-1|t-2}^{i}\right]. \label{eq: error Basline_Euler_CAC_util_est}
\end{align}
Then, grouping some terms from above we get%
\begin{align}
\Delta\widetilde{i}_{t}  &  =\frac{\rho}{1+\rho\beta}\Delta\widetilde{i}%
_{t-1}+\frac{\beta}{1+\rho\beta}\Delta\widetilde{i}_{t+1}+\dfrac{\beta\bar
{r}^{k}}{\sigma\delta\left(  1+\rho\beta\right)  }\zeta\widetilde{u}_{t+1}%
-\dfrac{1}{\sigma\delta\left(  1+\rho\beta\right)  }\tilde{r}_{t}%
^{p}
\nonumber\\
& -\dfrac{\beta\bar{r}^{k}\left(  1-\delta+\rho\right)  }{\sigma
\delta\left(  1+\rho\beta\right)  }\zeta\widetilde{u}_{t} +\dfrac{1-\delta+\rho}{\sigma\delta\left(  1+\rho\beta\right)  }\tilde
{r}_{t-1}^{p}+\dfrac{\rho(1-\delta)\beta\bar{r}^{k}}{\sigma\delta\left(
1+\rho\beta\right)  }\zeta\widetilde{u}_{t-1}
\nonumber \\
& -\dfrac{\rho(1-\delta)}%
{\sigma\delta\left(  1+\rho\beta\right)  }\tilde{r}_{t-2}^{p}+\overline
{\varepsilon}_{t-1}.\label{eq: Basline_Euler_CAC_util_est}
\end{align}
It can be gauged from looking at the equation (\ref{eq: error Basline_Euler_CAC_util_est})
that we need to use at least the second lag of endogenous variables as
instruments in order to ensure exogeneity.

\begin{figure}[htbhp]
\centering
\adjustbox{min width=1.9\textwidth, height=.4\textwidth}{
\begin{tabular}
[c]{cccccc}\hline\hline\\[-5pt]
&  & {\large{SW Investment}} &  & {\large{JPT Investment}} & \\\cline{3-3}\cline{5-5}\\
&  &{\small (a)}     &  & {\small (b)}     & \vspace{-.5cm}\\
& \raisebox{+12.7ex}{\rotatebox[origin=lt]{90}{S sets }}
&
\hspace{+.1cm}
{\includegraphics[angle=0,width=.35\textwidth,trim=120 280 140 230 ,
totalheight=.4\textwidth]{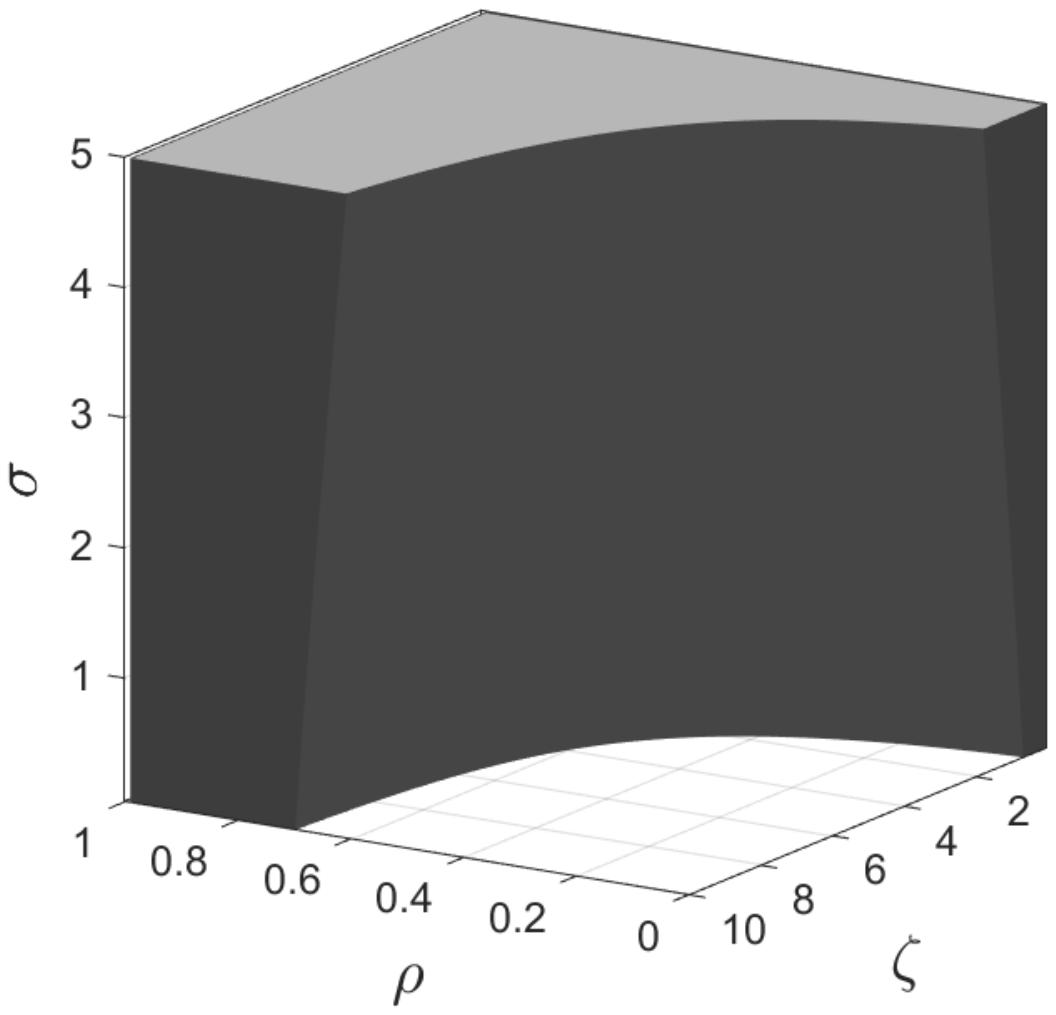}}
&
&
\hspace{+.3cm}
{\includegraphics[angle=0,width=.35\textwidth, trim=120 280 140 230 ,
totalheight=.4\textwidth]{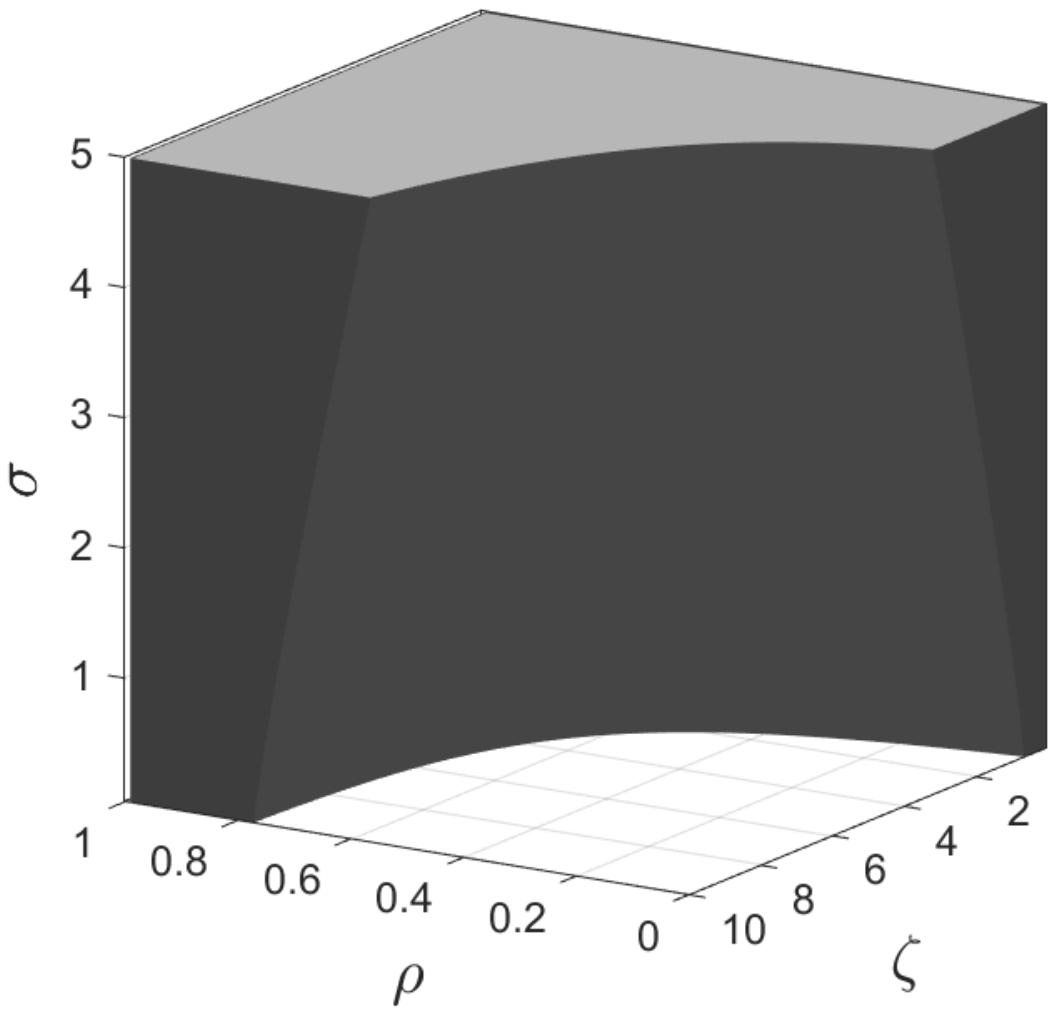}} & \\[10pt]
&  &{\small (c)}     &  & {\small (d)}     & \vspace{-.5cm}\\
& \raisebox{+7.7ex}{\rotatebox[origin=lt]{90}{qLL-S sets }}
&
\hspace{+.1cm}
{\includegraphics[angle=0,width=.35\textwidth,trim=120 280 140 230 ,
totalheight=.4\textwidth]{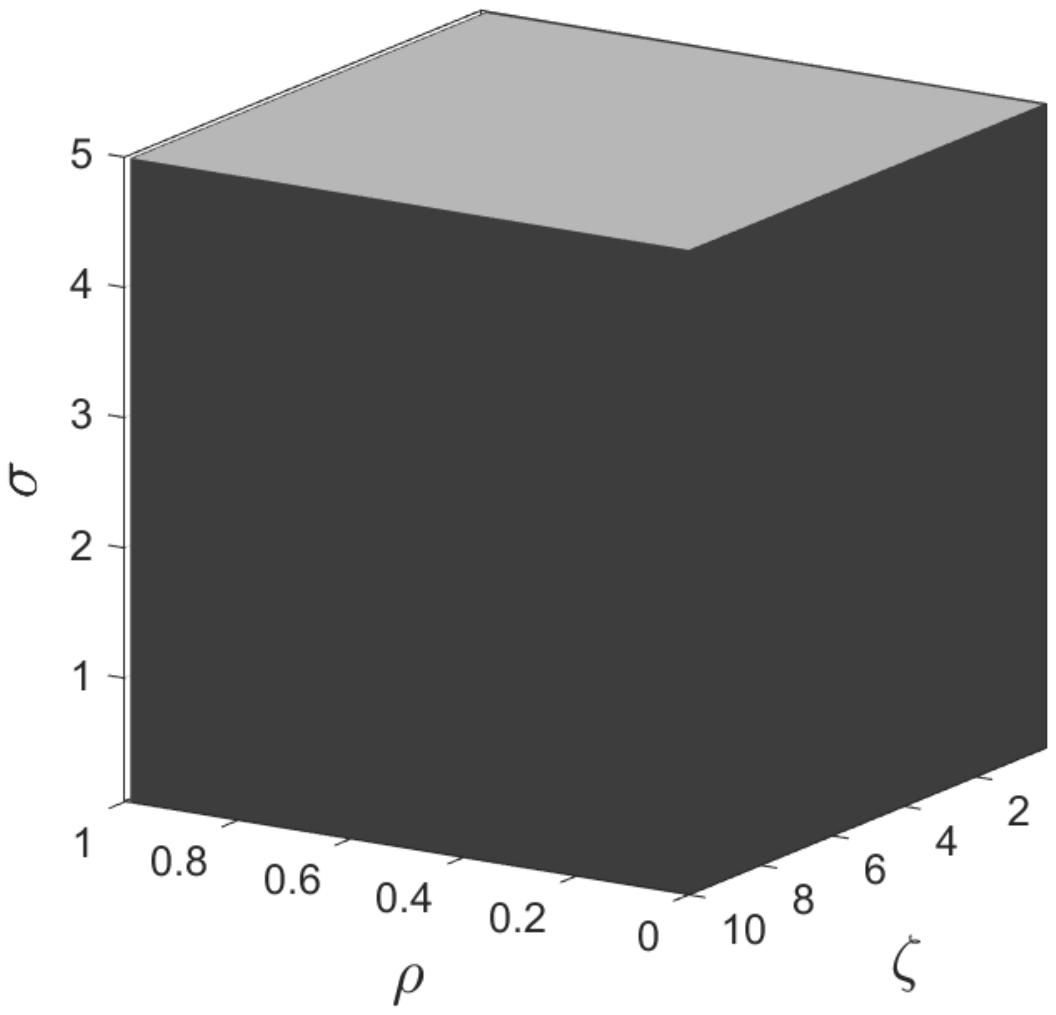}}
&
\hspace{+.5cm}
&
\hspace{+.3cm}
{\includegraphics[angle=0,width=.35\textwidth,trim=120 280 140 230, totalheight=.4\textwidth]
{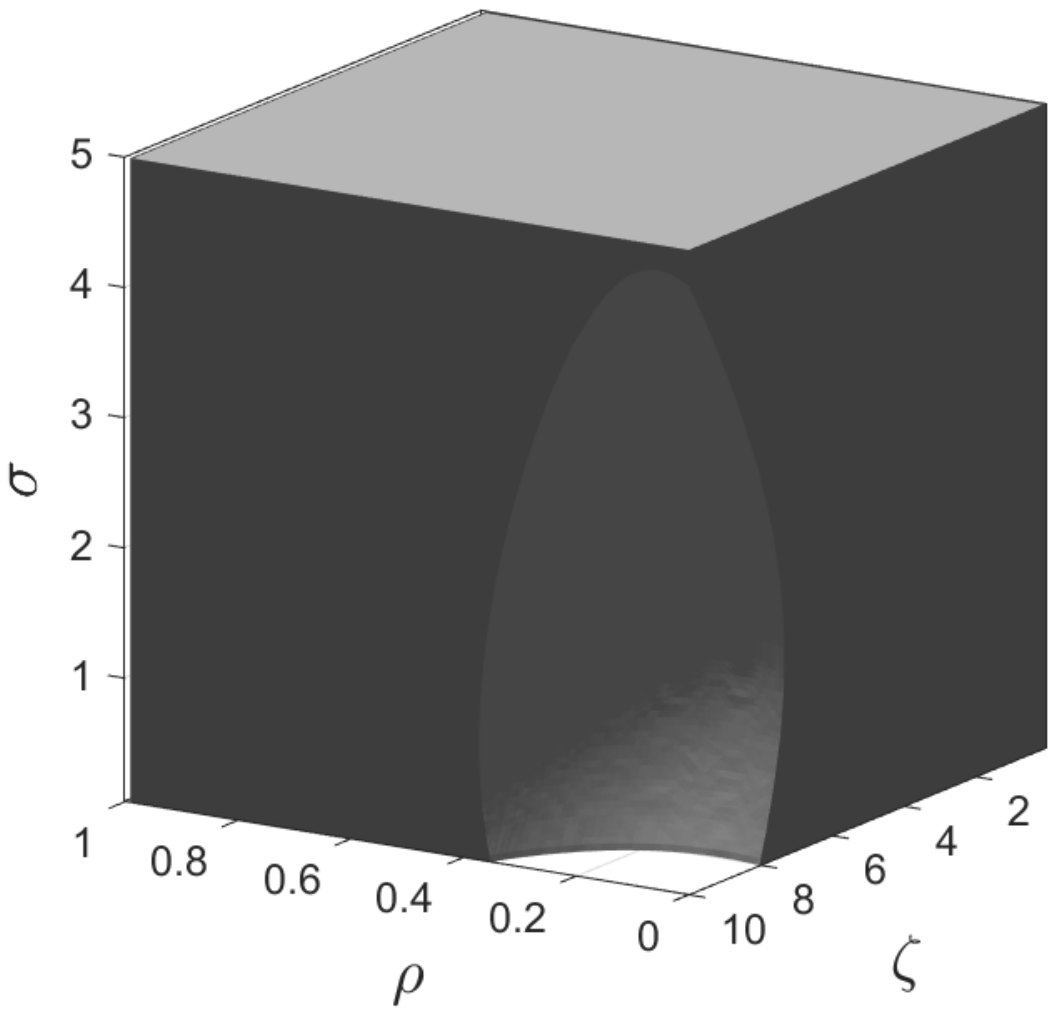}} &
\\[+3pt]
\hline\hline
\end{tabular}}
\caption{90\% S and qLL-S confidence sets for $\theta
=(\rho,\sigma,\zeta)$ derived from the investment Euler equation model
(\ref{eq: Basline_Euler_CAC_util_est}). Instruments: constant, $\Delta i_{t-2}$, $r_{t-3}^{p}$,
$u_{t-2}$. The investment proxies are Fixed Private Investment (left column) and the sum of Gross Private
Domestic Investment and Personal Consumption Expenditure on Durable Goods (right column). \cite{Newey_West_1987}
HAC. Period: 1967Q1-2019Q4.}%
\label{appfig: baseline}%
\end{figure}

Figure \ref{appfig: baseline} shows that confidence sets are large suggesting weak identification of the structural parameters also in the capital adjustment cost model.

\section{Derivation of Equation (\ref{eq: misspecified regression})}\label{app: misspecified derivation}
Suppose $x_{t}$ follows (\ref{eq: x AR}). Then, its first autocorrelation is
$\gamma.$ The pseudo true value of the coefficient $\theta$ in the MA(1)
specification (\ref{eq: x MA}) is obtained by solving the equation (see
Hamilton, 1995, p. 49)%
\begin{equation}
\gamma=\frac{\theta^{\ast}}{1+\theta^{\ast2}}, \label{eq: theta*}%
\end{equation}
and we can choose the (unique) invertible solution that satisfies $\left\vert
\theta_{i}\right\vert <1$. Given $\theta^{\ast},$ the corresponding estimate
of the shock in (\ref{eq: x MA}), $\omega_{t}^{\ast}$, can be solved from
$\left\{  x_{t}\right\}  $ using the backward recursion%
\begin{equation}
\omega_{t}^{\ast}=\sum_{j=0}^{\infty}\left(  -\theta^{\ast}\right)
^{j}x_{t-j}, \label{eq: filtered omega}%
\end{equation}
while the true structural shock in (\ref{eq: x AR}) is simply%
\[
\omega_{t}=x_{t}-\gamma x_{t-1}.
\]
So, we have%
\begin{align}
cov\left(  z_{t}^{\ast},\left(  z_{t}-z_{t}^{\ast}\right)  \right)   &
=cov\left(  z_{t}^{\ast},z_{t}\right)  +\theta^{\ast2}var\left(  \omega
_{t}^{\ast}\right) \nonumber\\
&  =\frac{\sigma_{\omega}^{2}\gamma\theta^{\ast}}{1-\gamma^{2}}\frac
{1}{1-\left(  \gamma\theta^{\ast}\right)  ^{2}}+\theta^{\ast2}\frac{\left(
1-\gamma\theta^{\ast}\right)  \sigma_{\omega}^{2}}{\left(  1+\gamma
\theta^{\ast}\right)  \left(  1-\gamma^{2}\right)  \left(  1-\theta^{\ast
2}\right)  }\nonumber\\
&  =\sigma_{\omega}^{2}\theta^{\ast}\frac{\gamma\left(  1-\theta^{\ast
2}\right)  +\theta^{\ast}\left(  1-\gamma\theta^{\ast}\right)  ^{2}}{\left(
1-\left(  \gamma\theta^{\ast}\right)  ^{2}\right)  \left(  1-\gamma
^{2}\right)  \left(  1-\theta^{\ast2}\right)  }, \label{eq: cov error}%
\end{align}
because%
\begin{align*}
cov\left(  z_{t}^{\ast},z_{t}\right)   &  =cov\left(  \theta^{\ast}\sum
_{j=0}^{\infty}\left(  -\theta^{\ast}\right)  ^{j}x_{t-j},\gamma x_{t}\right)
\\
&  =\frac{\sigma_{\omega}^{2}\theta^{\ast}\gamma}{1-\gamma^{2}}\sum
_{j=0}^{\infty}\left(  -\theta^{\ast}\gamma\right)  ^{j}=\frac{\sigma_{\omega
}^{2}\gamma\theta^{\ast}}{1-\gamma^{2}}\frac{1}{1-\left(  \gamma\theta^{\ast
}\right)  ^{2}}%
\end{align*}
and%
\[
x_{t}=\omega_{t}^{\ast}\left(  1+\theta^{\ast}L\right)  =\omega_{t}\left(
1-\gamma L\right)  ^{-1},
\]
so $\omega_{t}^{\ast}$ is an AR(2) process, $\omega_{t}^{\ast}\left(
1-a_{1}L-a_{2}L^{2}\right)  =\omega_{t},$ with $a_{1}=\gamma-\theta^{\ast},$
and $a_{2}=\gamma\theta^{\ast},$ and therefore, its variance is given by (see
Hamilton, 1995, p. 58)
\begin{align*}
var\left(  \omega_{t}^{\ast}\right)   &  =\frac{1-a_{2}}{1+a_{2}}\frac
{\sigma_{\omega}^{2}}{\left(  1-a_{2}\right)  ^{2}-a_{1}^{2}}\\
&  =\frac{1-\gamma\theta^{\ast}}{1+\gamma\theta^{\ast}}\frac{\sigma_{\omega
}^{2}}{\left(  1-\gamma\theta^{\ast}\right)  ^{2}-\left(  \gamma-\theta^{\ast
}\right)  ^{2}}\\
&  =\frac{\left(  1-\gamma\theta^{\ast}\right)  \sigma_{\omega}^{2}}{\left(
1+\gamma\theta^{\ast}\right)  \left(  1-\gamma^{2}\right)  \left(
1-\theta^{\ast2}\right)  }.
\end{align*}
Substituting for $\gamma$ using $\theta^{\ast}$ from (\ref{eq: theta*}) in
(\ref{eq: cov error}), we get%
\[
cov\left(  z_{t}^{\ast},\left(  z_{t}-z_{t}^{\ast}\right)  \right)
=\frac{\sigma_{\omega}^{2}\theta^{\ast2}\left(  2-\theta^{\ast4}\right)
\left(  \theta^{\ast2}+1\right)  ^{2}}{\left(  1-\theta^{\ast2}\right)
\left(  1-\theta^{\ast}+\theta^{\ast2}\right)  \left(  1+\theta^{\ast}%
+\theta^{\ast2}\right)  \left(  1+2\theta^{\ast2}\right)  }\neq0.
\]

\newpage
\bibliographystyle{chicago}
\bibliography{References}

\end{document}